\pdfoutput=1 
\documentclass[10pt,rmp,preprintnumbers,nofootinbib]{revtex4}

\usepackage{bm}
\usepackage{graphicx}
\usepackage{amsmath}
\usepackage{amssymb} 
\usepackage{color}
\usepackage{txfonts}
\usepackage{hyperref}
\usepackage{dcolumn}
\usepackage{subfigure}
\usepackage{multirow}
\usepackage{natbib}

\newcommand{\ie}{{\it i.e.~}}
\newcommand{\eg}{{\it e.g.~}}
\newcommand{\eqn}[1]{(\ref{#1})}
\newcommand{\fig}[1]{fig.~\ref{#1}}
\newcommand{\Fig}[1]{Fig.~\ref{#1}}

\newcommand{\xv}{{\bf x}}
\newcommand{\kv}{{\bf k}}

\newcommand{\qv}{{\bf q}}

\newcommand{\skyp}[1]{}

\def\be{\begin{equation}}
\def\ee{\end{equation}}
\def\bea{\begin{eqnarray}}
\def\eea{\end{eqnarray}}

\def\Mpc{\, h^{-1} \, {\rm Mpc}}

\def\cGpc{\, h^{-3} \, {\rm Gpc}^3}
\def\kMpc{\, h \, {\rm Mpc}^{-1}}
\def\icMpc{\, h^3 \, {\rm Mpc}^{-3}}
\def\nng{n_{\rm NG}}

\def\fnl{f_{\rm NL}}

\def\fnll{f_{\rm NL}^{\rm loc.}}
\def\dfnll{\Delta f_{\rm NL}^{\rm loc.}}
\def\fnle{f_{\rm NL}^{\rm eq.}}

\def\fnlo{f_{\rm NL}^{\rm orthog.}}

\def\dnng{\Delta n_{\rm NG}}

\def\pd{\partial}
\def\la{\langle}
\def\ra{\rangle}
\def\kp{k_p}
\def\lmax{l_{max}}

\def\lsim{\lesssim}

\def\a{\alpha}
\def\b{\beta}
\def\D{\Delta}
\def\d{\delta}
\def\e{\epsilon}

\def\l{\lambda}
\def\n{\eta}
\def\O{\Phi}

\def\t{\tau}

\def\W{\Omega}
\def\w{\omega}

\def\z{\zeta}

\def\px{\approx}
\def\pr{\prime}

\def\={\nonumber &=}
\def\nn{\nonumber}

\def\({\left(}
\def\){\right)}
\def\[{\left[}
\def\]{\right]}
\def\<{\left\langle}
\def\>{\right\rangle}

\def\uk{{\bf \hat{k}}}
\def\un{{\bf \hat{n}}}

\def\ux{{\bf \hat{x}}}
\def\bk{{\bf k}}

\def\curl{\mathcal}

\def\eq{\begin{eqnarray}}
\def\qe{\end{eqnarray}}

\def\and{\quad \mbox{and} \quad}

\def\fnl{f_\textrm{NL}}
\def\fnlest{\hat{f}_\textrm{NL}}

\def\fnlequil{f_\textrm {NL}^\textrm{equil}}

\def\fnlmodel{f_\textrm {NL}^\textrm{model}}
\def\bfnl{\kern2pt\overline{\kern-2ptf}_\textrm{NL}}

\def\lmax{\ell_\textrm{max}}
\def\lall{\ell_1,\ell_2,\ell_3}
\def\lsum{\ell_1+\ell_2+\ell_3}

\def\Blll{B_{\ell_1\ell_2\ell_3}}
\def\blll{b_{\ell_1\ell_2\ell_3}}
\def\bllllocal{b_{\alll}^\textrm{local}}

\def\kmax{k_\textrm{max}}
\def\kall{k_1,k_2,k_3}
\def\ksum{k_1+k_2+k_3}

\def\Bkkk{B(\kall)}
\def\Skkk{S(\kall)}

\def\Vtetra{{{\cal V}_{\cal T}}}

\def\Qn{\curl{Q}_n}

\def\Qp{\curl{Q}_p}

\def\Rn{\curl{R}_n}
\def\Rm{\curl{R}_m}

\def\barQ{\kern2pt\overline{\kern-2pt\curl{Q}}}
\def\barQn{\barQ_n}

\def\barR{\kern2pt\overline{\kern-2pt\curl{R}}}
\def\barRn{\barR_n}

\def\aR{\alpha^{\scriptscriptstyle{\cal R}}}
\def\aRn{\aR_n}
\def\aRp{\aR_p}
\def\baR{\bar{\alpha}^{\scriptscriptstyle{\cal R}}}
\def\baRn{\baR_n}

\def\aQ{\alpha^{\scriptscriptstyle{\cal Q}}}
\def\aQn{\aQ_n}

\def\baQ{\bar{\alpha}^{\scriptscriptstyle{\cal Q}}}
\def\baQn{\baQ_n}

\def\bbR{\bar{\beta}^{\scriptscriptstyle{\cal R}}}
\def\bbRn{\bbR_n}

\def\sumalllm{\sum_{\{\ell_i,m_i\}}}

\def\alll{\ell_1 \ell_2 \ell_3}
\def\allm{m_1 m_2 m_3}

\def\alm{a_{\ell m}}
\def\almone{a_{\ell_1 m_1}}
\def\almtwo{a_{\ell_2 m_2}}
\def\almthree{a_{\ell_3 m_3}}
\def\almfour{a_{\ell_4 m_4}}
\def\almfive{a_{\ell_5 m_5}}
\def\almsix{a_{\ell_6 m_6}}
\def\Ylm{Y_{\ell m}(\hat{\gamma})}
\def\Ylmone{Y_{\ell_1 m_1}(\hat{\gamma})}
\def\Ylmtwo{Y_{\ell_2 m_2}(\hat{\gamma})}
\def\Ylmthree{Y_{\ell_3 m_3}(\hat{\gamma})}

\def\bispvar{C_{\ell_1} C_{\ell_2} C_{\ell_3}}

\def\Gaunt{{\cal{G}}_{\alll}^{\allm}}

\def\ordpar{{\cal{O}}\( \fnl \left \langle \Phi^2_L(\mathbf{x}) \right \rangle \)}

\def\balpha{\mbox{\boldmath$\alpha$}_l}
\def\bbeta{\mbox{\boldmath$\beta$}_l}
\def\bgamma{\mbox{\boldmath$\gamma$}_l}
\def\bdelta{\mbox{\boldmath$\delta$}_l}

\def\Ylm{Y_{\ell m}(\hat{\bf n})}
\def\Ylmone{Y_{\ell_1 m_1}(\hat{\bf n})}
\def\Ylmtwo{Y_{\ell_2 m_2}(\hat{\bf n})}
\def\Ylmthree{Y_{\ell_3 m_3}(\hat{\bf n})}

\begin{document}

\title{Primordial non-Gaussianity and Bispectrum Measurements in the Cosmic Microwave Background and Large-Scale Structure}

\author{Michele Liguori}\email{ml453@cam.ac.uk}
\affiliation{Centre for Theoretical Cosmology\\
Department of Applied Mathematics and Theoretical Physics\\
University of Cambridge\\
Wilberforce Road, Cambridge, CB3 0WA, UK}

\author{Emiliano Sefusatti}
\email{emiliano.sefusatti@cea.fr}
\affiliation{Institut de Physique Th\'eorique\\
Commissariat \` a l'\'Energie Atomique\\
 F-91191 Gif-sur-Yvette, France}

\author{James R. Fergusson}\email{jf334@damtp.cam.ac.uk}
\author{E.P.S. Shellard}\email{E.P.S.Shellard@damtp.cam.ac.uk}
\affiliation{Centre for Theoretical Cosmology   \\
Department of Applied Mathematics and Theoretical Physics  \\
University of Cambridge  \\
Wilberforce Road, Cambridge, CB3 0WA, UK  }

\begin{abstract}

The most direct probe of non-Gaussian initial conditions has come from bispectrum measurements of temperature fluctuations in the Cosmic Microwave Background and of the matter and galaxy distribution at large scales.
Such bispectrum estimators are expected to continue to provide the best constraints on the non-Gaussian parameters in future observations. We review and compare the theoretical and observational problems, current results and future prospects for the detection of a non-vanishing primordial component in the bispectrum of the Cosmic Microwave Background and large-scale structure, and the relation to specific predictions from different inflationary models. 

\end{abstract}

\maketitle

\tableofcontents
 
\clearpage

\section{Introduction}

The standard inflationary paradigm predicts a flat Universe perturbed by nearly Gaussian and scale invariant primordial perturbations. These predictions have been verified to a high degree of accuracy by Cosmic Microwave Background (CMB) and Large-Scale Structure (LSS) measurements, such as those provided by the Wilkinson Microwave Anisotropy Probe  \citep[WMAP;][]{KomatsuEtal2009B}, the 2dF Galaxy Redshift Survey \citep[2dFGRS;][]{PercivalEtal2002} and the Sloan Digital Sky Survey \citep[SDSS;][]{TegmarkEtal2004}.   Despite this success, it has proved to be difficult to discriminate 
between the vast array of  inflationary scenarios that have been proposed by high-energy theoretical investigations, or even to rule-out alternatives to inflation.. Since most of the present constraints on the Lagrangian of the inflaton field have been obtained from measurements of the two-point function, or power spectrum, of the primordial fluctuations, a natural step is to extend the available information is to look at non-Gaussian signatures in higher order correlators.
 
The lowest order additional correlator to take into account is  the three-point function or its counterpart in Fourier space, the {\it bispectrum}. Every model of inflation is characterized by specific predictions for the bispectrum of the primordial  perturbations in the gravitational potential $\Phi(\kv)$. The bispectrum $B_\Phi(\kall)$ of these perturbations is defined as
\be
\langle\Phi(\kv_1)\,\Phi(\kv_2)\,\Phi(\kv_3)\rangle\equiv (2\pi)^3 \d_D(\kv_{123}) ~B_\Phi(\kall)\;,
\ee
where we have introduced the notation $\kv_{ij}\equiv\kv_1+\kv_2$ so that the Dirac delta function here is $\d_D(\kv_{123})\equiv\d_D(\kv_1+\kv_2+\kv_3)$.  Together with the assumption of statistical homogeneity and isotropy for the primordial perturbations, this implies that the bispectrum is a function of the triplet defined by the magnitude of the wavenumbers $k_1$, $k_2$ and $k_3$ forming a closed triangular configuration. The current constraints that we are able to derive on the
bispectrum $B_\Phi(\kall)$ provide additional information about the early Universe; the possible detection of a non-vanishing primordial bispectrum in future observations would represent a major discovery, especially as it is predicted
to be negligible by standard inflation. 

The cosmological observable most directly related to the initial curvature bispectrum is given by the bispectrum of the CMB temperature fluctuations, which provide a map of the density perturbations at the time of decoupling, the earliest information we have about the Universe. Current measurements of individual triangular configurations of the CMB bispectrum are, however, consistent with zero. Studies of the primordial bispectrum, therefore, are usually characterized by constraints on a single {\it amplitude parameter}, denoted by $\fnl$, once a specific model for $B_\Phi$ is assumed. Since most models predict a curvature bispectrum obeying the hierarchical scaling $B_\Phi(k,k,k)\sim P_\Phi^2(k)$, with $P_\Phi(k)$ being the curvature power spectrum, the non-Gaussian parameter roughly quantifies the ratio $\fnl \sim B_\Phi(k,k,k)/P_\Phi^2(k)$, defining the ``strength'' of the primordial non-Gaussian signal. In addition, we can write 
\be
B_\Phi(\kall) \equiv \fnl F(\kall)\;,
\ee
where $F(\kall)$ encodes the functional dependence of the primordial bispectrum 
on the specific triangle configurations. For brevity, the characteristic shape-dependence of a given bispectrum is often referred to simply as the {\it bispectrum shape} (a precise definition of the bispectrum shape function will be given in section~\ref{sec:bispshape}). Inflationary predictions for both the amplitude $\fnl$ and the shape of $B_\Phi$ that are
strongly model-dependent. Notice that the subscript ``NL" stands for ``nonlinear", since a common phenomenological model for the non-Gaussianity of the initial conditions can be written as a simple nonlinear transformation of a Gaussian field. 
Generically, of course, non-Gaussianity is associated with nonlinearities, such as nontrivial dynamics during 
inflation, resonant behaviour at the end of inflation (`preheating'), or nonlinear post-inflationary evolution. 
At the very least, future CMB and LSS observations are expected to be able to eventually detect the small last contribution.

Perturbations in the CMB provide a particularly convenient test of the primordial density field because CMB temperature and polarization anisotropies are small enough to be studied in the {\it linear regime} of cosmological perturbations. Once the effects of foregrounds are properly taken into account, a non-vanishing CMB bispectrum at large scales would be a direct consequence of a non-vanishing primordial bispectrum. As we will see, while other CMB probes of primordial non-Gaussianity are available, such as tests of the topological properties of the temperature map based on Minkowski Functionals or measurements of the CMB trispectrum, the estimator for the non-Gaussian parameter $\fnl$ has been shown to be optimal.  We will focus mostly on this bispectrum estimator in the section of this review dedicated to the CMB.

In the standard cosmological model, the large-scale structure of the Universe, that is,  the distribution of matter and galaxies on  large scales, is the result of the nonlinear evolution due to gravitational instability of the same initial density perturbations responsible for the CMB anisotropies. This is, perhaps, the most important prediction of the inflationary framework which provides a common origin for the CMB and large-scale structure perturbations as the result of tiny quantum fluctuations stretched over cosmological scales during a phase of accelerated expansion. The large-scale structure we observe at low redshift, however, is characterized by large voids and small regions with very large matter density, and it is therefore a much less direct probe of the initial conditions. The  {\it distribution of matter becomes a highly non-Gaussian field} precisely as a result of the nonlinear growth of structures, {\it even for Gaussian initial conditions}. This non-Gaussianity is expressed, in particular, by a non-vanishing matter bispectrum at {\it any} measurable scale, including the largest scales probed by current or future redshift surveys. In this context, the effect of primordial non-Gaussianity, \ie of an initial component in the curvature bispectrum, will constitute a {\it correction} to the galaxy bispectrum. It follows that the possibility of constraining or detecting this initial component is strictly related to our ability to {\it distinguish} it from other, primary sources of non-Gaussianity, that is the nonlinear gravitational evolution, and, in the case of galaxy surveys, nonlinear bias. 

The study of non-Gaussian initial conditions for large-scale structure has a relatively long history, with important contributions going back to the mid eighties. The standard picture that has been developed over the years, assumed that, at large scales, the effect of primordial non-Gaussianity on the galaxy distribution is simply given in terms of an additional component to the galaxy bispectrum. This is obtained, in perturbation theory, as the  linearly evolved and linearly biased initial matter bispectrum, related to the curvature bispectrum $B_\Phi(\kall)$ by the Poisson equation. Such component becomes subdominant as the gravity-induced non-Gaussian contribution grows in time. In this framework, as one can expect, high-redshift and large-volume galaxy surveys would constitute the best probes of the initial conditions. It has been shown, in fact, that proposed and planned redshift surveys, such as Euclid \citep{RefregierEtal2010}, should be able to provide constraints on the primordial non-Gaussian parameters comparable, if not better, than those expected from  CMB missions such as Planck. What is more important, in the event of a detection by Planck, is that confirmation by large-scale structure observations will be required. 

Recent results from N-body simulations with non-Gaussian initial conditions, however, have revealed a more complex picture. The effect of primordial non-Gaussianity at large scales is not limited to an additional contribution to the galaxy bispectrum, but it quite dramatically affects the galaxy bias relation itself, that is, the relation between the matter and galaxy distributions. A surprising consequence is that it induces a large correction even for the galaxy power spectrum. 
Such an effect has attracted considerable recent attention and, remarkably, have placed constraints on the non-Gaussian parameter from current LSS data-sets which already appear to marginally improve on CMB limits.  
However, from a theoretical point of view, a proper understanding of the phenomenon remains to be properly developed and, for example, reliable predictions for the galaxy bispectrum are not yet available. Most importantly, as for general cosmological parameter estimation, a complete likelihood analysis aimed at constraining, or detecting, primordial non-Gaussianity in large-volume redshift surveys should involve {\it joint} measurements of the galaxy power spectrum and bispectrum, as well as possibly higher-order correlation functions. While we are still far from a proper assessment of what such analysis would be able to achieve, current results in this direction are very encouraging. 

This review is divided in four parts. In section~\ref{sec:initialcond} we will first discuss initial conditions as defined in terms of the primordial curvature bispectrum and its phenomenology. We will then review the observational consequences of primordial non-Gaussianity on the CMB bispectrum, section~\ref{sec:CMB}, and on the large-scale structure bispectrum as measured in redshift surveys, section~\ref{sec:LSS}. In both cases we will discuss theoretical models for the observed bispectra and technical problems related to the estimation of the non-Gaussian parameters, with the differences that naturally characterize such distinct observables.   We also give an example of joint analysis using both CMB and large-scale
structure when we consider the possibility of constraining a strongly scale-dependent non-Gaussian parameter $\fnl(k)$, emerging in some recently proposed inflationary models. 


\section{Initial conditions and the primordial bispectrum}
\label{sec:initialcond}

\subsection{The primordial bispectrum and shape function}
\label{sec:bispshape}

The starting point for this discussion is the primordial gravitational potential perturbation $\Phi({\bf x},\,t)$ which was seeded by quantum fluctuations during inflation or by some other mechanism in the very early universe ($t\ll t_{\rm dec})$. When characterizing the fluctuations $\Phi$ we usually work in Fourier space with the (flat space) transform defined through
\eq
\Phi({\bf x}, \, t) = \int \frac {d^3{\bf k}}{(2\pi)^3} \, e^{-i{\bf k\cdot x} }\, \Phi({\bf k},t)\,.
\qe
The primordial power spectrum $P_\Phi(k)$ of these potential fluctuations is found using
an ensemble average,
\eq\label{eq:powerspect}
\langle \Phi({\bf k})\Phi^*({\bf k}') \rangle = (2\pi)^3 \d_D ({\bf k} -{\bf k}')P_\Phi(k)\,,
\qe
where we have assumed that physical processes creating the fluctuations are statistically isotropic so that only the dependence on the wavenumber remains, $k = |{\bf k}|$. Recall that for nearly scale-invariant perturbations,  the fluctuation variance on the horizon scale $k\approx H$ is almost constant   $\Delta_{k\sim H}^2 \approx  k^3 P_\Phi(k)/2\pi^2 \approx \mbox{const.}$, implying $P_\Phi(k) \sim k^{-3}$.

\begin{figure}[t]
\centering
\includegraphics[width=.35\linewidth]{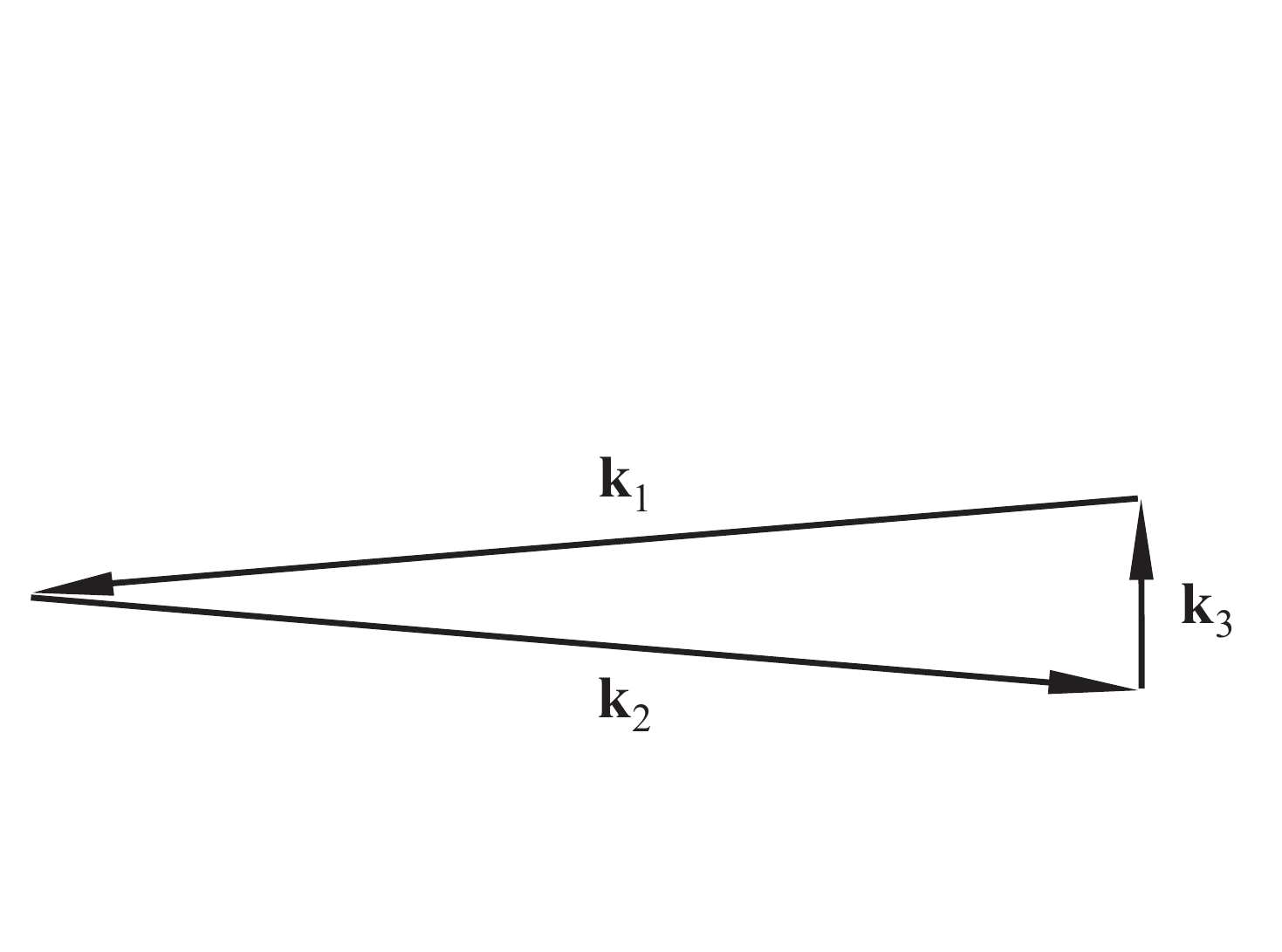}
\includegraphics[width=.27\linewidth]{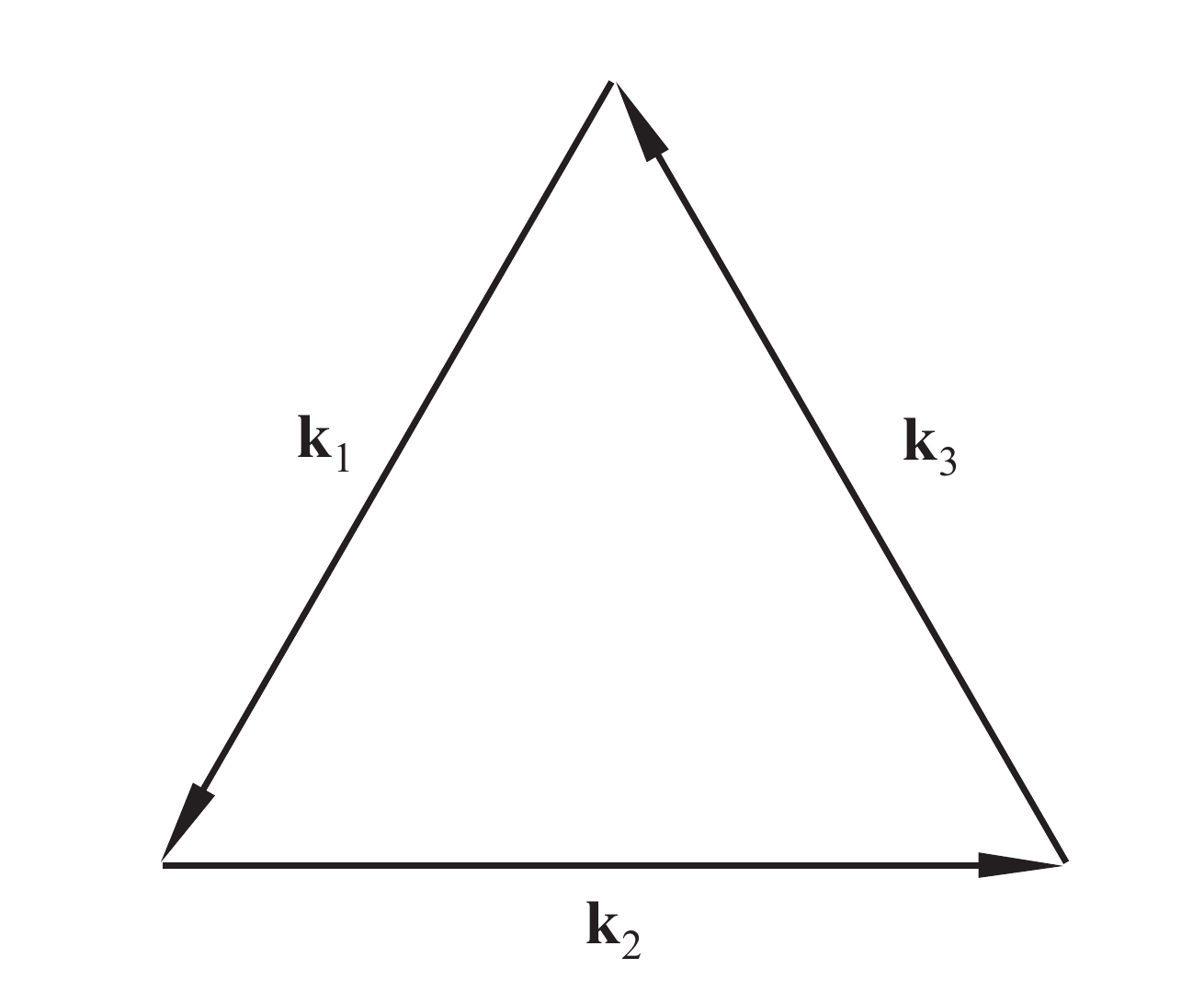}
\includegraphics[width=.35\linewidth]{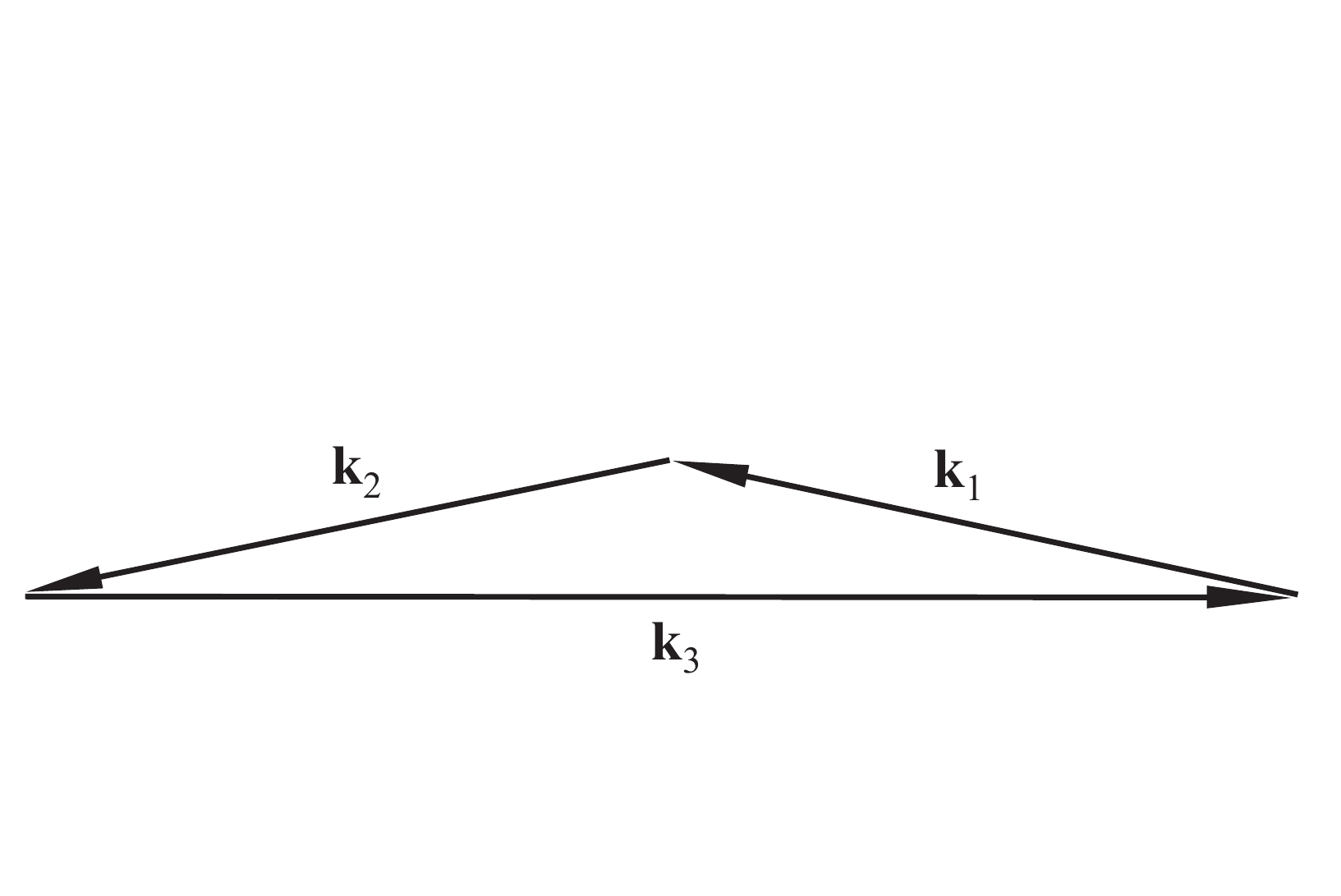}
\caption[Triangles ]{\small  Triangle types contributing to the bispectrum corresponding to `squeezed' or local configurations with $k_3\ll k_1,\,k_2$ (left),
equilateral configurations  with $k_3\approx k_1\approx k_2$ (centre) and flattened configurations with $k_3\approx k_1+ k_2$ (right).} 
\label{fig:triangles}
\end{figure}

\begin{figure}[t]
\centering
\includegraphics[width=.45\linewidth]{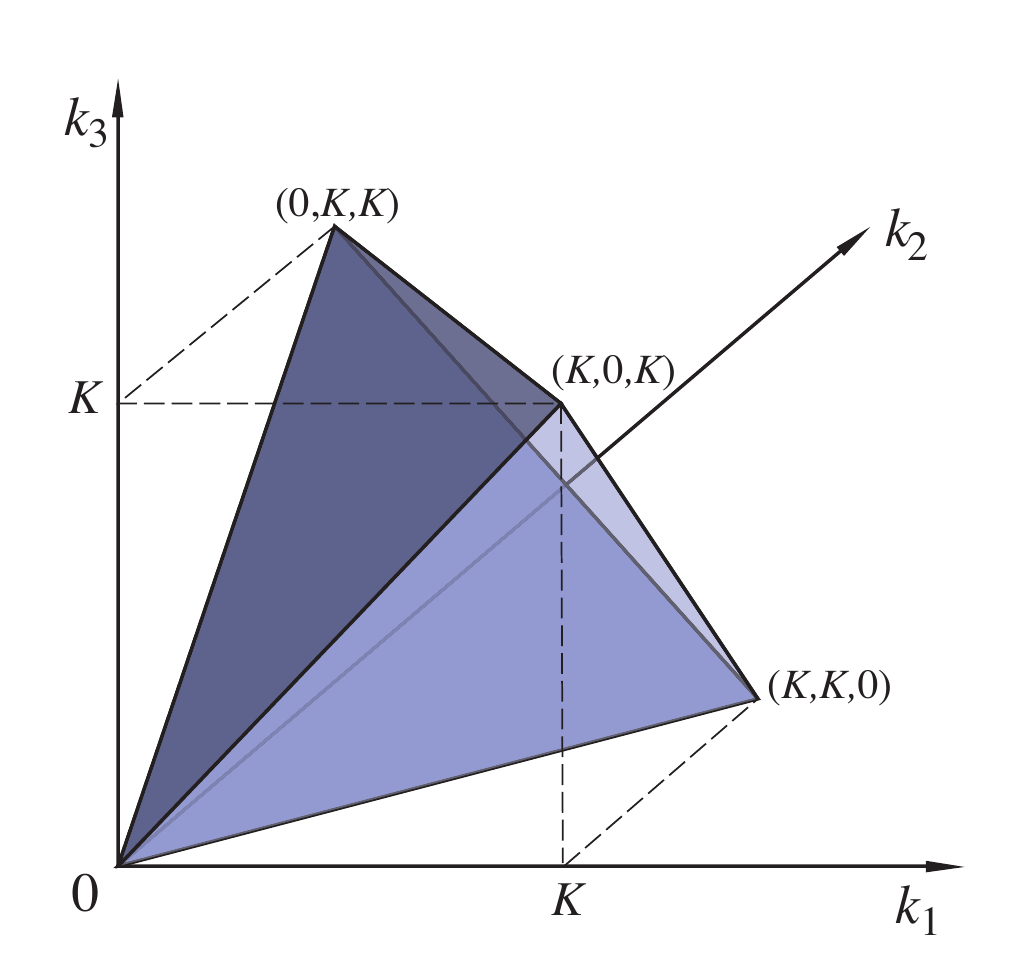}
\caption[Tetrahedral domain]{\small  Tetrahedral domain for allowed wavenumber configurations $\kall$ contributing to the primordial bispectrum $\Bkkk$).
A regular tetrahedron is shown satisfying $\ksum \le 2\kmax\equiv 2K$.} 
\label{fig:tetrapydk}
\end{figure}

The primordial bispectrum $B_\Phi(k_,\,k_2,\,k_3)$ is found from the Fourier transform of the three-point correlator as 
\begin{align}\label{eq:primbispect}
\<\O(\bk_1)\O(\bk_2)\O(\bk_3)\> = (2\pi)^3\,\d_D(\kv_{123})\, B_\Phi(\kall)\,.
\end{align}
Here, the delta function enforces the triangle condition, that is, the constraint that the wavevectors in Fourier space must close to form a triangle, ${\bf k}_1+{\bf k}_2+{\bf k}_3=0$. Examples of such triangles are shown in \fig{fig:triangles},
illustrating the basic squeezed, equilateral and flattened triangles to which we will refer later. 
Note that a specific triangle can be completely described by the three lengths of its sides and so, in the isotropic case, we are able to describe the bispectrum using only the wavenumbers $\kall$. The triangle condition restricts the allowed wavenumber configurations $(\kall)$ to the interior of the tetrahedron illustrated in \fig{fig:tetrapydk}.

The most studied primordial bispectrum is the {\it local model} in which contributions from `squeezed' triangles are dominant, that is, with e.g.\ $k_3\ll k_1,\, k_2$ (as illustrated in the left of \fig{fig:triangles}).   This is well-motivated physically as it encompasses `superhorizon' effects during inflation when a large scale mode $k_3$ (say) which has exited the Hubble radius  exerts a nonlinear influence on the subsequent evolution of smaller scale modes $k_1, \,k_2$.   Although this effect is small in single field slow-roll inflation, it can be much larger for multifield models.  In a weakly coupled regime,  the potential can be split into two components, the linear term $\Phi_{\rm L}$, representing a Gaussian field, giving the usual perturbation results plus a small local non-Gaussian term $\Phi_{\rm NL}$ \cite{SalopekBond1990},
\bea \label{eq:fnldefn}
\Phi({\bf x}) & = & \Phi_{\rm L}({\bf x}) + \Phi_{\rm NL}({\bf x})\nonumber\\
& = & \Phi_{\rm L}({\bf x}) + \fnl \big[\Phi^2_{\rm L}({\bf x}) - \langle \Phi^2_{\rm L}({\bf x})\rangle \big]\,,
\eea
where $\fnl$ is called the nonlinearity parameter. In Fourier space, the nonlinear term is then given by the convolution
\eq \label{eq:nlFT}
\Phi_{\rm NL}({\bf k}) = \fnl\left[ \int \frac {d^3{\bf k}}{(2\pi)^3} \, \Phi_{\rm L}({\bf k}+{\bf k}') \Phi_{\rm L}({\bf k}')  - 
(2\pi)^3 \d_D({\bf k})\langle \Phi^2_{\rm L}\rangle\right]\,.
\qe
From this we can infer, using (\ref{eq:powerspect}), that the only non-vanishing contributions to the bispectrum (\ref{eq:primbispect}) take the form  
\eq 
\<\O(\bk_1)\O(\bk_2)\O(\bk_3)\> = 2(2\pi)^3 \d_D(\kv_{123})\, P_\Phi(k_1)\,P_\Phi(k_2)\,.
\qe
Accounting for permutations, the local bispectrum becomes
\eq\label{eq:localB}
\nn B_\Phi(\kall) &=&~2\fnl\left[ P_\Phi(k_1)P_\Phi(k_2)+P_\Phi(k_2)P_\Phi(k_3)+P_\Phi(k_3)P_\Phi(k_1)\right]\\
&\simeq &~ 2\fnl\frac{\Delta_\Phi^2}{(k_1k_2k_3)^2}\(\frac{k_1^2}{k_2k_3}+\frac{k_2^2}{k_1k_3}+\frac{k_3^2}{k_1k_2}\)\,.
\qe
Although this is a rather pathological function which diverges along the edges of the tetrahedron (\ie when any $k_i\rightarrow 0$), we can infer from it some basic properties of the bispectrum for any model which is nearly scale-invariant. For example, we can observe that the bispectrum at equal $k_i$ has the characteristic scaling, 
\eq 
B_\Phi(k,\,k,\,k) = 2\fnl \Delta_\Phi^2 /k^6\,.
\qe
If we remove this overall $k^{-6}$ scaling by multiplying (\ref{eq:localB}) by the factor $(k_1k_2k_3)^2$, we note that on transverse slices through the tetrahedron defined by $\tilde k \equiv (\ksum)/2 = \mbox{const.}$ (see \fig{fig:tetrapydk}) the bispectrum only depends on the ratios of the wavenumbers, say $k_2/k_1$ and $k_3/k_1$.  Indeed, it can prove convenient to characterize the bispectrum in terms of the following transverse parameters \cite{RigopoulosShellardVanTent2006B,FergussonShellard2007}
\eq\label{eq:transcoords}
\tilde k = {\textstyle \frac{1}{2}}(\ksum)\,,\qquad \tilde \alpha = (k_2-k_3)/\tilde k \,,\qquad \tilde \beta = (\tilde k - k_1)/\tilde k\,,
\qe
with the domains  $\tilde k\le \kmax$, $0 \le\tilde \b \le 1$ and $ -(1-\tilde\b)\le \tilde\a \le 1 - \tilde\b$.  The volume element on the regular tetrahedron of allowed wavenumbers then becomes $ dk_1 dk_2 dk_3 = k^2d\tilde k \,d\tilde\alpha \,d\tilde \beta$.    

These considerations lead naturally to the definition of the primordial shape function \cite{BabichCreminelliZaldarriaga2004}
\begin{align} \label{eq:shapefn}
S(k_1,k_2,k_3) \equiv \frac{1}{N} (k_1 k_2 k_3)^2 B_\O(\kall)\,,
\end{align}
where $N$  is a normalization factor which is often chosen such that $S$ is unity for the equal $k_i$ case,  that is,  $S(k,\,k,\,k) = 1$ (we shall discuss alternatives to this rather arbitrary convention later).   For example, the canonical `local' model (\ref{eq:localB}) has the shape 
\eq\label{eq:localS}
S^{\rm local}(\kall) = \frac{1}{3} \(\frac{k_1^2}{k_2k_3}+\frac{k_2^2}{k_1k_3}+\frac{k_3^2}{k_1k_2}\)\,.
\qe
Thus it is usual to describe the primordial bispectrum in terms of an overall amplitude $\fnl$ and a transverse two-dimensional shape $S(\kall)= S(\tilde \alpha,\,\tilde \beta)$, which incorporates any distinctive momentum dependence. Of course, if there is a non-trivial scale dependence, then the full three-dimensional dependence of $S(\kall)$ on the $k_i$ must be retained.      

\begin{figure}[t]
\centering
\includegraphics[width=0.49\linewidth]{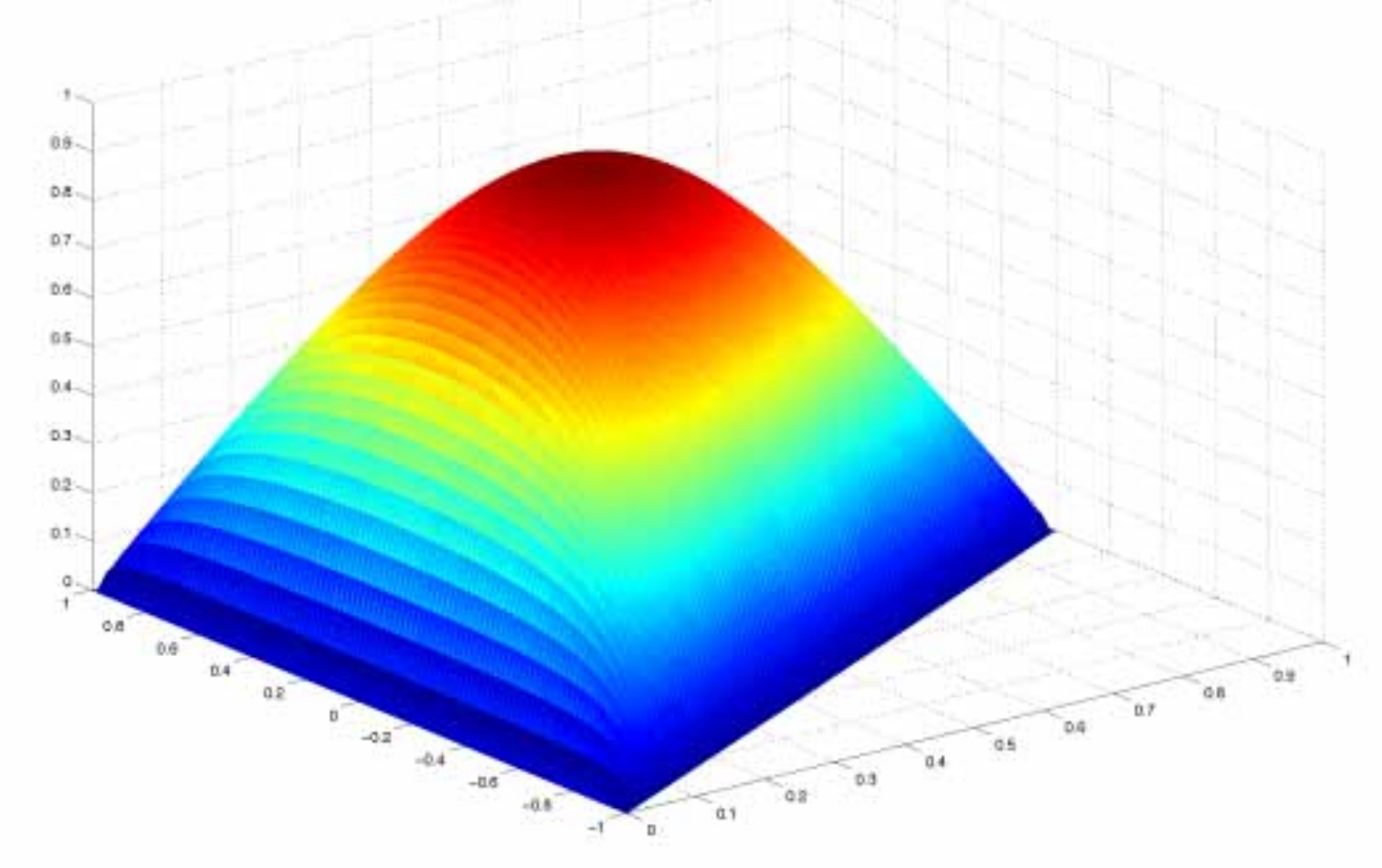} 
\includegraphics[width=0.49\linewidth]{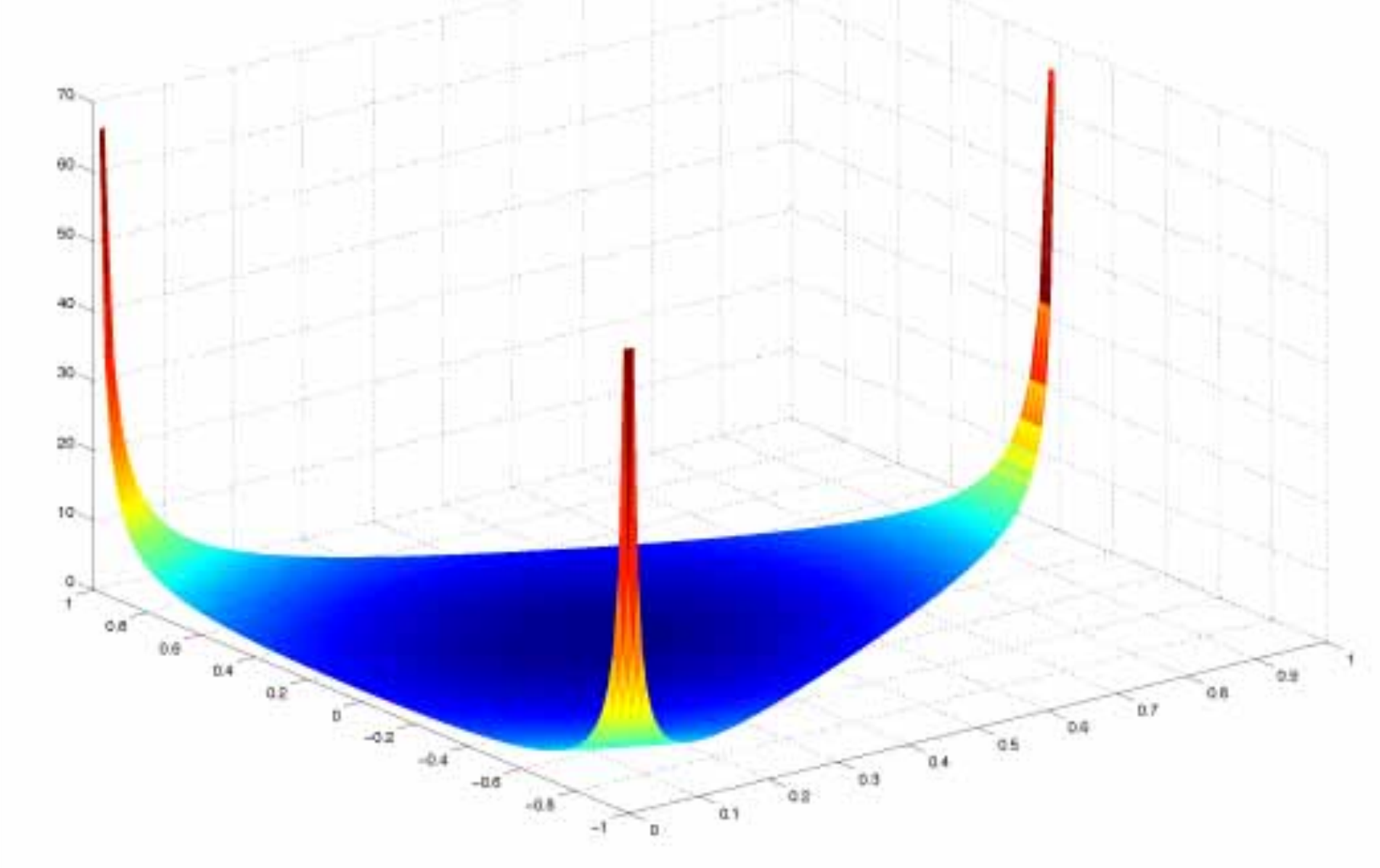} 
\caption[The shape function.]{\small Shape functions  for the scale-invariant equilateral (left) and local (right) models, $S(\kall) = S(\tilde \alpha,\,\tilde \beta)$ on transverse slices with $2\tilde k = \ksum = \mbox{const.}$\ }
\label{fig:shapes}
\end{figure}

There are other physically well-motivated shapes in the literature which have also been extensively studied. The simplest shape is the {\it constant model}
\eq \label{eq:constS}
S^{\rm const}(\kall) = 1\,,
\qe
which,  like the local model, has a large-angle analytic solution for the CMB bispectrum  \cite{FergussonShellard2009}.   
The local model tends to be the benchmark against which all other models are compared and normalized, but  for practical purposes the constant model is much more useful, given its regularity at both late and early times. The {\it equilateral shape} is another important case with \cite{BabichCreminelliZaldarriaga2004}
\eq\label{eq:equilS}
S^{\rm equil}(\kall) = \frac{(k_1+k_2-k_3)(k_2+k_3-k_1)(k_3+k_1-k_2)}{k_1k_2k_3}\,.
\qe
While not derived directly from a physical model, it has been chosen phenomenologically as a separable ansatz for higher derivative models \cite{Creminelli2003} and DBI 
inflation \cite{AlishahihaSilversteinTong2004}. The equilateral shape is contrasted with the local model in \fig{fig:shapes}. 

Another important early result was the primordial bispectrum shape for single-field slow roll inflation derived by \citet{Maldacena2003,AcquavivaEtal2003}
\eq \label{eq:MaldS}
\nn S^{\rm Mald}(k_1,k_2,k_3) &\propto& ~(3\e-2\n)  \left[\frac{k_1^2}{k_2k_3}+\frac{k_2^2}{k_1k_3}+\frac{k_3^2}{k_1k_2}\right]
+ \e\left[ (k_1 k_2^2 + \mbox{5 perm.})  + 4 \frac{k_1^2k_2^2 +k_2^2k_3^2 +k_3^2k_1^2}{k_1k_2k_3}\right] \\
&\simeq &~(6\e-2\n)S^{\rm local}(\kall) + \frac{5}{3}\e\, S^{\rm equil}(\kall)\,,
\qe
where $\e, \,\n$ are the usual slow roll parameters.  In the second line, we have noted that this shape can be accurately represented as the superposition of  local and equilateral shapes. The coefficients in (\ref{eq:MaldS}), which include  the scalar spectral index $n-1 = -6\e + 2\n \simeq -0.05$, confirm that $\fnl \ll 1$ and so standard single slow roll inflation cannot produce an observationally significant signal. Nevertheless, it is interesting to determine which shape is dominant in (\ref{eq:MaldS}) and to what extent 
other primordial shapes are independent from one another.   

Whether  two different primordial shapes can be distinguished observationally can be determined from the correlation between the corresponding two CMB bispectra weighted for the anticipated signal to noise,  as in the estimator (see next section) and the Fisher matrix analysis (see section~\ref{sec:forecasts}).   However, direct calculations of the CMB bispectrum can be very computationally demanding.   A much simpler approach is to determine the independence of the two shape functions $S$ and $S'$ from the
correlation integral \citep[][see also \citet{BabichCreminelliZaldarriaga2004}] {FergussonShellard2009}
\begin{align}
F_\e(S,S^\pr) = \int_{\curl{V}_k}\, S(k_1,k_2,k_3)\, S^\pr(k_1,k_2,k_3)\, \w_\e(k_1,k_2,k_3) d\curl{V}_k\,,
\end{align}
where we choose the weight function to be
\begin{align}\label{eq:weightcor}
\w_\e(k_1,k_2,k_3) = \frac{1}{k_1+k_2+k_3}\,,
\end{align}
reflecting the primary scaling of the CMB correlator. The shape correlator is then defined by 
\be\label{eq:shapecorrelator}
\bar{\curl{C}}(S,S^\pr) = \frac{F(S,S^\pr)}{\sqrt{F(S,S)F(S^\pr,S^\pr)}}\,.
\ee
Here, the integral is over the tetrahedral region shown in \fig{fig:tetrapydk}  taken out to a maximum wavenumber $k\lesssim \kmax$ corresponding to the experimental range $l \le \lmax$ for which forecasts are sought (with $\lmax \kern-2pt\approx \,\tau_0\, \kmax$).   The weight function $\w_\e(k_1,k_2,k_3) $ appropriate for mimicking the large-scale structure 
bispectrum estimator (see section~\ref{ssec:bsgBE}), would be different with varying scaling lawas introduced by the transfer functions for wavenumbers $k$ above and below $k_{\rm eq}$, the inverse comoving horizon at equal matter-radiation. 
Nevertheless, the $1/k$ weight given in (\ref{eq:weightcor}) provides a compromise between these scalings, and so 
shape correlation results should offer a useful first approximation. 

Below we will survey primordial models in the literature, showing how close the shape correlator comes to a full Fisher matrix analysis.  However, here we note that the local shape (\ref{eq:localS}) and the equilateral shape (\ref{eq:equilS}) have only a modest 46\% correlation. For the natural values of the slow roll parameters $\e\approx \n$ we find the somewhat surprising result that $S^{\rm Mald}$ is 99.7\% correlated with $S^{\rm local}$ (and it cannot be easily tuned otherwise because $3\e\approx\n$ is not consistent with deviations from scale-invariance favored observationally $n-1<0$).    Such strong correspondences are important in defining families of related primordial shapes, thus reducing the number of different cases for which separate observational constraints must be sought.

\subsection{General primordial bispectra and separable mode expansions}\label{sec:nonseparableshapes}

The three shape functions (\ref{eq:localS}),  (\ref{eq:constS}) and  (\ref{eq:equilS}) quoted above share the important property of {\it separability}, that is, they can be written in the form 
\eq \label{eq:separable}
S(\kall) = X(k_1) \,Y(k_2)\, Z(k_3) + \mbox{5 perms.}\,,
\qe
or as the sum of just a few such terms. As we shall see, if a shape $S$ is separable, then the computational cost of evaluating the corresponding CMB bispectrum $\Blll$ is dramatically reduced.  In fact, without this property, the task of estimating whether a non-separable bispectrum is consistent with observation appears to be intractable (for large $\lmax$).    Of course, the number of models which can be expressed directly in the form (\ref{eq:separable}) is very limited, despite the usefulness of approximate ans\"atze such as the equilateral shape (\ref{eq:equilS}). Indeed, approximating non-separable shapes by educated guesses for for the separable functions $X,\, Y,\, Z$ is neither systematic nor computationally 
efficient (because arbitrary non-scaling functions create numerical difficulties, as we shall explain later). 

Instead, we shall present a separable mode expansion approach for efficient calculations with any non-separable bispectrum, as described in detail in \citet{FergussonLiguoriShellard2009}  \citep[and originally proposed in][]{FergussonShellard2007}.  Our aim will be to express any shape function as an expansion in mode functions 
\eq \label{eq:separablebasis}
\Skkk = \sum_p\sum_r\sum_s \alpha _{prs} \,q_{\{p}(k_1) \, q_r(k_2) \, q_{s\}}(k_3) \equiv \sum_n \aQn \, \Qn(\kall)\,,
\qe 
where, here, for convenience we have represented the symmetrized products of the separable basis functions $q_p(k)$ as 
\eq\label{eq:Qdefn}
\curl{Q}_n(\kall) ={\textstyle{ \frac{1}{6}}}\left [q_p(x) q_r(y) q_s(z) +  \mbox{5 perms}\right]~\equiv~ q_{\{p}\,q_r\,q_{s\}}\,,
\qe
with a one-to-one mapping ordering the products as $ n \leftrightarrow \{prs\}$.  The important point is that the $q_p(k)$ must be an independent set of  well-behaved basis functions which can be used to construct complete and orthogonal three-dimensional eigenfunctions on the tetrahedral region $\Vtetra$  defined by (see fig.~\ref{fig:tetrapydk})
\eq \label{eq:tetrapydk}
\kall \leq \kmax\, ,\qquad 
k_1 \leq k_2+k_3~\hbox{for}~k_1 \geq k_2,\,k_3,  ~+~~ \mbox{2 perms}\,.
\qe
The introduction of the cut-off at $\kmax$ is motivated both by separability and the correspondence with the observational domain $l\le \lmax$.   In the shape correlator (\ref{eq:shapecorrelator}), we have already seen what is essentially an inner product between two shapes on this tetrahedral region, which we can define for two functions $f,\,g$ as 
\eq
\langle f,\, g\rangle = \int_{\curl{V}_\curl{T}}\, f(k_1,k_2,k_3)\, g(k_1,k_2,k_3)\, \w(k_1,k_2,k_3) d\curl{V}_\curl{T}\,,
\qe 
with weight function $w$.     

\begin{figure}[t]
\centering
\includegraphics[width=.75\textwidth, height = 5.5cm]{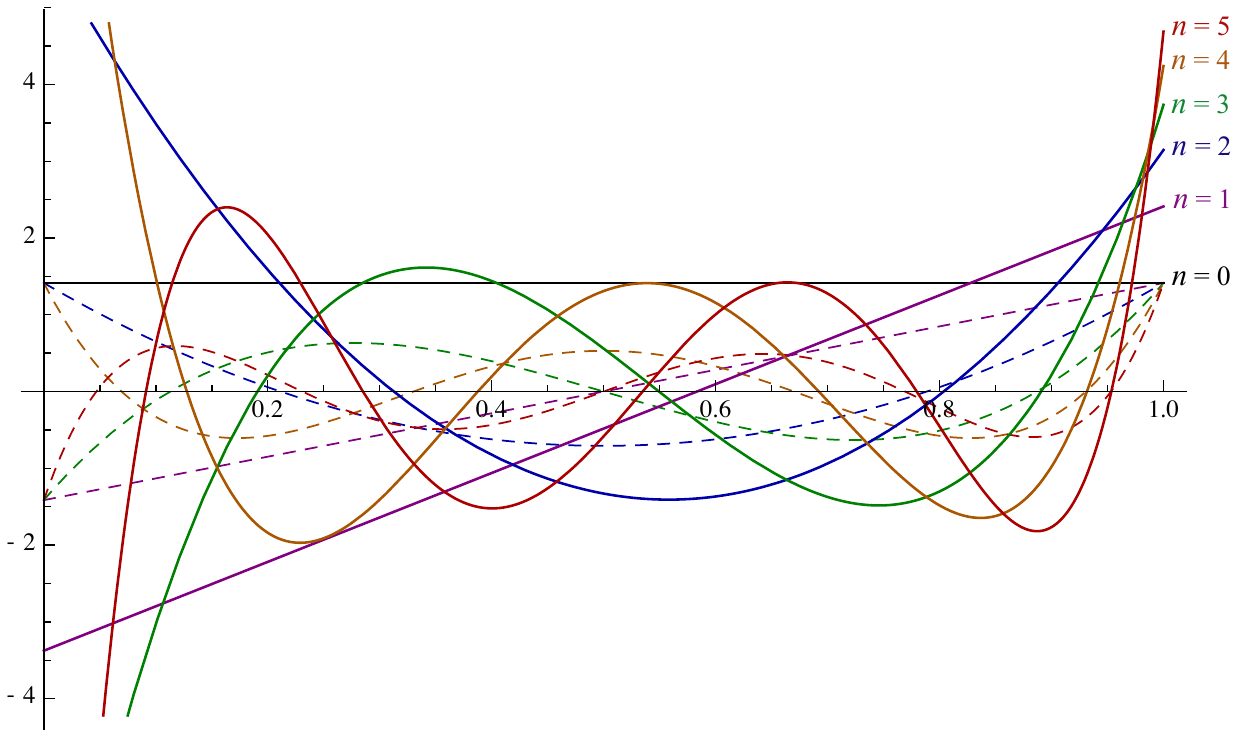}
\caption[Eigenfunctions in 1D on a tetrahedral domain]{\small The one-dimensional tetrahedral polynomials $q_n(k)$ on the domain (\ref{eq:tetrapydk}), rescaled to the unit interval for $n=0$--$5$. Also plotted are the shifted Legendre polynomials $P_n(2x-1)$ (dashed lines) which share qualitative features such as  $n$ nodal points.}
\label{fig:Qpolynomials}
\end{figure}
Satisfactory convergence for known bispectra can be found by using simple polynomials $q_p(k)$ in the expansion (\ref{eq:separablebasis}), that is, using analogues of Legendre polynomials on the domain (\ref{eq:tetrapydk}). With unit weight, the polynomials satisfying $\langle q_p(k_1),\, q_r(k_1)\rangle = \delta_{pr}$ can be found by generating 
functions with the first three  given by \cite{FergussonLiguoriShellard2009}
\eq \label{eq:Qpolys}
q_0(x) =\sqrt{2}\,,\qquad q_1(x) = 5.79 \,(-{\textstyle\frac{7}{12}} +x)\,,\qquad q_2(x) = 23.3 \left({\textstyle\frac{54}{215}} 
-{\textstyle\frac{48}{43}}x+ x^2\right)\,, ~...
\qe 
The first few polynomials $q_p(k)$ are plotted in \fig{fig:Qpolynomials}, where they are contrasted with Legendre polynomials.   

The three-dimensional separable basis functions $\Qn$ in (\ref{eq:Qdefn}) reflect the six symmetries of the bispectrum through the permuted sum of the product terms.   They could have been constructed directly from simpler polynomials, such as $1, \, k_1+k_2+k_3,\, k_1^2 +k_2^2 +k_3^2, \,...$, however, the $q_p$ polynomials have two distinct advantages.
First, the $q_p$'s confer partial orthogonality on the $\Qn$ and, secondly, these remain well-behaved when convolved with transfer functions.   

\begin{figure}[t]
\centering
\includegraphics[width=0.75\linewidth, height=5.25cm]{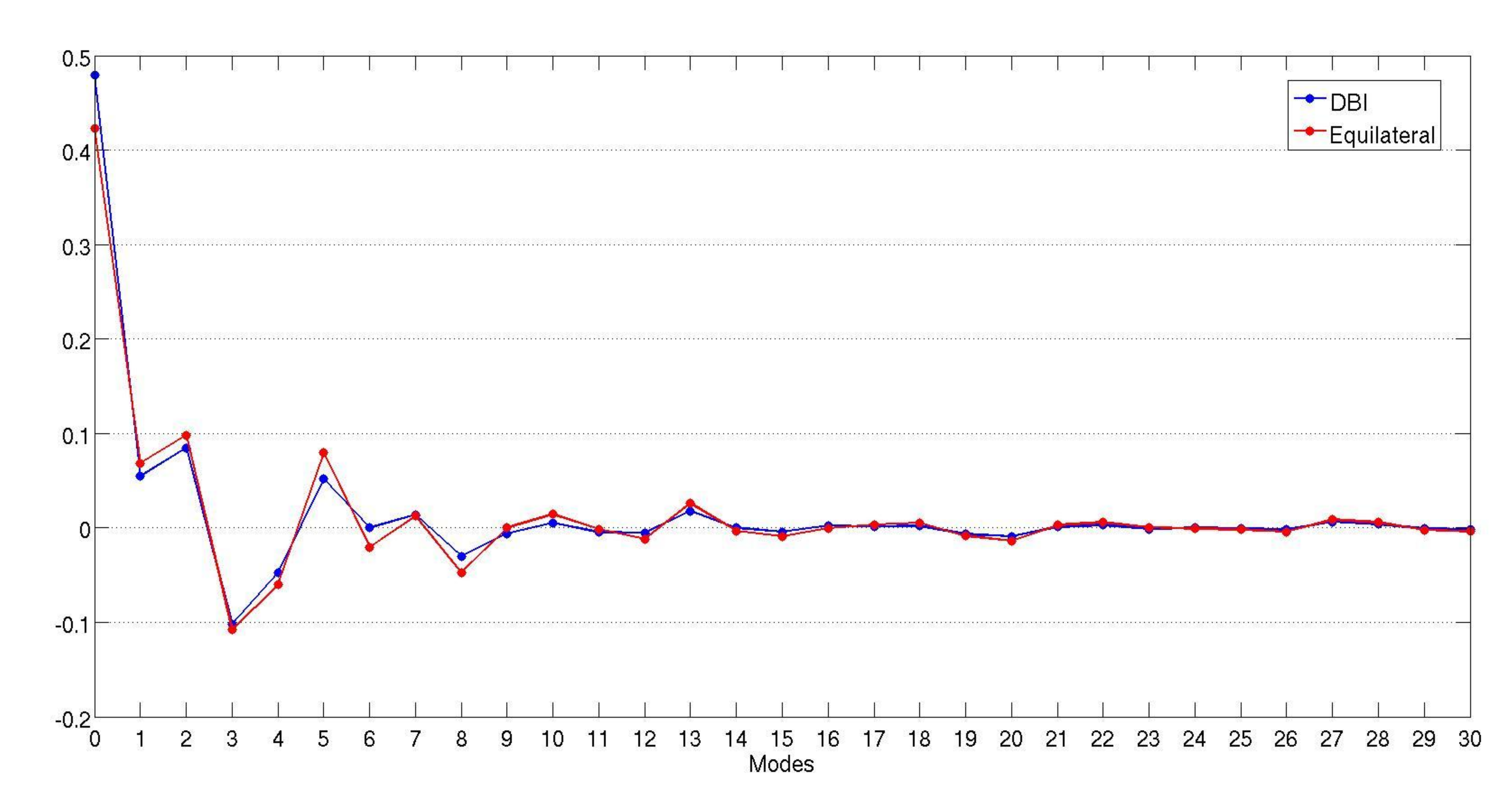}
\includegraphics[width=0.75\linewidth, height=5.25cm]{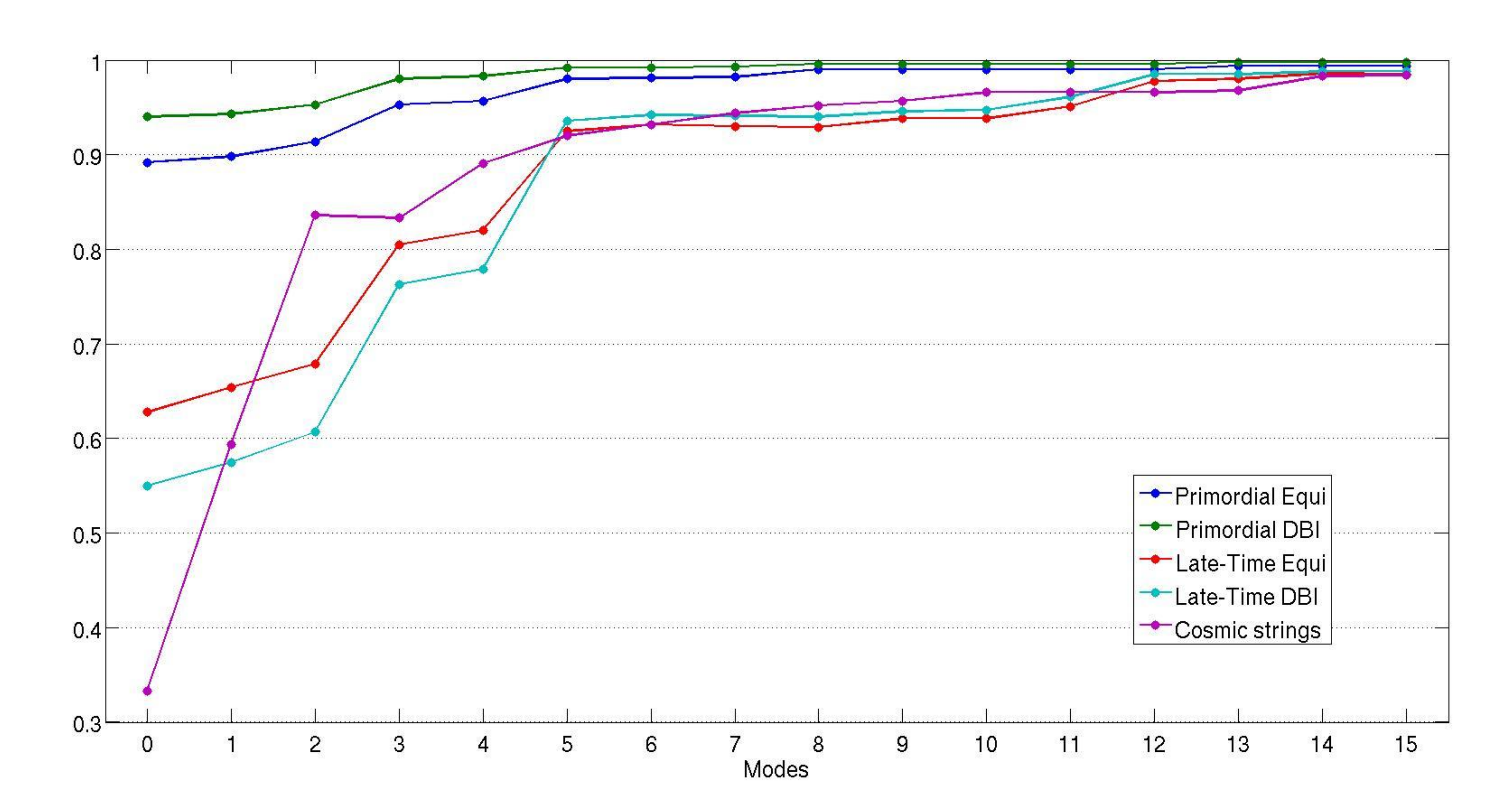}
\caption[Convergence]{\small Orthonormal eigenmode decomposition coefficients (\ref{eq:orthobasis}) for the equilateral and DBI models ({\it upper panel}) and shape correlations (\ref{eq:shapecorrelator}) of the original bispectrum against the partial sum up to a given mode $n$ ({\it lower panel}). The correlation plot includes both primordial and late-time CMB bispectra for the equilateral and DBI models,  as well as the late-time CMB bispectrum from cosmic strings (refer to section~\ref{sec:CMB}). In all cases, we find that we need at most 15 three-dimensional modes to obtain a correlation greater than 98\% (primordial convergence without the acoustic peaks requires only 6 modes).}
\label{fig:convergence}
\end{figure}

In order to rapidly decompose an arbitrary shape function $S$ into the coefficients $\aQn \leftrightarrow \alpha^{\scriptstyle\curl{Q}}_{prs}$, it is more convenient to work in a non-separable orthonormal basis $\Rn$ ($\langle \Rn,\,\Rm\rangle = \delta_{nm}$.  These can 
be derived directly from the $\Qn$ through Gram-Schmidt orthogonalization,  so that 
$\Rn = \sum_{p=0}^n \l _{mp} \Qp$ with $\l_{mp}$ a lower triangular matrix \citep[see][]{FergussonLiguoriShellard2009}.  Thus we can find the unique shape function decomposition 
\eq \label{eq:orthobasis}
\Skkk = \sum_n^N\aRn \, \Rn(\kall) = \sum_n^N \aQn \, \Qn(\kall) \,, \quad \mbox{with}~~ \aRn = \langle S,\, \Rn\rangle ~~ 
\mbox{and} ~~ \aQn =  \sum_p^N(\l ^\top)_{np}\, \aRp\,.
\qe
In the orthonormal $\Rn$ frame, Parseval's theorem ensures that the autocorrelator is simply $\langle S,\, S\rangle = \sum_n{ \aRn}^2$. Hence, with a simple and efficient prescription we can construct separable and complete basis functions on the tetrahedral domain (\ref{eq:tetrapydk})  providing rapidly convergent expansions for any well-behaved shape function $S$.  These eigenmode expansions will prove to be of  great utility in subsequent sections.   Examples of this bispectral decomposition and its rapid convergence for the equilateral and DBI models are shown in \fig{fig:convergence}.

\subsection{Families of primordial models and their correlations}
\label{sec:cmbcorrelator}

 We will now briefly survey the main categories of primordial models in the literature and
 their relative independence, closely following the discussion in \citet{FergussonShellard2009}.

\subsubsection{The constant model}

The constant model (\ref{eq:constS}) is the simplest possible primordial shape with triangles of every configuration contributing equally to the bispectrum $B(\kall)$; it is the equipartition model. The constant model was motivated initially 
by its simplicity  \cite{FergussonShellard2009} leading to an analytic solution for the large-angle CMB bispectrum,
as well as due to its close correlation with equilateral models.   However, the shape does have more explicit physical motivation in at least one context  \citep{ChenWang2009}, during multifield inflation for a slowly turning trajectory  (denoted quasi-single field inflation).  For multifield inflation, it is well known that the conversion of isocurvature fluctuations into curvature fluctuations during `corner-turning' can source significant non-Gaussianity \citep[see \eg][]{RigopoulosShellardVanTent2006B, VernizziWands2006}.   In the quasi-single field case with mass $m\sim H$ isocurvature modes,  a detailed investigation of the ongoing conversion into the curvature mode demonstrated that novel shapes could be generated \citep{ChenWang2009}, amongst them shapes which were very nearly constant.  Generically, these model-dependent shapes belonged to a one-parameter family which interpolated non-trivially between equilateral (\ref{eq:equilS}) and local (\ref{eq:localS}) shapes \citep[see also][]{SenatoreSmithZaldarriaga2009,RenauxPetel2009}.   This is an important caveat for the present discussion, because non-Gaussian searches could uncover shapes intermediate between the categories we will discuss below.

\subsubsection{Equilateral triangles -- centre-weighted models}

\begin{table}[b]
\centering
\begin{tabular}[t]{l||c|c||c|c||c|c}
 & \multicolumn{2}{c}{\bf DBI} & \multicolumn{2}{c}{\bf Ghost} & \multicolumn{2}{c}{\bf Single}    \\
\hline
 &$\curl{C}(S,S^\pr)$ & $\curl{C}(B,B^\pr)$ & $\curl{C}(S,S^\pr)$ &$\curl{C}(B,B^\pr)$ & $\curl{C}(S,S^\pr)$ &$\curl{C}(B,B^\pr)$   \\
\hline
{\bf Equilateral} &0.99 & 0.99 & 0.98 & 0.98 & 0.95 & 0.96   \\
{\bf DBI}  &  &  & 0.94 & 0.95 & 0.98 & 0.99   \\
{\bf Ghost} & & & & & 0.86 & 0.89  \\
\hline
\end{tabular}
\caption{Shape correlations (\ref{eq:shapecorrelator}) and CMB correlations (\ref{eq:cmbcor}) between the equilateral family of primordial models.} 
\label{tb:shapecorrelator}
\end{table}

Bispectra dominated by contributions from nearly equilateral triangle configurations, $k_1\approx k_2\approx k_3$ can be fairly easily characterized analytically and are the most amenable to CMB searches. However, equilateral non-Gaussianity requires the amplification of nonlinear effects around the time modes exit the horizon, which is not possible in a slow-roll  single field inflation. Instead, the kinetic terms in the effective action must be modified as in the Dirac-Born-Infeld (DBI) model \citep{AlishahihaSilversteinTong2004} or by explicitly adding higher derivative terms, such as in K-inflation \citep[see, for example,][]{ChenEtal2007}. The resulting corrections modify the sound speed $c_s$ and inflation is able to take place in steep potentials. For DBI inflation, this leads to non-Gaussianity being produced with a shape function of the form \citep{Creminelli2003, AlishahihaSilversteinTong2004}
\eq\label{eq:dbiS}
S(\kall) = \frac{1}{k_1 k_2 k_3 (k_1+k_2+k_3)^2} \[\sum_i k_i^5 + \sum_{i \neq j}\(2 k_i^4 k_j - 3 k_i^3 k_j^2\) 
+ \sum_{i \neq j \neq l}\(k_i^3 k_j k_l - 4 k_i^2 k_j^2 k_l\)\].
\qe
Another example of a model with non-standard kinetic terms is ghost inflation \cite{ArkaniHamedEtal2004} with a derivatively-coupled field driving inflation 
and a  trilinear term in the Lagrangian creating a non-zero equilateral-type shape $S^{\rm ghost}$ tending towards constant.  

General non-Gaussian shapes arising from modifications to single field inflation have been extensively reviewed in \citet{ChenEtal2007}. Using a Lagrangian that was an arbitrary function of the field and its first derivative, they were able to identify six distinct shapes describing the possible non-Gaussian contributions. Half of these had negligible amplitude being of the order of slow-roll parameters (with two already given in (\ref{eq:MaldS})). Of the remaining three shapes \citep[][see also \citealp{SeeryLidsey2005}]{ChenEtal2007}, one was believed to be subdominant, the second recovered the DBI shape (\ref{eq:dbiS}), leaving a third distinct single field shape which is the inverse of the local shape (\ref{eq:localS}), $S^{\rm single} \propto {S^{\rm local}}^{-1}$. Finally, we recall the original equilateral shape (\ref{eq:equilS}), noting that it was introduced not because of a fundamental physical motivation, but as a separable approximation to the DBI shape (\ref{eq:dbiS}) \cite{BabichCreminelliZaldarriaga2004}.

Despite the apparent visual differences between these shapes \citep[see][]{FergussonShellard2009}, particularly near the edges of the tetrahedral domain, the shape correlator (\ref{eq:shapecorrelator}) reveals at least a 95\% or greater correlation of the DBI, ghost and single shapes to the equilateral shape \eqn{eq:equilS} \citep[consistent with results in][]{BabichCreminelliZaldarriaga2004, SmithZaldarriaga2006} .   Comparative results between the shape correlator
are given in Table \ref{tb:shapecorrelator} (together with the corresponding CMB correlation results brought 
forward and showing the efficacy of these estiomates).  These particular centre-weighted shapes must be regarded as a single class which would-be extremely differentiate observationally, without a bispectrum detection of very high significance.

Finally, we comment on the `orthogonal' shape $S^{\rm orthog}$ proposed in \citet{SenatoreSmithZaldarriaga2009}, together with $S^{\rm equil}$, for characterizing single field inflation models with an approximate shift symmetry \citep[see also][]{ChenEtal2007}.   This shape is approximately $S^{\rm orthog} \propto S^{\rm equil} -2/3$, which means it is very  similar to an earlier study of flattened shapes \cite{MeerburgVanDerSchaarCorasaniti2009} which proposed an `enfolded' shape with $S^{\rm enfold} \propto S^{\rm equil} -1$.   From the eigenmode decomposition (\ref{eq:orthobasis}) of the equilateral model shown in fig.~\ref{fig:convergence}, it is clear how the degree of correlation can be altered by subtracting out the important constant term.

\subsubsection{Squeezed triangles -- corner-weighted models}

The local shape covers a wide range of models where the non-Gaussianity is produced by local interactions. These models have their peak signal in ``squeezed" states where one $k_i$ is much smaller than the other two due to non-Gaussianity typically being produced on superhorizon scales. We have already observed that single-field slow-roll inflation (\ref{eq:MaldS}) is dominated by the local shape \cite{BartoloEtal2004}, though $\fnll$ is tiny \cite{Maldacena2003,AcquavivaEtal2003,BartoloEtal2004,811001,820101}. The production of non-Gaussianity during multiple field inflation \cite{BernardeauUzan2002, BernardeauUzan2003, LythRodriguez2005, RigopoulosShellardVanTent2006A, SeeryLidsey2005, RigopoulosShellardVanTent2006B, VernizziWands2006, RigopoulosShellardVanTent2007} shows much greater promise through conversion of isocurvature  into adiabatic perturbations \citep[see, for example, recent work in][and references therein]{ByrnesChoiHall2008, NarukoSasaki2009, ChenWang2009, RenauxPetel2009}. The magnitude of the non-Gaussianity generated is normally around $\fnll \px O(1)$, which is at the limit for Planck detection, but models can be tuned to create larger signals. Significant $\fnll$ can be produced in curvaton models with $\fnll \px O(100)$ \citep{LindeMukhanov2006, LythUngarelliWands2003, BartoloMatarreseRiotto2004}. Large $\fnll$ can also be generated at the end of inflation from massless preheating or other reheating mechanisms \cite{EnqvistEtal2005A, EnqvistEtal2005B, ChambersRajantie2008}.

\begin{table}[t]
\centering
\begin{tabular}[t]{l||c|c||c|c||c|c||c|c}
\hline
 & \multicolumn{2}{c}{\bf Local} &  \multicolumn{2}{c}{\bf Warm} &  \multicolumn{2}{c}{\bf Flat} &   \multicolumn{2}{c}{\bf Feature}   \\
\hline
 &$\curl{C}(S,S^\pr)$ & $\curl{C}(B,B^\pr)$ & $\curl{C}(S,S^\pr)$ &$\curl{C}(B,B^\pr)$ & $\curl{C}(S,S^\pr)$ &$\curl{C}(B,B^\pr)$  & $\curl{C}(S,S^\pr)$ &$\curl{C}(B,B^\pr)$   \\
\hline
{\bf Equilateral} &0.46 & 0.51 & 0.44 & 0.42 & 0.30 & 0.39 & -0.36 & -0.43   \\
{\bf Local}  &  &  & 0.30 & 0.52 & 0.62 & 0.79 & -0.41 & -0.39   \\
{\bf Warm}  &  &  &  &  & 0.01 & 0.21  & -0.05 & -0.27   \\
{\bf Flat}  &  &  &  &  &  &  & -0.44 & -0.32   \\
\hline
\end{tabular}
\caption{Shape correlations (\ref{eq:shapecorrelator}) and CMB correlations (\ref{eq:cmbcor}) for  5 distinct families of primordial non-Gaussian models.} 
\label{tb:correlatorcomparison}
\end{table}

We note that local non-Gaussianity can also be created in more exotic scenarios.   Models based on non-local field theory, such as $p$-adic inflation, can have inflation in very steep potentials. Like single field slow-roll inflation, the predicted `non-local' shape function is a combination of a dominant local shape (\ref{eq:localS}) and an equilateral shape (\ref{eq:equilS}) \citep[see, for example,][]{SeeryMalikLyth2008, BarnabyCline2008, Zaldarriaga2004, Halliwell1993}.   The ekpyrotic model can also generate significant $\fnll$ \citep{CreminelliSenatore2007, KoyamaEtal2007, BuchbinderKhouryOvrut2008, LehnersSteinhardt2008A, LehnersSteinhardt2008B}. Here the density perturbations are generated by a scalar field rolling in a negative exponential potential, so non-linear interactions are important with  $\fnll \px O(100)$.

In using the shape correlator for the local model, we must introduce a small-wavenumber cut-off, taken to be a $k_{\rm min} = 2  / tau_0$,  otherwise the shape correlator $\bar{\cal C }(S^{\rm local},S^{\rm local})$ becomes infinite.   This logarithmic divergence does not afflict the CMB bispectrum $b_{l_1l_2l_3}$ because we do not consider contributions below the quadrupole $l=2$ (a threshold which is 
approximated by the primordial cut-off).
The local shape is modestly correlated at the 40-55\% level with the equilateral shapes, mainly through the constant term in the expansion \eqn{eq:orthobasis}.  As can be seen in table~\ref{tb:correlatorcomparison} this somewhat underestimates
the CMB correlator.   Nevertheless, a NG signal of only modest significance should be able to distinguish between 
these independent models.

Finally, warm inflation scenarios, \ie models in which dissipative effects play a dynamical role, are also predicted to produce significant non-Gaussianity \cite{GuptaEtal2002, GanguiEtal1994}. Contributions are again dominated by squeezed configurations but with a different more complex shape possessing a sign flip as the corner is approached (see fig.~\ref{fig:othershapes}).  This makes the warm  $S^{\rm warm}$ and local $S^{\rm local}$ shapes essentially orthogonal with only a  33\% correlation.  Again, in using the shape correlator, we need to introduce the same phenomenological 
cut-off $k_{\rm min}$ as for the local model, but we also note the more serious concern which is the apparent 
breakdown of the approximations used to calculate the warm inflation shape near the corners and edges.

\subsubsection{Flattened triangles -- edge-weighted models}

\begin{figure}[t]
\centering
\includegraphics[width=0.49\linewidth]{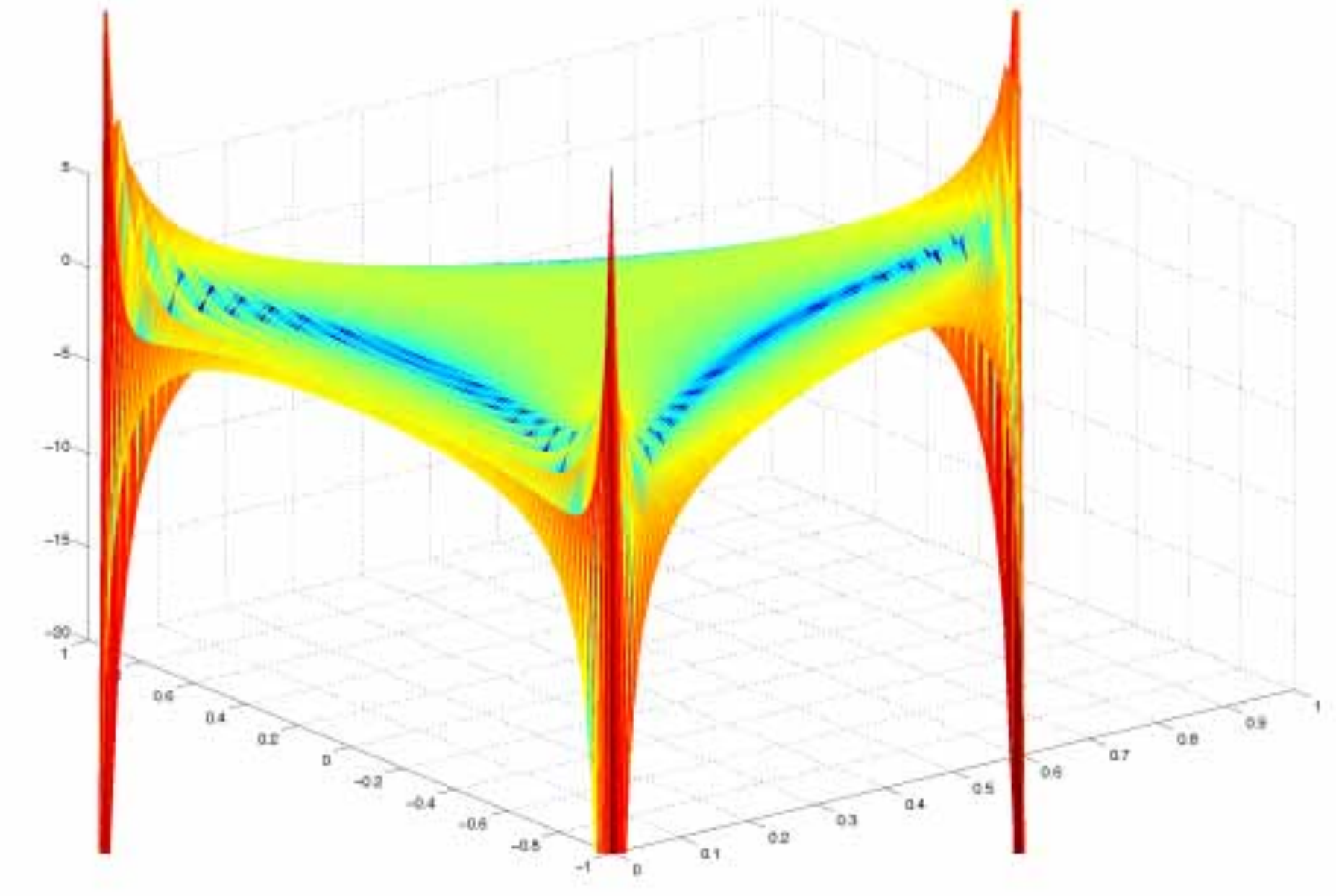} 
\includegraphics[width=0.49\linewidth]{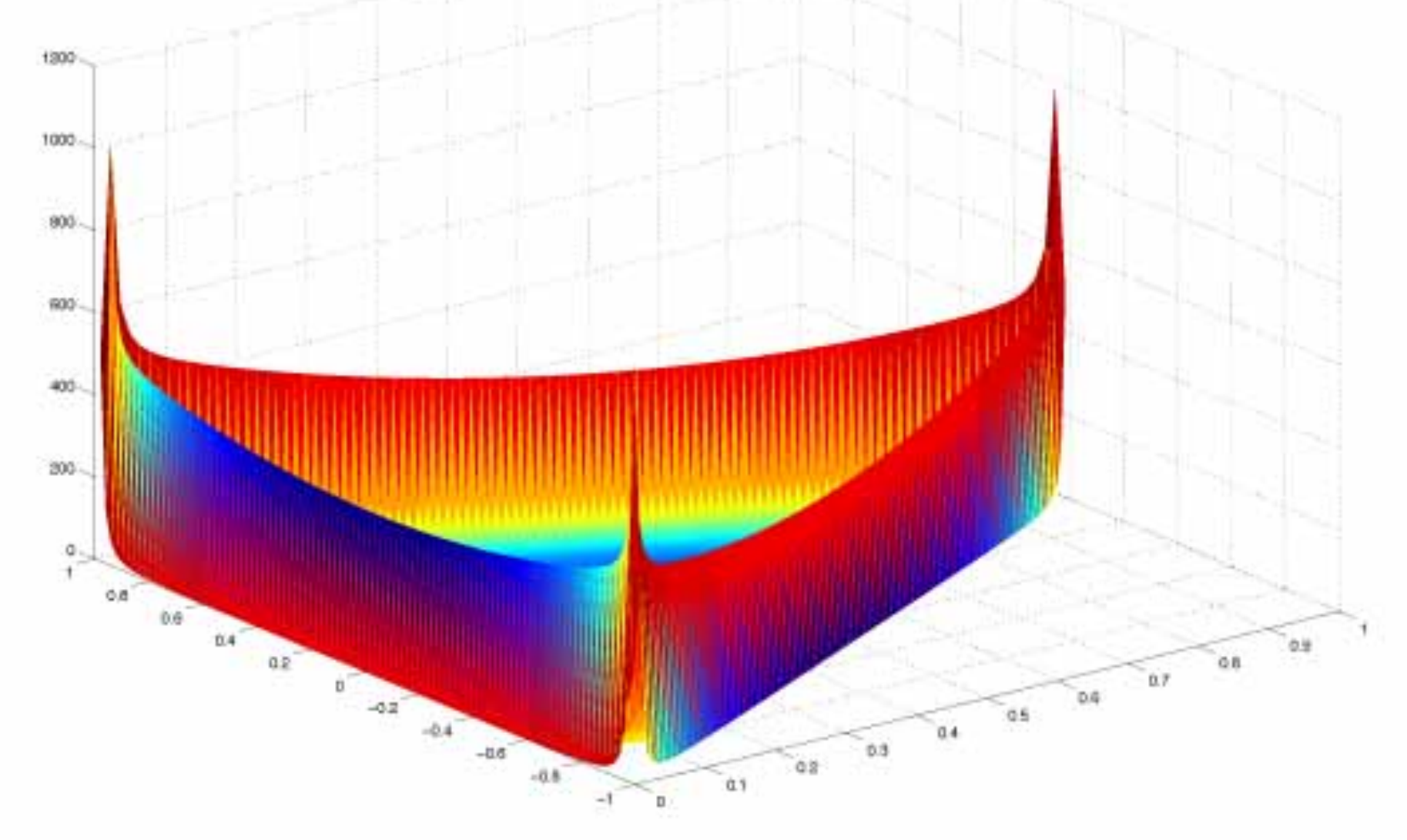} 
\caption[The shape function.]{\small Shape functions  for the nearly scale-invariant `warm' and `flat' NG models, $S(\kall) = S(\tilde \alpha,\,\tilde \beta)$ on transverse slices with $2\tilde k = \ksum = \mbox{const.}$\  These distinct and independent shapes prove to be largely uncorrelated
with each other and the local and equilateral models illustrated in fig.~\. From \citet{FergussonShellard2009}.}
\label{fig:othershapes}
\end{figure}

It is possible to consider inflationary vacuum states which are more general than the Bunch-Davies vacuum, such as an excited Gaussian (and Hadamard) state \citep[][see also discussions in \citealt{ChenEtal2007,MeerburgVanDerSchaarCorasaniti2009}]{HolmanTolley2008}.   Observations of non-Gaussianity in this case might provide insight into trans-Planckian physics. The proposed shape for the bispectrum is
\eq
\label{eq:flat}
S^{\rm flat}(k_1,k_2,k_3) ~\propto ~6\left\{\frac{k_1^2+k_2^2 -k_3^2}{k_2k_3} +\mbox{2 perms} \right\}
+ \frac{2(k_1^2+k_2^2+k_3^2)s}{(k_1+k_2-k_3)^2(k_2+k_3-k_1)^2(k_3+k_1-k_2)^2}\,.
\qe
The bispectrum contribution from early times is dominated by flattened triangles, with \eg $ k_3 \approx k_1+k_2$, and for a small sound speed $c_s\ll 1$ can be large. Unfortunately, as the divergent analytic approximation breaks down at the  boundary of the allowed tetrahedron, some form of cut-off must be imposed, as shown for the smoothed shape in \fig{fig:othershapes} where an edge truncation has been imposed together with a  Gaussian filter.
 The lack of compelling physical motivation and ill-defined asymptotics make predictions for this model uncertain.

\subsubsection{Features -- scale-dependent models}

There are also models in which the inflation potential has a feature, providing a break from scale-invariance. This can take the form of a either a step \citep{ChenEastherLim2006} or a small oscillation superimposed onto the potential \citep{BeanEtal2008B}. Analytic forms for both these three point functions have been presented in \citet{ChenEastherLim2008} with one approximation taking the form,
\begin{align}
\label{eq:feature}
S^{feat}(k_1,k_2,k_3) &~\propto~ \sin\(\frac{k_1+k_2+k_3}{k^*} + P\)\,, 
\end{align}
where $k^*$ is the associated scale of the feature in question and $P$ is a phase factor. Results for the shape correlator for a particular feature model (with $k^* \approx \ell^*/\tau_0$ and $\ell^*=50$), are given in table \ref{tb:correlatorcomparison}, showing that it is essentially independent of all the other shapes.   Clearly, scale dependent feature models form a distinct fifth category beyond equilateral, local, warm and flat shapes.

\section{Cosmic Microwave Background}
\label{sec:CMB}

\subsection{The CMB bispectrum}\label{sec:CMBbisp}

In this section we will study the connection between the primordial bispectrum at the end of inflation 
and the observed bispectrum of CMB anisotropies $\Blll$.
 Our work will be primarily concerned with the analysis of the three-point function induced by a NG primordial
gravitational potential $\Phi({\bf k})$ in the CMB temperature fluctuation field.
Temperature anisotropies are represented using the $a_{\ell m}$ coefficients of a spherical harmonic decomposition of 
the cosmic microwave sky, 
$$ 
\frac{\Delta T}{T}(\hat {\bf n}) = \sum_{\ell m} a^T_{\ell m} Y_{\ell m}(\hat {\bf n})\,.
$$
Analogous expansions are performed for the $E$ mode polarization field in order to produce 
polarization multipoles $\alm^E$. For simplicity and clarity, throughout most of this review we will focus on the temperature 
multipoles $\alm^T$, and omit the superscript T for convenience of notation. However we stress 
here that all of the considerations we make in the following can be readily applied 
to polarization multipoles and related bispectra. More discussion about this subject can be found in 
section \ref{sec:polarization}.

The primordial potential $\Phi$ is imprinted on the CMB multipoles $a_{lm}$ by a convolution with transfer
functions $\D_l(k)$ representing the linear perturbation evolution, through the integral: 
\begin{align} \label{eq:phi2alm1}
a_{\ell m} = 4\pi (-i)^l \int \frac{d^3 k}{(2\pi)^3}\, \D_{\ell}(k) \,\O(\bk)\, Y_{\ell m}(\uk)\,.
\end{align}
The radiation transfer functions encode all the typical effects observed in the CMB power spectrum at linear order, that is, 
the Sachs-Wolfe effect, Integrated Sachs-Wolfe effect, doppler peaks and silk damping \citep[see \eg][]{MaBertschinger1995,Dodelson2003}.  An equation identical to (\ref{eq:phi2alm1}) produces the E-mode polarization CMB multipoles starting from the primordial temperature fluctuation field, provided polarization transfer functions replace temperature transfer functions in the convolution above.
It is sometimes useful to rewrite equation (\ref{eq:phi2alm1}) in position, rather than Fourier, space. In this case 
it is straightforward to show that (\ref{eq:phi2alm1} becomes
\begin{align} \label{eq:phix2alm}
\alm = \int dr r^2 \alpha_{\ell}(r) \Phi_{\ell m}(r) \; ,
\end{align}
where, starting from the primordial potential $\Phi(\mathbf{x})$, we transform from Cartesian into polar coordinates
$\mathbf{x} = (r,\hat{x})$, and defined:
\begin{eqnarray}
\Phi_{\ell m}(r) & = & \int d \Omega_{\hat{x}} \Phi(\mathbf{x}) Y_{\ell m}(\hat{x}) \; , \\
\alpha_{\ell}(r) & = & \frac{2}{\pi}\int dk k^2 \Delta_{\ell}(k) j_{\ell}(kr) \, .
\end{eqnarray} 
In this expression $j_{\ell}$ is the spherical Bessel function of order $\ell$.
The CMB bispectrum is the three point correlator of the $a_{\ell m}$, so substituting we obtain 
\eq\label{eq:bispectrum1}
B^{\alll}_{\allm} &=& \<a_{\ell_1 m_1} a_{\ell_2 m_2} a_{\ell_3 m_3}\>\\
&=& (4\pi)^3 (-i)^{l_1+l_2+l_3} \int \frac{d^3 k_1}{(2\pi)^3} \frac{d^3 k_2}{(2\pi)^3} \frac{d^3 k_3}{(2\pi)^3} \D_{l_1}(k_1) \D_{l_2}(k_2) \D_{l_3}(k_3) \times\\
&&\qquad\qquad\qquad\qquad \<\O(\bk_1)\O(\bk_2)\O(\bk_3)\> Y_{\ell_1 m_1}(\uk_1)\, Y_{\ell_2 m_2}(\uk_2) \,Y_{\ell_3 m_3}(\uk_3) \\
&=& \(\frac{2}{\pi}\)^3 \int x^2d x \int  d k_1 d k_2 d k_3 (k_1 k_2 k_3)^2 B_\O(k_1,k_2,k_3) \,\D_{\ell_1}(k_1) \D_{\ell_2}(k_2) \D_{\ell_3}(k_3)\\
&&\qquad\qquad\qquad\qquad  \times j_{\ell_1}(k_1 x) j_{\ell_2}(k_2 x) j_{\ell_3}(k_3 x) \int d\W_{\ux} \,Y_{\ell_1 m_1}(\ux) Y_{\ell_2 m_2}(\ux) Y_{\ell_3 m_3}(\ux) \,,
\qe
where in the last line we have integrated over the angular parts of the three ${\bf k}_i$, having inserted the exponential integral form for the delta function in the bispectrum definition (\ref{eq:primbispect}). The last integral over the angular part of $\mathbf{x}$ is known as the Gaunt integral, which can be expressed in terms of Wigner-$3j$ symbols as (for more details on these functions and their properties \citep[see \eg][and references therein]{Komatsu2002}
\begin{align}\label{eq:Gaunt}
\nn \Gaunt &\equiv \int d\W_x Y_{\ell_1 m_1}(\ux) Y_{\ell_2 m_2}(\ux) Y_{\ell_3 m_3}(\ux) \\
&= \sqrt{\frac{(2\ell_1+1)(2\ell_2+1)(2\ell_3+1)}{4\pi}} \( \begin{array}{ccc} \ell_1 & \ell_2 & \ell_3 \\ 0 & 0 & 0 \end{array} \) \( \begin{array}{ccc} \ell_1 & \ell_2 & \ell_3 \\ m_1 & m_2 & m_3 \end{array} \)\,. 
\end{align}
Given that most theories we shall consider are assumed to be isotropic, the $m$-dependence can be factorized out of the physically relevant part of the bispectrum \cite{Luo1994}. It is then usual to work with the angle-averaged bispectrum,
\begin{align}
B_{\ell_1 \ell_2 \ell_3} = \sum_{m_i} \( \begin{array}{ccc} \ell_1 & \ell_2 & \ell_3 \\ m_1 & m_2 & m_3 \end{array} \)\<a_{\ell_1 m_1} a_{\ell_2 m_2} a_{\ell_3 m_3}\>\,.
\end{align}
or the even more convenient reduced bispectrum which removes the geometric factors associated with the Gaunt integral,
\begin{align}\label{eq:Gauntredbispect}
B^{\ell_1 \ell_2 \ell_3}_{m_1 m_2 m_3} = \curl{G}^{\ell_1 \ell_2 \ell_3}_{m_1 m_2 m_3} b_{\ell_1 \ell_2 \ell_3}\,.
\end{align}
From the previous two formulae we also derive the following useful relations between the full, averaged and reduced bispectra:
\be
B_{\alll} = \sqrt{\frac{(2 \ell_1 +1)(2 \ell_2 + 1)(2 \ell_3 +1)}{4 \pi}}
\( \begin{array}{ccc} \ell_1 & \ell_2 & \ell_3 \\ 0 & 0 & 0 \end{array} \) b_{\alll} \;\;\; , \;\;\;
B_{\alll}^{\allm} =  \( \begin{array}{ccc} \ell_1 & \ell_2 & \ell_3 \\ m_1 & m_2 & m_3 \end{array} \) B_{\alll} \; ,
\ee

The reduced bispectrum from (\ref{eq:bispectrum1}) then takes the much simpler form
\begin{align} \label{eq:redbispect}
\nn b_{\ell_1 \ell_2 \ell_3}= \(\frac{2}{\pi}\)^3 \int x^2dx \int & d k_1 d k_2 d k_3\, \( k_1 k_2 k_3\)^2\, B_\O(k_1,k_2,k_3)\\
&\times  \D_{\ell_1}(k_1) \,\D_{\ell_2}(k_2)\, \D_{\ell_3}(k_3)\, j_{\ell_1}(k_1 x)\, j_{\ell_2}(k_2 x)\, j_{\ell_3}(k_3 x)\,.
\end{align}
This is the key equation in this section, since it explicitly relates the primordial bispectrum, predicted by inflationary theories, to the reduced bispectrum observed in the cosmic microwave sky. This formula is entirely analogous to the well known relation linking the primordial curvature power spectrum $P_\Phi(k)$ and the CMB angular power spectrum $C_{\ell}$, that is, 
\be
C_\ell = \frac{2}{\pi} \int dk k^2 P_\Phi(k) \Delta^2_{\ell}(k)\,.
\ee 

Finally, it is important to note that the Gaunt integral in (\ref{eq:Gauntredbispect}) encodes several constraints on the angle 
averaged bispectrum $\Blll$ which are no longer transparent in the 
reduced bispectrum $\blll$.  These are:
\begin{enumerate}
\item The sum of the three multipoles $\ell_i$ must be even.
\item The $\ell_i$'s satisfy the 
triangle condition $|\ell_i-\ell_j| < \ell_k < \ell_i + \ell_j$.
\end{enumerate} 
Analogous to the wavenumber constraint (\ref{eq:tetrapydk}), the second condition 
is tells us that the only multipole configurations 
giving non-zero contributions to the bispectrum are those that form a closed triangle in harmonic ($\ell$-)space.  
For wavenumbers, the triangle condition is enforced through the $x$-integral over the three spherical Bessel functions $j_{\ell}(k_ix)$ which 
evaluates to zero if the $k_i$'s cannot form a triangle, 
whereas in multipole space it is enforced by the angular integration $d\Omega_x$ over the 
spherical harmonics $Y_{\ell_im_i}$ in (\ref{eq:Gaunt}).

\subsection{Separable primordial shapes and CMB bispectrum solutions}\label{sec:separableshapes}

 In terms of the shape function (\ref{eq:shapefn}), the reduced bispectrum (\ref{eq:redbispect}) 
 can be rewritten as
 \begin{eqnarray}\label{eq:redbispect2}
b_{\ell_1 \ell_2 \ell_3} &=& \frac{1}{N}\(\frac{2}{\pi}\)^3 \int  x^2 dx\int d k_1 d k_2 d k_3\,  S(k_1,k_2,k_3)\,
 \D_{\ell_1}(k_1) \D_{\ell_2}(k_2) \D_{\ell_3}(k_3)\, j_{\ell_1}(k_1 x) j_{\ell_2}(k_2 x) j_{\ell_3}(k_3 x).
\end{eqnarray}
The expression above can be simplified and simple analytic solutions can sometimes be obtained for the very important 
class of  separable shapes obeying the ansatz $ S=XYZ$,  as in (\ref{eq:separable}).   Substituting (\ref{eq:separable}) into (\ref{eq:redbispect2}), 
we find that 
\be\label{eq:separated}
b_{\alll} = \int dr r^2 X_{\ell_1}(r) \,Y_{\ell_2}(r) \,Z_{\ell_3}(r) + \mbox{ 5 perms} \; ,
\ee
where we have defined the quantities:
\bea
{{X}}_{\ell}(r) & \equiv & \int dk k^2 \,X(k) \,j_{\ell}(kr)\, \Delta_{\ell} \, , \nonumber \\     
{{Y}}_{\ell}(r) & \equiv & \int dk k^2\, Y(k) \,j_{\ell}(kr) \,\Delta_{\ell} \, ,  \\
{{Z}}_{\ell}(r) & \equiv & \int dk k^2\, Z(k) \,j_{\ell}(kr) \, \Delta_{\ell} \,.\nonumber
\eea
Instead of the three-dimensional integral of \eqn{eq:redbispect2} we now have to deal with a much more tractable product of three one-dimensional integrals. Moreover, if we work at large angular scales in the Sachs-Wolfe approximation, the transfer functions become $\D_l(k) = \frac{1}{3} j_{l}\[(\t_o - \t_{dec}) \, k\]$. The presence of a product of spherical Bessel functions in the integrals above can lead in some cases to simple analytic solutions. 

\begin{figure}[t]
\centering
\includegraphics[width=.48\linewidth]{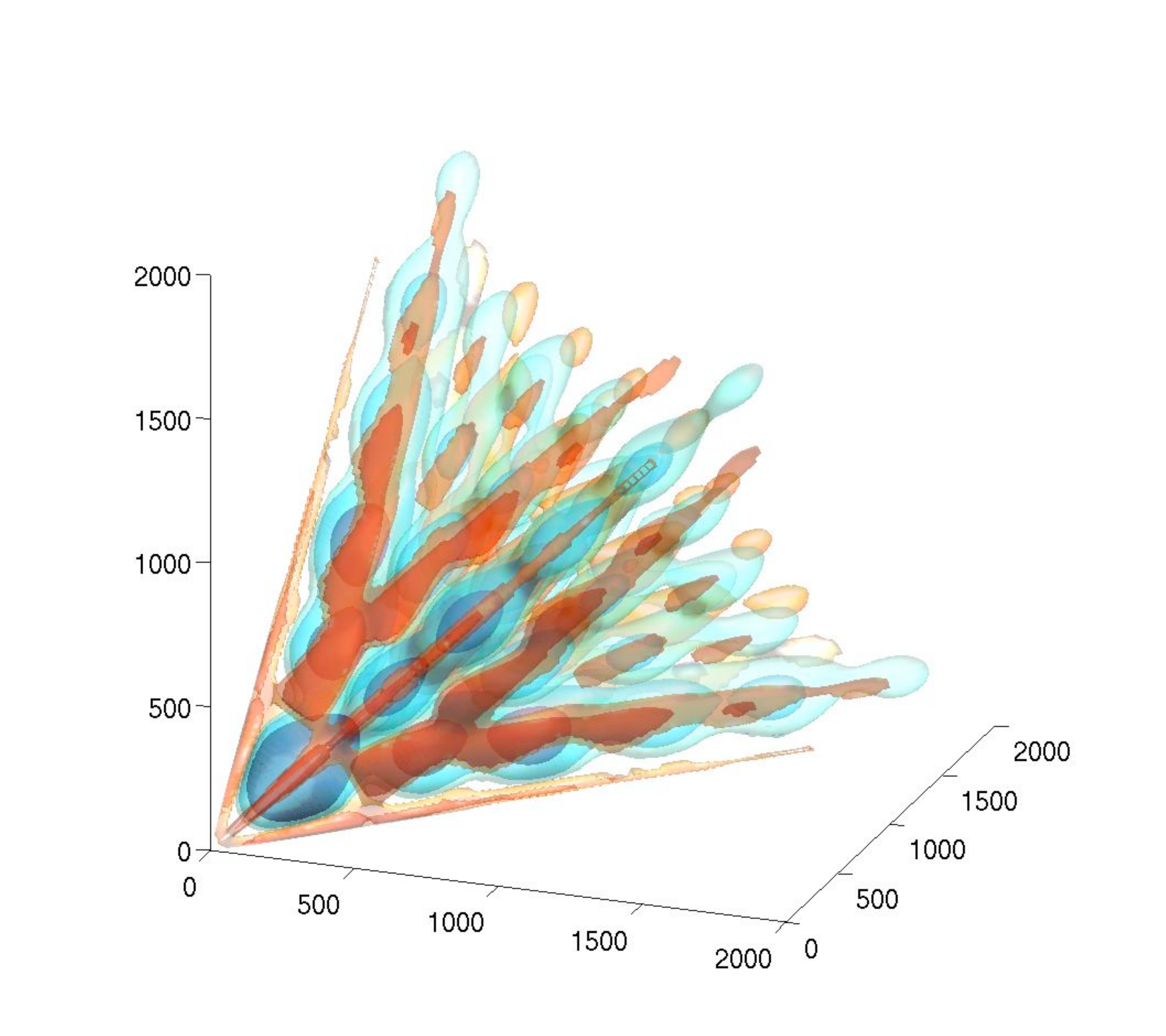}
\includegraphics[width=.48\linewidth]{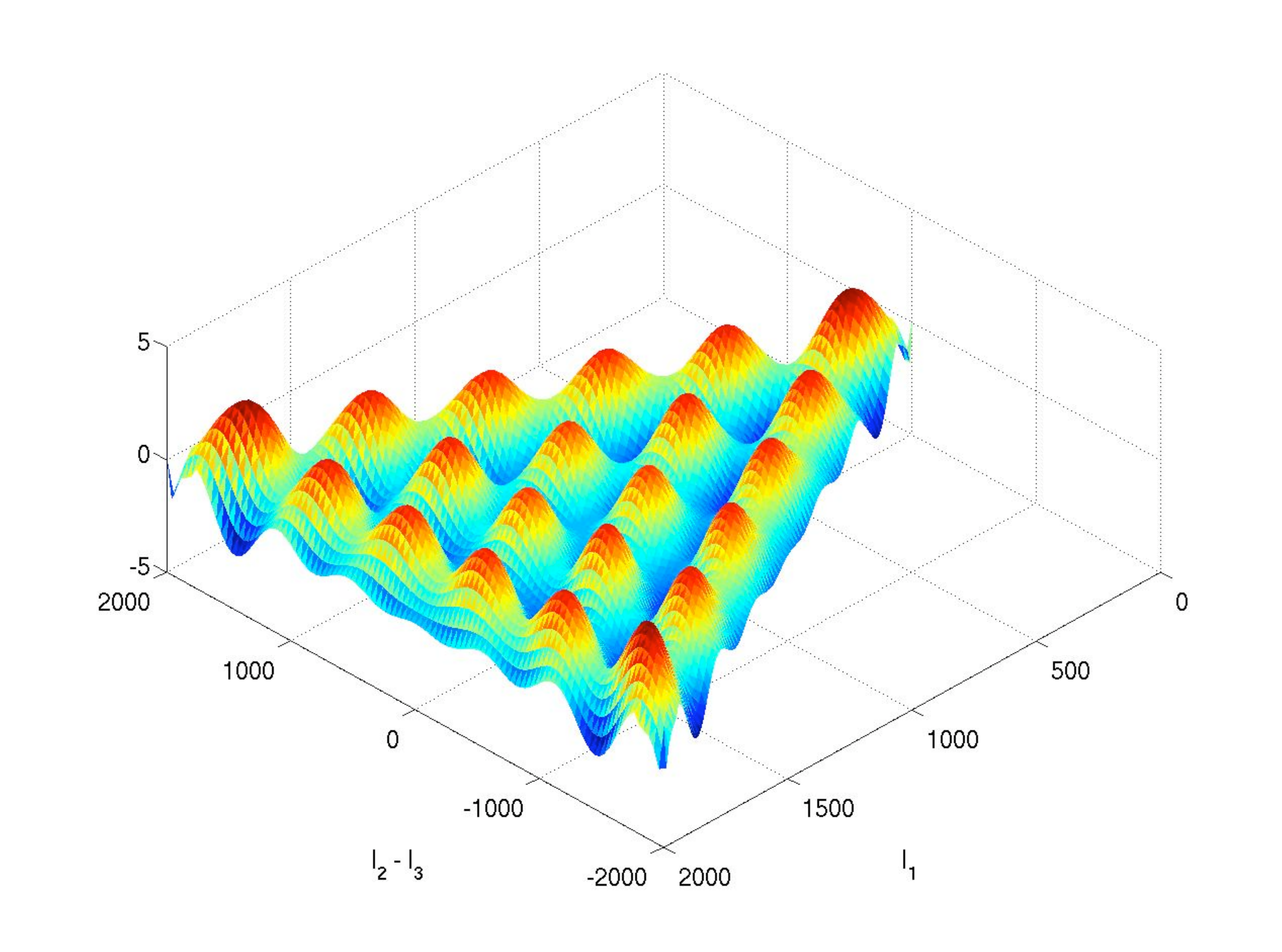}
\caption[Constant model]{\small The reduced CMB bispectrum  for the constant model $\blll^{\rm const}$ normalized relative to the large-angle constant solution \eqn{eq:constbispect}.   On the left, the bispectrum is plotted over the allowed tetrahedral region (see \fig{fig:tetrapydk}) using  several density contours  (light blue positive and magenta negative) out to 
$\ell_i\le 2000$ and, on the right, transverse triangular slices are shown at $\lsum= 2000$.   Note the coherent pattern of acoustic peaks induced by the transfer functions. From \citet{FergussonLiguoriShellard2009}.
}
\label{fig:constant}
\end{figure}

Let us demonstrate this for the separable primordial shapes considered in section \ref{sec:initialcond}. The simplest possible shape, the constant model (\ref{eq:constS}) with $S(\kall) = 1$, has a large-angle analytic solution 
for the reduced bispectrum \cite{FergussonShellard2009}, 
\be\label{eq:constbispect}
b_{\alll}^{const} = \frac{\D^2_\O}{27 N} \frac{1}{(2\ell_1+1)(2\ell_2+1)(2\ell_3+1)}\[\frac{1}{\ell_1+\ell_2+\ell_3+3} + \frac{1}{\ell_1+\ell_2+\ell_3}\]\,, \qquad (l\ll 200)\,. 
\ee
The large-angle solution (\ref{eq:constbispect}) is an important benchmark 
with which to compare the shape of late-time CMB bispectra from other models  $b_{\alll}$ (note the $l^{-4}$ scaling).
The more general constant solution does not have an analytic solution because 
the transfer functions cannot be expressed in a simple form, but  it can be evaluated 
numerically from the expression
\eq\label{eq:constbispectsep}\label{eq:Iterm}
b_{\alll}^{\rm const} = \frac{\D^2_\O}{N} \int  x^2 dx \, {\cal I}_{\ell_1}(x)\, {\cal I}_{\ell_2}(x)\, {\cal I}_{\ell_3}(x)\,,  \qquad \hbox{where} \quad   {\cal I}_\ell(x) = \frac{2}{\pi} \int d k\, \D_\ell(k) \, j_\ell(k x)\,.
\qe
The numerical solution is shown in fig.~\ref{fig:constant}, exhibiting the a regular pattern of acoustic peaks
introduced by the oscillating transfer functions. 

For the local shape (\ref{eq:localS}), the Sachs-Wolfe approximation also yields a large-angle analytic solution
\eq\label{eq:localbispect}
\bllllocal ~&=&~  \frac{2\D^2_\O}{27\pi^2}\(\frac{1}{\ell_1(\ell_1+1)\ell_2(\ell_2+1)} + \frac{1}{\ell_2(\ell_2+1)\ell_3(\ell_3+1)} + \frac{1}{\ell_3(\ell_3+1)\ell_1(\ell_1+1)}\)\,,
\qe
where the divergences for the squeezed triangles ($k_1\ll k_2,k_3 ...$) in the primordial 
shape (\ref{eq:localS}) are also reflected in $b_{\alll}^{local}$.   It is straightforward, in principle, to calculate the full bispectrum from the separable expressions arising from \eqn{eq:localS},
\be\label{eq:localbispectsep}
b_{\alll}^{\rm local}  =  \int x^2 dx \left [  \mbox{\boldmath$\alpha$}_{\ell_1} (x) \mbox{\boldmath$\beta$}_{\ell_2}(x)   \mbox{\boldmath$\beta$}_{\ell_3}(x) + \hbox{2 perm.}\right]\, ,
\ee 
where the separated integrals analogous to \eqn{eq:constbispectsep} become 
\be\label{eq:ABterm}
\mbox{\boldmath{$\alpha$}}_\ell (x) = \frac {2}{\pi} \int dk \, k^2\, \D_\ell(k) \, j_\ell(k x)\,, 
\qquad
\mbox{\boldmath{$ \beta$}}_\ell (x) = \frac {2}{\pi} \int dk \, k^2 P_\Phi(k) \, \D_\ell(k) \, j_\ell(k x)\,. 
\ee
However, we note that these highly oscillatory integrals must be evaluated numerically with considerable care. 

For the equilateral shape \eqn{eq:equilS} we first make its separability explicit by expanding the expression in the form, 
\eq\label{eq:equilS}
S(\kall) =-2 - \left(\frac{k_1^2}{k_2k_3} + \hbox{2 perm.}\right) +\left(\frac{k_1}{k_2} + \hbox{5 perm.}\right)\,.
\qe
While there is no simple large-angle analytic solution known for the equilateral 
model, it can be evaluated from the simplified expression 
\eq\label{eq:equilbispectsep}
b_{\alll}^{\rm equil} =  \int x^2 dx  \,\left[ 2  \mbox{\boldmath$\delta$}_{\ell_1}  \mbox{\boldmath$\delta$}_{\ell_2}   \mbox{\boldmath$\delta$}_{\ell_3} + \left( \mbox{\boldmath$\alpha$}_{\ell_1}  \mbox{\boldmath$\beta$}_{\ell_2}  \mbox{\boldmath$\beta$}_{\ell_3} + \hbox{2 perm.}\right)
+ \left(\mbox{\boldmath$\beta$}_{\ell_1}  \mbox{\boldmath$\gamma$}_{\ell_2}  \mbox{\boldmath$\delta$}_{\ell_3} + \hbox{5 perm.}\right) \right]\,,
\qe 
where $\balpha, \,\bbeta$ are given in (\ref{eq:ABterm}) and $\bgamma,\,\bdelta$ are defined by (compare with the local case)
\eq\label{eq:CDterm}
\bgamma(x) = \frac {2}{\pi} \int dk \, k^2\, P_\Phi(k) ^{1/3} \D_{l}(k) \, j_{l}(k x)\,, \qquad \bdelta(x) = 
\frac {2}{\pi} \int dk \, k^2\, P_\Phi(k) ^{2/3} \D_{l}(k) \, j_{l}(k x)\,.
\qe

\subsection{Non-separable bispectra revisited}
\label{sec:CMBnonseparable}

Recall the mode expansion \eqn{eq:separablebasis} of a general non-separable primordial shape.   If we substitute this into the  expression for the reduced bispectrum \eqn{eq:redbispect2}, then the separability of the expansion leads to the same efficient calculation route discussed in the previous section through \citep{FergussonLiguoriShellard2009}
\bea
\label{eq:sepblllprim}
\nn b_{\alll} &=& {\textstyle\(\frac{2}{\pi}\)^3} \D_\O^2 \fnl\int  x^2 dx\, d k_1 d k_2 d k_3\, \,6 \sum_n \aQn \,\Qn(\kall)
\,\D_{\ell_1}(k_1)\, \D_{\ell_2}(k_2)\, \D_{\ell_3}(k_3)\, j_{\ell_1}(k_1 x) \,j_{\ell_2}(k_2 x)\, j_{\ell_3}(k_3 x) \\
\nn & =& \D_\O^2 \fnl \sum_{n\leftrightarrow {prs}} \kern-6pt\alpha_{prs} \int  x^2 dx 
\left\{
\[{ \frac{2}{\pi} }\int d k_1\,  q_p(k_1) \, \D_{\ell_1}(k_1) \, j_{\ell_1}(k_1x) \]
\[ {\frac{2}{\pi}} \int d k_2 \,  q_r(k_2) \, \D_{\ell_2}(k_2) \, j_{\ell_2}(k_2x) \]
\right.\\
&& \qquad \qquad \qquad \qquad \qquad ~~\left.\times 
\[ {\frac{2}{\pi}} \int d k_3 \,  q_s(k_3) \, \D_{\ell_3}(k_3) \, j_{\ell_3}(k_3x) \] ~+ ~\mbox{5~ perm.}\right\}\\
\nn &=& \D_\O^2 \fnl \sum_{prs} \alpha_{prs} \int x^2dx \, q_{\{p}^{\,\ell_1} q_{r}^{\,\ell_2} q_{s\}}^{\,\ell_3}\,,
\eea
where the $q^{\ell}_p$ simply result from convolving the basis functions $q_p(k)$ with the transfer functions, 
\be\label{eq:qmultipole}
q_p^{\,\ell}(x) = \frac{2}{\pi} \int d k\,  q_p(k) \, \D_{\ell}(k) \, j_{\ell}(kx)\,.
\ee
The computationally costly 3D integrals have again reduced to a sum over products of 1D integrals; we note that this economy arises because the triangle condition is enforced in 
\eqn{eq:sepblllprim} through the product of Bessel functions, resulting in a manifestly separable form in which we can interchange orders of integration. With this mode expansion, all  non-separable theoretical CMB bispectra $\blll$ become efficiently calculable provided there is a convergent expansion for the shape function.

In the same way that we decomposed an arbitrary primordial shape $S(\kall)$ in section~\ref{sec:nonseparableshapes}, it is possible to construct analogous late-time separable basis functions $\barQn$ and orthonormal modes $\barRn$ with which to describe the CMB bispectrum $\b_{\alll}$  \cite{FergussonShellard2007, FergussonLiguoriShellard2009}.    The tetrahedral domain $\Vtetra$ defined by the triangle condition for multipole configurations $\{\lall\}$ is essentially identical to that for wavenumbers (\ref{eq:tetrapydk}), except that only even cases contribute $\sum \lsum= 2n,~ n\in \mathbb{N}$.  However, the appropriate weight function now incorporates Wigner-$3j$ symbols arising from bispectrum products, 
\eq\label{eq:lweightdiscrete}
w_{\alll} =\frac{1}{4\pi}\,{(2\ell_1+1)(2\ell_2+1)(2\ell_3+1)}
\( \begin{array}{ccc} l_1 & l_2 & l_3 \\ 0 & 0 & 0 \end{array} \)^2  \qquad \mbox{and} \qquad w^s_{\alll} =  \frac{w_{\ell_1 \ell_2 \ell_3}} {v_{\ell_1}^2v_{\ell_2}^2v_{\ell_3}^2} \,,
\qe
where in the second expression we have exploited the freedom to divide by a separable function $v_\ell = (2\ell+1)^{1/6}$  and use a weight which makes the bispectrum 
functions more scale-invariant (eliminating an $\ell^{-1/2}$ factor - see below).   The inner product between two functions $ f_{\alll}$ and $ g_{\alll}$ is altered from the primordial wavenumber integral (\ref{eq:innerprod}) into a sum over multipoles on
the tetrahedral domain, that is, 
\eq\label{eq:innerproductl}
\langle f,\,g\rangle ~\equiv~ \sum_{\lall\in\Vtetra } w^s_{\alll}\,  f_{\alll}\,  g_{\alll}\,.
\qe
But for the change in the weight (which only affects configurations near the edges of the tetrahedron), the 1D polynomials $\bar q_p(\ell)$ and 3D separable product basis functions $\barQn(\lall)=  \bar q_{\{p} q_{r}q_{s\}}$ ($n\leftrightarrow \{prs\}$), as well as the resulting orthonormal modes $\barRn$, are nearly identical to their primordial counterparts 
$q_p(k) $, $\Qn(\kall)$ and $\Rn(\kall)$ defined in section~\ref{sec:nonseparableshapes}.

We can now expand an arbitrary CMB bispectrum $\blll$ in both the separable and orthonormal mode expansions, which is achieved in the following form, 
\eq \label{eq:cmbestmodes}
\frac{v_{\ell_1}v_{\ell_2}v_{\ell_3}}{\sqrt{C_{\ell_1}C_{\ell_2}C_{\ell_3}}} \, \blll = \sum_n \baQn \barQn(\lall) = \sum_n \baRn \barRn(\lall)\,,
\qe 
where the variance term $\sqrt{C_\ell C_\ell C_\ell}$ reflects the signal-to-noise weighting expected in the CMB estimator (see section~\ref{sec:cmbestimator}).   Again, the coefficients in the expansions are determined, first, from the orthonormal inner products $\baRn = \langle\barRn,\,\cdot\,\rangle$ and, secondly,  the separable $\baQn$ are found with the transformation matrix analogous to (\ref{eq:orthobasis}).  Examples of the convergence
of these mode expansions for equilateral, DBI and cosmic string CMB bispectra are given in fig.~\ref{fig:convergence}.

\subsection{CMB bispectrum calculations and correlations}
\label{sec:cmbcorrelator}

Prior to the systematic mode expansion approach \eqn{eq:sepblllprim} being implemented, robust hierarchical schemes were developed to calculate any non-separable CMB bispectrum \eqn{eq:redbispect2} directly \citep{FergussonShellard2007, FergussonShellard2009}.   
These use the transverse coordinate system $(\tilde k,\,\tilde \alpha,\,\tilde \beta)$ given in (\ref{eq:transcoords}) and employ adaptive methods on a triangular grid to accurately determine the oscillatory 2D $\alpha\beta$-integrations, with important efficiencies also coming from the flat sky approximation, binning and interpolation schemes.   
Precision to greater than 1\% across the full Planck domain $\ell\le 2000$ was established by  direct comparison with analytic solutions such as \eqn{eq:constbispect} and \eqn{eq:localbispect}.    Examples of nonseparable (and separable) CMB bispectra found using these hierarchical coarse-graining methods are shown in figs~\ref{fig:constant} and \ref{fig:cmbbispectra}. While the CMB bispectra $b_{\alll}$ retain the qualitative features of the primordial shape functions $S(\kall)$, they are overlaid with the oscillatory transfer functions which give rise to a coherent pattern of acoustic peaks.   These direct bispectrum calculations revealed that typical primordial models could be described by eigenmode or other expansions using only a limited number of terms.   

Motivated by the form of the CMB estimator,   we can define the following correlator to determine whether or not two competing theoretical bispectra can be distinguished by an ideal experiment, 
 \begin{align}\label{eq:cmbcor}
\curl{C}(B,B^\pr) ~= ~\frac{1}{N}\sum_{l_i} \frac{B_{\alll}B^\pr_{\alll}}{C_{\ell_1} C_{\ell_2} C_{\ell_3}}~=~\frac{1}{N}\sum_{l_i} w_{\alll}  
\frac{b_{\alll}b^\pr_{\alll}}{C_{\ell_1} C_{\ell_2} C_{\ell_3}}
\end{align}
where  the normalization $N$ is defined as
\begin{align}
N = \sqrt{ \sum_{l_i}\frac{B^2_{\alll}}{C_{\ell_1} C_{\ell_2} C_{\ell_3}}} \sqrt{\sum_{\ell_i} \frac{{B^\pr}^2_{\alll}}{C_{\ell_1} C_{\ell_2} C_{\ell_3}}}\,.
\end{align}
The emergence of the inner product (\ref{eq:innerproductl}) in the expression (\ref{eq:cmbcor}) means that substitution of the mode expansions (\ref{eq:cmbestmodes}) for the theoretical bispectra reduces the correlator to
\eq
\curl{C}(B,B^\pr) = \sum_n \baRn\, \baRn{}'\,.
\qe
While the late time correlator (\ref{eq:cmbcor}) is the best measure of  whether two CMB bispectra are truly independent, it can be demonstrated that for the majority of models the shape correlator (\ref{eq:shapecorrelator}) introduced earlier is sufficient to determine independence.  

On the basis of the direct calculation of the bispectrum results and the CMB correlator, we can now quantitatively
check the forecasting accuracy of the primordial shape correlator proposed previously 
(again closely following the discussion in \citet{FergussonShellard2009}).  

\subsubsection{Nearly scale-invariant models}

\begin{figure}[t]
\centering
\includegraphics[width=.33\textwidth]{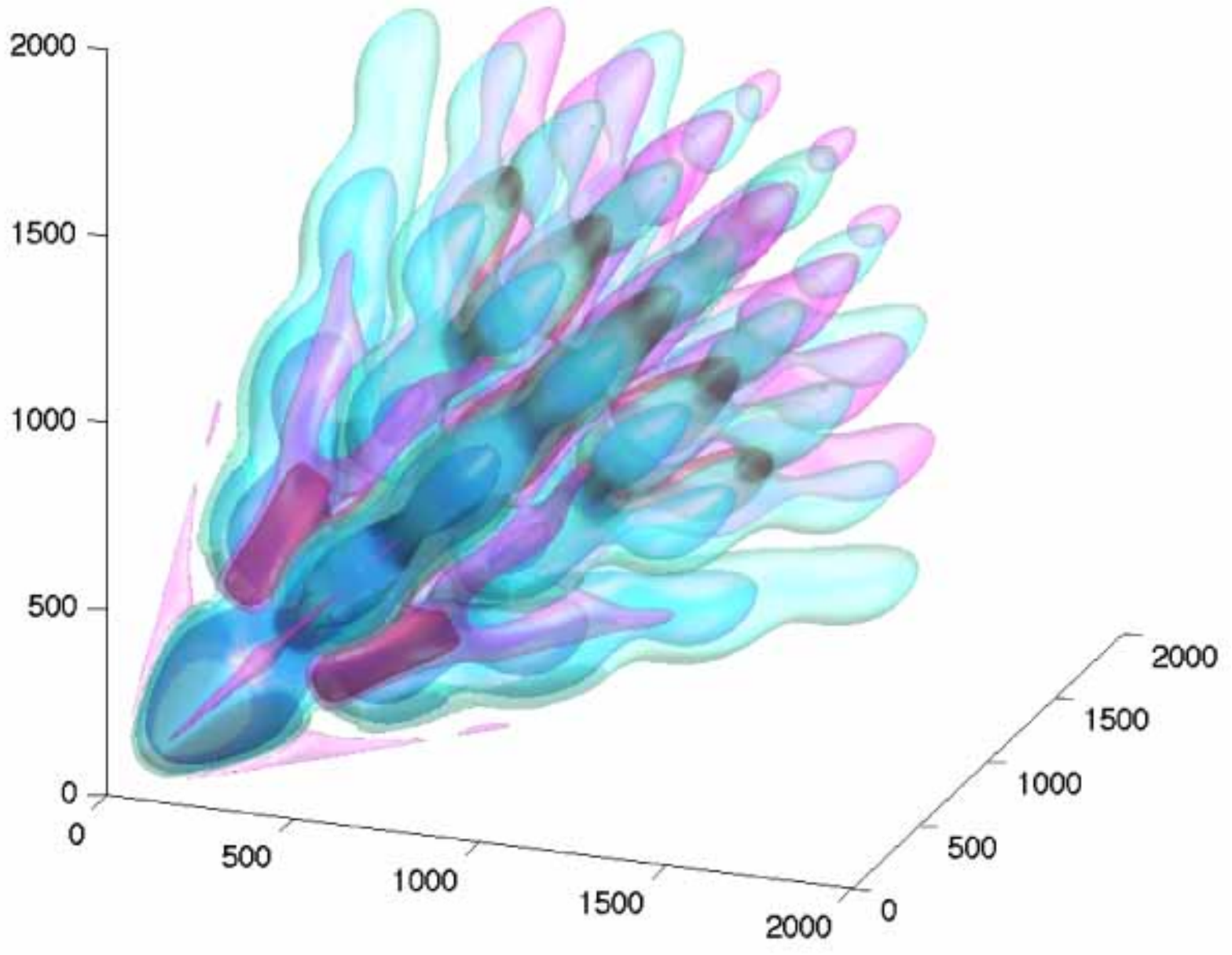}
\includegraphics[width=.33\textwidth]{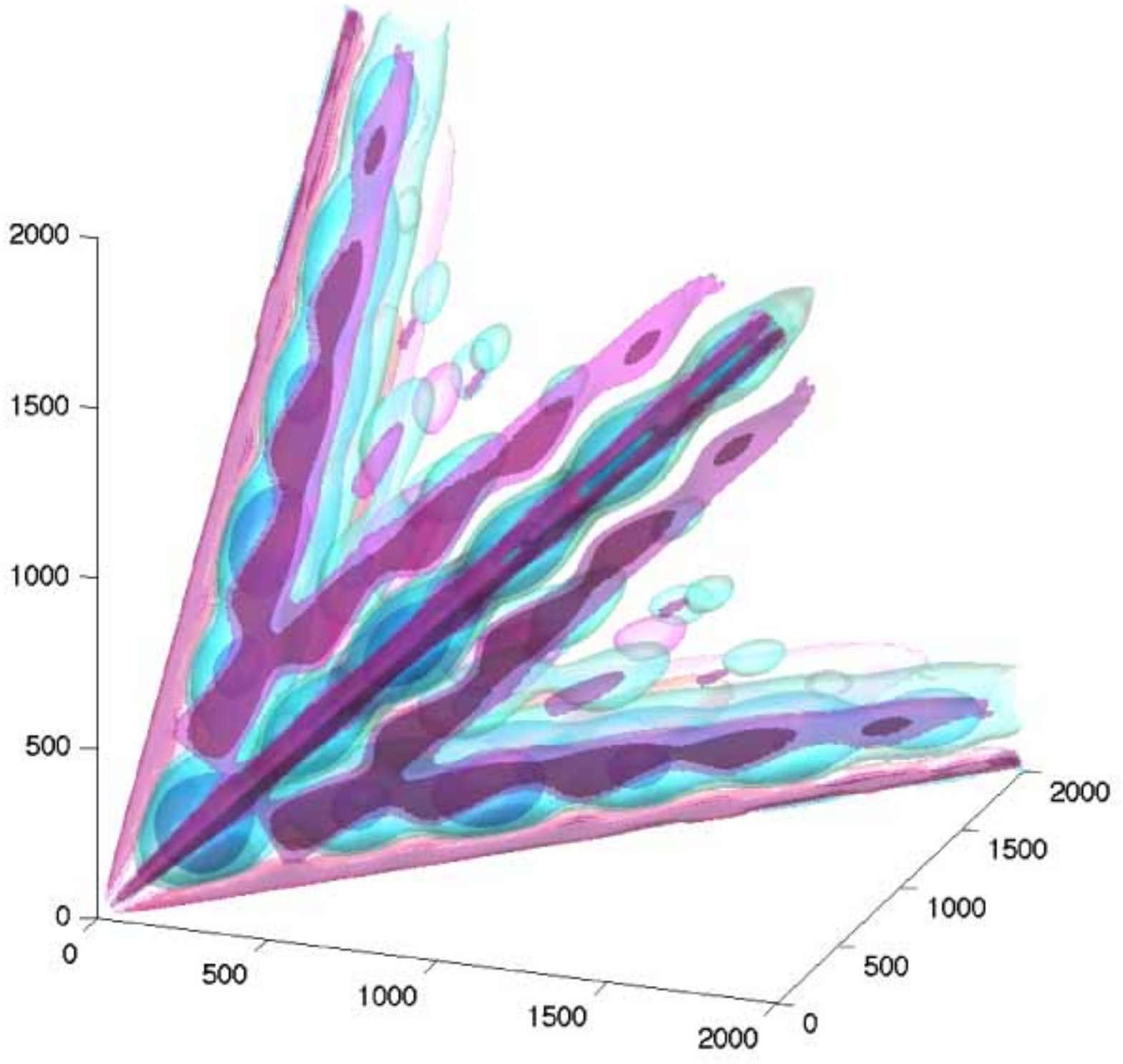}
\includegraphics[width=.33\textwidth]{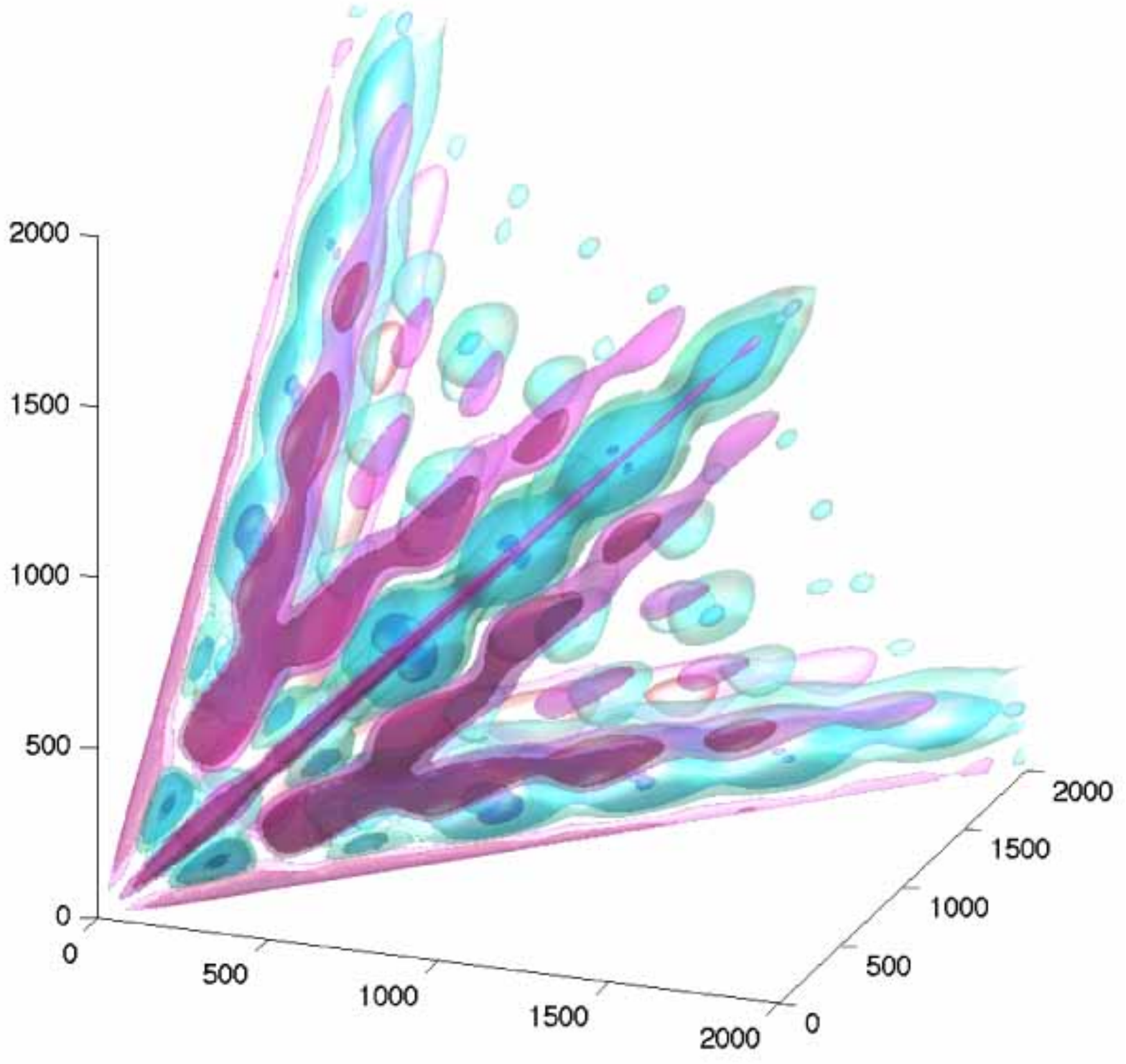}
\includegraphics[width=.33\textwidth]{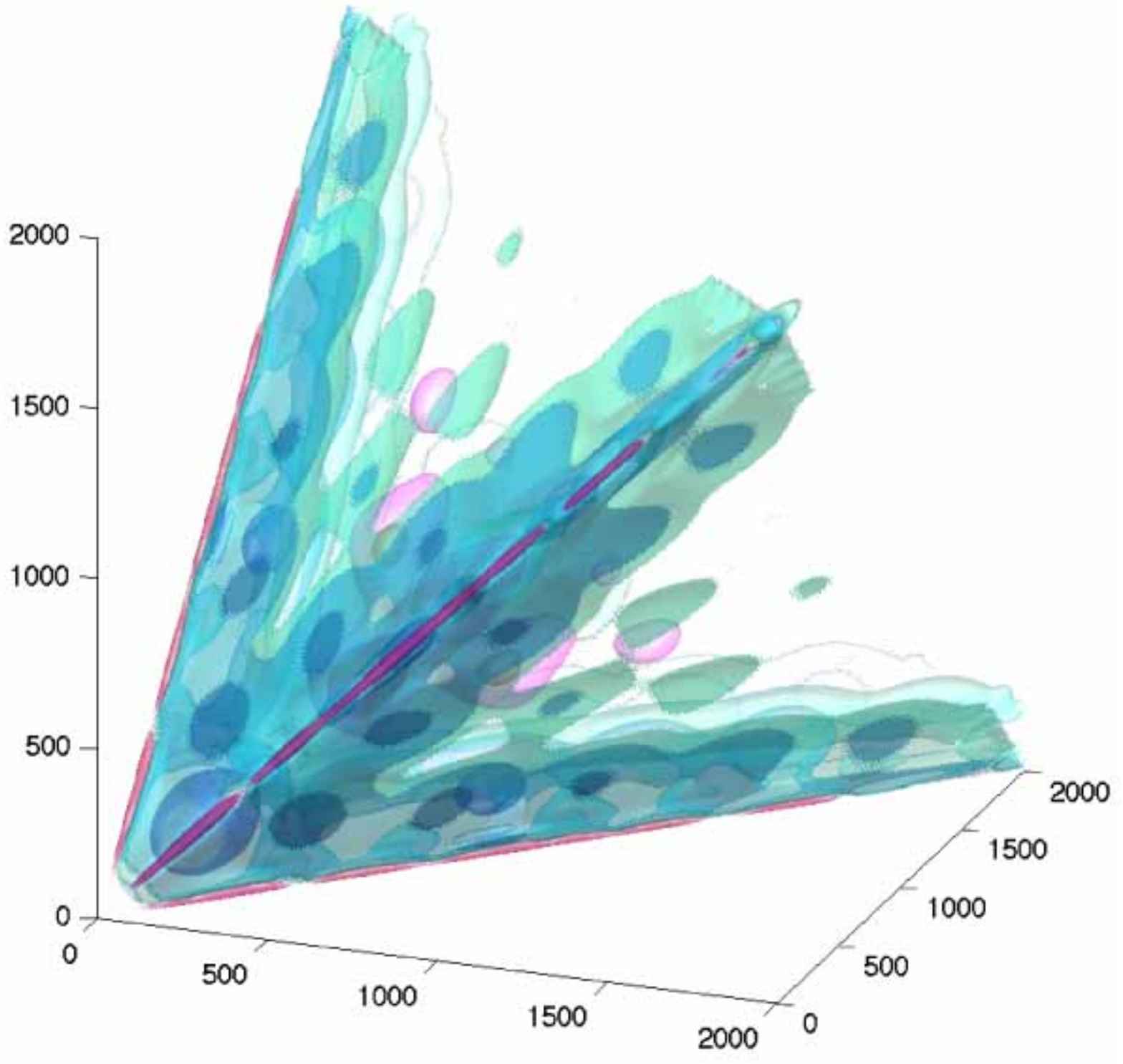} 
\includegraphics[width=.33\textwidth]{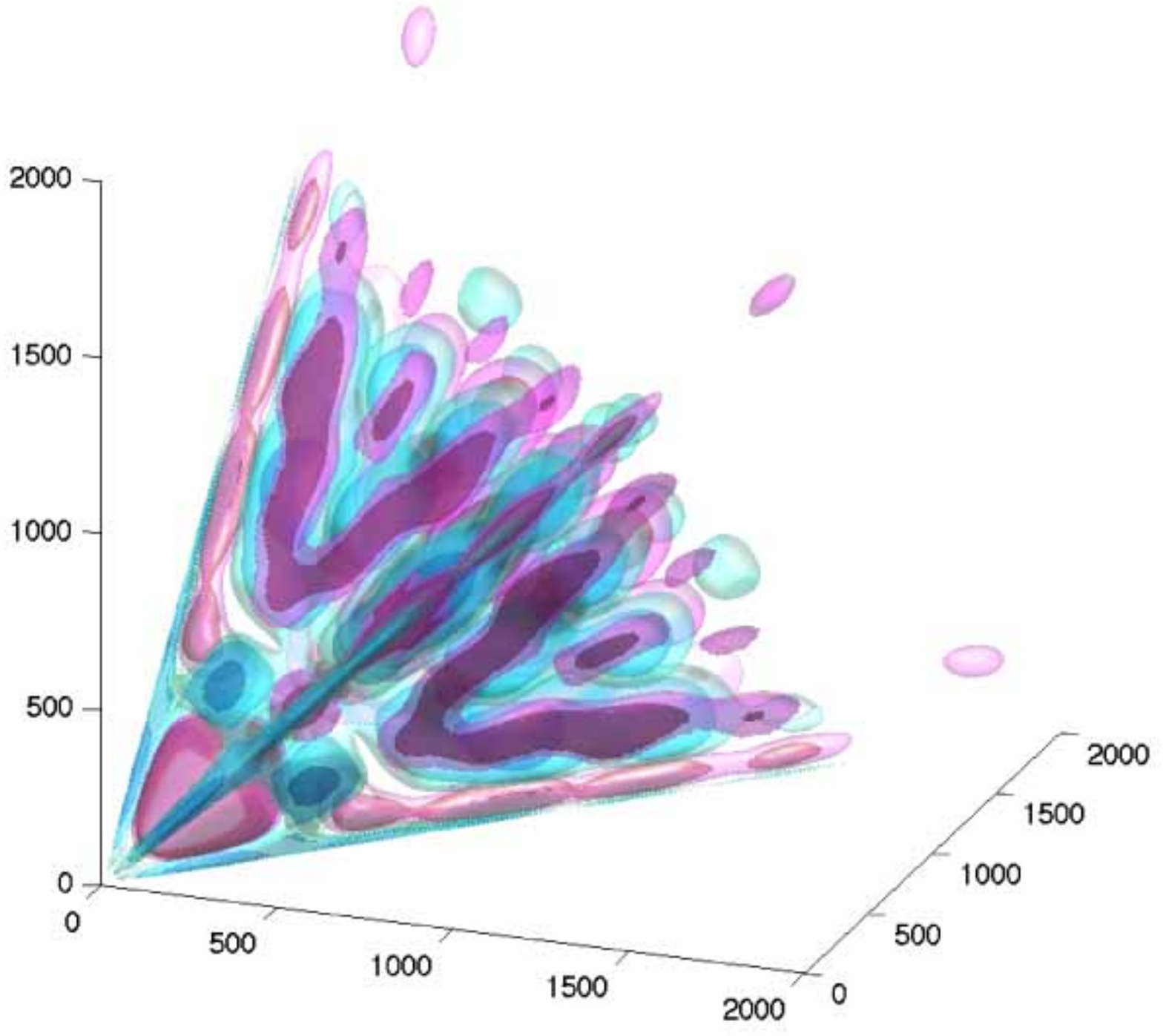} 
\includegraphics[width=.33\textwidth]{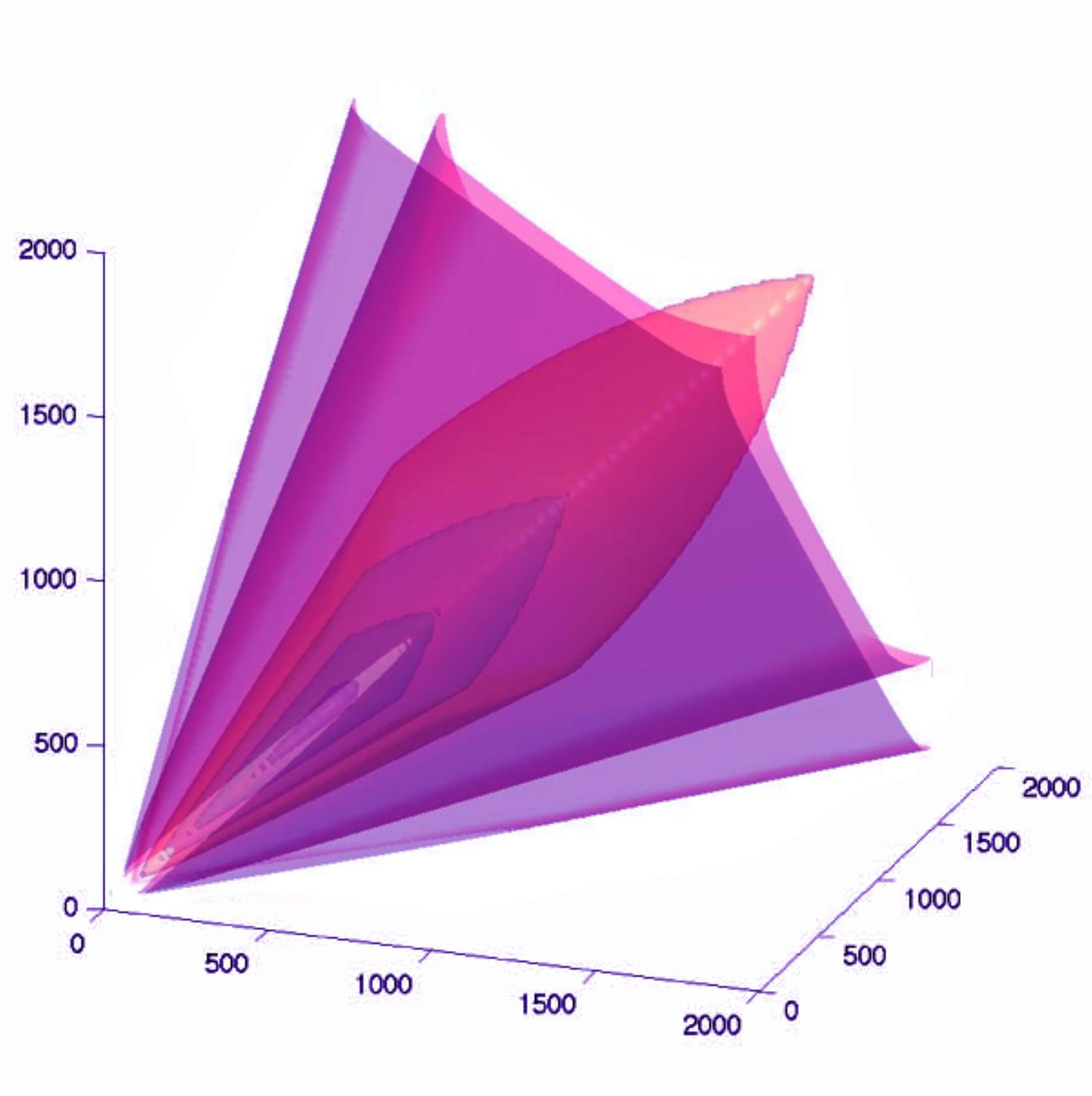} 
\caption[Equilateral and local non-Gaussianity]{The reduced CMB bispectra for several non-Gaussian models, including ({\it top panels, left to right}) equilateral, local, flattened models and ({\it bottom panels}) warm, feature, cosmic string models (see main text).   All five primordial models are normalised relative to the constant solution \eqn{eq:constbispect} and are taken from \citet{FergussonShellard2009}). The analytic cosmic string bispectrum \eqn{eq:stringbispect} is multiplied by $(\alll)^{4/3}$ and is taken from \citet{ReganShellard2009}.}
\label{fig:cmbbispectra}
\end{figure}

For nearly scale-invariant models, the centre values for the bispectrum $b_{lll}$ all have roughly the same profile but with different normalisations. As we see from (\ref{fig:cmbbipsectra}),  the oscillatory properties of the transfer functions for the CMB power spectrum, create a series of acoustic peaks for any combinations involving the following  multipole values, 
 $l = 200, 500, 800, ...$.    Of course, to observe the key differences between the scale invariant models we must study the bispectrum in the plane orthogonal to the $(l,l,l)$-direction, that is, the directions reflecting changes in the primordial shape functions. To plot the bispectrum (see figs~\ref{fig:constant} and \ref{fig:cmbbispectra}), we consistently 
 divide $b_{lll}$ by the large-angle CMB bispectrum solution for the constant 
model (\ref{eq:constS}).   This is analogous to multiplying the power spectrum $C_l$'s by $l(l+1)$, because it serves to remove the overall  $\ell^{-4}$ scaling of the bispectrum, flattening while preserving the  transverse momentum-dependence primordial shape and the effects of the oscillating transfer functions.  

The starting point is the constant model (\ref{eq:constS})
which, despite its apparent simplicity, has  a CMB bispectrum $\blll^{\rm const}$ revealing a non-trivial and coherent pattern of acoustic peaks that we have already noted (see fig.~\ref{fig:constant}).  Given that the constant model has no momentum dependence, we stress that the resulting bispectrum is the three-dimensional analogue of the angular power spectrum $\ell(\ell+1)C_\ell$ for a  scale-invariant model.  The largest (primary) peak, for example, is located where all three $\ell_i = 220$ (corresponding to the large blue region near the origin). We can interpret fig.~\ref{fig:constant}, therefore, as the pure window function or beam effect of convolving any model with the radiation transfer functions $\Delta_\ell(k)$ while transforming from Fourier to harmonic space.   

 The CMB bispectrum for the equilateral model is plotted in fig.~\ref{fig:cmbbispectra}, showing how the the centre-weighting from the primordial shape is well-preserved despite the convolution with the oscillating transfer functions.   For the full CMB correlator (\ref{eq:cmbcor}), the DBI, ghost and single shapes are generally even more closely correlated with equilateral, presumably because distinctive features are `washed out' by the transformation from Fourier to harmonic space. Comparative results between the shape correlator and Fisher matrix analysis are given in Table \ref{tb:shapecorrelator},
 establishing that these models are highly correlated and difficult to set apart observationally.  

The CMB bispectrum for the local model is also shown in \fig{fig:cmbbispectra}, demonstrating a marked contrast with 
equilateral which reflects their different primordial shapes shown in \fig{fig:shapes}. The dominance of the signal in the squeezed limit creates strong parallel ridges of acoustic peaks which connect up and emanate along the corner edges of the tetrahedron\citep[see][for further details]{BucherVanTentCarvalho2009}. The 51\% CMB correlation between the local and equilateral models is underestimated by the shape correlator at 41\%, presumably because of effective smoothing 
due to the harmonic analysis.     Reflecting their distinctive primordial properties, the CMB bispectra for the flat and
warm models are poorly correlated with most of the other models, though the flat shape could be susceptible to confusion with the local CMB bispectrum with which it has a larger correlation (see table~\ref{tb:correlatorcomparison}).    
It is clear that the local, equilateral, warm and flat shapes form four distinguishable categories among the 
scale-invariant models.

\subsubsection{Scale-dependent models, cosmic strings and other late-time phenomena}

Models which have a non-trivial scaling, such as the feature models, can have starkly constrasting 
bispectra as illustrated in fig.~\ref{cmbbispectra}.     For example, instead of having the same 
pattern of acoustic peaks which characterise the scale-invariant models, the feature model can become
entirely anticorrelated so that the primary peak has the opposite sign.   Later, for this particular choice 
of $k^*$ in (\ref{eq:feature})),  for increasing $l$ the phase of the oscillations becomes positively 
correlated by the second and third peaks.   This can lead to small correlation with the other 
primoridial shapes, all below 45\% as shown in table~\ref{tb:correlatorcomparison} for this 
$k^*$ and $\lmax$.   Clearly, these non-separable feature models form a distinct fifth category 
beyond the four scale-invariant shapes noted above and, of course, there are many possible 
model dependencies which can lead to further subdivision. 

By way of further illustration of the breadth of other possible non separable CMB bispectra,  
we present the late-time CMB bispectrum predicted analytically for cosmic strings \cite{ReganShellard2009} 
\begin{align}\label{eq:stringbispect}
\blll^{\rm string}  = \frac{A}{(\z \alll)^2} \[(\ell_3^2 - \ell_1^2 - \ell_2^2)\(\frac{L}{2\ell_3} + \frac{ \ell_3}{50L}\)\sqrt{\frac{\ell_*}{500}}\,\mbox{erf}(0.3 \z \ell_3) ~+~ \mbox{2 perm.}\] , \qquad (\ell \le 2000)\,,
\end{align}
where $\ell_{min} = \min(\lall)$,  $\ell_* = \min(500,\ell_{min})$, $\z = \min(1/500,1/\ell_{min})$ and 
\begin{align}
L = \z \sqrt{{\textstyle\frac{1}{2}}({\ell_1^2 \ell_2^2 + \ell_2^2 \ell_3^2 + \ell_3^2 \ell_1^2} )- {\textstyle\frac{1}{4}}{(\ell_1^4+\ell_2^4+\ell_3^4)}}\,.
\end{align}
Here, $A\sim (8\pi G\mu)^3$ is a model dependent amplitude with $G\mu = \mu/m_{\rm Pl}^2$ measuring the string tension $\mu$ relative to the Planck scale. The cutoffs around $\ell\approx 500$ in \eqn{eq:stringbispect} are associated with the string correlation length at decoupling (perturbations with $\ell\gtrsim500$ can only be causally seeded after 
last scattering).  Here, the non-separable nature and very different scaling  of the string CMB bispectrum are clear from a comparison with \eqn{eq:localbispect}.  Moreover, given the late-time origin of this signal from string metric perturbations,  the modulating effect of acoustic peaks from the transfer functions is absent, as is clear from \fig{fig:cmbbispectra}.  This is just one example of late-time phenomena such as gravitational lensing,  secondary anisotropies and contaminants which are accessible to analysis using the more general CMB mode expansions \eqn{eq:cmbestmodes}.

\subsection{The  estimation of $\fnl$ from CMB bispectra}
\label{sec:cmbestimator}

In light of the previous discussion, it is  evident how measurements of the bispectrum from CMB experimental data-sets are able to provide information about the primordial three point function of the cosmological curvature perturbation field at the end of inflation. This in turn allows us to put significant constraints on inflationary models, or on alternative models for the generation of cosmological perturbations. We will now start dealing with the problem of bispectrum estimation in the CMB, as a test of primordial non-Gaussianity.

Let us assume that we have measured the three point function of a given CMB dataset. There are now two general ways to exploit this information:

\begin{enumerate}

\item {\em Tests of the Gaussian hypothesis}. By comparing the measured three point function to its expected distribution obtained from Gaussian simulations we can detect if some configurations present a significant deviation from Gaussian expectations.
The issue with this approach is that it is sensitive not only to primordial non-Gaussianity, but also to any other possible source of NG, including those of non-cosmological origin. Original bispectrum tests of this kind on COBE maps \cite{FerreiraMagueijoGorski1998} revealed significant deviations from Gaussianity in the data. This NG signature in the three point function seemed to be localized in harmonic space around multipoles $\ell = 16$ and was object of much scrutiny  \citep[\eg][]{BromleyTegmark1999, MagueijoFerreiraGorski1999, MagueijoFerreiraGorski2000,   Magueijo2000, Magueijo2000Err}. It was then finally ascertained that the detected signal was not cosmological in origin, but due to a systematic artifact \citep{BandayZaroubiGorski2000}. Moreover, the overall statistical significance of the result disappeared in a later analyisis involving the measurement of all the bispectrum modes available in the map \citep{KomatsuEtal2002} (only a subset of all the configurations had been studied before).
General tests of Gaussianity are very useful to identify unexpected effects in the data, and to monitor systematics. However, as long as we are interested in a {\em primordial} NG signal, it is better to follow the approach of making an ansatz for the bispectrum we expect from the theory under study and obtain a {\em quantitative} constraint on a given model. This approach is outlined in the point that follows.

\item  {\em $\fnl$ estimation}. In this case we choose the primordial model that we want to test, characterizing it through its bispectrum shape. We then estimate the corresponding amplitude $\fnlmodel$ from the data. If the final estimate is consistent wih $\fnlmodel = 0$, we conclude that no siginificant detection of the given shape is produced by the data,
 but we still determine important constraints on the allowed range of $\fnlmodel$. Note that ideally we would like to do more than just constrain the overall amplitude, and reconstruct the entire shape from the data by measuring single configurations of the bispectrum. However, the expected primordial signal is too small to allow the signal from a single bispectrum triangle to emerge over the noise. For this reason we study the cumulative signal from all the configurations that are sensitive to $\fnlmodel$.

\end{enumerate}

Since in this review we are concerned with the study of the primordial bispectrum, we will take the latter approach, and deal with the problem of $\fnl$ estimation from measurements of the bispectrum in CMB maps. We will first present a cubic estimator that optimally extracts the $\fnl$ information from the data contained in the bispectrum (section~\ref{sec:cubicest}). We will then address the issue of understanding if this optimal cubic statistic extracts all the possible information available on $\fnl$ {\em in the data}, or if there is enough additional information beyond the three point function to allow more precise $\fnl$ measurements using non bispectrum-based estimators of $\fnl$ (section~\ref{sec:optimality}). We will then discuss concrete numerical implementations of bispectrum estimators (section~\ref{sec:numericalimplementation}) and review the experimental constraints on $\fnl$ obtained from bispectrum analysis of WMAP data (section~\ref{sec:constraints}). Using a standard Fisher matrix analysis, forecasts on the $\fnl$ error bars are achievable for future CMB surveys (section~\ref{sec:forecasts}). 
Following, we will study the NG signals in the map that could contaminate the primordial NG measurement and how they are dealt with when analyzing the data. Finally we will describe algorithms for the simulation of primordial NG CMB maps that are useful for testing and validation of estimators before applying them to real data.

In the following, we assume that the reader is familiar with essential concepts in statistical estimation theory, such as the definition of a statistical estimator, the role played by maximum likelihood estimators in statistics, the definitions of unbiasedness and optimality, and the definition and main applications of the Fisher information matrix. The reader unfamiliar with these concepts can consult the appendix of this review and references therein.

\subsubsection{Bispectrum estimator of $\fnl$}
\label{sec:cubicest}

In this section we are concerned with the statistical inference of $\fnl$ from measurements of the bispectrum of the CMB anisotropies. We recall that we defined $\fnl$ earlier as 
the amplitude of the bispectrum of the primordial potential. In principle, we can include both temperature and polarization multipoles $\alm^{T,E}$ in the analysis, in order to maximize the available data. However, for clarity we will consider only temperature multipoles in the following, and omit the superscript $T$ in $\alm$, for simplicity of notation. 
The extension to polarization is conceptually straightforward, and will be discussed in a following paragraph. We will start by considering a simple cubic\footnote{The estimator is dubbed cubic due to the fact that it contains the third power of the random variable $\alm$.} statistic written in the form:
\be\label{eq:purecubicestimator}
\fnlest = \frac{1}{{\cal{N}}} \sumalllm W_{\alll}^{\allm} \almone \almtwo \almthree \; .
\ee
In the previous equation $\fnlest$ represents the statistical estimate of $\fnl$ from the data, $\alm$ are the multipoles of the observed CMB temperature fluctuations, $W_{\alll}^{\allm}$ are some weight functions, and ${\cal{N}}$ is a normalization factor that has to be chosen to make the estimator {\em unbiased} \ie to ensure that:
\be
\langle \fnlest \rangle = \fnl \; .
\ee

We now want to find the weights $W_{\alll}^{\allm}$ that provide the best estimator
(\ie the minimum error bar estimator) {\em within the class of cubic statistics} written in the form 
(\ref{eq:purecubicestimator}). It is a well known result (see Appendix A) that the
 best unbiased estimator of a parameter 
from a given data-set is the maximum-likelihood estimator. In order to answer our question we then have to write 
the bispectrum likelihood as a function of the parameter $\fnl$, and maximize with respect to $\fnl$. 

In the assumption that the bispectrum configurations are characterized by a Gaussian 
distribution\footnote{This is not strictly true, but it is a good approximation. The same approach applies 
to most cosmological observables}, maximizing the likelihood is equivalent to minimizing the following $\chi^2$:
\be\label{eq:chi2}
\chi^2 = \sum_{\alll}  \frac{\( \fnl B^{\fnl=1}_{\alll} - B^{obs}_{\alll} \)^2}{\sigma^2} \; ,
\ee
where $B^{obs}_{\alll}$ is the observed angular averaged bispectrum \ie, by definition:
\begin{equation} 
B^{obs}_{\alll} = \sum_{\allm} \( \begin{array}{c c c} \ell_1 \; \ell_2 \;  \ell_3 \\ m_1 m_2 m_3 \end{array} \) 
\almone^{obs} \almtwo^{obs} \almthree^{obs} \; ,
\end{equation}
and $\sigma^2$ is the bispectrum variance \ie the $\alm$ six-point function: 
\begin{equation}\label{eq:sigma2}
\sigma^2 =  \langle \almone \almtwo \almthree \almfour \almfive \almsix \rangle \; .
\end{equation} 
We will now make the assumption that we are working in the {\em weak non-Gaussian limit} \ie $\fnl$ is small and the distribution of $\alm$ can be approximated as Gaussian in the calculation of the variance. The implications of this approximation will be discussed in greater detail in the following sections; for the moment it will suffice to point out that the weak non-Gaussian approximation is generally a good one since most inflationary models predict $\fnl$ to be small, and because the level of primordial non-Gaussianity is already constrained to be small by WMAP measurements \cite{KomatsuEtal2009B, SmithSenatoreZaldarriaga2009}. After restricting indeces so that $\ell_1 \leq \ell_2 \leq \ell_3$ and $\ell_4 \leq \ell_5 \leq \ell_6$, the six point function above can be calculated using Wick's theorem, yielding \citep{Luo1994}:
\begin{equation}\label{eq:sixpoint}
\langle \almone \almtwo \almthree \almfour \almfive \almsix \rangle = \Delta \, \bispvar 
\d_{\ell_1}^{\ell_4} \d_{\ell_2}^{\ell_5} \d_{\ell_3}^{\ell_6} \d_{m_1}^{m_4} \d_{m_2}^{m_5} \d_{m_3}^{m_6}    \; .
\end{equation}
In the last formula $\Delta$ is a permutation factor that takes the value of $1$ when all $\ell$'s are different, $2$ when two $\ell$'s are equal and $6$ when all $\ell$'s are equal.
We can now substitute (\ref{eq:sixpoint}) into (\ref{eq:chi2}), and differentiate with respect to $\fnl$ to get an explicit expression for the optimal cubic statistic we were looking for:
\bea
\label{eq:optimalcubicestnorm}
 \fnlest & = & \frac{1}{{\cal{N}}} \sumalllm \frac{{\cal{G}}_{\alll}^{\allm} b_{\alll}^{\fnl = 1}}{\bispvar} 
               \almone \almtwo \almthree  \\
 {\cal{N}} & = & \sumalllm \frac{ \({\cal{G}}_{\alll}^{\allm} b_{\alll}^{\fnl=1}\)^2}{\bispvar} \; ,
\eea
where $b_{\alll}$ is the reduced bispectrum and $\Gaunt$ is the Gaunt integral defined by equation (\ref{eq:Gaunt}); ${\cal{N}}$ is the normalization factor mentioned at the beginning of the paragraph, that guarantees the unbiasedness of the estimator. 

Note that the noise and window function of the experiment are included in the $C_l$ and $b_{\alll}$ that appear in the formula above, with the following replacements:
\be
C_\ell \rightarrow C_\ell {\cal{W}}_\ell^2 + N_\ell \; ,  \;\;\;\;\;\;\;\;\;
b_{\alll} \rightarrow b_{\alll} {\cal{W}}_{\ell_1} {\cal{W}}_{\ell_2} {\cal{W}}_{\ell_3} \; ; 
\ee
${\cal{W}}$ is the window function (not to be confused with the weights $W$) and 
$N_{\ell}$ is the noise power spectrum (constant for uncorrelated white noise). The noise is assumed to be Gaussian, thus characterized by a vanishing three point function. Comparing our result (\ref{eq:optimalcubicestnorm}) to the initial ansatz (\ref{eq:purecubicestimator}), we then see that the optimal weights are:
\be
W_{\alll}^{\allm} = \frac{\Gaunt b_{\alll}}{\bispvar} \; ;
\ee
in other words we are weighting the observed bispectrum by its expected signal-to-noise ratio.

We have now constructed a statistic that optimally extracts the information about $\fnl$ from the bispectrum of the map. The question now is: is there additional information about $\fnl$ in the map that is not contained in the bispectrum? This issue will be investigated in the following sections. For the impatient reader we anticipate that the answer is no: the bispectrum statistic built here is actually the minimum error bar estimator of $\fnl$ from CMB data.

\subsubsection{Optimality of the cubic estimator}
\label{sec:optimality}

In this section we address the issue of whether the cubic statistic \eqn{eq:optimalcubicestnorm} optimally extracts  {\em all} the $\fnl$ information contained in the $\alm$, or if other statistical estimators (\eg four-point function, 
or pixel space statistics such as the Minkowski functionals, or again wavelet estimators, just to make a few among many possible examples) are able to produce smaller error bars and are thus more efficient than the bispectrum.

In a non-Gaussian primordial CMB map, the $\alm$ likelihood depends on the NG parameter $\fnl$. We will indicate it with $p(\mathbf{a}|\fnl)$, where $\mathbf{a}$ indicates a vector including all the $\alm$'s (we will assume all other cosmological parameters are fixed, and concentrate on $\fnl$). It is a well known result in parameter estimation theory that there is a {\em lower limit} on the error bars that can be assigned to a given parameter (in our case $\fnl$). Such lower limit, also known as the {\em Rao-Cramer bound}, is defined in term of the {\em Fisher matrix} $F$ as\footnote{We again refer the reader unfamiliar with these concepts to the brief summary provided in appendix 
\ref{sec:estimationbasics}}:
\be
\Delta \fnl \geq \frac{1}{\sqrt{F_{\fnl \fnl}}} \; .
\ee
We remind the reader that the Fisher matrix is defined as:
\be
F_{\fnl \fnl}(\mathbf{a}) = \left\langle  \frac{ \partial^2 \ln p(\mathbf{a}|\fnl) }{\partial^2 \fnl}  \right\rangle\; .
\ee
If we can show that the bispectrum estimator of the previous section saturates the Rao-Cramer bound for the $\alm$ Fisher matrix above, then we conclude that it provides the best (\ie minimum variance) estimate of $\fnl$ {\em from the data}, rather than just the best $\fnl$ estimate {\em from the bispectrum} of the data. In other words, no more information about $\fnl$ could be extracted from the $\alm$ than the information contained in the bispectrum. The aim of this section is to show that this is actually the case.

The issue of the optimality of bispectrum estimators of $\fnl$ was addressed in great detail in \citet{Babich2005}. In this section we will basically review the main results of that study, referring the reader to the original paper for their complete derivation.

As we mention in appendix \ref{sec:estimationbasics}, there is a sufficient and necessary condition for an estimator ${\curl{E}}$ to saturate the Rao-Cramer bound, expressed by formula \eqn{eq:optimalitycondition}. This condition, applied to our case, reads:
\be
\label{eq:RCfnl}
\frac{\partial{\ln p({\mathbf{a}}|\fnl)}}{\partial \fnl} = F_{\fnl \fnl} ({\curl{E}}({\mathbf{a}}) - \fnl) \; .
\ee
Our aim is to show that the bispectrum statistic (\ref{eq:optimalcubicestnorm}) satisfies this condition. We then need to start from a computation of the full likelihood $p(\mathbf{a}|\fnl)$ for a general primordial non-Gaussian model. Following \citet{Babich2005} 
we will  start by limiting ourselves to the particular case of the local model. 

We recall from a previous section that local NG is the only case for which an explicit expression for the primordial potential is provided. In real space:
\be
\Phi(\mathbf{x}) = \Phi_L(\mathbf{x}) + \fnll \[ \Phi^2_L(\mathbf{x}) - \left \langle \Phi^2_L(\mathbf{x}) \right \rangle \] \; .
\ee
Starting from this formula it is possible to obtain a likelihood function for $\Phi$, dependent on the parameter $\fnll$. This is done by means of an {\em expansion} in terms of the order parameter $\fnll  \langle \Phi^2_L(\mathbf{x})\rangle$. 
The full expression for the Probability Density Function (PDF) $P(\Phi|\fnll)$ \citep[see][]{Babich2005} can be expanded around its Gaussian expectation for $\fnll = 0$, and schematically written as:
\be\label{eq:likelihoodphi}
\ln P(\Phi|\fnll) = \ln P_G (\Phi|C) + \fnll \ln P_{NG} (\Phi|C) + {{\cal{O}}} \(\fnl^2 \left \langle \Phi^2_L(\mathbf{x}) \right \rangle^2 \) \; ,
\ee
where $C$ is the covariance matrix of the Gaussian part of the potential $\Phi$ \ie
\be
C \equiv \left \langle \Phi_L(\mathbf{x_1}) \Phi_L(\mathbf{x_2}) \right \rangle \; .
\ee
Formula (\ref{eq:likelihoodphi}) is then telling us that the logarithm of the full likelihood can be decomposed into the sum of a Gaussian likelihood $P_G$, plus a NG term that depends linearly on $\fnll$, and that this decomposition is accurate up to terms of 
order ${{\cal{O}}} \(\fnl^2 \left \langle \Phi^2_L(\mathbf{x}) \right \rangle^2 \)$, \ie\ {\em we are assuming  that NG is weak}, as we did in the previous section.

After computing $P(\Phi|\fnll)$, one has to account for 2D projection and radiative transfer in order to obtain the required likelihood $P(\mathbf{a}|\fnll)$. As shown in \citet{Babich2005}, this can be achieved by expanding the PDF \ref{eq:likelihoodphi} in spherical harmonics and performing the functional integration:
\be\label{eq:functionalint}
P(\mathbf{a} | \fnll) = \int d^N \, \Phi \, \d_D^{(M)} \[\alm - \int dr r^2 \alpha_\ell(r) \Phi_{\ell m}(r) \] P(\Phi|\fnll) \; ,
\ee 
where $\d_D^{(M)}$ is the Dirac delta function of dimension M, and $M<N$ due to the 
2D projection\footnote{As noted in \citet{Babich2005} the additional degrees of freedom do not affect the CMB anisotropies and can therefore be integrated out.} The previous formula can be derived by recalling equation \eqn{eq:phix2alm}, together with the well known formula in probability theory:
\be
P(\mathbf{y}) = \int d\mathbf{x} P(\mathbf{x}) \d_D\(\mathbf{y}-\mathbf{F}(\mathbf{x})\) \; ,
\ee
where $\d_D$ is again the Dirac delta function, $\mathbf{x}$ and $\mathbf{y}$ are random variables linked by the functional relation 
$\mathbf{y} = \mathbf{F}(\mathbf{x})$, and $P(\mathbf{x})$, $P(\mathbf{y})$ are the PDFs of $\mathbf{x}$ and $\mathbf{y}$ respectively.
Solving the functional integral (\ref{eq:functionalint}) yields \cite{Babich2005}:
\be\label{eq:localCMBlikelihood}
\ln P(\mathbf{a}|\fnll) = -\frac{1}{2} \sum_{\ell m} \frac{\alm^* \alm}{C_\ell} + 
\fnll \sumalllm \frac{ {\cal{G}}_{\alll}^{\allm} 
b^{\fnll=1}_{\alll}}{C_{\ell_1} C_{\ell_2} C_{\ell_3} } \almone \almtwo \almthree + I_2(\mathbf{a},\fnl) + 
 {\cal{O}} \( \fnl^3 \left \langle \Phi^3_L(\mathbf{x}) \right \rangle \)
 \; .
\ee
In the previous formula we can recognize the standard $\alm$ PDF, valid in the standard Gaussian case, in the first term on the r.h.s. Added to this we find a first order $\fnl$-correction proportional to the CMB angular bispectrum. Higher order correlators are not present at order ${\cal{O}} \( \fnl \left \langle \Phi_L(\mathbf{x}) \right \rangle \)$. For reasons that will become clear shortly, although we have not computed it, we have explicitly denoted the  ${\cal{O}} \( \fnl^2 \left \langle \Phi^2_L(\mathbf{x}) \right \rangle \)$ term 
in the expansion with $I_2(\mathbf{a},\fnl)$. Note that, besides assuming weak NG in this formula, we
are also assuming {\em rotational invariance} (this is evident from the fact that the $\alm$ covariance matrix appearing in the Gaussian piece of \eqn{eq:localCMBlikelihood} is diagonal and equal to $C_\ell$). Rotational invariance is a general property of the CMB sky but it is broken when we deal with real CMB measurement characterized by inhomogeneous noise patterns and sky cuts. We will investigate these effects in the following section. For the moment we consider the purely ideal case described by \eqn{eq:localCMBlikelihood}. Armed with the PDF expression for local NG, and recalling the necessary and sufficient condition \eqn{eq:RCfnl}, we can finally determine whether the estimator \eqn{eq:optimalcubicestnorm} is optimal or not. 
First of all we see that:
\be\label{eq:diffpdflocal}
\frac{\partial \ln p(\mathbf{a}|\fnll)}{\partial \fnll} = \sumalllm \frac{ {\cal{G}}_{\alll}^{\allm} 
b^{\fnll=1}_{\alll}}{C_{\ell_1} C_{\ell_2} C_{\ell_3} }  \almone \almtwo \almthree + \frac{\partial I_2(\mathbf{a},\fnl)}{\partial \fnl} 
+ {\cal{O}} \( \fnl^2 \left \langle \Phi^2_L(\mathbf{x}) \right \rangle \).
\ee
We then see from combining (\ref{eq:diffpdflocal}) and (\ref{eq:optimalcubicestnorm}) 
that:
\be\label{eq:proofoptimality1}
\frac{\partial \ln p(\mathbf{a}|\fnll)}{\partial \fnll} \propto 
\[\fnlest(a) +  \frac{\partial I_2(\mathbf{a},\fnl)}{\partial \fnl}
+ {\cal{O}} \( \fnl^2 \left \langle \Phi^2_L(\mathbf{x}) \right \rangle\) \]\,.
\ee
We now see that, in order for the necessary and sufficient condition 
for optimality (\ref{eq:RCfnl}) to be verified, we need the $({\partial I_2 / \partial \fnl})$ term to be exactly equal to $-\fnl$.
The second order quantity $I_2$ should then be calculated explicitly in the expansion (\ref{eq:localCMBlikelihood}) in order 
to complete the calculation and verify if, or under which conditions, this is true. 
However this turns out not to be necessary if we consider the following ``regularity 
condition for a PDF''.\footnote{The condition \eqn{eq:regularitycondition} can be easily derived remembering that, for a given random variable $\mathbf{x}$ with probability density $p(\mathbf{x})$, we have by definition: $\langle F(\mathbf{x}) \rangle \equiv \int d \mathbf{x} F(x) p(\mathbf{x})$, and substituting $F(\mathbf{x}) \rightarrow {\partial \ln p(\mathbf{x}|\l) / \partial \l}$ in the previous expression. One then finds 
that the regularity condition \eqn{eq:regularitycondition} holds, provided the order of integration and differentiation can be exchanged (hence the ``regularity condition'' qualification). } For a general PDF of a random variable $\mathbf{x}$ depending on a parameter $\l$ we have
\be\label{eq:regularitycondition}
\left \langle \frac {\partial \ln p(\mathbf{x}|\l)}{\partial \l} \right \rangle = 0 \; .
\ee 
Since this regularity condition must be valid for each value of the parameter $\l$ ($\l \rightarrow \fnll$ in our case), it is clear that it must hold {\em term-by-term}, \ie at each order, in the expansion \eqn{eq:localCMBlikelihood}. By taking 
the average value of equation \eqn{eq:proofoptimality1}, keeping in mind that the estimator is unbiased, and imposing \eqn{eq:regularitycondition}, we then find that 
the average value of ${\partial I_2(\mathbf{a},\fnl) / \partial \fnl}$ must be 
exactly equal to $-\fnl$. If we could then replace ${\partial I_2(\mathbf{a},\fnl) / \partial \fnl}$ in \eqn{eq:proofoptimality1} with its average value we would exactly obtain the condition for optimality and conclude that the cubic estimator \eqn{eq:optimalcubicestnorm} 
saturates the Rao-Cramer bound. For present CMB experiments, the terms in the expansion \eqn{eq:localCMBlikelihood} are evaluated summing over a large number of $\ell$-modes\footnote{$\ell_{max} \simeq 500$ for WMAP, $\ell_{max} \simeq 2000$ for Planck in the 
signal dominated regime.}, or equivalently in pixel space, averaging over a large number of pixels ($\sim 10^6$ and $10^7$ for WMAP and Planck respectively).  For this reason we expect that the error made by replacing the $\fnl$-order term in \eqn{eq:proofoptimality1}
with its average value will be very small. In \cite{Babich2005}, an estimate of this error has been done in the approximation of neglecting radiative transfer and projection effects (\ie working in 3D with the primordial potential, rather than with a CMB map). The conclusion was that for a number of observations $N>30$ the approximation above works very well. Moreover the variance of the $\fnl$-order term scales like ${1/N}$. In the full radiative transfer case we expect the scaling to be unchanged, although the coefficients in front of it that led to the $N>30$ estimate might change. However, as noted above, the number of pixels in present-day experiments is many orders of magnitude larger than $30$. That leads us to conclude that the approximation of replacing the average of the first order term in equation \eqn{eq:proofoptimality1} is a very good one. We then reach the following important conclusion:
\begin{quote}
{\em for a rotational invariant CMB sky, in the limit of weak NG, the 
cubic estimator defined by formula {\rm \eqn{eq:optimalcubicestnorm}} is the best unbiased CMB estimator of} 
$\fnll$
\end{quote}

Let us now move to problem of generalizing the last conclusion to shapes different from local. In this case a full expression of the primordial potential $\Phi(\mathbf{x})$ 
is not available. The steps that lead to the conclusion that the local $\fnl$ estimator is optimal can thus not be reproduced. However it was pointed out in \citet{Babich2005} that, in the limit of weak NG, the full CMB NG likelihood can still be expressed in terms 
of its power spectrum and bispectrum by mean of an Edgeworth expansion, regardless of its full expression. The Edgeworth expansion is basically a way to express a NG PDF as a series expansion around its Gaussian part \citep{BernardeauKofman1995,BernardeauEtal2002,TaylorWatts2001}. For CMB anisotropies one finds, at the end of the calculation:
\be\label{eq:edgeworth}
P(\mathbf{a}|\fnl) = \prod_{\ell m} \frac{e^{-\frac{\alm \alm^*}{2 C_{\ell}}}} {\sqrt{2 \pi C_{\ell}}}
 \[1+\sumalllm b_{\ell_1 \ell_2 \ell_3}  {\cal{G}}_{\alll}^{\allm}
\frac{\almone \almtwo \almthree}{C_{\ell_1} C_{\ell_2} C_{\ell_3}} \] \; .
\ee
It is easy to see that $\ln P(\mathbf{a}|\fnl)$ takes the same form as in equation 
\eqn{eq:localCMBlikelihood}. For this reason all the previous derivation applies also to the 
present case and the following conclusion holds:
\begin{quote}
{\em in the weak non-Gaussian limit and assuming rotational invariance of 
the CMB sky, the cubic estimator {\rm \eqn{eq:optimalcubicestnorm}} is the best 
unbiased CMB estimator of {\rm $\fnl$} for any non-Gaussian shape}.
\end{quote}

Before concluding this section, we would like to stress that, despite the technical complications arising in the detailed probe of the bispectrum estimator's optimality, the physical reason behind this result is quite clear. We can always expand the $\alm$ PDF in series of its momenta. The order parameter of this expansion is $(\fnl \left \langle \Phi \right \rangle)$. This parameter is the natural measure of the amplitude of primordial NG, and it is actually predicted by inflation to be very small. For this reason higher order momenta in the primordial non-Gaussian $\alm$ PDF are suppressed with respect to the bispectrum. Basically, the information on $\fnl$ in a CMB map is entirely contained in the three point function.

\subsubsection{Breaking rotational invariance}

So far the assumption of rotational invariance of the CMB sky has been made when probing the optimality of the cubic estimator \eqn{eq:optimalcubicestnorm}. In an ideal situation, the CMB sky is clearly rotationally invariant. However, two elements break rotational invariance in a CMB map derived from a real experiment: an anisotropic distribution of noise in pixel space, and a galactic mask. Anisotropic noise comes from the fact that the CMB sky is generally scanned in a non-uniform way: regions that are less contaminated by astrophysical foreground emission are generally observed more times, and are thus characterized by a lower 
noise level (see \fig{fig:cutnoise} for an example). A sky cut has also to be introduced in order to remove the regions on the galactic plane that are most contaminated by foregrounds. When rotational invariance is broken the considerations of the previous two sections do not strictly apply anymore and the estimator (\ref{eq:optimalcubicestnorm}) become sub-optimal. However, the same Edgeworth expansion approach that was adopted in the previous section can still be applied, but this time keeping rotation invariance breaking terms in the calculation, 
in order to find the new more general form of the optimal estimator. The general estimator turns out to be the sum of two terms: the first term is cubic in $\alm$ and is analogous to the one appearing in the rotationally invariant case, while the second term is linear in $\alm$ and accounts for breaking of rotational invariance. The explicit expression of this general optimal $\fnl$ estimator is \cite{CreminelliEtal2006}:
\bea\label{eq:cubicestlinear}
{\curl{E}}(a) & = &  \frac{1}{{\cal{N}}} \sumalllm \left( \Gaunt b_{\alll} \(C^{-1}_{\ell_1 m_1, \ell_4 m_4} \almone\) 
                   \(C^{-1}_{\ell_2 m_2, \ell_5 m_5} \almtwo \) \(C^{-1}_{\ell_3 m_3, \ell_6 m_6} \almthree \) \right.
                   \nonumber \\
            & &    \left. -3 \left \langle \almone \almtwo \almthree \right \rangle C^{-1}_{\ell_1 m_1, \ell_2 m_2} 
                   C^{-1}_{\ell_3 m_3, \ell_4 m_4} \almfour \right) \\
& & \nonumber \\
{\cal{N}} & = & \sumalllm \left \langle \almone \almtwo \almthree \right \rangle 
                C^{-1}_{\ell_1 m_1, \ell_4 m_4} C^{-1}_{\ell_2 m_2, \ell_5 m_5} 
                C^{-1}_{\ell_3 m_3, \ell_6 m_6} \left \langle \almfour \almfive \almsix \right \rangle \; .
\eea

In the rotationally invariant case the $\alm$ covariance matrix $C_{\ell_1 m_1, \ell_2 m_2}$ is diagonal and equal to $C_\ell$, while the linear term is proportional to a monopole. We then recover the form of the cubic estimator $\ref{eq:optimalcubicestnorm}$ as expected.
Note that in the signal dominated regime of the experiment under study (\eg $\ell \lesssim 300$ for WMAP and $\ell \lesssim 1000$ for Planck), and if the mask is not too large, then the simple cubic estimator (\ref{eq:optimalcubicestnorm}) is still basically optimal, since we are in a nearly rotationally invariant case. For small masks it has been shown in  \citet{KomatsuSpergel2001} that the bispectrum and power spectrum of the map are, to a good approximation, just rescaled by a factor $f_{sky}$, representing the fraction of the sky left free by the mask, \ie
\be\label{eq:fsky}
b_{\alll}^{mask} = f_{sky} b_{\alll}^{full \, sky} \;\; , \;\;\;   
C_{\ell}^{mask} =  f_{sky} C_{\ell}^{full \, sky} \,.
\ee
In this case one can then assume the covariance matrix to be diagonal, and account for the effects of the mask by correctly rescaling the normalization term in order to keep the estimator unbiased. This nearly rotationally invariant estimator then takes the form:
\bea\label{eq:cubicestnearlyrotinv}
\fnlest & = & \frac{1}{{\cal{N}}} \sumalllm {\cal{G}}_{\alll}^{\allm} b_{\alll}^{\fnl = 1} 
              \almone \almtwo \almthree  \, ,\\
{\cal{N}} & = & f_{sky} \sumalllm \frac{ \({\cal{G}}_{\alll}^{\allm} b_{\alll}^{\fnl=1}\)^2}{\bispvar} \; .
\eea

\subsubsection{Large $\fnl$ regime}
\label{sec:largefnl}

The approximation of weak non-Gaussianity is the basis for all the results derived so far. One can then ask at which point (\ie for which values of $\fnl$) this approximation breaks down. As we observed earlier, the Edgeworth expansion \eqn{eq:edgeworth} shows that the likelihood of a generic primordial NG distribution can be expanded in series of its momenta, with order parameter ${\cal{O}} \( \fnl \left \langle \Phi^2_L(\mathbf{x}) \right \rangle \)$. We know that $\Phi_L \sim 10^{-5}$, while WMAP observations already constrain $\fnl \lesssim 100$. That means that the order parameter of the PDF expansion is $\sim 10^{-3}$ and thus the weak NG approximation seems a very good one in the entire range of allowed 
and predicted $\fnl$. However a subtle effect has been pointed out by \citet{CreminelliEtal2007}, that change the previous conclusions in certain cases. Let us quickly summarize their main results. We already saw that, for the angular averaged bispectrum of a Gaussian temperature field:
\be
\left \langle B^2_{\alll} \right \rangle \propto \bispvar \; .
\ee
We then included this expression for the variance in the weights of the optimal estimator 
\eqn{eq:optimalcubicestnorm}, and in the normalization factor ${\cal{N}}$. It is easy to see that in this approximation the variance of the estimator can be predicted as:
\be\label{eq:fisher1}
\left \langle \(\Delta {\curl{E}}\)^2 \right \rangle = \sumalllm \frac{(\Gaunt b_{\alll})^2}{\bispvar} \; .
\ee
However the approximation of taking $\fnl = 0$ in the calculation of the estimator variance is not always a good one if $\Delta \fnl$ is dominated by squeezed configurations\footnote{We recall that by squeezed configurations we mean triangles in which one of the sides is much smaler than the other two, \ie $\ell_1 << \ell_2, \ell_3$.}, or more in general by configurations in which one of the $\ell$'s is small. It turns out that, in these cases, the 
$\fnl$-dependent corrections to the Gaussian expectation of the bispectrum variance become 
important when $\fnl$ gets large enough. This effect increases the variance of the estimator with respect to the expectation for $\fnl = 0$. There is a clear physical interpretation for this. One can see, for example by calculating \eqn{eq:fisher1} in the simple Sachs Wolfe case (see also section \ref{sec:forecasts}) that the variance of the estimator scales roughly like ${1/\ell_{max}}$, or equivalently like ${1/N_{pix}}$ in pixel space, $N_{pix}$ being the number of pixel in the observed map. This increase of the signal-to-noise ratio with the number of pixels is due to the fact that more and more bispectrum configurations are included into the sum over modes to estimate $\fnl$. However, if the signal is completely dominated by 
low-$\ell$ modes, as in the local case, there is an intrinsic large cosmic variance, due to the small number of low-$\ell$ configurations. Clearly, cosmic variance cannot be beaten by increasing the resolution of the map. \citet{CreminelliEtal2007} then found that 
for $N_{pix}$ getting large, the $S/N$ of the estimator for local NG grows asymptotically as $(\ln N_{pix})$ \ie much slower than the expected $N_{pix}$, that one would obtain by neglecting $\fnl$-dependent corrections in the calculation of the estimator variance. They carried out a calculation of the estimator variance in the flat-sky approximation, and neglected the transfer functions, to find the following expression:
\be\label{eq:largefnlcorrection}
\left \langle \(\Delta {\curl{E}}\)^2 \right \rangle = \frac{1}{4 A N_{pix} \ln N_{pix}}\(1+\frac{8 \fnl^2 A N_{pix}}{\pi \ln N_{pix}}\) \; ,
\ee
where $A$ is the bispectrum amplitude.
We clearly see from this formula what we were stating above \ie when $\fnl$ gets large, the variance starts scaling like ${1/\ln N_{pix}^2}$. The same formula also shows the technical point behind this behaviour: in the correction term, the order parameter is actually {\em not} $\fnl A^{{1/2}}$ anymore but rather $\fnl A^{{1/2}} N_{pix}$. This enhancement by a factor $N_{pix}$ can make the first order corrections non-negligible anymore. The natural 
question is now how large an $\fnl$ we need to make the correction term important in (\ref{eq:largefnlcorrection}). Following \citet{CreminelliEtal2007} let's call $\sigma_0$ the standard deviation of the estimator computed for $\fnll = 0$. Let's say that for an experiment at a given angular resolution (defined by $\ell_{max}$ in harmonic space or by $N_{pix}$ in pixel space) a value of $\fnll$ is measured, equal to $n \sigma_0$. Substituting this value into formula \eqn{largefnlcorrection}behavior, and calling $\sigma^2_{\fnl}$ the real estimator variance, one finds the following relative
correction to $\sigma_0$:
\be
 \frac{\sigma_{\fnl}^2}{\sigma_0^2} - 1 = \frac{2 n^2}{\pi \ln^2 N_{pix}} \; .
\ee
For an experiment like WMAP the r.h.s. term becomes $\sim 1$ when $\fnl$ is about $6 \sigma_0$ away from the origin. For an experiment like Planck $\fnl$ has to be about $3.5 \sigma_0$. The definition of an high-$\fnl$ regime is thus dependent 
on the experiment under study, as a consequence of the fact that the enhancement of first order terms in the variance expression $\left \langle \( \Delta {\curl{E}} \) \right \rangle$ depends on $N_{pix}$. In other words we can conclude the following:
\begin{quote}
{\em if $\fnll$ will be detected at several $\sigma$ (in terms of the Gaussian expectation for the standard deviation), then {\rm $\fnl$}-dependent correction terms in the estimator variance will have to be taken into account, and the simple expansion (\ref{eq:localCMBlikelihood}) of the CMB likelihood does not constitute a valid approximation anymore}.
\end{quote}

 One caveat in all this discussion is that formula \eqn{eq:largefnlcorrection} was obtained in flat sky and neglecting the transfer functions and it should be checked how dependent the final results are on these approximations. Since the scaling argument is based on a very general physical reason, \ie the weight of squeezed configurations in the local $\fnl$ estimator discussed earlier in this section, one expects that the general scaling with $N_{pix}$ obtained in \eqn{eq:largefnlcorrection} does not depend on the details of radiative transfer and 2D projection. \citet{LiguoriEtal2007} actually checked the results of this section numerically, by applying an implementation of the optimal cubic estimator \eqn{eq:optimalcubicestnorm} to full-sky simulations of CMB local NG maps with different $N_{pix}$ and $\fnll$, including the full radiative transfer\footnote{For details about the numerical implementation of the optimal cubic estimator, and about the generation of NG CMB maps, see sections \ref{sec:numericalimplementation} and \ref{sec:mapmaking}}. Although, as expected, the coefficients in formula (\ref{eq:largefnlcorrection}) change with respect to the simple flat sky no radiative transfer approximation, the scaling of the error bars with $N_{pix}$ follows very well the expectations, going from $\sim {1/{N_{pix}\ln N_{pix} }}$ at low $\fnl$ to ${1/\ln N_{pix}}$ when $\fnl$ is detected at high significance.
Since in the large $\fnl$ regime the variance starts to scale very slowly, like ${1 / \ln^2 N_{pix}}$, one is led to wonder whether the estimator discussed in the previous sections
 becomes suboptimal at this point, and whether a better one can be found. The answer to this question is not immediate. In order to check for the optimality of an estimator, as we have seen, one has to see if it saturates the Rao-Cramer bound.   However, also the local $\fnl$ likelihood and Rao-Cramer bound estimated in the previous sections have to be recomputed,
since they were obtained neglecting higher order terms in $\fnl A^{1/2}$. In order to account for the $\fnl A^{1/2} N_{pix}$ enhanced terms it is necessary to produce an exact expression of the full likelihood. This can be extremely challenging in the full radiative transfer case, but it is feasible in the flat sky no radiative transfer approach that we are considering (and that we showed earlier to be a good approximation as long as scaling arguments are involved). \citet{CreminelliEtal2007} proceed to calculate the full likelihood in this approximation and conclude that the optimal cubic estimator of weak local NG indeed {\em does not} saturate the Rao-Cramer bound in the high-$\fnl$ regime. The estimator (\ref{eq:optimalcubicestnorm}) is thus no longer optimal in this case. They then proceed (always in the flat sky no transfer function case) to derive a cubic estimator that saturates the Rao-Cramer bound also for large $\fnll$. We won't enter into the details of this derivation here, referring the reader to \citet{CreminelliEtal2007} for a complete discussion. The main aim of this section was to show under which conditions the optimality of the cubic estimator that we discussed in previous sections is valid.   Since current bispectrum analysis of WMAP data \citep{KomatsuEtal2009B, SmithSenatoreZaldarriaga2009} find $\fnll \sim 2\sigma_0$, the weak NG approximation applies, and the cubic estimator we derived earlier is indeed an optimal estimator in this case. However, if future Planck measurement will produce a detection of $\fnll$ at high significance the estimator will have to be modified in order to account for the enhanced variance in the high-{$\fnl$} regime. This is not necessarily a remote possibility if one considers that the present central value of $\fnl$ from WMAP estimates would produce a $\sim 8 \sigma_0$ detection with Planck.
Before concluding this section we would like to remark once again that the variance enhancement discussed here applies {\em only} to non-Gaussianity of the {\em local} type, whose bispectrum is dominated by squeezed configurations, affected by a large cosmic variance. For the other shapes the cubic estimator (\ref{eq:optimalcubicestnorm}) is optimal in both the small and high-$\fnl$ regimes.

\subsection{Numerical implementation of the bispectrum estimator}\label{sec:numericalimplementation}

In this section we turn to the problem of finding an efficient numerical implementation of the optimal bispectrum estimator \eqn{eq:cubicestlinear}. Let us repeat its expression here for convenience:
\bea\label{eq:cubicestlinear2}
{\curl{E}}(a) & = &  \frac{1}{{\cal{N}}} \sumalllm \left[ \Gaunt b_{\alll} \(C^{-1}_{\ell_1 m_1, \ell_4 m_4} \almone\) 
                   \(C^{-1}_{\ell_2 m_2, \ell_5 m_5} \almtwo \) \(C^{-1}_{\ell_3 m_3, \ell_6 m_6} \almthree \) \right.
                   \nonumber \\
            & &    \left. -3 \left \langle \almone \almtwo \almthree \right \rangle C^{-1}_{\ell_1 m_1, \ell_2 m_2} 
                   C^{-1}_{\ell_3 m_3, \ell_4 m_4} \almfour \right] \, ,\\
& & \nonumber \\
{\cal{N}} & = & \sumalllm \left \langle \almone \almtwo \almthree \right \rangle 
                C^{-1}_{\ell_1 m_1, \ell_4 m_4} C^{-1}_{\ell_2 m_2, \ell_5 m_5} 
                C^{-1}_{\ell_3 m_3, \ell_6 m_6} \left \langle \almfour \almfive \almsix \right \rangle \; .
\eea

We remind the reader that this is the full expression, valid for the general non-rotationally invariant case. For a rotationally invariant CMB sky the linear term in the formula above vanishes, and the covariance matrix is diagonal and reduces to $C_\ell$, giving the simplified expression \eqn{eq:optimalcubicestnorm} that we reproduce again here for convenience:
\bea\label{eq:optimalcubicestnorm2}
\curl{E}(a) & = & \frac{1}{{\cal{N}}} \sumalllm \frac{{\cal{G}}_{\alll}^{\allm} b_{\alll}^{\fnl = 1}}{\bispvar} 
              \almone \almtwo \almthree \, , \\
{\cal{N}} & = & \sumalllm \frac{ \({\cal{G}}_{\alll}^{\allm} b_{\alll}^{\fnl=1}\)^2}{\bispvar} \; .
\eea                 

In a schematic way, the full estimator can be written as:
\be
\curl{E}(a) = \frac{{\curl{E}}^{cubic}(a) + {\curl{E}}^{linear}(a)}{\curl{N}} \; ,
\ee
where the ``cubic'' term is the one containing the product $\almone \almtwo \almthree$, while the linear term is the one dependent on a single $\alm$ and vanishing in the rotationally invariant case, where it is proportional to a monopole. It was shown before in a formal way that a pure cubic estimator becomes suboptimal when rotational invariance is broken, and adding the linear term is necessary to restore optimality. It is useful to try to understand the reason of this effect qualitatively and in a more intuitive way. Let's assume 
that we have a map characterized by non-stationary noise, and we are observing a region of the sky that was sampled many times so that the noise level in this area is low. That implies that the level of {\em small} scale power in this {\em large} region is lower than average. Now, for a specific realization of the CMB sky, this modulation of small scale power on a large region can look like a non-Gaussian signal sourcing a {\em squeezed} configuration of the bispectrum. On average this effect must cancel if the underlying noise model is Gaussian. However, this ``confusion'' between signal and noise increases the variance of any estimator of a primordial NG signal that is peaked on squeezed configurations. We know that this happens for the local model. This heuristic argument thus shows that, even though in principle a linear term must always be included when rotational invariance is broken, for a realistic noise model only local non-Gaussian estimates will be affected.

\subsubsection{Primary cubic term for $\fnl$}

Let us focus for the moment on the rotationally invariant case, where the linear term vanish, and the covariance matrix is simply $C = C_{\ell} \delta_{\ell_1 \ell_2} \delta_{m_1 m_2}$. We immediately see that a brute force implementation of equation \eqn{eq:optimalcubicestnorm2}, consisting in computing and summing over all the bispectrum configurations, would take ${\cal{O}} \, \(\ell_{max}^5\)$ operations, where $\ell_{max}$, the maximum multipole in the calculation, depends on the resolution of the experiment. As mentioned earlier, $\ell_{max} \sim 500$ for WMAP and $\ell_{max} \sim 2000$ for Planck in the signal dominated regime. At these resolutions a brute force approach would be absolutely unfeasible for a general shape. If however we assume that the primordial shape under study is 
{\em separable} then the dimensionality of the problem can be reduced and the overall number of operations scaled down significantly, making the computational cost affordable. Let us illustrate this point more in detail. Substituting \eqn{eq:separated} into the estimator expression \eqn{eq:optimalcubicestnorm}, and remembering the identity (\ref{eq:Gaunt}),
\be
\Gaunt = \int d \Omega_{\hat{n}} \Ylmone\, \Ylmtwo\, \Ylmthree \; ,
\ee
it is possible to rewrite \eqn{eq:optimalcubicestnorm2} as follows (or more in general as a linear combination of terms of the following kind):
\be
\curl{E}(a) = \frac{1}{{\cal{N}}} 
\int dr r^2 \int d \Omega_{\hat{n}} \sum_{\ell_1 m_1} \frac{\almone {{X}}_{\ell_1}(r) \Ylmone}{C_{\ell_1}} 
 \sum_{\ell_2 m_2} \frac{\almtwo {{X}}_{\ell_2}(r) \Ylmtwo}{C_{\ell_2}}  
\sum_{\ell_3 m_3} \frac{\almthree {{X}}_{\ell_3}(r) \Ylmthree}{C_{\ell_3}}  + {\rm perm.} \; .
\ee
From an inspection of previous formula we see how, as a direct consequence of separability, 
the initial sum over the indeces $\alll, \allm$ has been factorized in the product of three sums, each running over two indices $\ell,m$.  This greatly reduces the computational cost from ${\cal{O}} \, \( \ell_{max}^5 \)$ to ${\cal{O}} \, \(\ell_{max}^3\)$ operations. If we define the new quantities 
\bea
M_{X}(r,\hat{\bf n}) & \equiv & \sum_{\ell m} \frac{\alm {{X}}_{\ell}(r)}{C_{\ell}} \Ylm \; , \nonumber \\     
M_{Y}(r,\hat{\bf n}) & \equiv & \sum_{\ell m} \frac{\alm {{Y}}_{\ell}(r)}{C_{\ell}} \Ylm \; , \nonumber \\     
M_{Z}(r,\hat{\bf n}) & \equiv & \sum_{\ell m} \frac{\alm {{Z}}_{\ell}(r)}{C_{\ell}} \Ylm \; ,
\eea
we can recast the estimator expression above in the following form:
\be\label{eq:positionspace}
\curl{E}(a) = \frac{1}{{\cal{N}}} \int dr \, r^2 \int d \Omega_{\hat{\bf n}} M_X(r,\hat{\bf n}) 
M_Y(r,\hat{\bf n}) M_Z(r,\hat{\bf n})  + {\rm perms.} \; ,
\ee
where it is evident that we are now calculating our statistic in position space rather than in pixel space. Note how the 
filtered maps $M_X,\, M_Y,\,M_Z $ can be efficiently calculated using a fast harmonic transform algorithms such as those included in the HEALPix package.
This fast position space algorithm was initially introduced in \citet{KomatsuSpergelWandelt2005} in the context of local $\fnl$ estimation, and applied to the estimation of WMAP 1-year data by the WMAP team in \citet{KomatsuEtal2003}. It was then applied to equilateral $\fnl$ estimation for the first time in \citet{CreminelliEtal2006}. An alternative numerical implementation with respect to the one used by the aforementioned authors was introduced in \citet{SmithZaldarriaga2006}. Although different under many technical aspects, this second algorithm is still based on the calculation of the position space statistic \eqn{eq:positionspace}; we refer the reader to the original work for additional details. This second implementation has been used to produce alternative estimates of $\fnll$, and $\fnle$ from WMAP data, and to estimate the amplitude of the orthogonal shape, recently introduced in \citet{SenatoreSmithZaldarriaga2009}. 

Let us now discuss the possible limitations of this numerical approach. As noted in section \ref{sec:initialcond}, the separability condition is in principle quite restrictive: the only separable shape arising directly from primordial models of inflation is  the local one. On the other hand, it is still possible to study non-separable models by finding separable shapes that are highly correlated to the primordial one. As observed in \citet{CreminelliEtal2006,SmithZaldarriaga2006,FergussonShellard2009}, the $\fnl$ limits obtained from a highly correlated separable shape in this way will be very close to those that would have been obtained using the original non-separable model (see again sections \ref{sec:initialcond}, \ref{sec:separableshapes} and \ref{sec:nonseparableshapes} for a detailed discussion of this issue). We know from earlier sections that the other two shapes mentioned so far in this section besides local, namely the equilateral and orthogonal shape, have actually been derived as separable approximations of theoretical inflationary shapes. These approximations were obtained in an heuristic way \ie an educated guess of a good separable approximation of the shape under study was made, and the correlation was checked {\it a posteriori}. There is obviously no a priori guarantee that this approach would be easily repeatable for all the shapes of interest. The eigenmode expansion method introduced in \cite{FergussonLiguoriShellard2009}, and summarized by equation \eqn{eq:orthobasis}, however, provides a general and rigorous method to find separable approximations of any shape, thus enabling the estimation of {\em any} possible primordial model.   In this case, recall that we expand our (non-separable) primordial 
shape function in terms of the separable basis functions $\Qn$ (see (\ref{eq:orthobasis}), constructed from symmetric polynomical products $q_p(k)$, as 
\eq 
\Skkk = \sum_{prs} \alpha_{prs} \, q_p(k_1) \, q_r(k_2) \, q_s(k_3) \, , \quad \longrightarrow \quad 
b_{l_1 l_2 l_3} = \D_\O^2 \fnl \sum_{prs} \alpha_{prs} \int x^2dx \, q_{\{p}^{\,\ell_1} q_{r}^{\,\ell_2} q_{s\}}^{\,\ell_3}\,,
\qe 
where the second expression for the reduced bispectrum $\blll$ (\ref{eq:sepblllprim}) expands in convolved basis
functions (\ref{eq:qmultipole}) in harmonic space with
\eq\label{eq:convolved}
q_p^{\,l}(x) = \frac{2}{\pi} \int d k\,  q_p(k) \, \D_{l}(k) \, j_{l}(kx)\,.
\qe
In the mode expansion approach, then, the $\fnl$-estimator for a specific model generalises to the following
\eq
\curl{E}(a) = \frac{1}{{\cal{N}}} \sum_{prs} \alpha_{prs} \int dr r^2 \int d \Omega_{\hat{n}} \,
M_{\{p}(r,\hat{\bf n}) \,
M_r(r,\hat{\bf n}) \, M_{s\}}(r,\hat{\bf n} )\, ,
\qe
where the filtered maps or shells $M_p(r,\hat{\bf n})$ are defined by 
\eq\label{eq:filteredmaps}
 M_p(r,\hat{\bf n}) = \sum_{lm} q_p^{\,l}\,\, \frac { a_{lm} Y_{lm} }{C_{l}}\,.
\qe
Defining the integral $\beta_{prs} \equiv  \int dr r^2 \int d \Omega_{\hat{n}} \, M_{\{p}\, M_r\, M_{s\}}$, the estimator
collapses into the compact form 
\eq \label{eq:sepestimator} 
\curl{E}(a) = \frac{1}{{\cal{N}}}  \sum_{prs} \alpha_{prs} \, \beta_{prs}\,,
\qe
where it is possible to show a precise relationship between the theoretical bispectrum expansion 
coefficients $\alpha_{prs}$ and expectations for the observed coefficients $\beta_{prs}$.   

It was also pointed out in \citet{FergussonLiguoriShellard2009} (see also \citet{FergussonShellard2007}), and summarized in formula \eqn{eq:cmbestmodes}, that the separation can be performed directly in harmonic space on the reduced bispectrum $b_{\alll}$, rather than on the primordial shape $\Skkk$. This provides an alternative, but equivalent, late-time $\fnl$-estimation pipeline with respect to the primordial shape separation approach given above (\ref{eq:sepestimator}).  In fact, since orthonormality is more direct 
on the harmonic domain without the intervention of transfer functions, the approach is considerably more 
straightforward conceptually.   In this case, expectations for the observational expansion coefficients in the orthonormal frame $\barRn$
(with $n\leftrightarrow \{prs\}$ see (\ref{eq:cmbestmodes}))  become simply
$\langle \bbRn \rangle = \baRn$,  that is, for an ensemble of maps possessing the theoretical bispectrum described
by the coefficients $\baRn$.  This means that for a NG bispectrum signal of sufficient significance we can 
consider directly and efficiently reconstructing the bispectrum from the observed coefficients $\bbRn$ using (\ref{eq:cmbestmodes}).   
We also note that the harmonic space separation scheme also allows for the estimation of non-inflationary late-time bispectra, such as the bispectrum of cosmic strings, as well as other secondary anisotropies. 

We can then conclude, in light of these developments, that the fast cubic statistic \eqn{eq:positionspace} can be applied in complete generality to any model of primordial NG, as well as to any other potential source of CMB NG. We also point  out that alternative approaches have been considered for harmonic space $\fnl$ analysis using wavelets and binning. 
For example, \citet{BucherVanTentCarvalho2009} recently proposed using a suitable binning scheme in which the full expression for $b_{\alll}$ is calculated in a subset of all the triples $\lall$, small enough to make the calculation feasible  while maintaining calculation accuracy. Approaches based on a harmonic space separation scheme, of course, require the full calculation of the reduced bispectrum $b_{\alll}$ in order to determine the correlation between the theoretical prediction
 and the final expanded or binned bispectrum. 
The calculation of $b_{\alll}$ implies the necessity of numerically solving the 
radiative transfer integral \eqn{eq:phi2alm1} for all the configurations $\lall$ which appears to be 
intractable  in the non separable case, since the dimensionality of the problem cannot be reduced.  
However, this can be achieved efficiently in the general case either using the separable mode expansion 
integral (\ref{eq:sepblllprim}) or else the hierarchical adaptive approach  \citet{FergussonShellard2007}
discussed in section~\ref{sec:cmbcorrelator}.

\subsubsection{Linear correction term for $\fnl$}

\begin{figure}[t]
\centering
\includegraphics[width=0.45\textwidth,height=0.25\textheight]{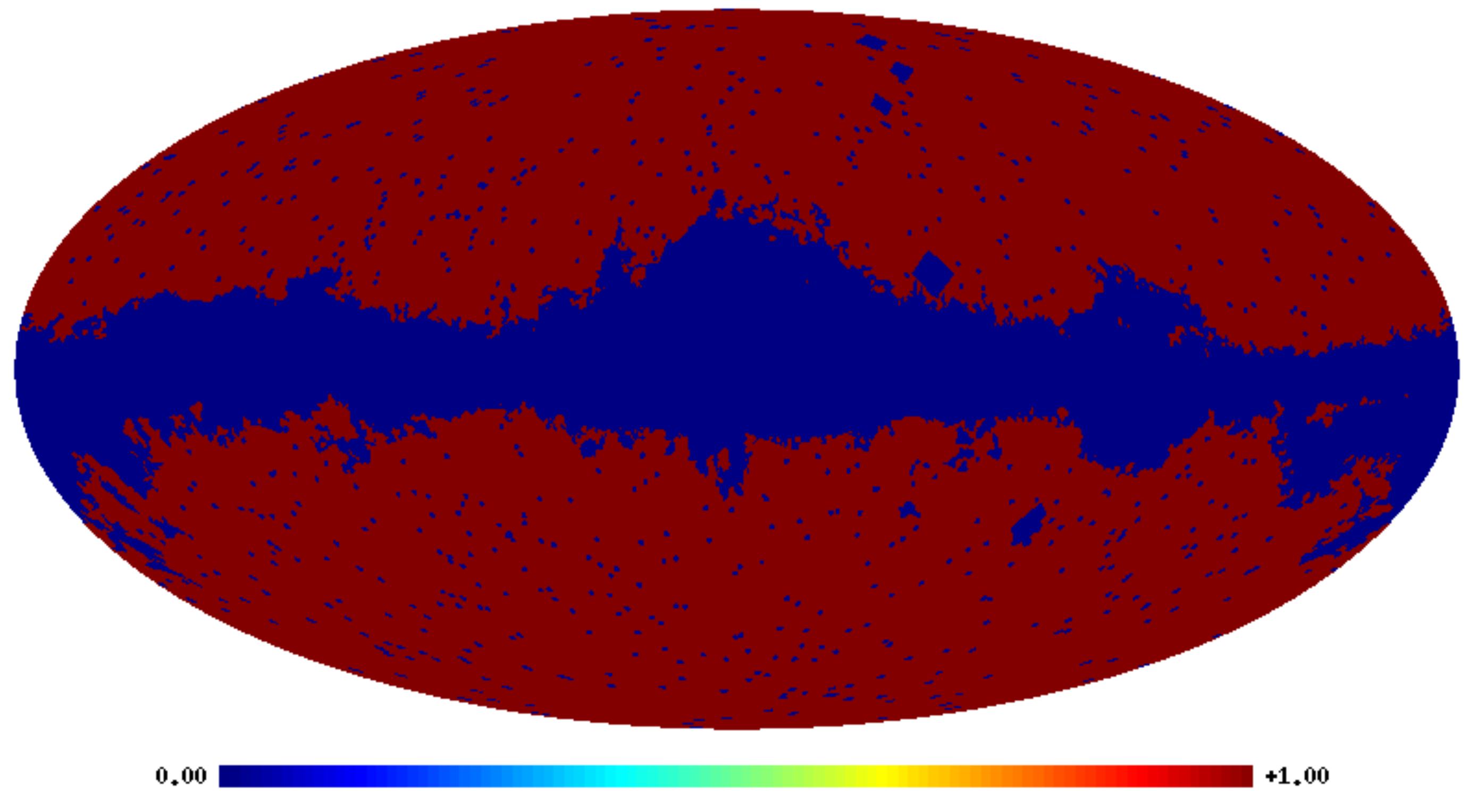}
\hspace{0.2cm}
\includegraphics[width=0.45\textwidth,height=0.25\textheight]{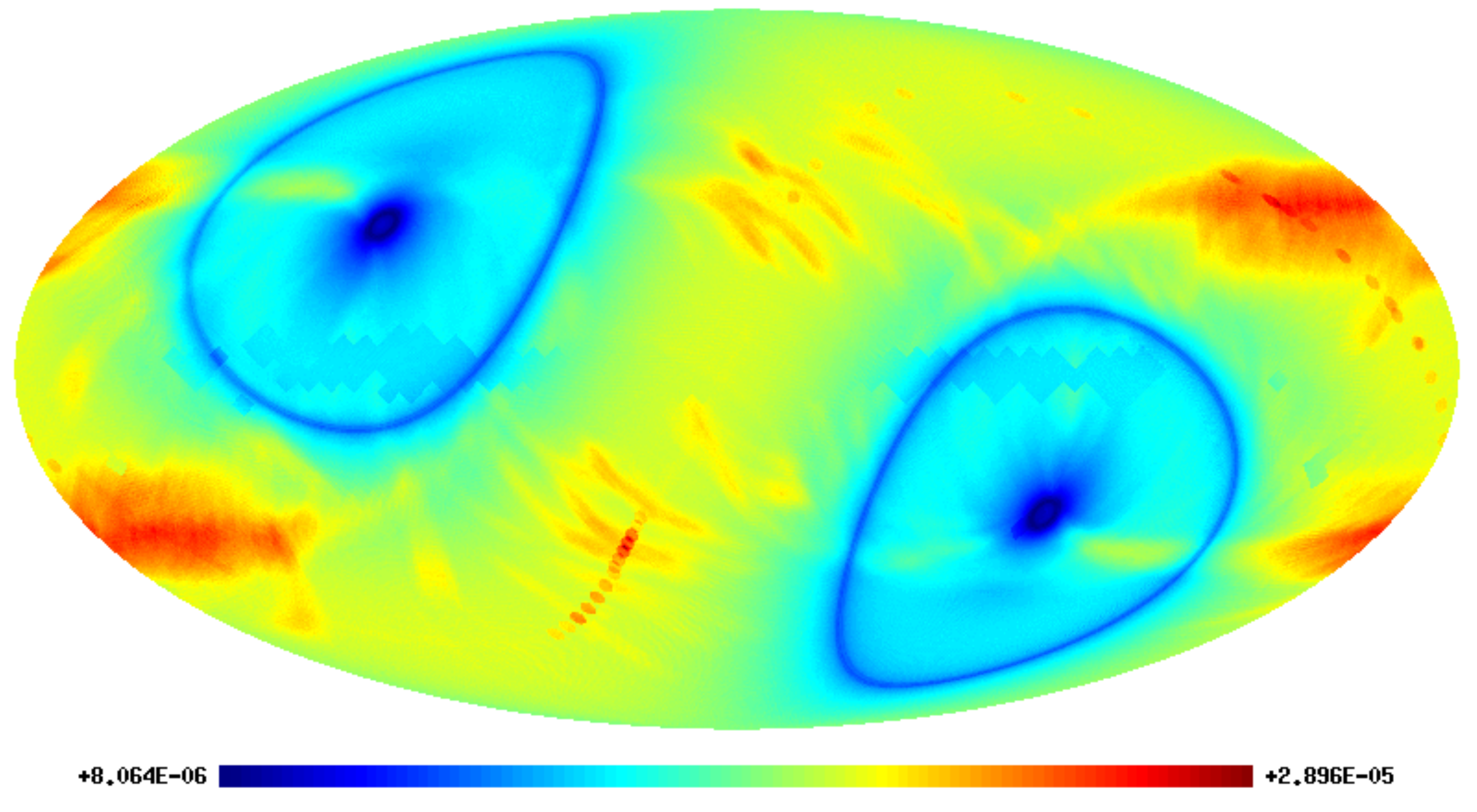}
\caption{{\it Left panel}: the KQ75 galactic and point source mask used for non-Gaussian analysis of WMAP data. {\it Right panel}: anisotropic distribution of the noise for a coadded map of the WMAP V and W frequency channels. These two features break rotational invariance of the observed CMB sky, and spoil the optimality of the standard cubic statistic \eqn{eq:optimalcubicestnorm}, unless an additional ``linear term'' is included, as explained in the main text. Data from the LAMBDA website, \href{http://lambda.gsfc.nasa.gov/index.cfm}{http://lambda.gsfc.nasa.gov/index.cfm}.}
\label{fig:cutnoise}
\end{figure}

Let us now consider the realistic situation in which inhomogeneous noise and a sky-cut break rotational invariance (see \fig{fig:cutnoise}). In this case two complications arise:
\begin{enumerate}
\item A linear term in $\alm$ has to be added
\item The $\alm$ covariance matrix is now no longer diagonal. The inverse covariance weighting $C^{-1} a$ that appears in expression (\ref{eq:cubicestlinear2}) is hard to compute numerically for high angular resolution experiment, since its size makes a brute force numerical inversion impossible.
\end{enumerate}
A first approach,  introduced in \citet{CreminelliEtal2006}, is to simplify the problem by assuming that the covariance matrix is diagonal in the cubic term of the estimator, and then finding the linear term that minimize the variance under this assumption. In other words, we keep the cubic term in the form \eqn{eq:optimalcubicestnorm2} and compute the variance of this term, relaxing the assumption of isotropy at this point\footnote{This means that, when we apply the Wick's theorem to the $\alm$ six point function in the calculation of the cubic term variance, we take $\left \langle \almone \almtwo \right \rangle = C_{\ell_1 m_1, \ell_2 m_2}$ instead of $\left \langle \almone \almtwo \right \rangle = C_{\ell_1 m_1} \d_{\ell_1 \ell_2} \d_{m_1 m_2}$}

\begin{figure}[t]
\centering
\includegraphics{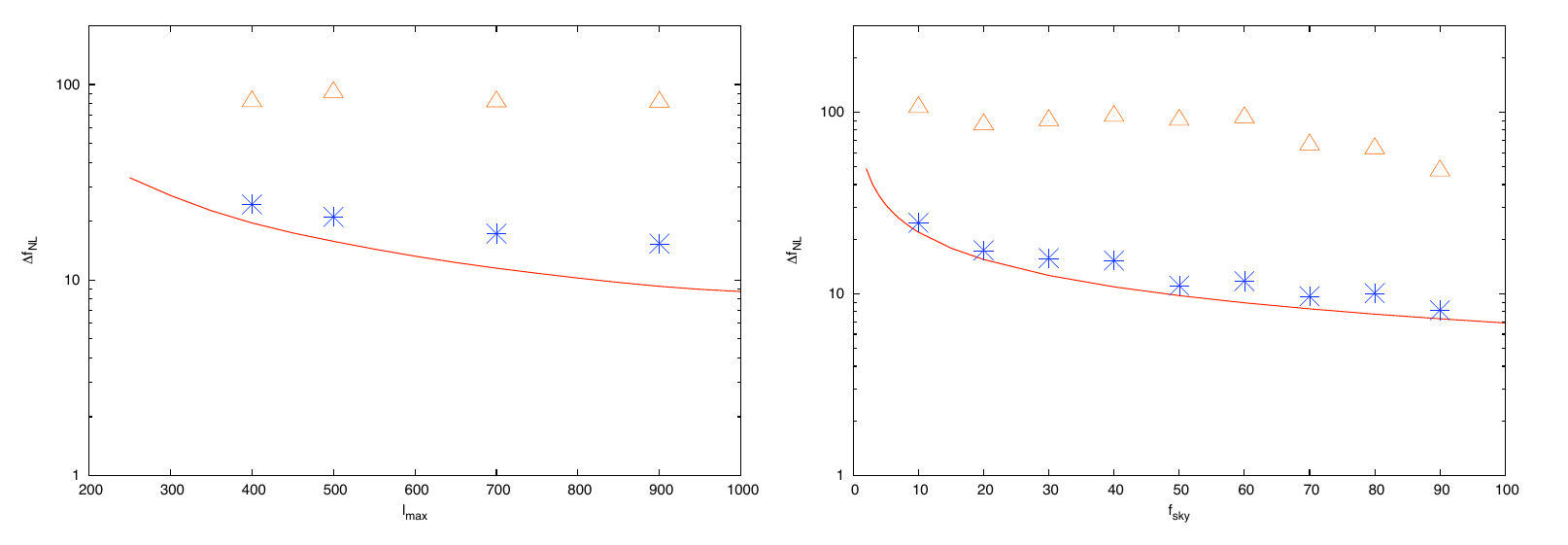}
\caption{Error bars (obtained from Gaussian simulations) for the pure cubic ({\em triangles}) 
and pseudo-optimal ({\em stars}) implementations of the bispectrum estimator, to be compared to the solid red line, representing the Fisher matrix (Rao-Cramer) bound, saturated by the full optimal statistic described in the text. {\it Left panel}: error bars as a function of the maximum multipole included in the analysis. {\it Right panel}: error bars as a function of the fraction of the sky considered in the analysis. This analysis included both temperature and polarization data. From \citet{YadavEtal2008}.}
\label{fig:errorbars}
\end{figure}

It turns out that the variance is minimized (while leaving the estimator unbiased) for the following choice of the linear term:
\be\label{eq:lineartermapprox}
{\curl{E}}_{lin} = -\frac{3}{ {\curl{N}} } \sumalllm \frac{\Gaunt b^{\fnl=1}_{\alll}}{\bispvar} C_{\ell_1 m_1, \ell_2 m_2} \almthree \; ,
\ee
where $ {\curl{N}} = f_{sky} \sum_{\alll} \({B_{\alll}^2 / \bispvar}\)$ is the normalization term. Despite being sub-optimal with respect to a full implementation of equation \eqn{eq:cubicestlinear2}, this choice of linear term has been shown to significantly improve the error bars with respect to the simple cubic statistic \eqn{eq:optimalcubicestnorm2} in presence of anisotropic noise. At the same time, the simplicity of this implementation in comparison to the full optimal statistic (\ref{eq:cubicestlinear2}) is manifest, since no $C^{-1}$ terms appear in equation \eqn{eq:lineartermapprox}. Let us consider again a separable primordial bispectrum shape that can be written $S(k_1,k_2,k_3) = X(k_1)Y(k_2)Z(k_3) + {\rm perm.}$. Applying the same procedure as we did for the cubic term, the linear term can be recast in the form:
\bea
{\curl{E}}_{lin} & = & -\frac{3}{ {\curl{N}} } \int dr r^2 \int d \Omega_{\hat{n}} 
\( \sum_{\ell_1 m_1} \frac{X_{\ell_1}(r)}{C_{\ell_1}}Y_{\ell_1 m_1}(\hat{\bf n}) 
\sum_{\ell_2 m_2} \frac{Y_{\ell_2}(r)}{C_{\ell_2}}Y_{\ell_2 m_2}(\hat{\bf n}) 
\sum_{\ell_3 m_3} \frac{Z_{\ell_3}(r)}{C_{\ell_3}}Y_{\ell_3 m_3}(\hat{\bf n})   \right. \nonumber \\
& & \left.  \times \left \langle \almone \almtwo \right \rangle \almthree + {\rm perm.} \) \; ,
\eea 
where we explicitly wrote $C_{\ell_1 m_1, \ell_2 m_2}$ as $\left \langle \almone \almtwo \right \rangle$. This last formula can be rewritten as:
\bea\label{eq:linearsub}
{\curl{E}}_{lin}& =& -\frac{6}{ {\curl{N}} } \int dr r^2 \int d \Omega_{\hat{n}} \sum_{\ell m} \[ 
\frac{X_{\ell}(r)}{C_{\ell}} \left \langle M_Y(r,{\bf n}) M_Z(r,{\bf n}) \right\rangle + 
\frac{Y_{\ell}(r)}{C_{\ell}} \left \langle M_X(r,{\bf n})M_Z(r,{\bf n}) \right\rangle  + \right.\nonumber\\
& & \left.\qquad
+\frac{Z_{\ell}(r)}{C_{\ell}} \left \langle M_X(r,{\bf n}) M_Y(r,{\bf n}) \right\rangle \] \; .
\eea
Like for the cubic part of the estimator, we have rewritten the linear term as a fast position space integral. The ensemble averages appearing in the last formula can be computed as Monte Carlo averages over a large number of Gaussian realization of the CMB sky, characterized by the same beam, mask and noise properties as the experiment under study. This pseudo-optimal, but relatively straightforward, implementation of the linear term has been adopted by a number of groups in order to estimate $\fnll$ from WMAP data \citep{KomatsuEtal2009B, CreminelliEtal2006, CreminelliEtal2007, YadavWandelt2008}. The full optimal estimator \eqn{eq:cubicestlinear} was implemented only quite recently in \citet{SmithSenatoreZaldarriaga2009}, where the authors developed an efficient conjugate gradient inversion \citep[\eg][]{Shewchuk1994} algorithm based on earlier results from \citet{SmithZahnDore2007}, in order to compute the $C^{-1} a$ pre-filtering in reasonable CPU-time. Note that after the inverse covariance matrix pre-filtering is calculated, the numerical implementation of the estimator is very similar to the one outlined above for the pseudo-optimal case. The new position space statistic is obtained from formulae \eqn{eq:positionspace}, \eqn{eq:linearsub}, by making the following replacements, wherever the corresponding quantitites appear:
\bea
\alm & \rightarrow & \alm^{filtered} \equiv \( C^{-1} a \)_{\ell m} \nonumber \;, \\
& & \nonumber \\
M_X(r,\hat{\bf n}) & \rightarrow &  
\tilde{M_X}(r,\hat{\bf n}) \equiv \sum_{\ell m} \alm {{X}}_{\ell}(r) \Ylm \nonumber \; , \\
\frac{X_{\ell}(r)}{C_{\ell}} & \rightarrow & X_{\ell}(r) \; ,
\eea
with analogous substitutions to be made for the $Y,{\cal{Y}},Z,{\cal{Z}}$ terms appearing in the same equations. The improvement in error bars from the pure cubic sub-optimal estimator, to the pseudo-optimal and optimal statistics is shown on \fig{fig:errorbars}.

\subsection{Experimental constraints on $\fnl$}
\label{sec:constraints}

In order to obtain an estimate of $\fnl$ from a given data-set one has first to generate sets of Gaussian CMB maps and obtain the MC averages that appear in the linear term expression \eqn{eq:linearsub},  after an inverse covariance pre-filtering of the full optimal estimator is implemented. The normalization term ${\cal{N}}$ can be pre-computed using formula (\ref{eq:separated}) to evaluate numerically the theoretical bispectrum shape 
for the model we want to estimate. The statistic \eqn{eq:cubicestlinear2} can then be computed for the experimental data $a_{obs}$ to get our result:
\be
\fnlest(a_{obs}) \equiv \frac{{\curl{E}}^{cubic}(a_{obs}) + {\curl{E}}^{linear}(a_{obs})}{{\cal{N}}}\,.
\ee
The error bars are then obtained by running the estimator on simulated Gaussian maps\footnote{The error bars can be obtained from Gaussian simulations as long as the weak NG approximation applies. As we saw earlier, this works at any $\fnl$ for any shape, except for the local shape when a large $\fnll$ makes the error bars $\fnl$-dependent. In this case the error bars would need to be calculated from NG simulations of $\fnll$. So far no high-significance detection of $\fnll$ has been reported, so working with G maps is at this stage sufficient to get accurate error bars.}:
\be\label{eq:estvar}
\sigma_{\fnlest} = \sqrt{\left \langle \(\fnlest(a_{sim})\)^2 \right \rangle_{MC}} \; ,
\ee
where $\left \langle . \right \rangle_{MC}$ indicates the MC average and $a_{sim}$ a vector of simulated multipoles (obviously including mask, beam and noise features of the experiment). For an accurate step-by-step description of an $\fnl$-analysis of WMAP data, including details about channel coadding, noise model, beams and pixel weighting schemes, we refer the reader to the explanations contained in \citet{KomatsuEtal2009B}. The most stringent limits so far have been obtained by applying the bispectrum estimator to the WMAP datasets. Constraints have been put on the local, equilateral and orthogonal shapes. The best constraints come from the full implementation of the optimal estimator done in \citet{SmithSenatoreZaldarriaga2009} and \citet{SenatoreSmithZaldarriaga2009}. They are, at 95\% C.L.,
\begin{eqnarray}\label{eq:constraints}
-4 & < & \fnll \;\, < 80 \; , \\
-125 & < & \fnl^{equil.} < 435 \; , \\ 
-369 & < & \fnl^{ortho.} < 71
\end{eqnarray}
Since the first release of WMAP data, different groups have used the cubic statistic described in the previous paragraph, either in its pure cubic form \eqn{eq:positionspace}, or in the improved version including the pseudo-optimal linear term implementation \eqn{eq:linearsub}. The results of different analysis of the WMAP 1-year, 3-year and 5-year 
datasets are summarized and commented in table \ref{tab:constraints}, where just the local and equilateral shapes have been included since the only constraint on the orthogonal shape to date has been already mentioned in \eqn{eq:constraints}.

\begin{table}[t]
\begin{tabular}{c |cl | cl }
       & \multicolumn{2}{c}{\bf Local} & \multicolumn{2}{c}{\bf Equilateral} \\ 
       \hline
   \multirow{2}{*}{\bf Pure cubic} & $-58 < \fnl < 134$ & \cite[W1]{KomatsuEtal2003} & $-366 < \fnl < 238$ & \cite[W1]{CreminelliEtal2006}  \\ &  $-54 < \fnl < 114$ & \cite[W3]{SpergelEtal2007} &  $-256 < \fnl <332$  &\cite[W3]{CreminelliEtal2006} 
   \\
  \hline 
  \multirow{5}{*}{\bf Pseudo-optimal}  & $-27 < \fnl < 121$ & \cite[W1]{CreminelliEtal2006}  &  $-151 < \fnl < 253$ & \cite[W5]{KomatsuEtal2009B}  \\ & $-36 < \fnl < 100$  & \cite[W3]{CreminelliEtal2006} &  \\  & $\,\,27\, < \fnl < \, 147$ & \cite[W3]{YadavWandelt2008}   &  \\ & $9\,\, < \fnl < 129\,$ & \cite[W3]{SmithSenatoreZaldarriaga2009} &
   \\ & $\,-9\, < \fnl < \, 111$ &  \cite[W5]{KomatsuEtal2009B} &  \\
  \hline
\multirow{2}{*}{\bf Optimal} & $12<\fnl<104$ & \cite[W3]{SmithSenatoreZaldarriaga2009}  & $-125<\fnl<435$&  \cite[W5]{SmithSenatoreZaldarriaga2009}  \\ & $-4<\fnl<80$ & \cite[W5]{SmithSenatoreZaldarriaga2009}  &        \\
   \hline
\end{tabular}
\caption{Constraints on $\fnl^{local}$,$\fnl^{equil.}$, obtained by different groups on the one-year (W1), three-year (W3), and five-year (W5) 
WMAP data releases. Different rows correspond to the different implementations of the $\fnl$ estimator described in the text: the ``pure cubic'' implementation \eqn{eq:optimalcubicestnorm2} in which no linear term is included, the ``pseudo-optimal'' implementation (\ref{eq:linearsub}) in which a linear term is added but the covariance matrix is assumed diagonal in the cubic term, and the fully ``optimal'' implementation (\ref{eq:cubicestlinear2}). As we noted in the text the linear term is important 
mostly for estimates of local NG, since anisotropic noise ``mimic'' squeezed configuration. For this reason ``pure cubic'' estimates of equilateral NG in the table above are nearly optimal, while local ones are significantly sub-optimal, especially because they have to be confined to the pure signal dominated region $l \lesssim 300$, where the assumption of rotational invariance is correct. There is a certain degree of friction between some of the results shown. In particular the $27  < \fnll <  147$ WMAP 3-year estimate obtained in \citet{YadavWandelt2008}, corresponding to a ``nearly 3-$\sigma$" detection of local NG, seems not to agree well with the $9 < \fnll < 129$, $\sim 2.3 \sigma$ result obtained on the {\em same} data set in \citet{SmithSenatoreZaldarriaga2009}. The origin of the discrepancy is unclear, although it is argued in \citet{SmithSenatoreZaldarriaga2009} that it might be due to differences in the coadding scheme of different data channels, or analogous differences in the choice of some weights. As pointed out in \citet{SmithSenatoreZaldarriaga2009}, one additional advantage of the fully optimal implementation of the estimator is actually that all the ambiguity related to the use of different coadding schemes disappears, since the optimal coadding strategy is automatically selected in the inverse covariance filtering process. Another discrepancy is that between the two equilateral constraints on WMAP 5-year data. It seems that the pseudo-optimal estimator produces better constraints than the optimal one. This is clearly not possible. \citet{SmithSenatoreZaldarriaga2009} claim that their numerical pipeline calculates the theoretical ansatz for the bispectrum shape more accurately than it was done before. That is due to a subtlety that went unnoticed in previous works, consisting in the necessity to extend above the horizon the upper integration limit in the calculation of the equilateral shape related quantities $\beta_{\ell}(r)$, $\gamma_{\ell}(r)$, $\delta_{\ell}(r)$ (see equation \eqn{equilbispectsep}). This is required in order to obtain stable numerical solutions, and it calls for a reasessment of the expected and measured error bars, that actually increase with respect to previous calculations.}
\label{tab:constraints}
\end{table}

\subsection{Fisher matrix forecasts}
\label{sec:forecasts}
 
The fisher matrix, defined as the curvature of the likelihood function calculated in its peak
reassessment (see equation \eqn{eq:fishermatrixdef} in Appendix \ref{sec:estimationbasics}), 
plays a very important and well-known role in parameter estimation theory, not only because it defines the optimality of estimators through the Rao-Cramer bound, but also because it  
allows us to estimate {\em a priori} what the smallest error bars attainable will be for 
a given parameter (see again Appendix \ref{sec:estimationbasics}). In other words, using the Fisher matrix we can forecast how well  a parameter will be measured by a given experiment. This is very useful in order to optimize the experimental design to the detection of the parameters of interest. In our specific case, a Fisher matrix analysis will help us 
to understand what is the smallest $\fnl$ detectable in principle using different CMB datasets, and which experimental features can be improved in order to increase 
the sensitivity to $\fnl$.

\subsubsection{A general derivation}

Formula ({\ref{eq:fisher3}}) from appendix \ref{sec:estimationbasics}, when applied to our case yields:
\be\label{eq:fisherbisp}
F_{\fnl \fnl} = \frac{1}{6} \sum_{\alll=2}^{\ell_{max}} \frac{\(B^{\fnl=1}_{\alll}\)^2}{\bispvar} \; ,
\ee
where $B_{\alll}$ is the angular averaged bispectrum (\ie the measured quantity). This can be rewritten in terms of the reduced bispectrum as:
\be\label{eq:fisherbisp2}
F_{\fnl \fnl} = \frac{1}{6} \sum_{\alll=2}^{\ell_{max}} I_{\alll}^2 \frac{\(b^{\fnl=1}_{\alll}\)^2}{\bispvar} \; ,
\ee
where we have defined (see also $w_{l_1l_2l_3}$ in (\ref{eq:lweightdiscrete})):
\be
I_{\alll} = \sqrt{\frac{(2 \ell_1 +1)(2 \ell_2 + 1)(2 \ell_3 +1)}{4 \pi}}
\( \begin{array}{ccc} \ell_1 & \ell_2 & \ell_3 \\ 0 & 0 & 0 \end{array} \) \; .
\ee 
Note how the features of the experiment enter the Fisher matrix through the parameter $\ell_{max}$, defining the angular resolution, and in the angular power spectrum expression in the denominator, that contains the angular beam and experimental noise:
\be
\tilde C_\ell = {\cal{C}}_{\ell} W_{\ell} + N_{\ell} \; ,
\ee
where ${\cal{C}}_{\ell}$ is the theoretical power spectrum for a given set of cosmological parameters, $W_{\ell}$ is the beam of the experiment, and $N_{\ell}$ is the experimental noise. $N_{\ell}$ is a constant for uncorrelated noise. Likewise, the theoretical bispectrum will be convolved by the experimental beam.
\be
B_{\alll} = {\cal{B}}_{\alll} W_{\ell_1}W_{\ell_2}W_{\ell_3} \; .
\ee
Note that, since the noise is generally Gaussian, its three point function vanishes. 
The experimental noise thus  only enters in the denominator of the Fisher matrix expression. The effects of partial sky coverage can be easily accounted for. From \eqn{eq:fsky} it follows that if only a fraction $f_{sky}$ of the full sky is covered then the Fisher matrix takes a $f_{sky}$ factor in front, that produces a degradation of the error bars of $\sqrt{f_{sky}}$.

We saw previously that for separable shapes the reduced bispectrum can be calculated either analytically, under some simplifying assumptions on the transfer functions (\eg the Sachs-Wolfe approximation), or numerically through formula  (\ref{eq:redbispect}). It is then possible to evaluate numerically the fisher matrix and the corresponding error $\D \fnl \equiv \sqrt{1/F}$. In the context of $\fnl$ estimation, the first calculation of this kind was done for $\fnll$ in \citet{KomatsuSpergel2001}, where it was found that WMAP could reach a sensitivity $\Delta \fnl = 20$ (note how this bound is actually saturated by the optimal estimator results presented in table \ref{tab:constraints}), while Planck
\cite{Planck2006} could go down to $\D \fnl = 5$\footnote{Note that all the errors quoted in this section are at 1-$\sigma$.}. What allows Planck to improve on WMAP is that it has a much better angular resolution and that it is cosmic variance dominated in a very large range of scales \ie the power spectrum signal $C_{\ell}b_{\ell}$ is larger than the noise $N_{\ell}$
  up to $\ell_{max} = 2000$. Angular resolution and sensitivity are the two factors that increase the ability of a CMB experiment to constrain $\fnl$. This information is provided by the Fisher matrix expression \eqn{eq:fisherbisp2}. Looking at such expression, we notice how the signal-to-noise ratio is obtained by adding over all the bispectrum configurations up to $\ell_{max}$, weighted by their variance. Thus, the higher $ \ell_{max}$ is, the more configurations are included in the sum and the larger is the final sensitivity to $\fnl$. 
On the other hand we see that, if the power spectrum of the instrumental noise appearing in the variance term in the  denominator dominates from a certain $\ell_{S=N}$, then the signal 
contribution is suppressed above that threshold by the noise power spectra appearing in the denominator of \eqn{eq:fisherbisp2}. 
So what determines the sensitivity of a CMB experiment to $\fnl$ is the range of 
$\ell$ over which the instrumental noise is low, so the experiment is cosmic variance dominated. 
This range is $\ell \lesssim 2000$ for Planck and $\ell \lesssim 500$ for WMAP, 
hence Planck can obtain tighter constraints than WMAP. This is shown in \fig{fig:fisher}, where Fisher matrix forecasts of $\fnl$ are plotted for different CMB experiments: the predicted error bars decrease with $\ell$ up to the angular scale at which the measurements start to be noise dominated, after which the $\fnl$ signal-to-noise ratio saturates. 
A simple calculation done in \citet{BabichZaldarriaga2004} taking the Sachs Wolfe approximation, and working in flat sky, showed that before noise dominates the signal-to-noise ratio for the local shape grows as:
\be
\frac{S}{N} \propto \ell_{max} \ln\(\frac{\ell_{max}}{\ell_{min}}\) \; ,
\ee 
where the ($\ln$) is dictated by the coupling between large and small scales introduced by squeezed configurations, from which most of the local signal comes.

Note also how, in absence of experimental noise, the beams  in the numerator and in the denominator of \eqn{eq:fisherbisp2} cancel each other out. An ideal noiseless CMB experiment would then have a signal to noise ratio indefinitely growing.  However, this would not imply infinite sensitivity to $\fnl$, because above a certain $\ell_{max}$ secondary anisotropies would start to dominate. Fisher matrix analysis of the equilateral shape \citep[\eg]{SmithZaldarriaga2006, SefusattiEtal2009} showed that the minimum achievable error bars in this case are $\D \fnl \simeq 100$ and $\D \fnl \simeq 60$, for WMAP and Planck respectively\footnote{Note how the larger error bars in this case with respect to the local constraints does not reflect a higher sensitivity of CMB measurement to $\fnll$, but only the conventional choice of the normalization of the bispectrum amplitude in the definition of $\fnl$. The normalizations are in fact chosen in such a way that the bispectra have the same value for equilateral configurations $\ell_1 = \ell_2 = \ell_3$, where the local bispectrum is suppressed, and the equilateral bispectrum is peaked.}. Additional shapes are studied in \citet{SmithZaldarriaga2006}.

\subsubsection{Polarization}
\label{sec:polarization}

\begin{figure*}[t]
\centering
\includegraphics[width=0.48\textwidth]{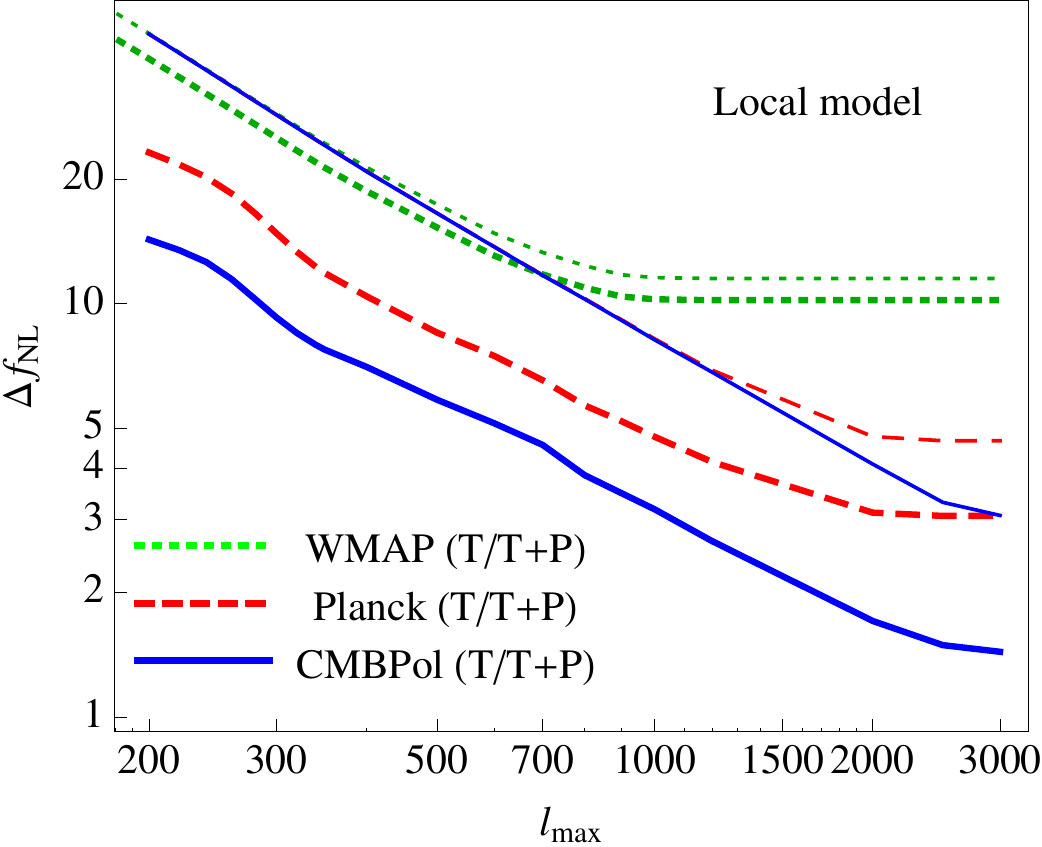}
\includegraphics[width=0.48\textwidth]{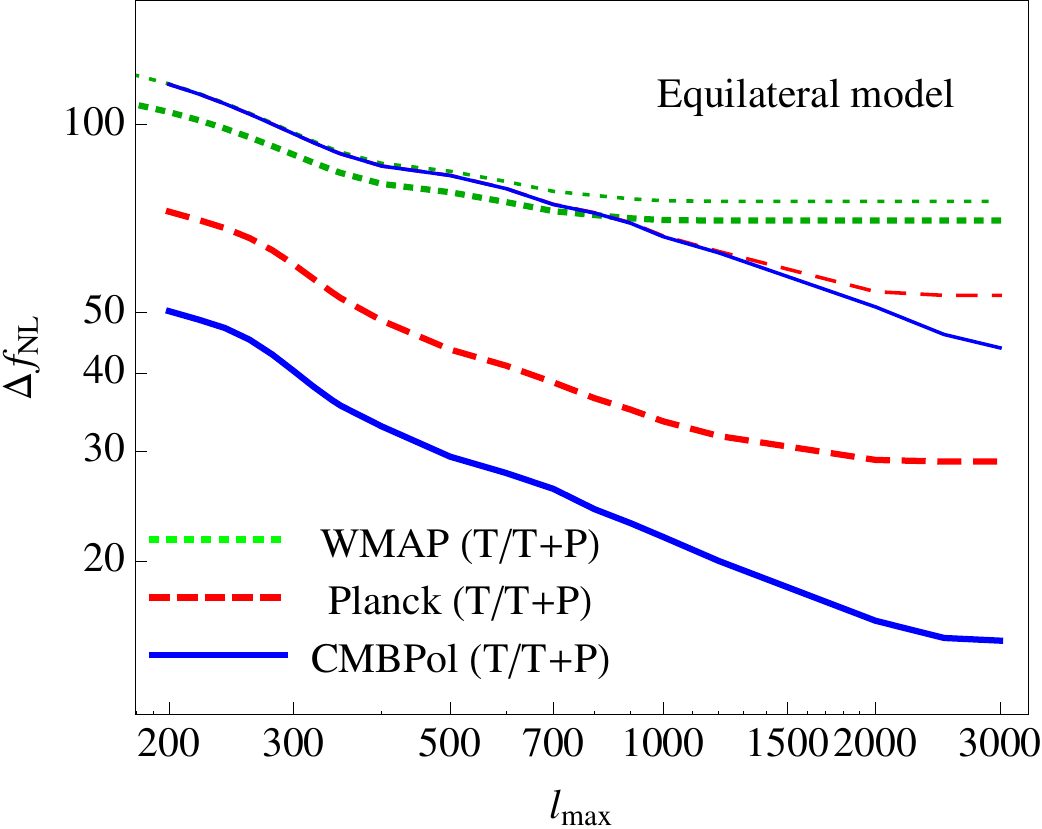}
\caption{Fisher matrix forecasts on $\Delta \fnl$, featured for different experiments: WMAP ({\em green, dotted lines}), Planck ({\em red, dashed lines}) and the proposed CMBpol \cite{BaumannEtal2009} survey ({\em blue, solid lines}). The left panel shows results for the local shape, while the right panel refers to the equilateral shape. Thin 
lines are obtained from temperature data only, and thick lines show the improvement in the error bars coming from adding polarization datasets to the analysis for the various experiment.}
\label{fig:fisher}    
\end{figure*}

\citet{BabichZaldarriaga2004} showed with a Fisher matrix analysis that the CMB E-mode polarization measurements can be used to improve the sensitivity to $\fnl$. Although we have dealt so far only with temperature bispectra and related estimators, including polarization is fairly straightforward. As usual the calculation starts from the formula \eqn{eq:phi2alm1} linking the multipoles of CMB anisotropies to the primordial potential $\Phi$, but this time including the polarization radiative transfer $\D^E_{\ell}(k)$ in the convolution integral:
\be\label{eq:phi2almE}
a^E_{\ell m} = 4\pi (-i)^l \int \frac{d^3 k}{(2\pi)^3}\, \D^E_{\ell}(k) \,\O(\bk)\, Y_{\ell m}(\uk)\,. \; ,
\ee
The bispectrum is then defined in the usual way, but this time more configurations can be built 
by correlating temperature and polarization multipoles:
\begin{eqnarray} 
B_{\alll}^{TTT} & \equiv & \left \langle \almone^T \almtwo^T \almthree^T \right\rangle \nonumber \; , \\
B_{\alll}^{TTE} & \equiv & \left \langle \almone^T \almtwo^T \almthree^E \right\rangle \nonumber \; , \\
B_{\alll}^{TET} & \equiv & \left \langle \almone^T \almtwo^E \almthree^T \right\rangle \nonumber \; , \\ 
& \vdots & \nonumber \\
B_{\alll}^{EET} & \equiv & \left \langle \almone^E \almtwo^E \almthree^T \right\rangle  \nonumber \; , \\
& \vdots & \nonumber \\
B_{\alll}^{EEE} & \equiv & \left \langle \almone^E \almtwo^E \almthree^E \right\rangle \; .
\end{eqnarray}
The point to emphasize is that the polarization signal is generated on scales where the temperature signal is suppressed by Silk damping. The polarization bispectra thus open a window over a new $\mathbf{k}$-range in the $3D \rightarrow 2D$ 
projection $\mathbf{k} \rightarrow \ell$, and increase the overall information available. In other words, since the new configurations $TTE$, $TEE$, etc. including polarization are partially independent of the pure temperature ($TTT$) bispectrum, adding those additional 
configurations to the Fisher matrix (and to the actual $\fnl$ estimation from data) increases the total signal available. The Fisher matrix 
expression now becomes:
\be\label{eq:fisherpol}
F = \sum_{pqr}\sum_{ijk} \sum_{\alll} B_{\alll}^{pqr} [{\rm cov}^{-1}]_{pqr|ijk}^{\alll} B_{\alll}^{ijk} \; ,
\ee
where $i,j,k,p,q$ and $r$ run over the $T$ and $E$ superscripts. We still work in the 
assumption that all the quantities involved are Gaussian, but now the 
different bispectra of temperature and polarization are correlated for a 
given configuration $\lall$, thus defining a multivariate Gaussian distribution. The full covariance matrix between bispectra (indicated by ${\rm cov}$ in the formula above)
has then to be evaluated. A numerical evaluation of \eqn{eq:fisherpol} shows \citep{BabichZaldarriaga2004} that, for an ideal (\ie noiseless) experiment, adding the polarization signal produces an improvement of a factor $\sim 2$ on $\fnl$ constraints. For WMAP, adding polarization bispectra produces very little improvement, since polarization data are mostly noise dominated. For Planck, however, including polarization does generate a significant improvement, bringing the forecasted error bars from $\Delta \fnl \simeq 5$ to $\Delta \fnl \simeq 3.5$ . Some error bar forecasts from temperature and polarization bispectra as a function of $\ell_{max}$ for different experimental designs including WMAP and Planck are shown in \fig{fig:fisher}. Motivated by this analysis, \citet{YadavKomatsuWandelt2007,YadavEtal2008} have implemented a bispectrum estimator of $\fnl$ including both temperature and polarization bispectra. All the general considerations about optimality and the numerical implementation techniques described in previous sections apply in an analogous way to the temperature $+$ polarization case, although the presence of additional bispectra with a non-trivial covariance matrix introduces a few additional technical complications. We refer the reader to \citet{YadavKomatsuWandelt2007,YadavEtal2008} for further discussion.

\subsection{Non-Gaussian contaminants}
\label{sec:NGcontaminants}

So far we have considered only primordial non-Gaussianity as a source of the three-point function of the CMB. However many other astrophysical and cosmological effects can produce an observable angular bispectrum. Among these, diffuse astrophysical foreground emission \citep[see \eg][and references therein]{BennettEtal2003B, LeachEtal2008}, unresolved point sources \citep[\eg][]{KomatsuEtal2003} and secondary anisotropies are probably the most important NG sources. Since the main focus on this review is on the primordial bispectrum, we will not describe this NG sources in great detail. We will however outline in this section their main effects in order to understand if, and how,  they could contaminate an estimate of primordial NG. Let us consider a number $N_s$ of sources of a  CMB bispectrum signal and call $B_{\alll}^i$ the bispectrum produced by the $i$-th source. Let us also indicate with $B_{\alll}^{\fnl=1}$ the primordial component of the bispectrum calculated for $\fnl=1$.  For our purposes $B_{\alll}^{\fnl=1}$ is the signal that we want to measure, while the other bispectra are contaminants, that we would like to eliminate. The total bispectrum of the map in presence of these contaminants is then:
\be
B_{\alll} = \fnl B^{\fnl=1}_{\alll} + \sum_{i=1}^{N_s} A_i B_{\alll}^i \; ,
\ee
where $A_i$ is the amplitude of the $i$-th bispectrum. If we have a precise 
prediction of the bispectra generated by the contaminants, we can then think of extending our $\fnl$ estimator to a joint-estimator of all the amplitude parameters. The optimal cubic $\fnl$ estimator defined in \eqn{eq:optimalcubicestnorm} would then be generalized to the multi-parameter case by minimizing the following $\chi^2$:
\be\label{eq:joint}
\chi^2\(\fnl,A_i\) = \sum_{\alll} \frac{\( \fnl B^{\fnl=1}_{\alll} + \sum_{i=1}^{N_s} A_i B_{\alll}^i - B_{\alll}^{obs}\)}{\bispvar} \; .
\ee
The new errors on $\fnl$ in this case can be forecasted as usual by means of a Fisher matrix analysis. The Fisher matrix described in the previous paragraph can be generalized straightforwardly to the multi-parameter case. In this case $F$ becomes an array whose entries are defined as:
\be 
F_{i j} = \sum_{\alll} \frac{B^{i}_{\alll} B^{j}_{\alll}}{\bispvar} \; .
\ee
The optimal errors on a given amplitude $A_i$ (including $\fnl$) then become, according to the multi-dimensional generalization of the Rao-Cramer bound: 
\be
\Delta A_i = \sqrt{\(F^{-1}\)_{ii}} \; ,
\ee
where the crucial point to notice is that we now {\em first} invert the Fisher matrix and then we take the square root of the diagonal elements to find the errors. This is the error that is obtained when the full joint-parameter likelihood is calculated and then the 1-dimensional likelihood for a given parameter is obtained by integrating out all the other degrees of freedom: a process defined in statistics as {\em marginalization}. One can see that the inverse of the Fisher matrix defines the {\em covariance matrix} of the parameters under study.  If the various parameters are completely uncorrelated, then the Fisher matrix is diagonal and we would have $F_{ii}^{-1} = \(F^{-1}\)_{ii}$, showing that the parameters can obviously be estimated independently and the marginalization process doesn't change the error bars on a given parameter of interest (in our case $\fnl$). If the different parameters are  correlated, however, then off-diagonal terms appear in the Fisher matrix and the error bars after marginalization (\ie the ``real'' error bars to quote in the results) are larger
 than those that would have been obtained by naively neglecting contaminants. An obvious but useful observation is that two bispectral amplitudes will be strongly correlated when the respective shapes are similar. To make a practical example, the bispectrum generated 
by correlating weak lensing of CMB anisotropies with the Integrated Sachs-Wolfe (ISW) effect can be shown to be peaked on squeezed configurations. For this reason the presence of this effect can be a significant contaminant for estimates of local non-Gaussianity.

So far in this section we have described the degradation effects on the error bars if an hypothetical joint-estimator of all the CMB bispectrum amplitudes was built, and the amplitudes of contaminants were marginalized over to estimate $\fnl$. However a joint-estimation might be difficult, due to factors like the presence of theoretical uncertainties on the shapes of contaminant bispectra or possible practical difficulties in finding an efficient implementation of this full bispectrum-likelihood estimator (\eg if the additional 
secondary bispectra are non-separable). As a result, the practical approach so far has been to estimate only $\fnl$ using the techniques described in previous sections, and neglect possible non-primordial contaminants. In this case the possible effect of contaminants would not show up as a degradation of the error bars but in an even worse 
way, by introducing a {\em bias} in the $\fnl$ measurements. Let us see this by assuming that the CMB three point function takes contributions both from a primordial NG component and from a contaminant bispectrum with amplitude $A_i$. Let us also assume that we can produce a set of NG Monte Carlo simulations of CMB maps including both bispectra. We assign a given $\fnl$ in input to the primordial component of our simulated maps. Finally we estimate the average $\fnl$ obtained from the simulations by applying the usual optimal cubic statistic described so far. The result of our MC average will be:
\bea
\left \langle \fnlest \right \rangle & = & \frac{1}{{\cal{N}}} \sum_{\alll} \frac{B^{\fnl=1}_{\alll} B^{observed}_{\alll}}{\bispvar} 
\nonumber \\
& = & \fnl + \frac{1}{{\cal{N}}} \sum_{\alll} \frac{B^{\fnl=1}_{\alll} B^{cont.}_{\alll}}{\bispvar}\label{eq:bias1} \; ,
\eea
where $B^{observed}$ is the averaged bispectrum extracted from the map. The $\fnl$ term on the r.h.s. of the second line comes from the fact that the normalization ${\cal{N}}$ is chosen in such a way as to obtain an unbiased estimator of the primordial component. However a second term is present, which accounts for the fact that a contaminant bispectrum, $B^{cont.}_{\alll}$ is in the map; this term clearly biases the estimator\footnote{Let us remind that by definition an estimator of a parameter $\l$ (in our case $\fnl$) is unbiased if $\left \langle \hat{\l} \right \rangle = \l$, $\hat{\l}$ being the estimate from data, and $\l$ being the true parameter of the underlying model.}. The magnitude of the bias will depend again on how similar the shape of the contaminant bispectrum is to the primordial one. If, for example, the contaminant bispectrum is strongly peaked on equilateral configurations and suppressed on squeezed ones, a local estimator of NG will then not be significantly biased by it, since the second term in equation \eqn{eq:bias1} will cancel-out. However, an estimate of $\fnl^{equil}$ will in this case be significantly biased.

In general we can define the correlation coefficients between two bispectra, labeled $i$ and $j$, as:
\be
r_{ij} = \frac{\sum_{\alll} \frac{B^{(i)}_{\alll} B^{(j)}_{\alll}}{\bispvar}}
{\sqrt{\frac{\sum_{\alll} B_{\alll}^{(i) 2}}{\bispvar}} 
\sqrt{ \frac{\sum_{\alll} B_{\alll}^{(j) 2}}{\bispvar}}}  \; .
\ee
The definition of ``correlation coefficient'' becomes completely transparent if we rewrite the previous formula in terms of the Fisher matrix, and keep in mind that $F^{-1}$ define the covariance matrix of the bispectrum amplitudes:
\be\label{eq:correlation}
r_{ij} = \frac{\(F^{-1}\)_{ij}}{\sqrt{{\(F^{-1}\)_{ii}} {\(F^{-1}\)_{jj}}}} \; .
\ee
The correlation coefficient vary by definition from $0$, for totally uncorrelated shapes, to $1$, for identical shapes. The more a given contaminant bispectrum is correlated to the primordial bispectrum that we want to measure, the larger will be the induced bias. At this point we distinguish between three possibilities. The first is that the contaminant bispectrum shape and amplitude are perfectly known. In that case we can compute the expected bias from formula \eqn{eq:bias} and subtract from our estimate. The second possibility is that the shape of the contaminant bispectrum is known, but its amplitude is defined with a given uncertainty. In this case we can propagate this uncertainty by quoting it in addition
 to the statistical error bars on $\fnl$ obtained in the usual way. The third and worst possibility is that we are unaware of the presence of some contaminant effect, or we know nothing about its bispectrum. In this case we might obtain a biased estimate of $\fnl$ without knowing it and thus eventually misinterpret a spurious NG effect as primordial NG. Contaminants are then very dangerous, because if not properly taken into account can lead to spurious claim of detection of primordial NG. For this reason, if a positive detection of $\fnl$ were to be made at some point for a certain model, all possible tests for the presence of contaminant effects should be performed. Moreover, since we cannot be absolutely sure that we are considering all possible source of NG contamination, {\em cross-validation} of the result using other non-bispectrum based estimators will be very important. These other 
estimators (Minkowski Functionals, wavelets, needlets, higher order correlators are just some examples among those considered in the literature) are by construction sub-optimal estimators of the primordial component. However, in principle they are expected to produce a 
totally different response to NG contaminants than the primordial bispectrum. A cross-detection of $\fnl$ with many different statistics would then be much less 
likely due to some unknown spurious effect. Another way to test the primordial origin of an observed NG signal, recently proposed in \citet{MunshiHeavens2009}, is to modify the optimal 
bispectrum estimator in order to evaluate a function of $\ell$ rather than a single amplitude $\fnl$. The point is that if a clear detection of $\fnl$ is achieved at several $\sigma$, then the signal is large enough to allow a less radical data compression. \citet{MunshiHeavens2009} have then recently proposed to estimate the ``bispectrum related power spectrum'', $C_{\ell}^{skew}$ defined as:
\be
C_{\ell}^{skew} = \frac{1}{2 (\ell+1)} \sum_{\ell_1 \ell_2} \frac{B_{\alll}^{\fnl=1} B_{\alll}^{obs}}{\bispvar} \; .
\ee
Like in the usual $\fnl$ estimator, the optimal $S/N$ weighting is included, and the observed bispectrum from the map is correlated to the theoretical shape. However in this case we don't measure the overall amplitude, but rather the amplitude for each $\ell$-bin. Note that
\be
\fnlest \equiv \frac{1}{{\cal{N}}} \sum_\ell C_\ell^{skew} \; ; 
\ee
by construction of $C_{\ell}^{skew}$, the usual $\fnl$ estimator is then retrieved by summing the bispectrum-related power spectrum over all the $\ell$. The general idea is now that the functional dependence of this skew power spectrum on $\ell$ will show significant variation between different sources of NG, allowing a clearer test of the hypothesis that the origin of the observed signal is primordial. A number of investigations of WMAP data have already been performed using this statistic in order to look for primordial and secondary signals \citep{CalabreseEtal2009,SmidtEtal2009A}, and related pseudo-$C_l$ statistics have been developed in \citet{MunshiEtal2009}.

In any case, as long as bispectrum estimators are considered, independently of the specific statistic or implementation, the best way to deal with NG contaminants is to make sure to list all of them and study their bispectra, or at least find ways to assess their potential impact on the final results. In the following paragraphs we will then turn our attention to a classification of the most important potential sources of spurious NG, and see how they are treated in the primordial bispectrum analysis. Finally, we will consider some effects that interact with the $\fnl$ measurement not necessarily by directly producing a secondary bispectrum, but rather by changing the normalization of the estimator or by increasing the error bars without producing any bias.

\subsubsection{Diffuse foreground emission}

There are three main astrophysical effects producing a galactic microwave emission from our galaxy in the typical frequency range of a CMB experiment \citep{BennettEtal2003B,Dodelson2003}: free-free emission from electron-ion scattering, synchrotron emission from acceleration of cosmic ray electrons in magnetic fields, and thermal dust emission.

Since these sources produce signals with a peculiar spectral and spatial distribution, multifrequency observations allows the separation of them from the primordial component of the CMB signal by suitable {\em component separation} algorithms. In the resulting 
``cleaned'' map the foreground contribution to the $\alm$ is minimized, although obviously 
it can never be completely eliminated. The remaining foreground contamination after cleaning is called the {\em foreground residual}. Note that the emission from the galactic plane of the CMB map is so strong that a clean separation of the primordial CMB component from the foregrounds is impossible. The galactic regions that are too contaminated to produce a clean component separation have to be masked out in the analysis. 
The size of the galactic mask will depend on the choice of the foreground flux level above which the pixel is considered too contaminated to be included in the analysis.
 The choice of the cut-off will depend on the specific analysis that one wants to perform on the data. Since the primordial NG signal is much smaller than the Gaussian component, more conservative masks (i.e.\ larger) need generally to be used for $\fnl$ estimates than those applied to $C_{\ell}$ estimation. Direct information about the spatial distribution of foreground emission in the sky (\ie free-free, synchrotron or dust) is provided in the form of templates, obtained either from the most foreground contaminated channels of the CMB experiment itself, or from external astrophysical surveys (\eg observations of radio-emission, maps of H$\alpha$ emission). Templates are affected by several sources of uncertainties and errors \citep[see \eg][]{BennettEtal2003B} and using them in assessing the possible impact on $\fnl$ of foreground emission or residuals has both advantages and disadvantages. The safest approach is probably to combine internal consistency tests 
on the data with analysis involving the use of templates.

The first extensive tests of possible foreground contamination in $\fnl$ measurements were performed in \citet{YadavWandelt2008}, where a detection of a primordial local signal at above $99.5 \%$ level on WMAP 3-years data was claimed. 
As explained earlier (see caption of table \ref{tab:constraints}), further analysis on more recent datasets and/or using more optimal estimators have led to an updated $\fnll$ estimate that is about 2-$\sigma$ away from the origin, \ie just a ``hint'' of a possible local signal, rather than a detection. However, as long as a detection was claimed in \citet{YadavWandelt2008}, tests to exclude a possible contamination from diffuse foregrounds 
had to be carried out. In this case the authors relied mostly on the ``internal consistency test'' approach. Their analysis included:
\begin{enumerate}
\item Expanding the original galactic mask in order to see if the estimated value of $\fnl$ is stable for different choice of the mask. A significantly lower value of $\fnl$ for a larger mask might mean that some unmasked noise contribution is affecting the measurement with the original mask.
\item Comparing $\fnl$ estimates from foreground reduced maps to estimates from ``raw'' maps that include a galactic mask, but have not gone through a component separation process. If foregrounds have a significant impact on $\fnl$ then one expects the measurements from raw and reduced maps to differ significantly.
\item Comparing different frequency channels. If foregrounds significantly contaminate measurements at given frequencies, then different channels should produce different results.
\end{enumerate}
Analysis involving some kind of prior information about foreground emission were carried on both in \citet{YadavWandelt2008} and \citet{SmithSenatoreZaldarriaga2009}. The two approaches adopted in this case were:
\begin{enumerate}
\item Producing simulations including both a Gaussian primordial CMB signal and the foreground emission. The latter has in this case to be generated according to a model that allows for a good reconstruction of the observed templates. The $\fnl$ estimator can then be 
applied to these simulations in order to check if the measured $\fnl$ is consistent with 0 (as it should be, in absence of significant foreground contamination, since the primordial input is Gaussian).
\item For an optimal estimator including full $C^{-1}$ pre-filtering \citep{SmithSenatoreZaldarriaga2009}, adding the foreground templates to the 
noise covariance, by assigning {\em infinite} variance to each template $T^{i}_{templ}(\hat{\bf n})$. In this way the estimate is ``blind'' to the template amplitudes. This produces a loss of information that in turn determines an increase of the variance. 
The larger the contamination from foreground is, the more the variance increase. For negligible contamination, the variance stays the same. In any case, the effect of foregrounds is entirely included in the error bars, provided the assumed templates are accurate enough. This method of analysis, called {\em template marginalization}, is adopted in \citet{SmithSenatoreZaldarriaga2009}. A complete mathematical derivation of this method is provided in \citet{RybickiPress1992}.
\end{enumerate}
In addition to the methods outlined above, there is also the possibility of using the foreground templates for a joint-estimation of $\fnl$ and of the templates amplitudes (see equation \eqn{eq:joint}). This approach has been recently used in \citet{CabellaEtal2009} 
for a needlet estimator. It could be obviously reapplied in the same form to a bispectrum estimator.

In conclusion, all the tests above have been applied to WMAP 3-years and 5-years data releases. No evidence for the presence of a significant contamination of the local $\fnl$ measurement from diffuse foreground was produced. Other shapes of $\fnl$ were not considered since the only type of non-Gaussianity that has produced a marginal detection is so far the local one. Although diffuse foregrounds and foreground residuals do not seem to contaminate primordial NG measurements in WMAP, this is not guaranteed to hold true for Planck, due to its much higher sensitivity.

\subsubsection{Unresolved point sources}  

Extragalactic point sources are the most important foreground at small angular scales \citep[see][]{WrightEtal2009}. Sources are identified by searching the maps for bright spots that fit the beam profile, and then masked out. However not all the sources can be resolved and eliminated in this way. Unresolved point sources contaminate the map and are a source of a NG signal that can potentially interfere with primordial NG measurements. Unclustered extra-galactic point sources have a Poisson distribution and their bispectrum is then 
simply a constant:
\be
b^{ps}_{\alll} = b^{ps} \; ,
\ee
with an amplitude that has to be estimated from the data and depends on the level of contamination from unresolved sources. We can now use equation \eqn{eq:correlation} to estimate the correlation between primordial shapes and the point source bispectrum. For a given choice of the amplitude we can also estimate the expected bias on the $\fnl$ estimator.  Simulations of NG maps including the bispectrum from point sources can also be produced and the primordial $\fnl$ estimator for different shapes applied to them in order to estimate the bias. Finally, since $b^{ps}_{\alll}$ is manifestly separable, an estimator of $b^{ps}$ can be built. All of these analysis were performed in \citet{KomatsuEtal2003,KomatsuEtal2009B} on local and equilateral shapes to conclude that point sources do not contaminate significantly the estimate of $\fnll$. On the other hand they have a larger impact on $\fnlequil$: their induced bias from MC simulations is $\D\fnlequil = 22 \pm 4$, to be compared to the statistical error bar $\D\fnlequil \sim 100$.  Additional tests were performed in \citet{SmithSenatoreZaldarriaga2009} to account for the possible presence of clustered unresolved point sources. No significant contamination on $\fnll$ was found in this case. As for the diffuse foreground case, the enhanced $\fnl$ sensitivity that Planck can achieve with respect to WMAP might increase the impact of these  effects.

\subsubsection{Secondary anisotropies}

One big advantage of using CMB anisotropies to test primordial NG is that they are small and can then be treated in the linear regime. The CMB temperature fluctuation field is thus linked to the primordial potential through a {\em linear} convolution with radiation transfer functions, as we saw earlier. At this level, the Gaussianity of the primordial potential is conserved in the CMB temperature fluctuation field. If, however, we work at second order in perturbation theory, the initial conditions are propagated {\em non-linearly} into the observed CMB anisotropies, and the resulting CMB fluctuations are mildly non-Gaussian
 even starting from a Gaussian primordial curvature field. 
Second order effects are clearly very small. However they may well be of the same order 
of magnitude as primordial NG, since the NG component of the primordial potential is $\ordpar$. In conclusion, {\em secondary anisotropies are a potential source of CMB NG, at a level that could in principle contaminate estimates of primordial non-Gaussianity}. 
To fully account for these effects, it is necessary to obtain a relation analogous to equation \eqn{eq:phi2alm1}, but to second order in perturbation theory. Radiation transfer functions are obtained at first order by solving the linearized system of Boltzmann-Einstein
 equations \citep[see \eg][]{Dodelson2003, MaBertschinger1995}. The same equations will then have to be expanded and  numerically integrated at second order in this case. Having obtained second order transfer functions, the full angular bispectrum of secondary anisotropies can be calculated and correlated to the primordial one in order to check for the presence of contaminant effects. 
A full second order Boltzmann code is actually not yet available, although much progress has been made over the past few years in that direction. The full system of second order Einstein-Boltzmann equations has been derived in \citep{BartoloMatarreseRiotto2006A, BartoloMatarreseRiotto2006B,PitrouUzanBernardeau2008,Pitrou2009} and partially integrated numerically in \citep{NittaEtal2009} including only the source terms that can be written as product of first order perturbations. These terms have been shown to produce a totally negligible NG contamination. No numerical evaluation is available to date of the ``genuine'' second order source terms. Although a full solution of the relevant equations hasn't been obtained yet, a significant number of secondary effects are known, and have been modeled for some time. Among these there are for example weak lensing, Sunyaev-Zeldovich (SZ) effect, Rees-Sciama (RS) effect, and so on. Therefore, a natural approach that was adopted in the literature consisted in studying the bispectra arising from these well-known effects 
and from their correlations (\eg ISW-lensing correlation, SZ-lensing correlation and so on). 
It goes beyond the purpose of this review to discuss in detail these results and their implications. Let us just mention them briefly. A fisher matrix analysis in \citet{SerraCooray2008} showed that the combination of bispectra arising from ISW-lensing, SZ-lensing and unresolved point sources produced a negligible contamination at the angular resolution and sensitivity of WMAP, but a significant one for an experiment with the characteristics of Planck. It was in particular shown that estimates of local NG would be biased, especially by ISW-lensing correlation, with $\fnl^{bias} \sim 10$ for local NG. A similar result on ISW-lensing was obtained in another Fisher matrix analysis by \citet{SmithZaldarriaga2006}, and a similar level of contamination was found in \citet{MangilliVerde2009} for the analogous RS-lensing bispectrum. A bispectrum estimator of local and equilateral NG was applied to simulated lensed primordial NG CMB maps by 
\citet{HansonEtal2009}, and three main effects were studied: a possible bias induced by neglecting the lensing of primordial bispectrum in the normalization and weights of the estimator, an increase of the variance due to lensing-produced higher order correlators, and 
ISW-lensing bias. The only significant effect turned out to be the ISW-lensing bias on $\fnll$, at a level confirming Fisher matrix predictions. Note that this bias, being well-known and expected, can be simply calculated and subtracted from future Planck estimates. 
The reason why the coupling between lensing and ISW tend to bias the local estimate can be understood physically: {\em large} scale potential fluctuations source the ISW effect and produce a lensing signal on {\em small} scales, generating a NG signal on squeezed triangles. 
Although both the primordial local bispectrum and the ISW-lensing bispectrum are peaked on squeezed triangles, the presence of acoustic oscillations in the primordial configurations reduce the overall correlation between the two shapes, thus making the final bias significant, but not too large. In order to conclude our brief survey of studies of secondary bispectra, let us mention the work done in \citet{BabichPierpaoli2008}, where point source density modulation bispectra induced by lensing magnification and selection effects, as well as SZ modulation from lensing magnification were studied. The conclusion was again that these effect are negligible for WMAP but close to the sensitivity level of Planck. Despite the great attention received so far in the literature, much work still has to be done in the area of assessing NG contamination from secondary sources. It is clear that a complete and accurate description of secondary bispectra will be crucial for analysis of the future Planck data set.

\subsubsection{Non-Gaussian noise}

Systematics are another potential cause of contamination beyond astrophysical and cosmological sources. The noise in the experiment is generally well described as Gaussian. 
However possible non-Gaussian properties have to be tested in our context. This was done in 
\citet{YadavWandelt2008} by taking differences of yearly WMAP data in order to create jackknife realizations of WMAP noise maps for different detectors, including instrument systematics. The estimator can then be applied to these realizations in order to check that a negligible $\fnl$ is measured. This was the result obtained on the WMAP 3-year data-set.

\subsubsection{Other effects}

In this section we quickly summarize other effects that could interfere with estimates of primordial non-Gaussianity, but did not fit the classification above in the sense that they do not correspond to NG effects contaminating the CMB sky or the instrument noise. 

One of this effects is ${1/f}$ noise, expected to affect especially the low frequency channels of Planck. The ${1/f}$ noise component is generally removed from the map using ``destriping'' algorithms \citep[see \eg][]{MainoEtal2002, Efstathiou2005}. The unsubtracted ``destriping residuals'' form a Gaussian {\em correlated} random field in pixel space. Their non-trivial covariance matrix should in principle be included in the inverse covariance pre-filtering of the optimal estimator. If not included in the pre-filtering, this effect could in principle enhance the estimator error bars (although it cannot generate any bias, since it is Gaussian). Unfortunately, a full numerical evaluation of this covariance matrix is quite challenging. \citet{DonzelliEtal2009} applied the estimator in its pseudo-optimal 
implementation to maps of Gaussian CMB signal $+$ noise, accounting only for anisotropic noise in the linear term, but including destriping residuals in the noise model adopted for the simulations. The final result shows that the error bars do not increase when ${1/f}$ noise effects are included in the simulations, even though they are neglected in the covariance matrix appearing in the estimator.  

Another effect to take into account for Planck is that of an asymmetric beam. The beam in the estimator normalization term is approximated as a circular beam. However Planck optical simulations \citep[\eg][]{SandriEtal2004} show that in reality we have to deal with elliptic beams, characterized by a non-trivial azimuthal dependence. If the circular beam approximation in the normalization of the estimate is not accurate enough, a bias could be introduced. Moreover the anisotropy of the beam could cause an increase of the variance if neglected in the inverse covariance pre-filtering. Again, these effects were found to be negligible in tests on realistic simulations performed by  \citet{DonzelliEtal2009}.

Finally, the estimate of $\fnl$ is done assuming a given cosmological model, \ie\ by fixing all the other cosmological parameters to their best-fit value obtained from a likelihood analysis of the angular power spectrum. Since they are themselves the product of a statistical estimation process, these values obviously present uncertainties that should be propagated into the final $\fnl$-error bars\footnote{In particular, since we are not doing a joint-likelihood estimation of all the parameters and marginalizing to get $\fnl$ (that would be the optimal but time consuming approach), the effect of uncertainties in the parameters propagate onto the $\fnl$ measure as a {\em bias}. This bias has to be evaluated and quoted in addition to the usual statistical $\fnl$ error bar}. This calculation was done in \citet{LiguoriRiotto2008}, where it was found that the propagated error is $\fnl$ 
dependent and it can become important only if a large $\fnl$ will be detected in the data at some point.

\subsection{Generation of simulated non-Gaussian CMB maps}
\label{sec:mapmaking}

\begin{figure*}[t]
\centering
\subfigure{\includegraphics[width=0.47\textwidth]{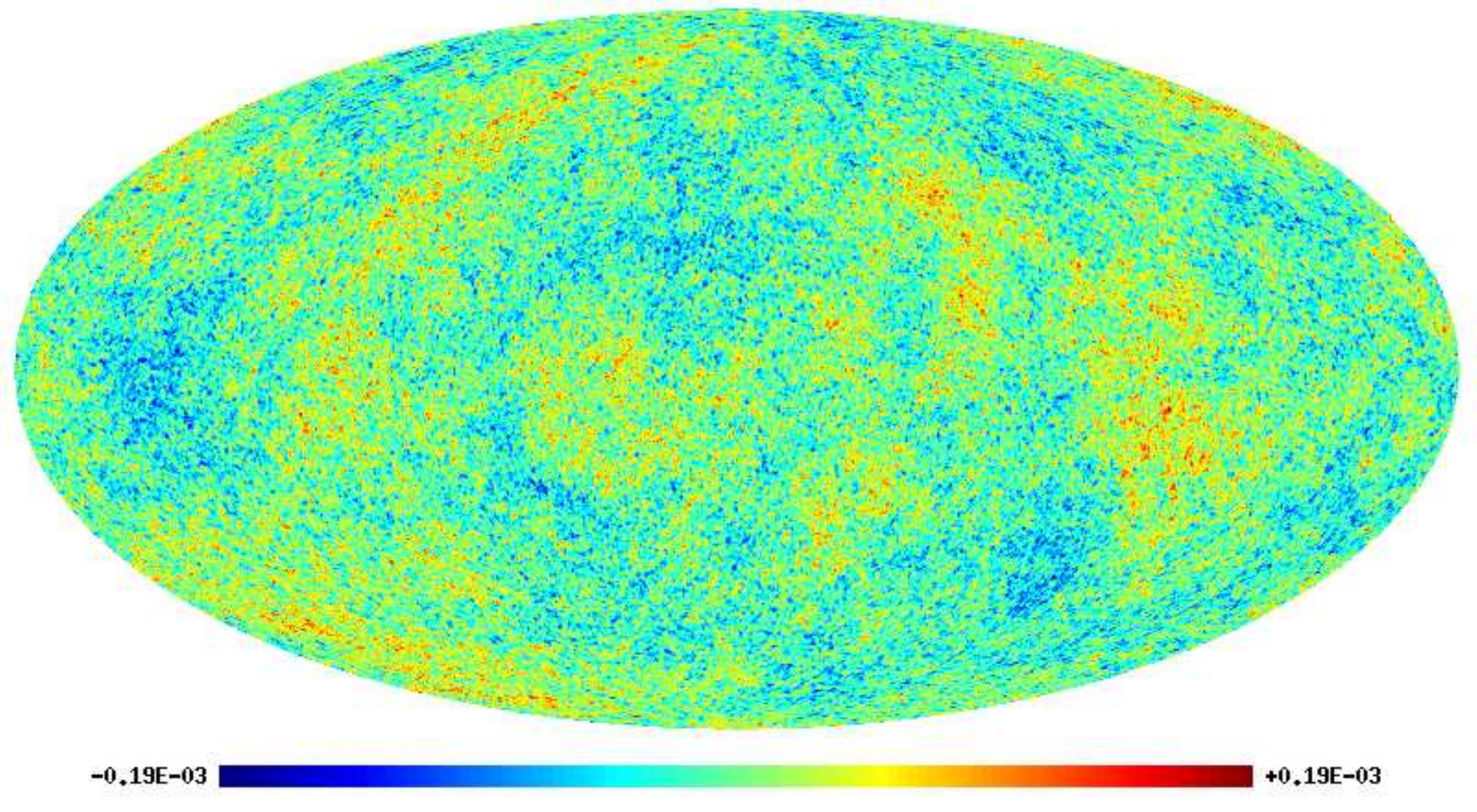}} \qquad
\subfigure{\includegraphics[width=0.47\textwidth]{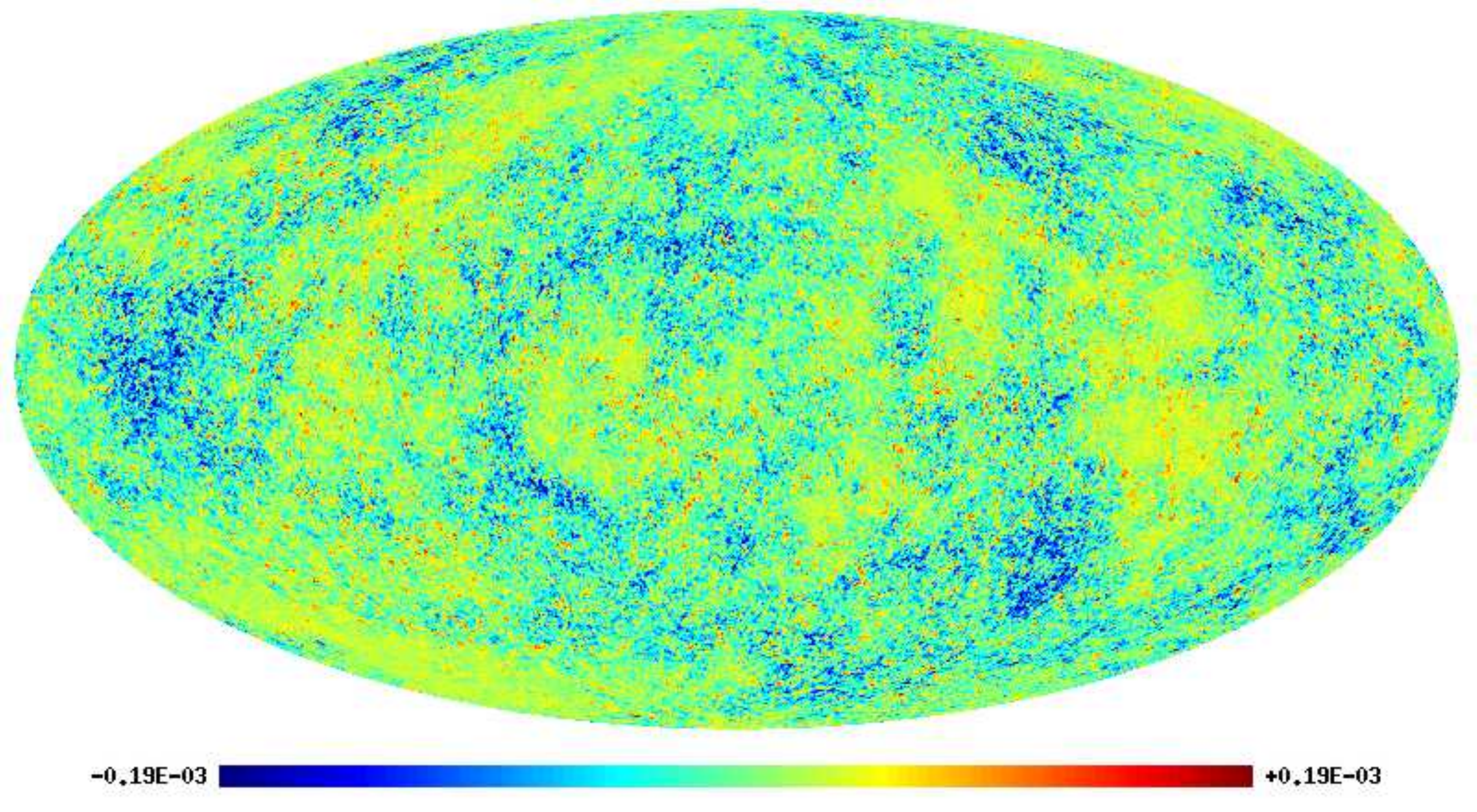}} \\
\subfigure{\includegraphics[width=0.47\textwidth]{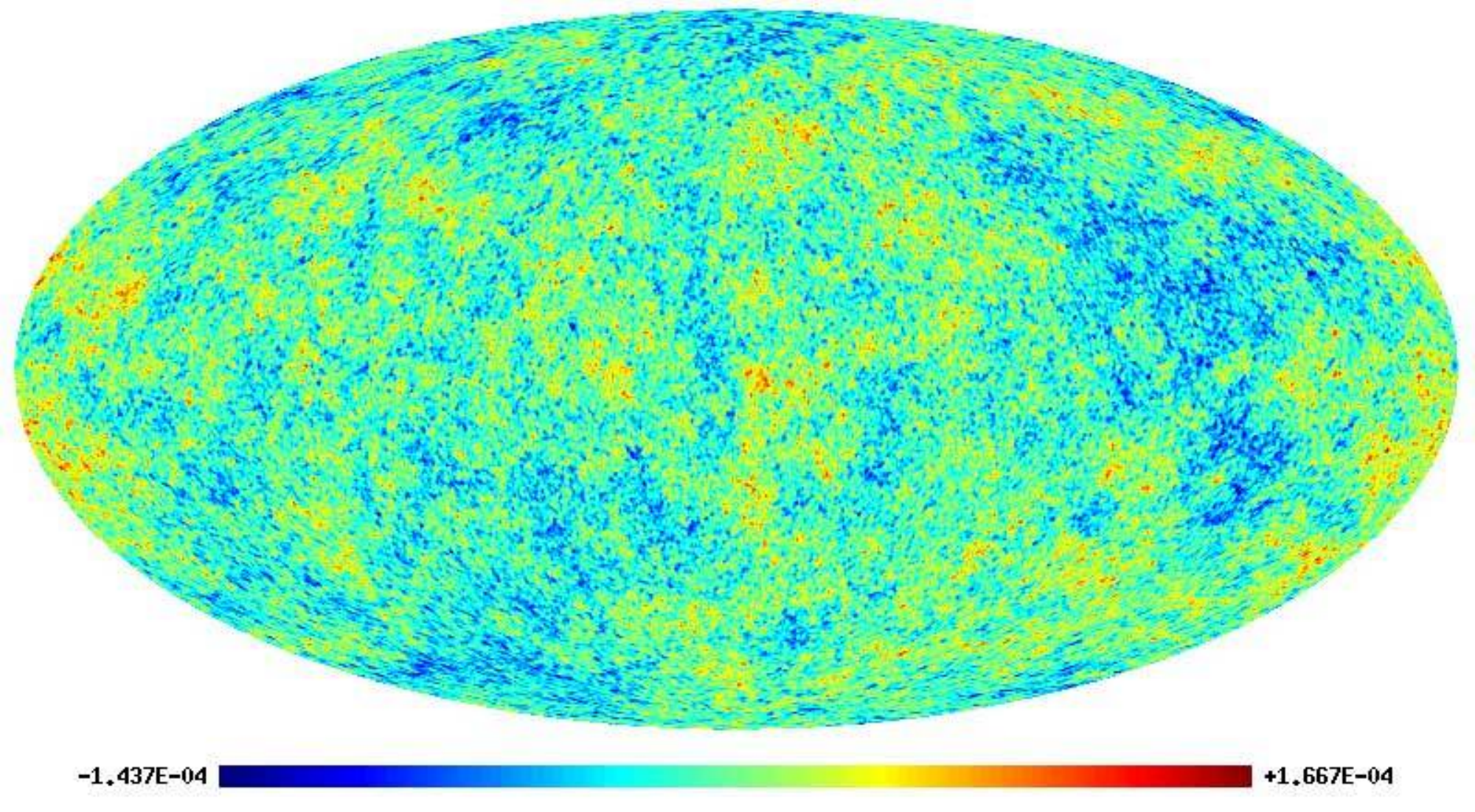}} \qquad
\subfigure{\includegraphics[width=0.47\textwidth]{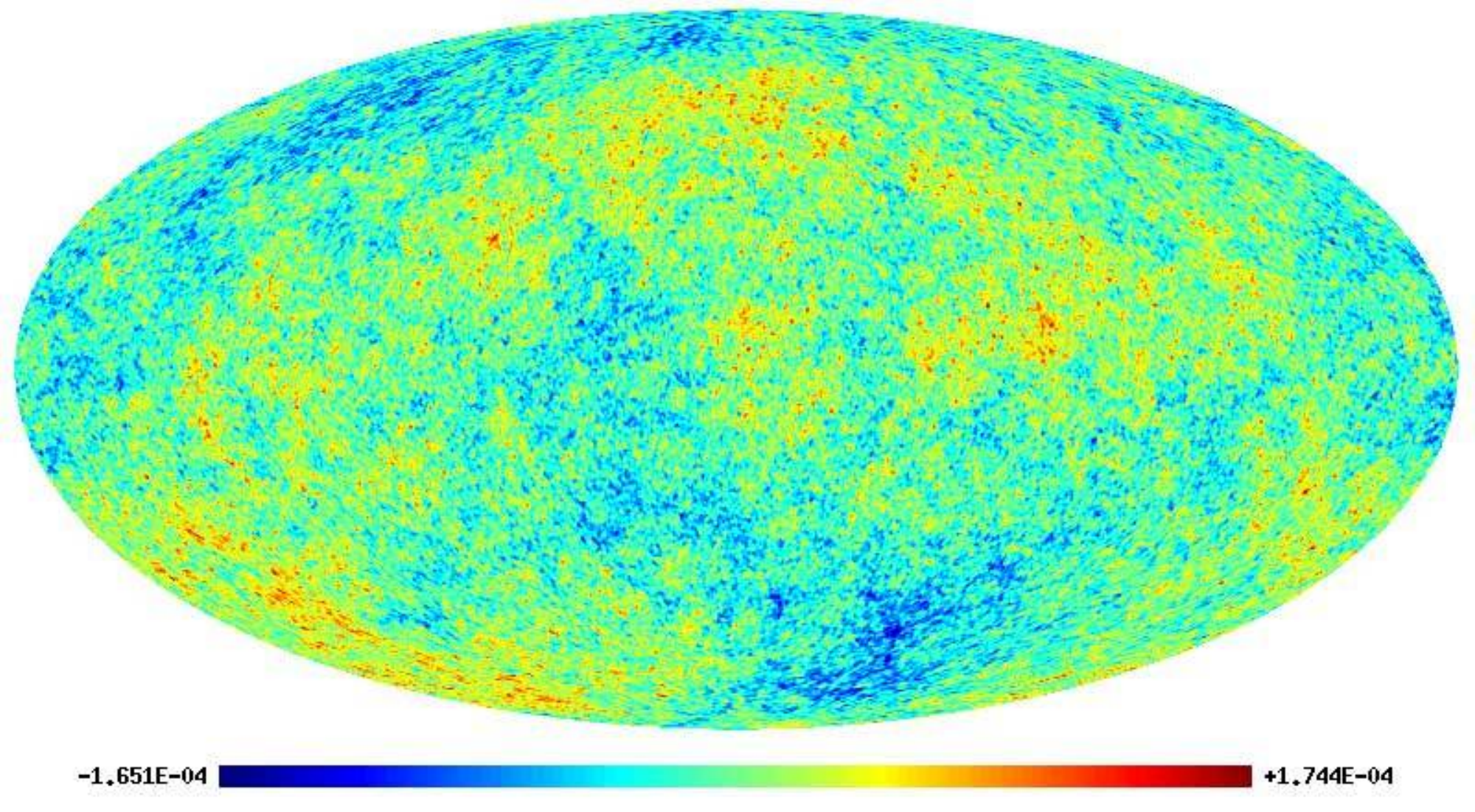}} \\
\caption{{\it Top panels}: a Gaussian realization of the CMB sky ({\it left}) and a non-Gaussian local CMB map ({\it right}), obtained by adding to the Gaussian one a NG component with $\fnll=3000$. From \citet{LiguoriMatarreseMoscardini2003}. {\it Lower panels}:  a Gaussian CMB map ({\it left}) and a non-Gaussian DBI map ({\it right}) with $\fnl^{DBI}=4000$. From \cite{FergussonLiguoriShellard2009}. The maps in the upper panel have been obtained using the local algorithm described in section \ref{sec:localmaps}. The maps in the lower panel have been produced with the bispectrum algorithm of section  \ref{sec:bispectrummaps}, after having separated the primordial DBI shape using the eigenmode expansion defined in \eqn{eq:orthobasis}. }
\label{fig:maps}    
\end{figure*}

In this section we will describe algorithms for the generation of non-Gaussian CMB maps with a given bispectrum. There are three main reasons why primordial NG simulations of the CMB are useful in the context of bispectrum estimation of $\fnl$:
\begin{enumerate}
\item To test the unbiasedness of the $\fnl$ bispectrum estimator (by checking that the Monte Carlo average of the recovered $\fnl$ reproduces the $\fnl$ set in input). 
\item To study how the expected primordial NG signal imprinted in the CMB is modified by the presence of other effects, like those considered in section \ref{sec:NGcontaminants}. 
For example, weak lensing of primordial NG might in principle change the observed bispectrum and affect the estimates. This can be studied again by testing the estimator on NG lensed simulations, as it was done in \citet{HansonEtal2009}
\item For local NG, to obtain the error bars of the $\fnl$ estimator if a large $\fnl$ is detected at several $\sigma$ (see section \ref{sec:largefnl}). We have previously seen that for a several sigma detection of local NG the bispectrum variance is $\fnl$-dependent. The Monte Carlo average \eqn{eq:estvar} thus has to be evaluated on NG simulations with the measured $\fnl$ in input.
\end{enumerate} 
Unless we are in the situation described at point 3. of the list above all we need to produce is then maps with given power spectrum and bispectrum, since higher order correlators can be neglected. In the large local $\fnl$ case higher order correlators are instead important and have to be included. Fortunately the local case is the only one for which we have a full expression of the primordial potential $\Phi(\mathbf{x})$ that allows us to produce exact simulations.

We will divide this section in two parts. In the first we will describe exact simulation algorithms of local NG, while in the second we will describe methods to generate maps with given power spectrum and bispectrum, starting from an arbitrary primordial shape.

\subsubsection{Algorithms for local non-Gaussianity}
\label{sec:localmaps}

First of all, let us recall that the CMB multipoles $a_{\ell m}$ are related to the primordial gravitational potential $\Phi$ through the well known formula:
\be\label{eq:phi2alm} 
a_{\ell m} = \int \frac{d^3 k}{(2 \pi)^3} \Phi(\mathbf{k}) Y_{\ell
  m}(\hat{k}) \Delta_\ell(k) \; ,
\ee
where $\Delta_\ell(k)$ are the radiation transfer function and the potential is written in Fourier space. We already met this formula when we calculated the relation between the primordial and CMB bispectrum in section \ref{sec:CMBbisp}.  We also recall that the local non-Gaussian primordial potential takes a very simple expression in real space, where:
\be
\Phi(\mathbf{x}) = \Phi_L(\mathbf{x}) + \fnl \left[\Phi_L^2(\mathbf{x}) - \left\langle \Phi_L^2(\mathbf{x}) \right \rangle \right] \; .
\ee
In the previous expression $\Phi_{\rm L}$ is a Gaussian random field, characterized by a primordial power spectrum $P_\Phi(k) = Ak^{n-4}$; in the following we will refer to $\Phi_{\rm L}(\mathbf{x})$ as the Gaussian part of the primordial potential. The remaining non-Gaussian part of the potential is simply the square of the Gaussian part point by point 
(modulo a constant term, necessary to enforce the condition $\langle \Phi(\mathbf{x}) \rangle = 0$; however it is clear that this term only affects the CMB monopole). It is then convenient to work directly in real space and recast formula \eqn{eq:phi2alm} in the following form
\be\label{eq:phi2alm_real}
a_{\ell m} = \int d^3 r \, \Phi(\mathbf{r}) Y_{\ell m}(\hat{r}) \alpha_\ell(r) \; ,
\ee
where $\alpha_{\ell}(r) \equiv \int dk \, k^2 j_\ell(kr) \D_{\ell}(k)$, also used in \eqn{eq:phix2alm}, is the real space counterpart of the radiation transfer functions $\D_{\ell}(k)$, $j_\ell(kr)$ is a spherical Bessel function, and $r$ is a look-back conformal distance. This formula suggests to structure an algorithm for the generation of local CMB NG maps in the following steps
\begin{enumerate}
\item Generate the Gaussian part $\Phi_L$ of the potential in a box whose side is the present cosmic horizon.
\item Square the Gaussian part point by point to get the non-Gaussian part. 
\item Expand in spherical harmonics the Gaussian and non-Gaussian parts of the potential for different values of the radial coordinate $r$ in the simulation box.
\item Convolve the spherical harmonic expansions of $\Phi_{\rm L}$ and $\Phi_{\rm NL}$ with the radiation transfer function $\D_\ell(r)$ in order to obtain the Gaussian and non-Gaussian part of the multipoles of the final NG CMB simulation. For a given choice of the non-Gaussian parameter $\fnl$ a CMB map is then obtained simply through the linear combination $a_{\ell m} = a_{\ell m}^L + \fnl a_{\ell m}^{\rm NL}$ (the superscripts L and NL always indicating Gaussian and non-Gaussian respectively).
\end{enumerate}

The most difficult and time consuming part in this process is actually the generation of the Gaussian part of the potential $\Phi$. One possibility is to generate the Gaussian part of the potential in a cubic box in Fourier space (where different modes are uncorrelated
and have variance given by the primordial power spectrum $P_\Phi(k)$, then apply a Fast Fourier Transform (FFT) algorithm to go to real space. Cartesian coordinates are then transformed into spherical coordinates by means of an interpolation algorithm in order to transform $\Phi_L(\mathbf{x})$ into $\Phi_L(r,\hat{n})$. Finally, the Gaussian potential in spherical coordinates is squared point by point to get the NG part and the spherical harmonic expansion and radiation transfer function convolution at point 4 of the list above are performed in order to obtain the multipoles of the final CMB map. The aforementioned algorithm was implemented in \citet{KomatsuEtal2003} to generate NG local CMB maps at the resolution of the Planck satellite. 

The difficulty with this approach arises from the fact that we are working in a box of the size of the present cosmic horizon, (about $15$ Gpc in conformal time), but at the same time a cell in this box must have a side no bigger than $20$ Mpc in order to resolve the last
scattering surface, where most of the CMB signal is generated. A more convenient and accurate way to produce the local NG $\alm$ was found in \citep{LiguoriMatarreseMoscardini2003, LiguoriEtal2006}: the idea is to work directly in spherical coordinates, use a non uniform 
discretization of the simulation box (since no sample points are needed in a large region of the box where photons are just free streaming, while many sample points are needed at last scattering, as we just pointed out above) and generate the multipoles of the expansion of $\Phi_{\rm L}(\mathbf{x})$ through the following two step approach:

\begin{enumerate}

\item Generated uncorrelated radial multipoles $n_{\ell m}(r)$, gaussianly distributed and characterized by the following spectrum:
\be
\label{eq:whitenoise} 
\left \langle n_{\ell_1 m_1}(r_1)
n^*_{\ell_2 m_2}(r_2) \right \rangle = 
\frac{\d_D(r_1-r_2)}{r^2}\d_{\ell_1}^{\ell_2} \d_{m_1}^{m_2}\; ;  
\ee 
where $\d_D$ is the Dirac delta function. 

\item Filter the multipoles $n_{\ell m}$ with suitable functions in order to produce a Gaussian random field with the properties of the multipole expansion of the primordial Gaussian potential $\Phi_L$ . It can be shown that the expression of the filter functions
is:
\be
\label{eq:filter} 
W_\ell(r,r_1) =
\frac{2}{\pi} \int \! dk \, k^2 \, \sqrt{P_\Phi(k)} \, j_\ell(kr)
j_\ell(kr_1) \; ,  
\ee
where $P_{\Phi}$ is the primordial curvature power spectrum, and the filtering operation takes the form 
\be
\label{eq:nlm2phil} 
\Phi^{\rm L}_{\ell m}(r) = \int \! dr_1 \, r_1^2 \, n_{\ell m}(r_1) 
W_\ell(r,r_1) \; .
\ee 
In the last expression $\Phi^{\rm L}_{\ell m}(r)$ are the desired quantities, \ie the multipoles of the expansion of the Gaussian part of the primordial potential for a given $r$.

\end{enumerate}

\begin{figure}[t] 
\begin{center} 
\includegraphics[height=0.45\textheight,width=0.6\textwidth]{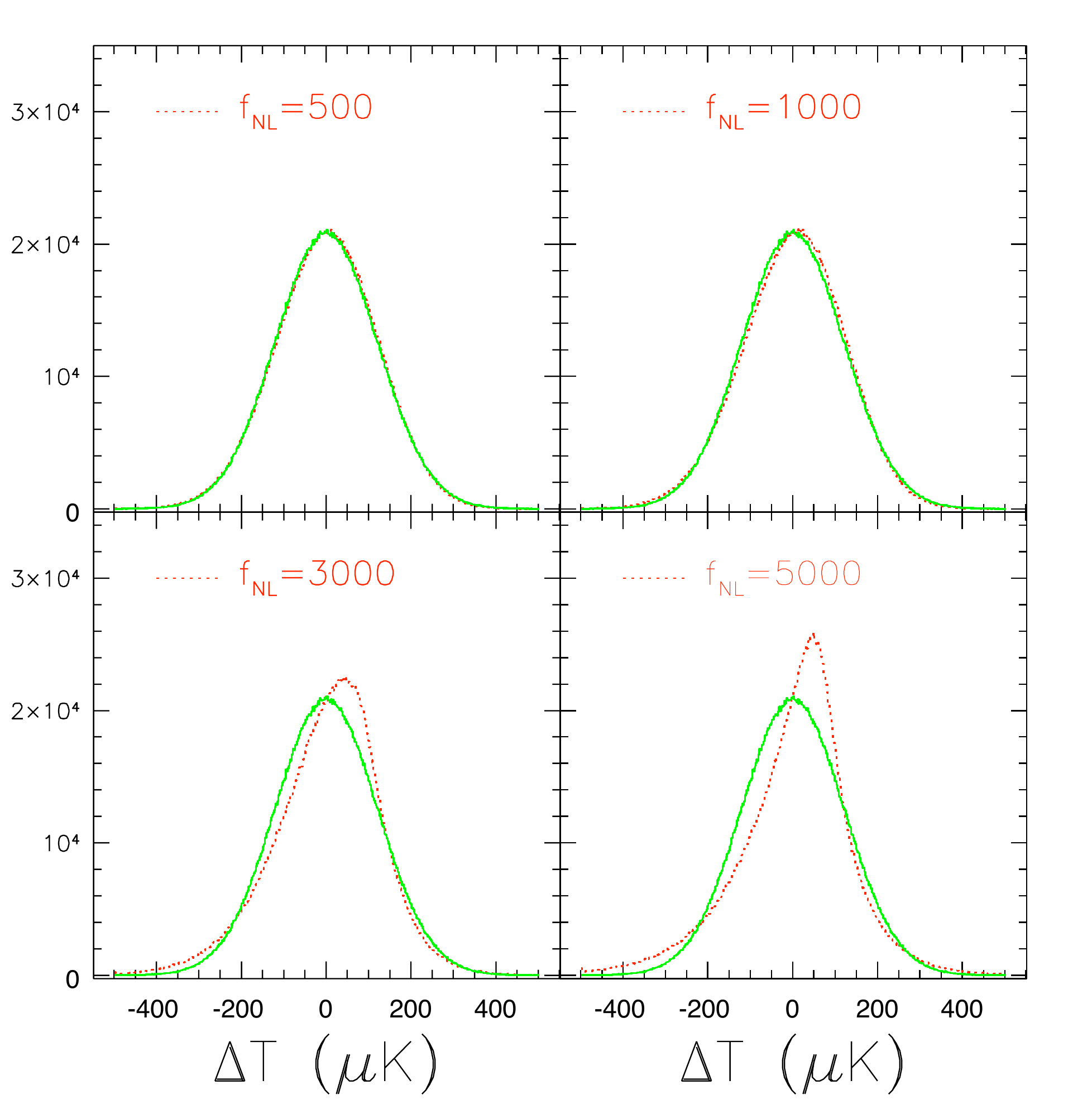} 
\caption{Probability density function of temperature pixel from local primordial non-Gaussian CMB maps, obtained with the ``exact'' simulation algorithm described in section \ref{sec:localmaps}. Different panels show the result for different values of $\fnll$, in order to give an idea of the order of magnitude of the signal that one wants to detect. For $\fnl < 1000$ the non-Gaussianity is too small to be seen this this plots. Note that WMAP constrain $\fnll$ to be $\lesssim 100$.}
\label{fig:PDF} 
\end{center}
\end{figure}

This algorithm, recently improved in \citet{ElsnerWandelt2009}, was used to produce NG local 
maps at the resolution of WMAP and Planck in temperature and polarization. An example of its results is shown in the upper panels of \fig{fig:maps}. \Fig{fig:PDF} shows 1-point PDFs of temperature anisotropies for different values of $\fnll$, extracted from these simulations.

\subsubsection{Algorithms for arbitrary bispectra}\label{sec:bispectrummaps}

In the limit of weak non-Gaussianity, an algorithm to produce non-Gaussian CMB simulations with a given power spectrum and bispectrum  for separable primordial shapes was described in \citet{SmithZaldarriaga2006}. In this algorithm the non-Gaussian components of the CMB multipoles are obtained using the following formula:
\begin{align}\label{eq:almNG}
a_{\ell m}^{\rm NG} = 
\frac{1}{6} \sum_{\ell_i m_i} B_{\ell_1 \ell_2 \ell_3} 
\(  \begin{array}{ccc} \ell & \ell_2 & \ell_3 \\ m & m_2 & m_3 \end{array} \)
  \frac{a_{\ell_2 m_2}^{\rm G *}}{C_{\ell_2}} \frac{a_{\ell_3 m_3}^{\rm G *}
  }{C_{\ell_3}} \; ,
\end{align}
where $a_{\ell m}^G$ is the Gaussian part of the CMB multipoles, generated using the angular power spectrum $C_{\ell}$, while $B_{\alll}$ is the given bispectrum of the theoretical model for which simulations are required. Note that alternative algorithms to generate CMB maps with given bispectrum have been proposed in the literature \citep{ContaldiMagueijo2001, RochaEtal2005} but they are less general that the one introduced by equation \eqn{eq:almNG}.
Although \eqn{eq:almNG} is completely general, as before its numerical evaluation is
only computationally affordable for bispectra that can be written in separable form. We have emphasized already that separability results in a reduction of the computational cost of the estimator (\ref{eq:optimalcubicestnorm2}) from ${\mathcal O}(\ell_{max}^5)$ to ${\mathcal O}(\ell_{max}^3)$ operations; the same argument applies here, and allows to rewrite  \eqn{eq:almNG} into an equivalent form in pixel space. Starting from formula \eqn{eq:separated}, and substituting it in \eqn{eq:almNG}, we find:
\be\label{eq:pixelspacemap}
a_{\ell m}^{\rm NG} = \int dr r^2 \int d \Omega_{\hat{n}}  \( 2 X_\ell(r) M_Y(r,\hat{\bf n}) M_Z(r,\hat{\bf n}) +
                          2 Y_\ell(r) M_X(r,\hat{\bf n}) M_Z(r,\hat{\bf n}) + 
                          2 Z_\ell(r) M_X(r,\hat{\bf n}) M_Y(r,\hat{\bf n}) \)\,.
\ee
As already discussed in the $\fnl$-estimator section, the limitation dictated by separability is clearly overcome by using the eigenfunction representations for the bispectrum \eqn{eq:orthobasis} and \eqn{eq:cmbestmodes} introduced in \citet{FergussonLiguoriShellard2009}. 
As usual, the basic idea is to start by expanding an arbitrary bispectrum shape $S$ (either primordial or in the CMB) using a separable polynomial decomposition until a good level of convergence is achieved and then to substitute the mode decomposition into \eqn{eq:almNG} to get a linear combination of numerically tractable terms written in the form \eqn{eq:pixelspacemap}.   Using the separable mode coefficients $\alpha_{prs}$ for the reduced bispectrum (\ref{eq:sepblllprim}) and 
the filtered map expressions $M_p(r,\hat{\bf n})$ (\ref{eq:filteredmaps}) as the starting point,  we find that the
expression (\ref{eq:pixelspacemap}) generalises to
\begin{align}\label{eq:almNGsplit}
a_{lm}^{NG} = \frac{1}{18} \sum_{prs} \alpha_{prs} \int dx x^2 q_{p}^l(x)  \int d\Omega_{\hat{n}}\,
Y^{m*}_l(\hat{\mathbf{n}}) \,M^G_r(r,\un) \, M^G_{s}(r,\un)\,,
\end{align}
where the $M_p^G(r,\un)$ are found by summing using a set of Gaussian $\alm^G$'s convolved with the 
$q^l_p$'s (refer to eqn (\ref{eq:convolved})),
\eq\label{eq:Gmapfilter}
M^G_p(\un,x) =  \sum_{lm} q_p^{\,l}\,\, \frac { a^G_{lm} Y_{lm} }{C_{l}} \,.
\qe
 Here, the accuracy of convergence with the $\alpha_{prs}$ is parametrized in terms of the
correlation $\bar {\cal C}(S,S_N)$ between the original non-separable shape and the eigenmode
expansion, as defined previously (\ref{eq:shapecorrelator}). Note that this convergence can also be checked more accurately using the full Fisher matrix correlation on the CMB bispectra ${\cal C}(\blll,\blll^N)$, described in sections \ref{sec:forecasts} and \ref{sec:NGcontaminants}.

In addition to the bispectrum separability requirement, there is an important further caveat which can prevent the straightforward implementation of the algorithm \eqn{eq:almNG}.
By construction, terms $\curl{O}(f^2_{\rm NL})$ and higher are not explicitly controlled. Following the discussion in \cite{HansonEtal2009} we can write the connected N-point functions as:
\bea
\langle a^{*}_{\ell_1 m_1} a_{\ell_2 m_2} \rangle  & = & \left[ C_{\ell_1}
  + \fnl^2 C_{\ell_1}^{NG} \right] \label{eq:clsim} \\ 
\langle a_{\ell_1 m_1} a_{\ell_2 m_2} a_{\ell_3 m_3} \rangle & = &
\left[\fnl B_{\ell_1 \ell_2 \ell_3} + \curl{O} (\fnl^3) \right] \\
\langle a_{\ell_1 m_1} a_{\ell_2 m_2} a_{\ell_3 m_3} \dots
a_{\ell_N m_N} \rangle & = & \curl{O} \left( \fnl^3 \right) \; .
\eea 
Thus the condition that the map has the power spectrum $C_l$ specified in input will only be satisfied if the power spectrum of the non-Gaussian component in \eqn{eq:clsim} remains small. Since this method does not control $\curl{O}(f^2_{\rm NL})$ terms, one has to ascertain that spuriously large $C_l^{NG}$ contributions do not affect the overall
power spectrum significantly. It turns out that this effect plagues current map
simulations if the standard separable expressions for the local and equilateral bispectra are
directly substituted into \eqn{eq:almNG}. However a slight modification of equation \eqn{eq:almNG}, described in \citet{HansonEtal2009} and \citet{FergussonLiguoriShellard2009} allows us to overcome this problem at no computational cost. Moreover, it was shown by  \citet{FergussonLiguoriShellard2009} that maps obtained from the eigenmode expansions \eqn{eq:orthobasis} and \eqn{eq:cmbestmodes} are stable independently of the shape under study, thus making this map-making generating algorithm 
robust and fully general. 
Examples of DBI NG maps produced by combining the eigenmode expansion method with the map making algorithm described in this section are shown in the lower panels of \fig{fig:maps}.

\clearpage

\section{Large-Scale Structure}
\label{sec:LSS}

In the standard scenario, early perturbations produced during inflation are responsible for the common origin of the CMB temperature fluctuations and the large-scale matter and galaxy distributions in the Universe, \ie the large-scale structure. The Cosmic Microwave Background provides a remarkable example of a Gaussian random field in nature. Information on cosmological parameters is in fact derived from measurements of its power spectrum, the $C_l$'s, while bispectrum measurements from WMAP data remain consistent with zero. The distribution of matter, as we can infer today from shear or galaxy observations, unlike the CMB, can be described as a highly non-Gaussian random field, {\it even} for Gaussian initial conditions.  

The matter overdensity $\d(\xv)$ is defined in terms of the matter density $\rho(\xv)$ and its mean value $\bar{\rho}$ by
\be
\d(\xv)\equiv\frac{\rho(\xv)-\bar{\rho}}{\bar{\rho}}\,,
\ee
with zero mean by construction.  Here, at late times, the limiting value $\d=-1$ in voids, accounting for a large fraction of the volume of the Universe, while achieving values $\d\gg 1$ in collapsed objects such as dark matter halos. Its probability distribution function is therefore expected, at low redshift, to depart strongly from a Gaussian distribution centred at $\d=0$,  even though it could be well approximated by it at decoupling, when perturbations around $\d= 0$ were of the order of $\d\sim 10^{-5}$. Such non-Gaussianity is the result of the nonlinear evolution of structures subject to {\it gravitational instability}. 

In addition, nonlinearities in the bias relation between the galaxy and matter distributions constitute a {\it second source of non-Gaussianity} in the large-scale structure  mapped out by redshift surveys. Non-Gaussian initial conditions would therefore provide a third component in the non-Gaussianity of the galaxy distribution. The question regarding the detection of effects due to {\it primordial} non-Gaussianity, is therefore strictly related to our ability to {\it distinguish} between these different contributions and, ultimately, it will depend on the robustness of our theoretical predictions in the linear and mildly nonlinear regime. From this respect, cosmological Perturbation Theory (PT), and its more recent developments, is very important for providing the tools to study the evolution of non-Gaussianities  and how to differentiate their origin. 

Considering only the {\it matter} distribution, the leading order prediction in standard PT for the matter bispectrum at large scales is given by the {\it sum} of a primordial component and a component due to gravitational instability, which is present also for Gaussian initial conditions. Until fairly recently it was assumed that this picture could be easily extended to the galaxy distribution, with the galaxy bispectrum receiving an additional contribution due to nonlinear bias. Following the historical development
of the subject, in section~\ref{ssec:skewness} we will discuss early work on higher-order {\it moments} of the matter and galaxy distribution, starting with the {\it skewness}. Here, most of the theoretical results on higher-order correlation functions are developed. We will then consider in \ref{ssec:lssm} the matter bispectrum and its description in {\it Eulerian} perturbation theory, with specific attention given to effects at large scales due to a primordial component,  as well as at small-scales, nonlinear corrections in presence of non-Gaussian initial conditions. In section~\ref{ssec:lssg}, we will deal with the galaxy bispectrum. We will first introduce the simple model based on local bias and discuss problems related to bispectrum measurements in redshift surveys with specific attention given to the detection of primordial non-Gaussianity. We will see how early results indicated that the galaxy bispectrum could be used as a tool to constrain non-Gaussian initial conditions which is, in principle, {\it competitive} with the CMB, illustrating this with actual results from current data-sets. We will then consider the outcome of recent N-body simulations with non-Gaussian initial conditions showing that the simple prediction for the galaxy bispectrum assumed in most of the previous  literature on the subject fails to describe not only the measured halo bispectrum, but even the halo {\it power spectrum}, even {\it at large scales}! We now know that correlators of biased populations such as galaxies and dark matter halos, receive large corrections, at large scales, from local primordial non-Gaussianity. These results opened  up new and promising opportunities for detection in future large-scale structure observations. Although, in our view, a proper understanding of these effects remains to be adequately developed at the time of writing, particularly with respect to higher-order galaxy correlation functions, we will describe the different descriptions proposed so far in the literature and the prospects for detection of primordial non-Gaussianity in measurements of the galaxy bispectrum.

From an historical perspective, non-Gaussian initial conditions have been studied for quite a long time. For instance, early works on the clustering of density peaks and rare objects can be found in \citet{GrinsteinWise1986, MatarreseLucchinBonometto1986, LucchinMatarrese1988}, while early N-body simulations with non-Gaussian initial conditions go back to the early eighties \citep{MessinaEtal1990, MoscardiniEtal1991, WeinbergCole1992, ColesEtal1993, White1999}. In the early days, a large variety of non-Gaussian models, often defined in terms of a nonlinear transformation of a Gaussian field were considered. In some cases, a large non-Gaussian component were studied because, on the one hand, they could be used falsify some models and, on the other,  as a way to reconcile contradictory observational results with  theoretical frameworks. In this review, however, we will consider only models predicting small departures from Gaussian initial conditions which are consistent with CMB observations. 

Finally, while we focus in this review on direct bispectrum measurements, it should be stressed that the effects of primordial non-Gaussianity on large-scale structure are not limited to corrections to its higher-order correlation functions. Aside from the recent results on the galaxy power spectrum mentioned above, significant departures from Gaussian initial conditions are expected to have important effects on the halo mass function and therefore on the observed cluster number density. See section 2.1 in \citet{SefusattiEtal2007} for an brief overview of previous work and\citet{LoVerdeEtal2008, DalalEtal2008, PillepichPorcianiHahn2008, AfshordiTolley2008, MaggioreRiotto2009C, FedeliMoscardiniMatarrese2009, Oguri2009, DesjacquesSeljakIliev2009, Valageas2009, GrossiEtal2009, LamSheth2009} for recent theoretical and N-body results. In addition, the corresponding effect on the abundance of voids has been studied by \citet{KamionkowskiVerdeJimenez2009}, while the possibility of constraining primordial non-Gaussianity from measurements of Minkowski Functionals in large-scale structure has been explored by \citet{HikageKomatsuMatsubara2006,HikageEtal2008A}. Further effects on the intergalactic medium and reionization \citep{VielEtal2009,CrocianiEtal2009} or on future 21cm observations \citep{Cooray2006,PillepichPorcianiMatarrese2007} have also been investigated. We refer the reader to other 
reviews in this issue for a more complete discussion of these alternative approaches.

\subsection{The skewness}
\label{ssec:skewness}

\begin{figure}[t]
\begin{center}
{\includegraphics[width=0.4\textwidth]{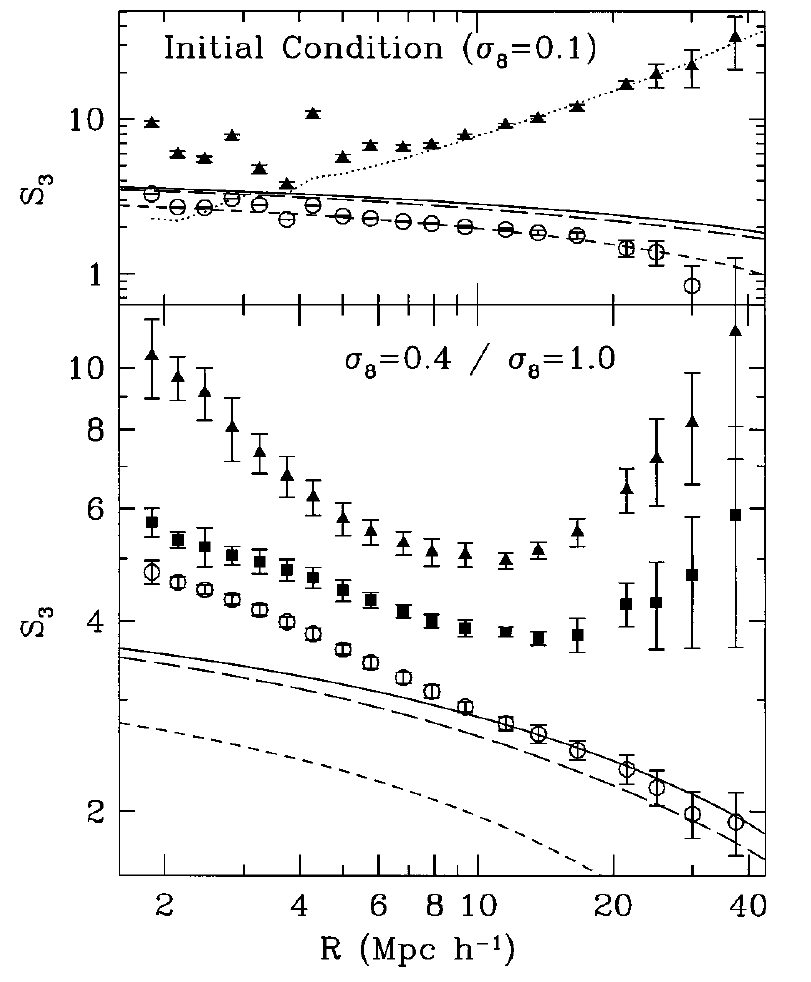}}
{\includegraphics[width=0.50\textwidth]{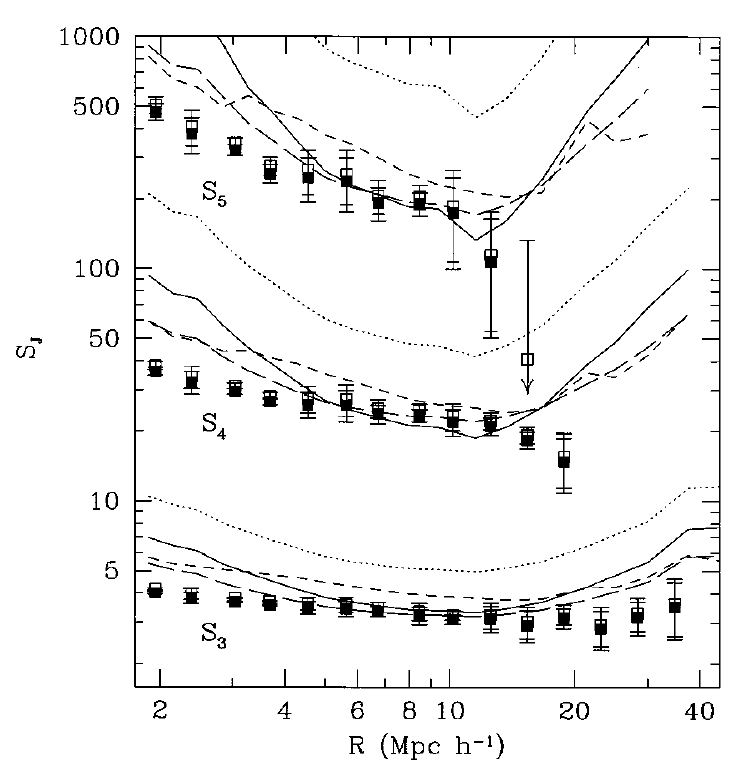}}
\caption{{\it Left panel}: measurements of the skewness of the matter distribution in N-body simulations as a function of the smoothing scale $R$ for the non-Gaussian texture model ({\it filled symbols}) and for Gaussian initial conditions ({\it open circles}). The time evolution is parametrized by the value of $\sigma_8$, with the non-Gaussian results shown at $\sigma_8=0.4$ ({\it triangles}) and $\sigma_8=1$ ({\it squares}). Lines show different theoretical predictions. {\it Right panel}: measurements of the third, fourth and fifth-order moments of the galaxy distribution in the APS Galaxy Survey, compared with the simulation results with non-Gaussian initial conditions with different bias assumptions. From \citet{GaztanagaMahonen1996} (see the reference for further details). }
\label{fig:GM96}
\end{center}
\end{figure}

Since the first large-scale observations did not allow an accurate determination of individual bispectrum or trispectrum configurations, most of the attention in the early literature focused on the moments of the galaxy distribution, and, in the first place on the third- and fourth-order moments, \ie the {\it skewness} and {\it kurtosis}, respectively. The  ``normalized'' moment of order $p$ can be defined in terms of the smoothed density field $\d_R(\xv)$ as 
\be
s_{p,R}\equiv\frac{\la\d^p_R(\xv)\ra_c}{\la\d_R^2(\xv)\ra^{p/2}}\,,
\ee
For Gaussian initial conditions, a perturbative treatment of the equations of gravitational instability predicts at leading order \citep{Peebles1980}
\be
s_{3,R}=\frac{34}{7}\sigma_R\,,
\ee
with $\sigma_R^2=\la\d_R^2\ra$, computed in linear theory. When non-Gaussian initial conditions are present, one expects an extra contribution to the skewness, typically with a different relation with $\sigma_R$, whose value depends on the non-Gaussian model. Comparisons between the second- and third-order moments, $S_{3,R}$ and $\sigma_R$, (as well as higher-order moments such as the kurtosis) measured in redshift surveys have been early recognized as a tool to test the Gaussianity of primordial perturbations, \citep{ColesFrenk1991,JuszkiewiczBouchet1992,BouchetEtal1992,JuszkiewiczBouchetColombi1993,LahavEtal1993,LuoSchramm1993,ColesEtal1993,LucchinEtal1994,FryScherrer1994}. These works recognized as well the importance of reliable predictions in the nonlinear regime and of a proper modeling of the effects of galaxy bias. In this respect, \citet{FryScherrer1994} proposed a more quantitative prediction for the contribution to the galaxy skewness due to galaxy bias based perturbation theory and on the local bias expansion of \citep{FryGaztanaga1993}. They derived, for the skewness of the galaxy distribution, an expression of the form
\be\label{eq:skewPT}
s_{3,R}= s_{3,R}^{(0)}+\frac{34}{7}\sigma_R+\frac{6b_2}{b_1}\sigma_R\,,
\ee
where we assumed non-Gaussian initial conditions described by a non-vanishing initial skewness $s_{3,R}^{(0)}$ (but vanishing higher-order moments) and where $b_1$ and $b_2$ represent constant bias parameters typical of the galaxy population (which we will discuss explicitly in section~\ref{ssec:lssg}). This relatively simple expression describe the skewness measured in galaxy surveys, as the sum of {\it three components corresponding to three sources of non-Gaussianity for the galaxy distribution}: one primordial, one due to gravitational instability and the last to nonlinear bias. Further studies in perturbation theory can be found in \citep{ChodorowskiBouchet1996, DurrerEtal2000} while an alternative derivation of the smoothed moments of the density field based on the spherical collapse model has been studied in \citep{GaztanagaFosalba1998}. The skewness predicted by texture models has been studied in simulations as a function of the smoothing scale $R$ by \citet{GaztanagaMahonen1996} and compared to measurements of the same quantities in the APS Galaxy Survey \citep{Gaztanaga1994}, see \fig{fig:GM96}. The differences between the $s_{3,R}$ in the non-Gaussian texture model with respect to the Gaussian case provides a qualitative example of the typical effects that we expect for non-Gaussian initial conditions as a function of the smoothing scale $R$ and redshift.

The measured skewness, as higher order moments, corresponds to a single number. Despite the possibility to study its peculiar dependence on the smoothing scale $R$, it is nevertheless difficult to separate the different components, particularly with respect to bias effects. However, this possibility is offered in principle by direct measurements of the galaxy bispectrum, relying on its dependence on the shape of triangular configurations. In the next sections we will discuss in details first the bispectrum of the {\it matter} distribution then the bispectrum of the {\it galaxy} distribution, a direct observable in redshift surveys.

\subsection{The matter bispectrum}
\label{ssec:lssm}

In this review we will focus on the predictions for correlation functions in Fourier space from Eulerian Perturbation Theory (PT). This approach solves perturbatively the equations for the matter density and velocity field evolution governed by gravitational instability. These are the continuity equation, the Euler equation and Poisson equation relating the matter density and the gravitational potential. In the PT framework, the relation between the results and the initial conditions, given in terms of the initial correlators of the density field is particularly transparent. Moreover, recent works have significantly extended, as we will discuss later, the predicting power of this specific tool.   Different approaches are also available: see, for instance, \citet{Scoccimarro2000B} for a comparison between bispectrum measurements in N-body simulations and predictions in Lagrangian Perturbation Theory. We refer the reader to \citet{BernardeauEtal2002} for a comprehensive review of cosmological perturbation theory of the large-scale structure. 

\subsubsection{Leading-order results in Perturbation Theory}
\label{sssec:lssmPT}

As mentioned before, we consider specifically models where non-Gaussian initial conditions are completely given in terms of the correlators of the curvature perturbations at early times, and the mechanism responsible for the {\it extra} non-Gaussian properties of the density field is not active during the subsequent evolution of matter perturbations, governed only by gravitational instability. In PT, the solution for the evolved matter density contrast is expressed as a series of corrections to the linear solution $\d^{(1)}$ \citep{Fry1984}
\be\label{eq:PT}
\d_\kv=\d_\kv^{(1)}+\d_\kv^{(2)}+\d_\kv^{(3)}+\ldots.\,,
\ee
where each term can be written formally as\footnote{From now on we will adopt a different convention for the Fourier transform with respect to the one used for the formulae in previous section. The present convention is more common in the large-scale structure literature and conforms with the one adopted in the classical review \citet{BernardeauEtal2002}.}
\be\label{eq:PTterm}
\d_\kv^{(n)}\equiv\int d^3q_1\ldots d^3q_nF_n(\qv_1,\ldots,\qv_n)~\d_{\qv_1}^{(1)}\ldots\d_{\qv_n}^{(1)}\,,
\ee
with $F_n(\qv_1,\ldots,\qv_n)$ representing the symmetrized $n$-order kernel in PT. The initial conditions in the Gaussian case are completely specified by the linear power spectrum $P_0(k)$, with $\la\d^{(1)}_{\kv_1}\d^{(1)}_{\kv_2}\ra=\d_D(\kv_{12})P_0(k_1)$, where we adopt the notation $\kv_{ij}\equiv\kv_i+\kv_j$. Non-Gaussian initial conditions are described, in the first place, by a non-zero expression for the three-point function of the linear solution, that is $\la \d_{\kv_1}^{(1)}\d_{\kv_2}^{(1)}\d_{\kv_3}^{(1)}\ra$. In turn, the {\it initial} matter correlators, \ie the correlators of the {\it linear} solution $\d^{(1)}$, are given in terms of the correlators of the curvature perturbations, as 
\be
\la\d_{\kv_1}\cdots\d_{\kv_n}\ra=M(k_1,z)\cdots M(k_n,z)~\la\Phi_{\kv_1}\cdots\Phi_{\kv_n}\ra,
\ee
where we introduce the function
\be
M(k,z)=\frac{2}{3}\frac{k^2~T(k)~D(z)}{\Omega_mH_0^2}\,,
\ee
with $T(k)$ being the matter transfer function and $D(z)$ the growth factor, expressing Poisson's equation in Fourier space as
\be\label{eq:Poisson}
\d_\kv(z)=M(k,z)~\Phi_\kv\,.
\ee
Notice that we denote with $\Phi$ the {\it primordial} curvature perturbations, \ie evaluated during the matter dominated era, not their value linearly extrapolated at present time\footnote{This choice, not unique in the literature, is particularly convenient since curvature perturbations are constant during matter domination. Also, it conforms to the definition of $\fnl$ in terms of $\Phi$ assumed in the CMB literature on observational constraints and specifically in \citet{KomatsuSpergel2001}.}. The linear, \ie {\it initial}, power spectrum is given by 
\be
P_0(k)=M^2(k,z)P_\Phi(k)\,,
\ee
while the initial bispectrum and trispectrum are
\bea
B_0(k_1,k_2,k_3) & = & M(k_1)M(k_2)M(k_3)~B_\Phi(k_1,k_2,k_3)\,,
\label{eq:B0}\\
T_0(k_1,k_2,k_3,k_4) & = & M(k_1)M(k_2)M(k_3)M(k_4)~T_\Phi(k_1,k_2,k_3,k_4)\,.
\label{eq:T0}
\eea
Notice that given these simple relations between curvature and primordial matter correlators, issues such as the property of separability discussed in section~\ref{sec:nonseparableshapes} for the CMB bispectrum are not present in the case of three-dimensional, large-scale structure observables.

The nonlinear power spectrum is obtained perturbatively from the expansion
\be
\la\d_{\kv_1}\d_{\kv_2}\ra=\la\d_{\kv_1}^{(1)}\d_{\kv_2}^{(1)}\ra+
\la\d_{\kv_1}^{(1)}\d_{\kv_2}^{(2)}\ra+
\la\d_{\kv_1}^{(2)}\d_{\kv_2}^{(2)}\ra+
\la\d_{\kv_1}^{(1)}\d_{\kv_2}^{(3)}\ra+\dots\,,
\ee
where the term $\la\d_{\kv_1}^{(1)}\d_{\kv_2}^{(1)}\ra$ corresponds to the linear solution, $P_0(k)$ while the other terms represent, in analogy with perturbation theory in quantum field theory, {\it one}- and higher {\it loop corrections} as they involve integrations over internal momenta. In particular, the term $\la\d_{\kv_1}^{(1)}\d_{\kv_2}^{(2)}\ra$ vanishes for Gaussian initial conditions as it depends on the initial bispectrum $B_0$ (see \citealt{TaruyaKoyamaMatsubara2008} for an analysis of nonlinear corrections to the matter power spectrum {\it due to} primordial non-Gaussianity). 

In a similar fashion, nonlinear corrections in \eqn{eq:PT} provide a perturbative expansion for the matter bispectrum,
\be
\la\d_{\kv_1}\d_{\kv_2}\d_{\kv_3}\ra=\la\d_{\kv_1}^{(1)}\d_{\kv_2}^{(1)}\d_{\kv_3}^{(1)}\ra+
\la\d_{\kv_1}^{(1)}\d_{\kv_2}^{(1)}\d_{\kv_3}^{(2)}\ra+
\la\d_{\kv_1}^{(1)}\d_{\kv_2}^{(2)}\d_{\kv_3}^{(2)}\ra+
\la\d_{\kv_1}^{(1)}\d_{\kv_2}^{(1)}\d_{\kv_3}^{(3)}\ra+\dots\,.
\ee
In this case, the leading order contributions are given by the {\it tree-level} terms $\la\d_{\kv_1}^{(1)}\d_{\kv_2}^{(1)}\d_{\kv_3}^{(1)}\ra$ and $\la\d_{\kv_1}^{(1)}\d_{\kv_2}^{(1)}\d_{\kv_3}^{(2)}\ra$, with the first being the initial component and the second corresponding to a contribution to the matter bispectrum due to {\it gravity alone}, of the form
\be\label{eq:bsmGtree}
B_G^{tree}(\kall)=2F_2(\kv_1,\kv_2)P_0(k_1)P_0(k_2) + 2{\rm ~perm.}\,.
\ee
Notice that this contribution is present even for Gaussian initial conditions as it depends only on the initial power spectrum $P_0$ and describe the emergence of non-Gaussianity due to gravitational instability. The leading order, {\it tree-level} expression of the matter bispectrum with non-Gaussian initial conditions is therefore given in terms of the {\it sum}
\be\label{eq:bsmTL}
B^{tree}(\kall)= B_0(k_1,k_2,k_3;z)+B_G^{tree}(k_1,k_2,k_3;z)\,.
\ee
This expression corresponds to the first two terms on the r.h.s. of \eqn{eq:skewPT} for the skewness, which can be obtained from \eqn{eq:bsmTL} by integration. 

\begin{figure}[t]
\begin{center}
{\includegraphics[width=0.48\textwidth]{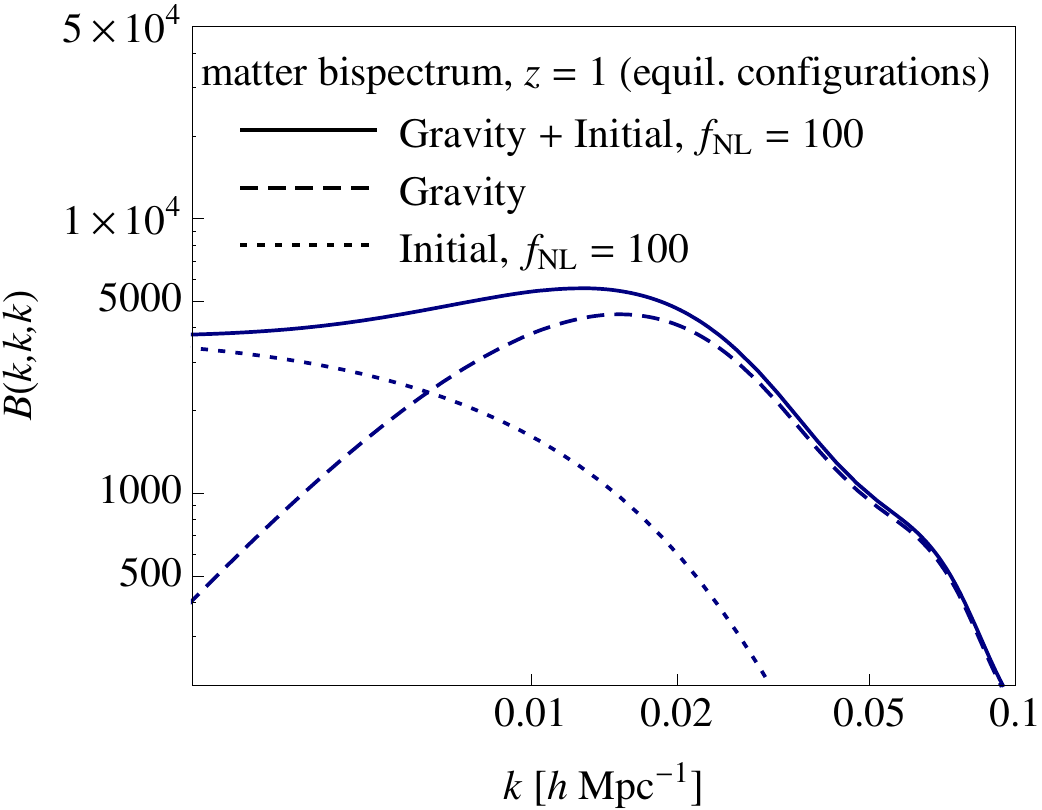}}
{\includegraphics[width=0.48\textwidth]{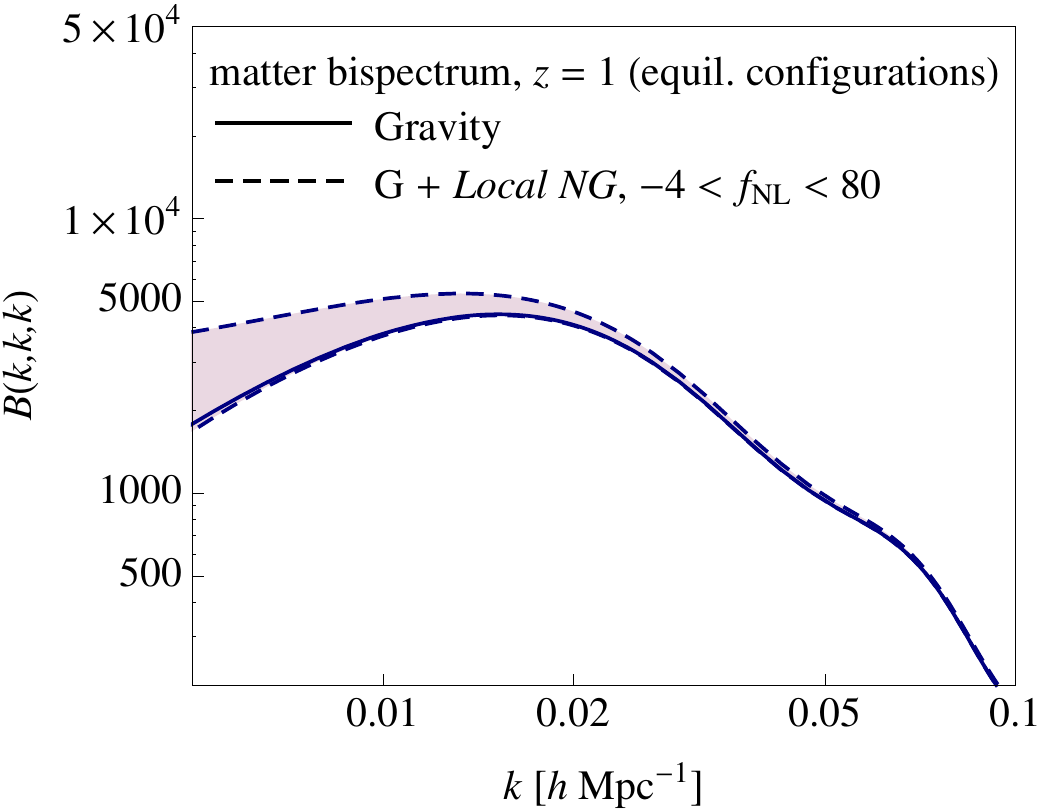}}
{\includegraphics[width=0.48\textwidth]{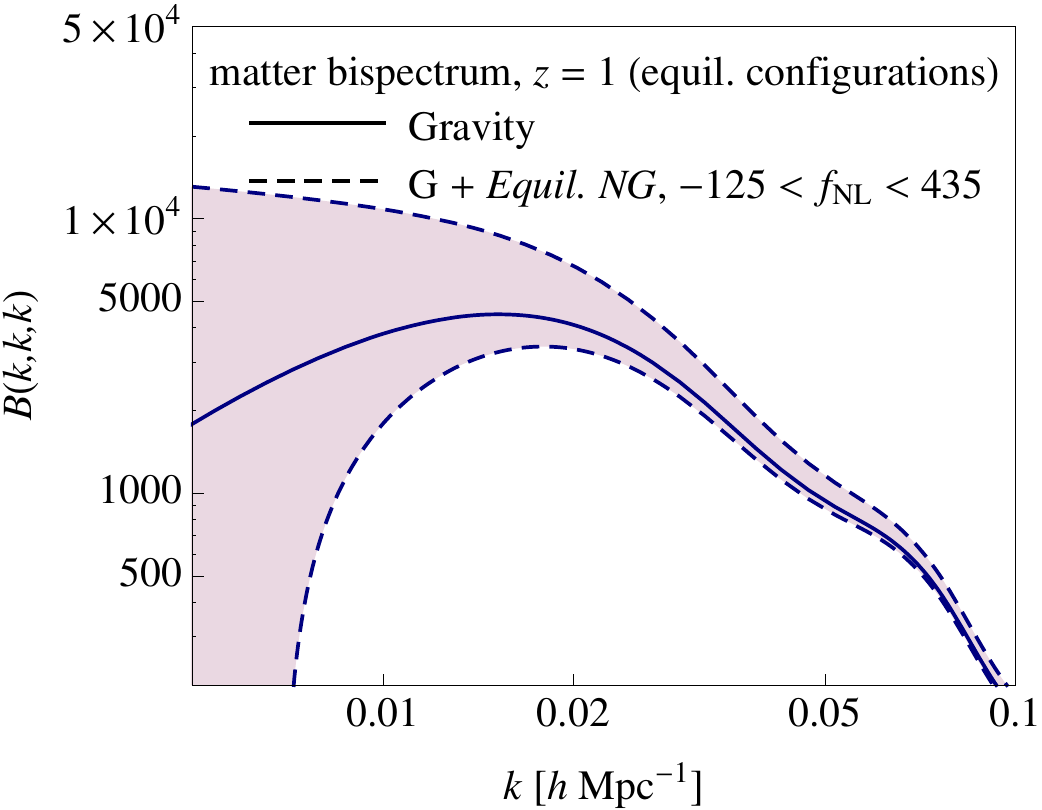}}
{\includegraphics[width=0.48\textwidth]{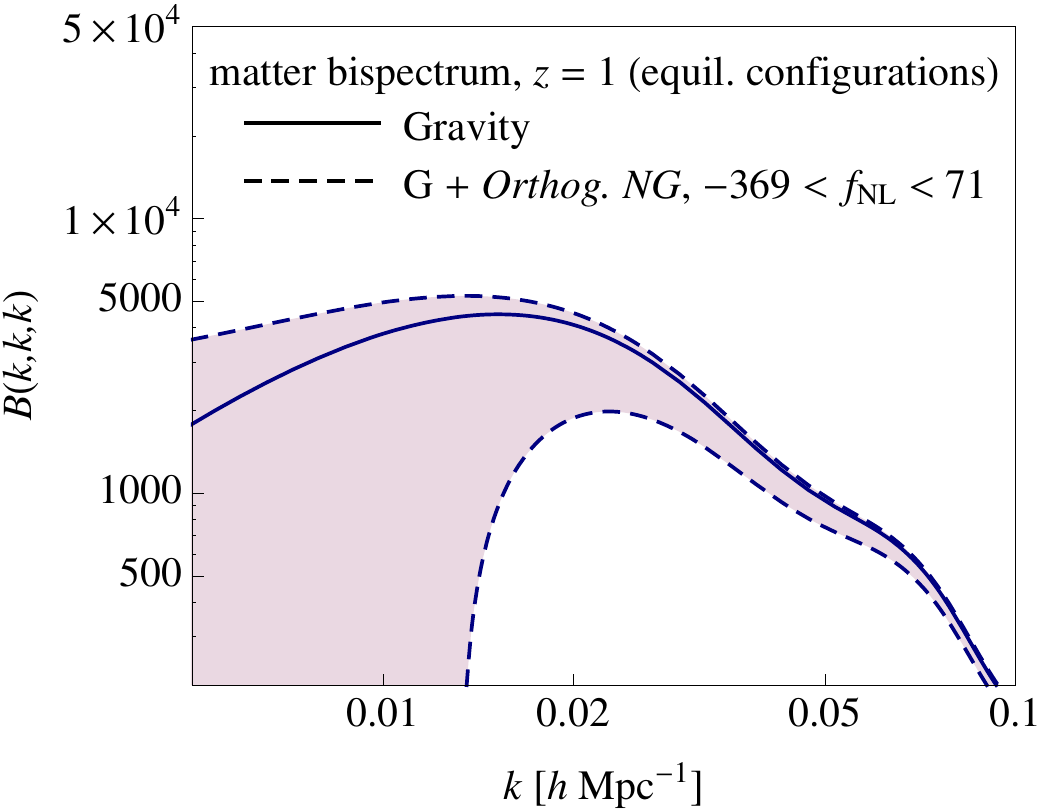}}
\caption{Effect of the primordial component for different non-Gaussian models on the equilateral configurations of matter bispectrum, $B(k,k,k)$, at redshift $z=1$, as a function of scale, at tree-level in PT. In the upper left panel the continuous line shows the initial component $B_0$({\it dotted line}), the gravity-induced component, $B_G^{tree}$ ({\it dashed line}) and their sum ({\it continuous line}). For equilateral configurations the initial component coincides for the local, equilateral and orthogonal models while it vanishes in the folded model. In the other panels, continuous lines show the gravity component alone while dashed lines show the tree-level bispectrum including the primordial component for the local ({\it upper right panel}), equilateral ({\it lower left panel}) and orthogonal ({\it lower right panel}) models assuming the values of $\fnl$ corresponding to the 95\% C.L. limits as determined by \citet{SmithSenatoreZaldarriaga2009} and \citet{SenatoreSmithZaldarriaga2009} from WMAP observations. The shaded area indicates the currently allowed region. }
\label{fig:bsmeq}
\end{center}
\end{figure}

The possibility of distinguishing the primordial component $B_0$ from the gravity-induced one $B_G$ relies on their specific and distinct dependence on {\it scale}, on the {\it triangular configuration shape} and on {\it redshift}. For a primordial non-Gaussianity described by a curvature bispectrum obeying the hierarchical scaling $B_\Phi\sim P_\Phi^2$, typical of weakly non-Gaussian models such as the local and equilateral ones, the different redshift and scale dependence of the two contributions is evident in their ratio for {\it equilateral triangles} ($k_1=k_2=k_3=k$), given by\footnote{The first equality is in fact {\it identical} for local, equilateral and orthogonal non-Gaussianity, simply by definition of the equilateral bispectrum, Eq.~\ref{eq:}, introduced in \citep{BabichCreminelliZaldarriaga2004} and of the orthogonal bispectrum, Eq.~\ref{eq:}, introduced in \citep{SenatoreSmithZaldarriaga2009}, where $\fnle$ and $\fnlo$ are precisely the amplitudes that provide the same value for the curvature bispectrum as the local model for equilateral configurations.}
\be
\frac{B_0(k,k,k;z)}{B_G^{tree}(k,k,k;z)}=\frac{7}{4}\frac{\fnl}{M(k;z)}
\stackrel{k\rightarrow 0}{\sim}\frac{\fnl}{k^2 D(z)}\,.
\ee
We therefore expect, for a wide range of non-Gaussian models, the initial contribution $B_0$ to be larger at large scales and at high redshift. The upper left panel of \fig{fig:bsmeq} shows the two contributions and their sum for equilateral configurations $B(k,k,k)$ as a function of $k$. The other panels show the effect of the primordial component for different non-Gaussian models, for values of the respective parameters $\fnl$ corresponding to the current 95\% C.L. limits \citep{SmithSenatoreZaldarriaga2009,SenatoreSmithZaldarriaga2009}, with the shaded area indicating the allowed region.  

\begin{figure}[t]
\begin{center}
{\includegraphics[width=0.48\textwidth]{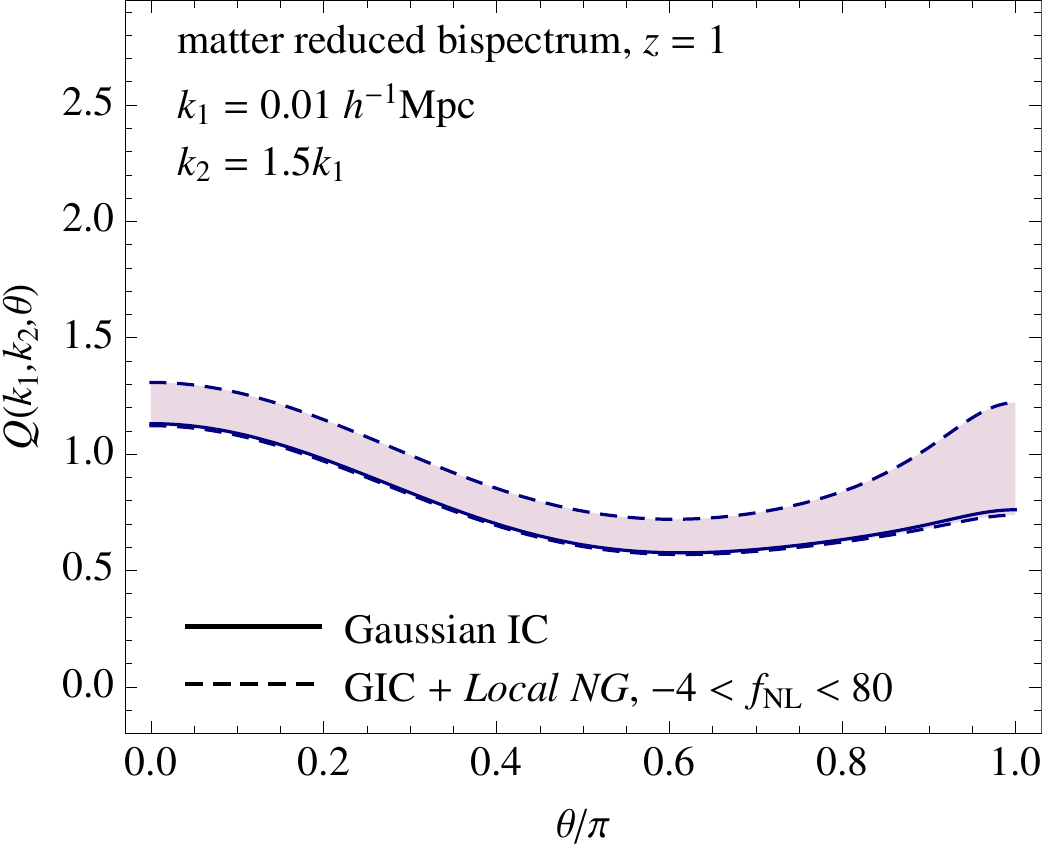}}
{\includegraphics[width=0.48\textwidth]{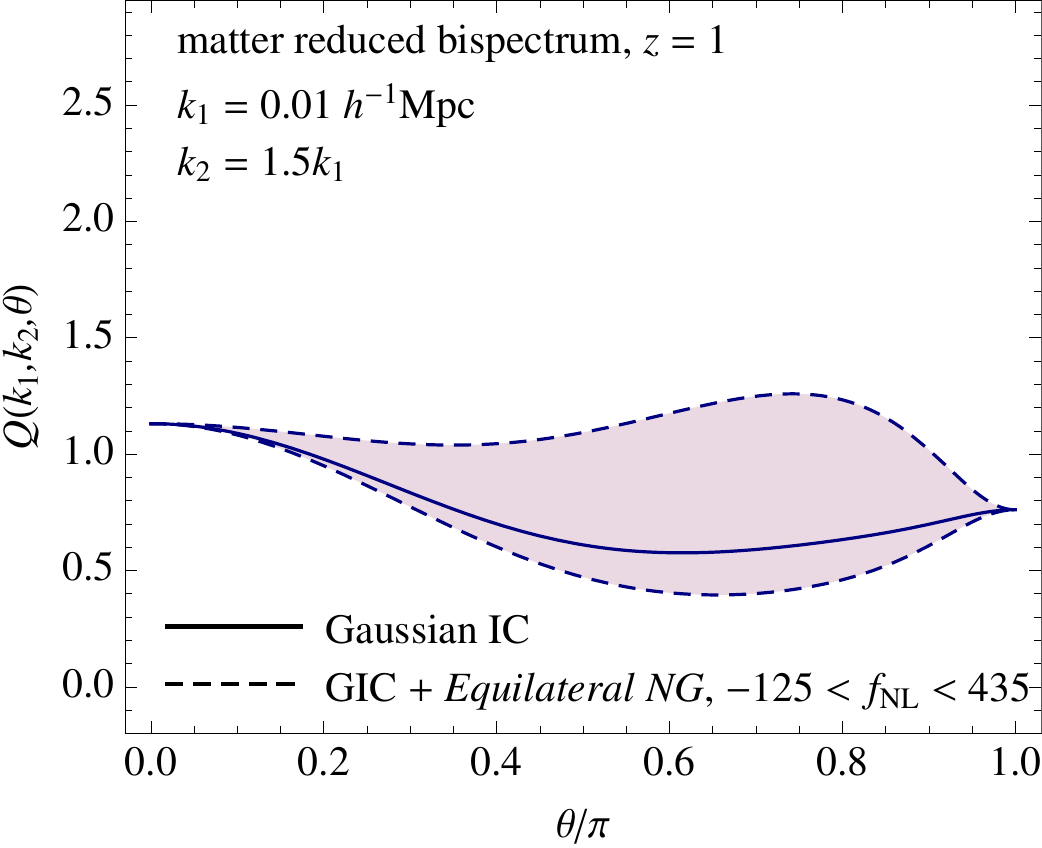}}
{\includegraphics[width=0.48\textwidth]{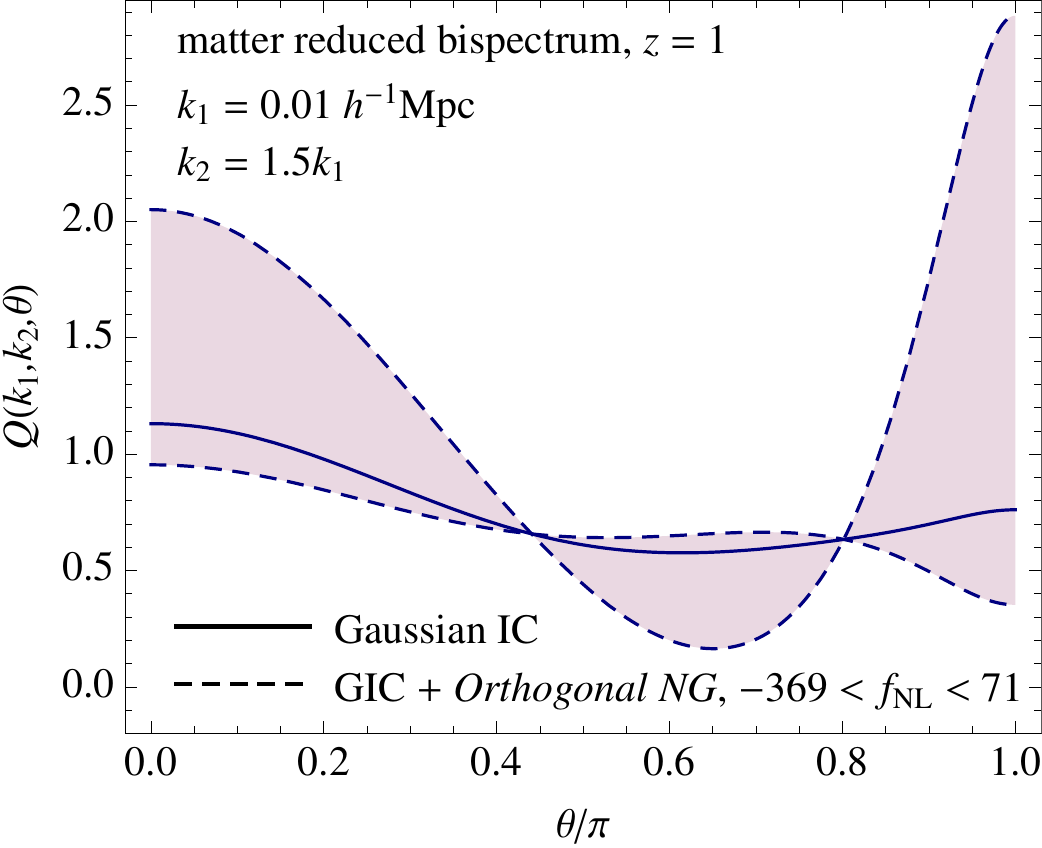}}
{\includegraphics[width=0.48\textwidth]{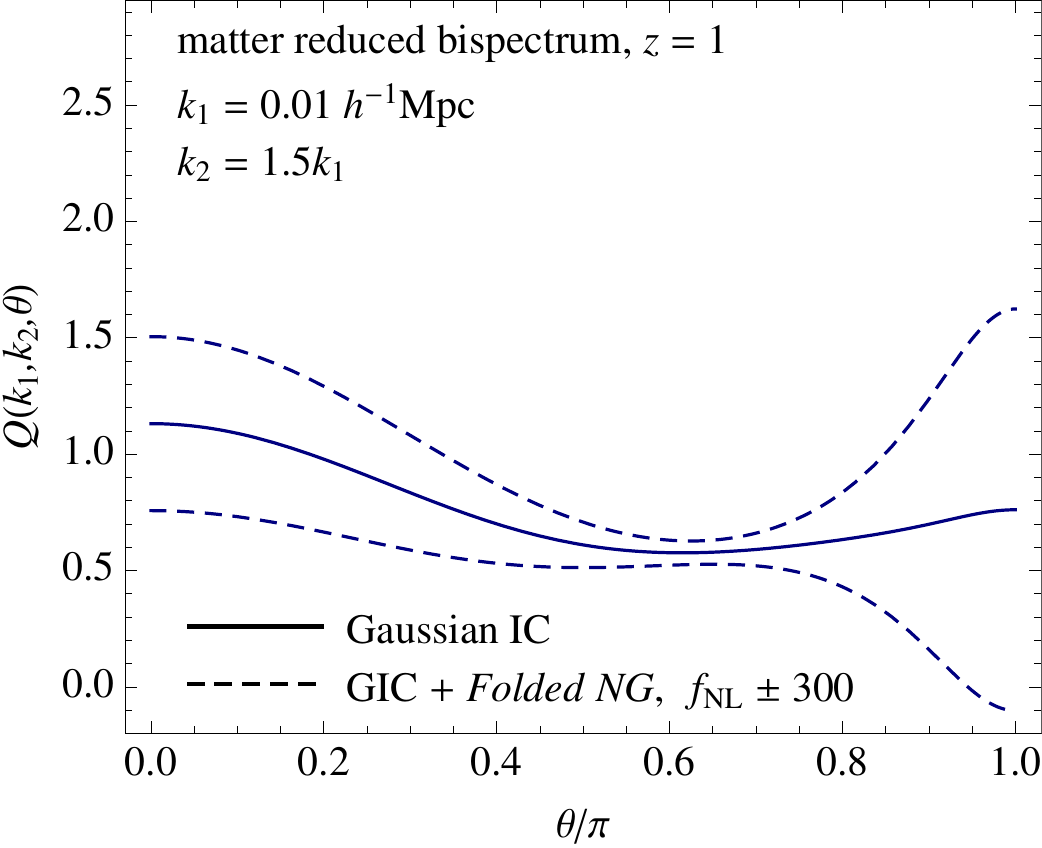}}
\caption{Effect of the primordial component for different non-Gaussian models on the matter reduced bispectrum, as a function of the triangle shape. The continuous line shows the reduced bispectrum $Q(\kall)$ at tree-level in PT for Gaussian initial conditions at redshift $z=1$ assuming $k_1=0.01\kMpc$, $k_2=1.5k_1$ as a function of the angle $\theta$ between $k_1$ and $k_2$. Dashed lines show the reduced bispectrum including the primordial component for the local ({\it upper left panel}), equilateral ({\it upper right panel}), orthogonal ({\it lower left panel}) and folded ({\it upper left panel}) models. For the local, equilateral and orthogonal models we assume the values of $\fnl$ corresponding to the 95\% C.L. limits as determined by \citet{SmithSenatoreZaldarriaga2009} and \citet{SenatoreSmithZaldarriaga2009} from WMAP observations. The shaded area indicates the currently allowed region. For the folded model, for which no observational constraints are available, the values $\fnl=\pm300$ are considered. }
\label{fig:bsmA2}
\end{center}
\end{figure}
In addition, $B_G^{tree}$ presents a specific dependence on triangle shapes, determined by gravitational instability and described by \eqn{eq:bsmGtree} at tree-level. The shape dependence of $B_0$, determined by the specific non-Gaussian model under consideration, is generically different. Such differences can be explicitly shown in plots of the {\it reduced} bispectrum, defined as
\be
Q(k_1,k_2,k_3)=\frac{B(k_1,k_2,k_3)}{P(k_1)P(k_2)+2{\rm ~perm.}}\,,
\ee
which removes the redshift and scale dependencies of the gravity contribution. \fig{fig:bsmA2} shows the reduced bispectrum $Q(k_1,k_2,k_3)$ at tree-level in perturbation theory, at $z=1$ for $k_1=0.01\kMpc$, $k_2=1.5 k_1$ as a function of the angle $\theta$ between $k_1$ and $k_2$. In all panels, the continuous line represents the gravity-induced term which assumes larger values for nearly collapsed triangles, \ie for $\theta\simeq 0$ or $\pi$. This indicates that the {\it probability} of finding larger values for the matter density in triplets of points forming a squeezed or folded triangle is larger than for nearly equilateral triangles. This prediction is confirmed by the typical {\it filamentary} nature of the large-scale structure, evident from snapshots of N-body simulations or images of redshift surveys, since along these filaments it is easier to form collapsed triangles than equilateral ones. It should be stressed that the bispectrum is, in fact, the lowest order statistic sensitive to the three-dimensionality of structures and that these features are not captured by the information contained in the power spectrum alone. The effects of the primordial component on the matter bispectrum are shown by the dashed lines which correspond, as in \fig{fig:bsmeq} to the 2-$\sigma$ limits from CMB observations, in the case of the local ({\it upper left panel}), equilateral ({\it upper right panel}) and orthogonal ({\it lower left panel}) models while they correspond to the values $\fnl=\pm300$ in the folded case, for which no experimental bounds are available. Although the large scales $k_1=0.01\kMpc$ and $k_2=0.015\kMpc$ and the relatively high redshift $z=1$ have been chosen to enhance the effect of the non-Gaussian component, these triangles are not completely out of reach for future, large-volume surveys. Primordial non-Gaussianity modifies, in very specific ways, the shape dependence of the matter bispectrum produced by gravitational instability. 

While the dependence of the matter bispectrum on scale and redshift is responsible for the specific behavior of the skewness of the matter density field on the smoothing scale $R$ and redshift, the sensitivity to the triangle shape is completely lost in analysis of the density higher-order moments. Instead, accurate measurements of the bispectrum, when achievable, offer in principle the possibility to disentangle the different contributions when triangles of different size {\it and shape} are included in the analysis. 

The matter bispectrum is not, unfortunately, a direct observable. While we will discuss later how the statistical properties of the matter distribution can be inferred from galaxy redshift surveys, we should mention that the shear field in weak lensing surveys is another observable directly related to the matter distribution.  The observational consequences on the {\it weak lensing} bispectrum, of a primordial non-Gaussian component (of the {\it local} type) such as the one in \eqn{eq:bsmTL}, have been explored in \citet{TakadaJain2004}. The authors find that, the primordial component {\it alone} (\ie {\it without} contamination from the gravitational one) could be detected if $\fnll>150 f_{sky}^{1/2}$, assuming $l_{max}\simeq 500$ and a tomography over four redshift bins for a galaxy number density of $\bar{n}_g=100$ arcmin$^{-2}$. The large cosmic variance for low $\ell$'s makes difficult the detection of the primordial component, prominent instead at larger scales. As we will see in the next section, primordial non-Gaussianity has some effect on small scales as well, due to the nonlinear evolution of structures.

\subsubsection{Second-order corrections}
\label{sssec:lssmPT1l}

The simple prediction of \eqn{eq:bsmTL} for the matter bispectrum is expected to be valid at the largest observable scales and at high-redshift, where nonlinear evolution is subdominant. Despite the fact that such conditions correspond as well to the regime where a detection of the initial component $B_0$ is favored, the effects of non-Gaussian initial conditions can be significant even at smaller scales and at low redshift. Since these effects are the result of nonlinear gravitational evolution {\it and} non-Gaussian initial conditions, it is no longer possible to identify distinct contributions resulting from distinct sources of non-Gaussianity, as it is the case for the tree-level expression of \eqn{eq:bsmTL}. Nevertheless, it is possible to distinguish individual corrections in PT to the matter bispectrum depending exclusively on the initial power spectrum $P_0$, and therefore present as well for Gaussian initial conditions, and corrections depending instead on higher-order initial correlators, such as the initial bispectrum $B_0$ and trispectrum $T_0$, which can be interpreted as small-scales effects due to non-Gaussian initial conditions. One-loop corrections in PT for Gaussian initial conditions have been studied in \citet{Scoccimarro1997}, while the extension of these results to non-Gaussian initial conditions is studied in \citet{Sefusatti2009}.

\begin{figure}[t]
\begin{center}
{\includegraphics[width=0.48\textwidth]{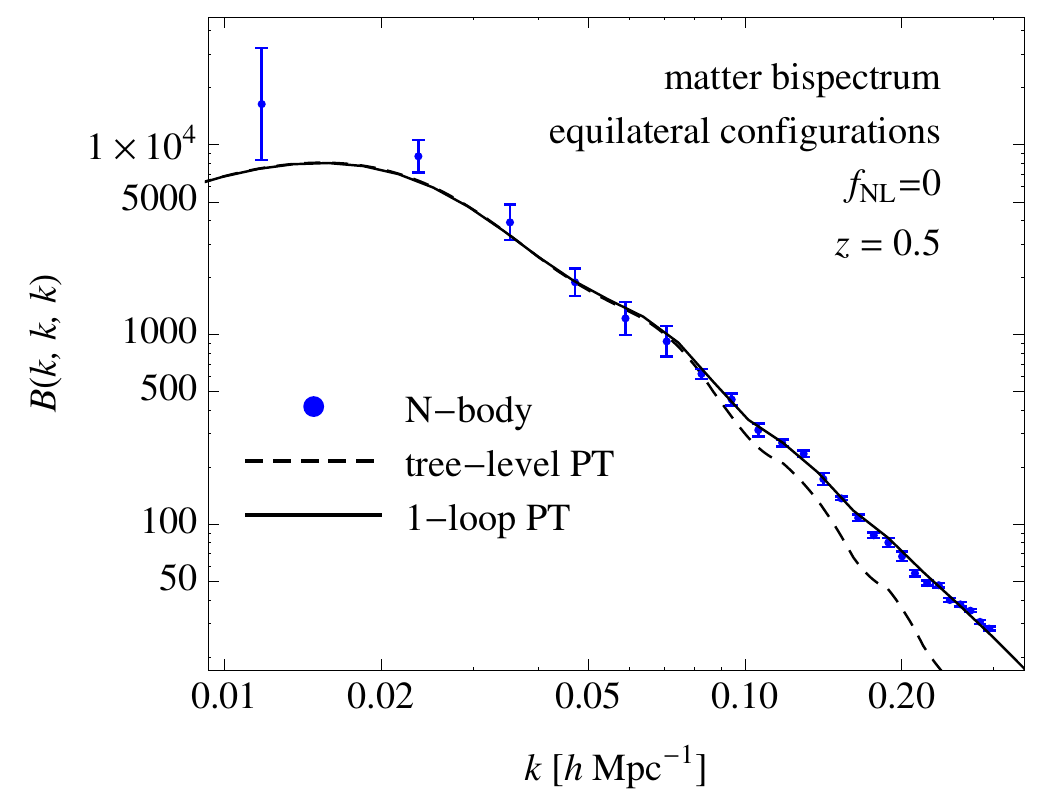}}
{\includegraphics[width=0.48\textwidth]{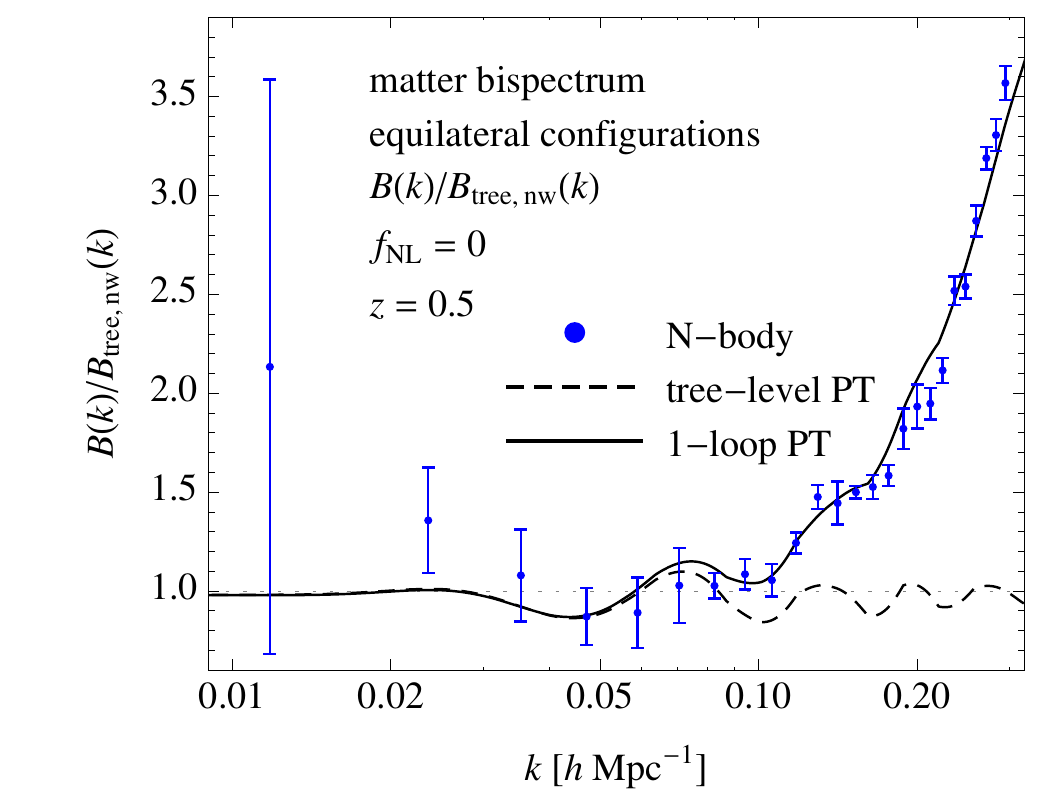}}
{\includegraphics[width=0.48\textwidth]{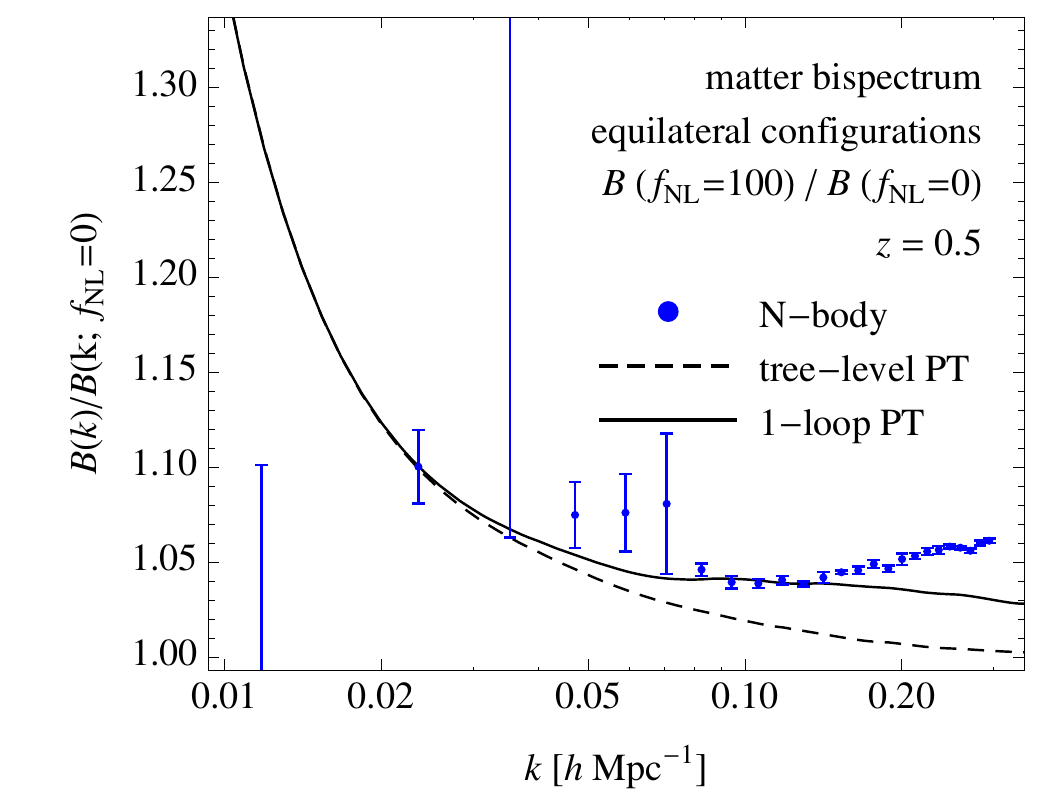}}
{\includegraphics[width=0.48\textwidth]{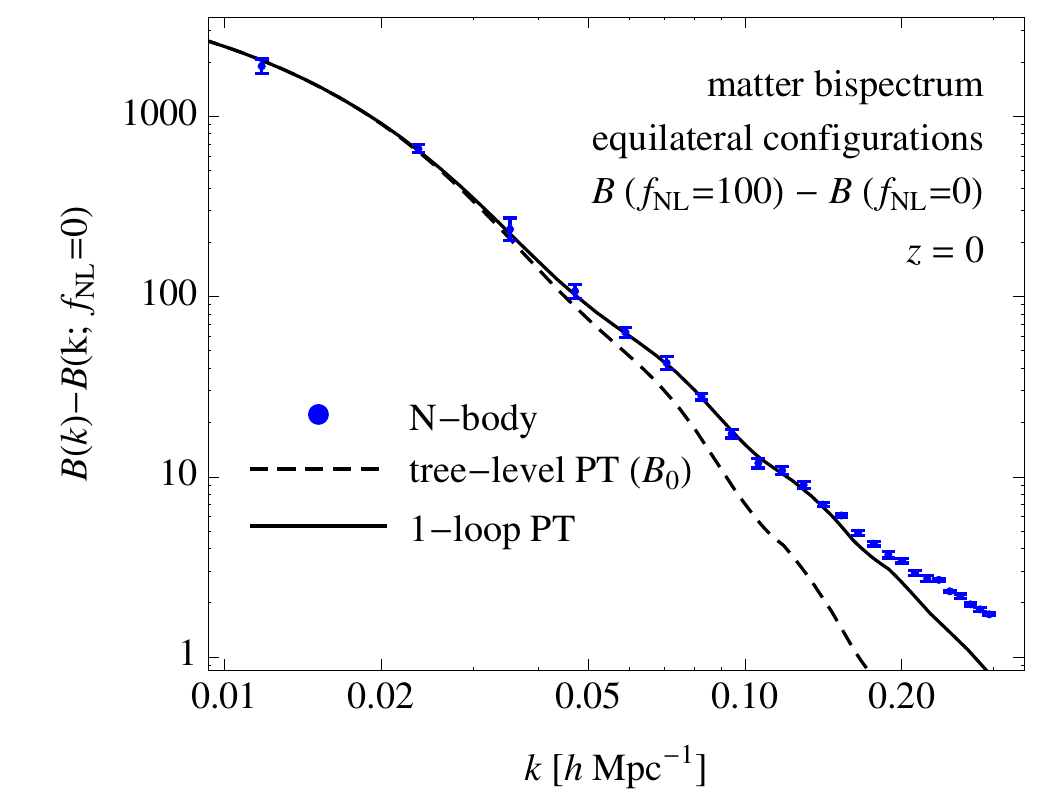}}
\caption{{\it Upper panels}: equilateral configurations of the matter bispectrum measured in N-body simulations with Gaussian initial conditions ({\it data points}) and tree-level ({\it dashed lines}) and one-loop ({\it continuous lines}) predictions in perturbation theory. The right panel shows the ratio to the tree-level prediction with acoustic oscillations removed. {\it Lower panels}): ratio ({\it left}) and difference ({\it right}) between the matter bispectrum measured in realizations with local non-Gaussian initial conditions ($\fnl=100$) and the Gaussian case, compared with PT predictions. From \citep{SefusattiCrocceDesjacques2010}.}
\label{fig:bsm}
\end{center}
\end{figure}

A comparison of these results with measurements of the matter bispectrum in N-body simulations \citep{DesjacquesSeljakIliev2009} with non-Gaussian initial conditions of the local kind can be found in \citet{SefusattiCrocceDesjacques2010}. \Fig{fig:bsm} shows the equilateral configurations of the matter bispectrum measured in N-body simulations together with predictions from perturbation theory at tree-level ({\it dashed line}) and one-loop ({\it continuous line}). In particular, the upper left panel considers $B(k,k,k)$ for Gaussian initial conditions while the upper right panels shows the same quantity divided by the tree-level prediction in PT to highlight the small-scales non-linear behavior. The lower left and right panels show, respectively, the ratio and the difference between the matter bispectrum with an initial local component corresponding to $\fnl=100$ and the Gaussian case. The agreement between one-loop predictions and the simulations results is quite remarkable, while we notice that the tree-level prediction fails to accurately describe the effect of primordial non-Gaussianity already at relatively large scales.  

The significance of these relatively small corrections to individual configurations is to be considered in relation to the much larger number of configurations that can be measured as we include smaller and smaller scales and they could lead to a measurable effect when considered in terms of the cumulative signal to noise ratio. On the other hand, these effects loose in part the shape dependence of the original initial bispectrum and require an accurate model (perhaps beyond standard perturbation theory) and strong priors on the underlying cosmological parameters to be distinguished from the nonlinear, "Gaussian" component. A step in the direction of improved predictions is offered by the promising results of Renormalized Perturbation Theory \citep{CrocceScoccimarro2006A, CrocceScoccimarro2006B, BernardeauCrocceScoccimarro2008} and of the Renormalization Group approach \citep{MatarresePietroni2007, Pietroni2008}. The extension of the latter to the case of non-Gaussian initial conditions has been recently considered in \citet{BartoloEtal2009}, which studies specific predictions for the matter power spectrum and bispectrum.

\subsection{The Galaxy Bispectrum}
\label{ssec:lssg}

From the discussion above, we could expect that future, {\it large-volume} and {\it high-redshift} galaxy surveys will be able to {\it directly} detect a possible, large primordial component to the matter bispectrum by measurements of the galaxy bispectrum, or at least provide constraints on the non-Gaussian parameters {\it comparable} to the constraints from measurements of the CMB bispectrum. Such an expectation is motivated by the simple observation that the number of {\it Fourier modes} available in a {\it three-dimensional}, ideal, all-sky galaxy survey is in principle much larger than the number of modes available in {\it two-dimensional} CMB maps. 

The galaxy distribution is, however, a less direct probe of the early Universe than the CMB temperature fluctuations. On top of the nonlinear evolution of structures and its contribution to higher-order correlation functions, one has to take into account the {\it nonlinear} nature of {\it galaxy bias}, itself responsible for additional non-Gaussianity. An analysis of the galaxy bispectrum should therefore be able to detect a small primordial component by separating it from these primary contributions. 

In this respect, an even more complex picture, due to additional and somehow unexpected  effects of primordial non-Gaussianity {\it on} galaxy bias, has been emerging in the last couple of years, following the results of \citet{DalalEtal2008}. N-body simulations have shown, in fact, that nonlinear bias and an initial bispectrum are not two distinct sources of non-Gaussianity for the galaxy bispectrum, {\it not even at large scales!} Instead a {\it local} initial component can significantly affect the bias relation precisely at large scales, adding extra corrections.  In the spirit of a review and since {\it we do not have}, at the time of writing, {\it a satisfactory model of the galaxy bispectrum in presence of non-Gaussian initial conditions}, in this section we will summarize earlier results, while in section~\ref{ssec:bsgNGBias} we will present the recent developments that radically changed our understanding of the effects of local non-Gaussianity on the large-scale structure and finally comment, in section~\ref{ssec:bsgNGBiasBisp} on some consequences for galaxy bispectrum measurements as far as current research provides.

\subsubsection{The galaxy bispectrum and local bias}
\label{ssec:bsgTL}

{\it Until recently}, it was commonly assumed, {\it even for non-Gaussian initial conditions}, that the galaxy overdensity $\d_g(\xv)$, defined in terms of the galaxy density $n_g(\xv)$ and its mean $\bar{n}_g$ as
\be
\d_g(\xv)\equiv\frac{n_g(\xv)-\bar{n}_g}{\bar{n}_g}\,,
\ee 
can be expressed, at large-scales, as a {\it local} function of the matter density contrast, $\d(\xv)$\footnote{Properly speaking we should consider here the {\it smoothed} matter density contrast, that is $\d_R(\xv)=\int d^3x' W_R(\xv-\xv')\d(\xv')$ with $W_R$ a top-hat filter function. For simplicity, we implicitly assume a {\it smooth} density field, so that, for large enough filtering scale, \eg $R\sim 10\Mpc$, matter perturbations are small, $\d\lsim 1$},  \ie
\be
\d_g(\xv)=f[\d(\xv)]\,.
\ee
Such a reasonable expectation is based on the fact that the physics of galaxy formation operates on much smaller scales, below the typical halo size, than those we are interested in. At large scales, where fluctuations are small, $\d_R\lesssim 1$, we can consider the Taylor expansion, \citep{FryGaztanaga1993}
\be\label{eq:bias}
\d_g(\xv)=b_1\,\d(\xv)+\frac{1}{2}\,b_2\,\d^2(\xv)+\frac{1}{3!}\,b_3\,\d^3(\xv)+\dots\,,
\ee
describing the bias relation between galaxy and matter in terms of a series of {\it constant bias parameters}, $b_i$. This expansion allows for a consistent extension of the perturbative expressions for the matter correlators to the galaxy ones. In fact, from \eqn{eq:bias} we can derive the galaxy three-point function in position space
\be
\la\d_g(\xv_1)\d_g(\xv_2)\d_g(\xv_3)\ra=b_1^3\,\la\d(\xv_1)\d(\xv_2)\d(\xv_3)\ra+
b_1^2b_2\,\la\d(\xv_1)\d(\xv_2)\d^2(\xv_3)\ra+{\rm perm.}+\dots\,,
\ee
and the {\it tree-level} expression for the {\it galaxy bispectrum} given by
\be
B_g(\kall)  =  b_1^3\,B^{tree}(\kall)
 + b_1^2b_2\,[P_0(k_1)P_0(k_2)+2{\rm ~perm.}]\,,
\ee
where the second term on the r.h.s., proportional to the {\it quadratic} bias parameter $b_2$, is of the same order of the gravity-induced contribution to the matter bispectrum $B_G^{tree}$, \eqn{eq:bsmGtree}. Relying on this simple result, measurements of the galaxy bispectrum has been considered in the first place, in the context of Gaussian initial conditions, as a way to determine the bias parameters, and break the degeneracy between linear bias ($b_1$) and the amplitude of matter fluctuations (\eg $\sigma_8$), otherwise affecting power spectrum measurements \citep{Fry1994B,MatarreseVerdeHeavens1997,Scoccimarro2000B,ScoccimarroCouchmanFrieman1999,SefusattiScoccimarro2005,SefusattiEtal2006,NishimichiEtal2007,JeongKomatsu2009A}. In this respect, the corresponding reduced galaxy bispectrum is
\be\label{eq:Qg}
Q_g(\kall)\equiv\frac{B_g(\kall)}{P_g(k_1)P_g(k_2)+2{\rm ~perm.}}=\frac{1}{b_1}Q(k_1,k_2,k_3)+\frac{b_2}{b_1^2}\,,
\ee
where $Q$ is the reduced {\it matter} bispectrum (including a possible initial contribution) and the effect of nonlinear bias is simply given by an additive constant term. 
As already mentioned, measurements of triangular configurations different in shape and size allow to disentangle the different sources of non-Gaussianity and determine independently $b_1$ and $b_2$, provided that accurate predictions for the matter bispectrum, from PT or N-body simulations,  are available \citep{GuoJing2009A,GuoJing2009B} and the effects of redshift distortions and the survey geometry are properly taken into account \citep{Scoccimarro2000B,SmithShethScoccimarro2008}.

In particular, if we allow the possibility of non-Gaussian initial conditions, then the matter bispectrum includes an initial contribution, so that we can rewrite \eqn{Qg} at tree-level explicitly as
\be\label{eq:QgNG}
Q_g^{tree}=\frac{1}{b_1}\left[Q_I(\fnl)+Q_G^{tree}\right]+\frac{b_2}{b_1^2}\,,
\ee 
and we can extend the analysis to obtain simultaneous constraints on the bias parameters {\it and} on the parameter determining the amplitude of the primordial bispectrum, \ie $\fnl$. A first conservative estimate of the possibilities offered by this method in measurements of the galaxy bispectrum in the 2dF Galaxy Redshift Survey \citep{CollessEtal2001} and in the Sloan Digital Sky Survey \citep[SDSS][]{YorkEtal2000} is given in \citet{VerdeEtal2000} as a simple extension of previous results for the bias alone \citep{MatarreseVerdeHeavens1997} suggesting that a primordial component could be detected for values of a {\it local} $\fnl$ of the order of $10^3$-$10^4$. As we will see in the next sections, a complete analysis of the galaxy bispectrum, including all measurable configurations can improve this estimate by more than an order of magnitude.

Among the various observational issues in analyses of galaxy correlators, \eg finite volume effects or completeness of the galaxy samples, we stress that particularly relevance has the problem of redshift distortions. Redshift distortions have in fact a significant impact on the shape dependence of the galaxy bispectrum, particularly at small scales \citep{ScoccimarroCouchmanFrieman1999, Scoccimarro2000B}. A recent treatment of redshift distortions in bispectrum predictions (with Gaussian initial conditions) can be found in \citet{SmithShethScoccimarro2008}.

\subsubsection{A bispectrum estimator}
\label{ssec:bsgBE}

In this section we define a simple estimator for the measurement of the galaxy bispectrum in N-body simulations as well as actual data. This allows us to derive an expression for the bispectrum variance and define a Fisher matrix for an analysis of the galaxy bispectrum in terms of the non-Gaussian (and bias) parameters. In the next section we will consider a proper likelihood analysis and the effects of the bispectrum covariance. Since what follows can be applied in general to bispectrum measurements we will consider, to simplify the notation, the case of the {\it matter} density field in Fourier space, described by the density contrast $\d_{\kv}$. We will point-out relevant differences in the application to the galaxy distribution.

For a cubic box of volume $V$, a bispectrum estimator can be defined as \cite{ScoccimarroEtal1998}
\bea
\label{Best}
\hat{B}(\kall) & \equiv & \frac{V_f}{V_{B}(\kall)}\int_{k_1}\!\!\!\!d^3 q_1\int_{k_2}\!\!\!\!d^3 q_2\int_{k_3}\!\!\!\!d^3 q_3 \;\delta_D(\qv_{123})~ \d_{\qv_1}~\d_{\qv_2}~\d_{\qv_3}\,,
\eea
where $V_f\equiv k_f^3=(2\pi)^3/V$ is the volume of the fundamental cell and where each integration is defined over the bin $q_i\in[k_i-\D k/2,k_i+\D k/2]$ centered at $k_i$ and of size $\D k$ equal to a multiple of the fundamental frequency $k_f$. The Dirac delta function $\d_D(\qv_{123})$ ensures that the wavenumbers $\qv_1$, $\qv_2$ and $\qv_3$ indeed form a closed triangle, as imposed by translational invariance, while the normalization factor $V_B(\kall)$, given by
\bea
V_B(\kall) & \equiv & \int_{k_1} \!\!\!\! d^3 q_1\int_{k_2} \!\!\!\! d^3 q_2 \int_{k_3} \!\!\!\! d^3 q_3 \,\delta_D(\qv_{123})
\simeq
8\pi^2\ k_1 k_2 k_3\ \D k^3\,,
\eea
represents the number of {\it fundamental} triangular configurations (given by the triplet $\qv_1$, $\qv_2$ and $\qv_3$) that belong to the triangular configuration {\it bin} defined by the triangle sizes $k_1$, $k_2$ and $k_3$ with uncertainty $\D k$. 

The leading contribution to the bispectrum variance following from this estimator, in analogy with the power spectrum case \citep{FeldmanKaiserPeacock1994}, is given by \citet{ScoccimarroEtal1998}~\footnote{This expression, \citep[see][]{ScoccimarroSefusattiZaldarriaga2004}, corrects a typo in equation~(A16) of~\citet{ScoccimarroEtal1998}.}
\be
\D B^2(\kall)=V_f \frac{s_{123}}{V_B(\kall)}\ P_{tot}(k_1)P_{tot}(k_2)P_{tot}(k_3)\,,
\label{eq:Berror}
\ee
with the factor $s_{123}=6,2,1$ respectively for equilateral, isosceles and general triangles and where
\be
P_{tot}(k)  \equiv P(k)+\frac{1}{(2 \pi)^3}\ \frac{1}{ \bar{n}}\,,
\ee
with the particle (or galaxy) number density $\bar{n}$ accounting for the shot noise contribution. In the case of a galaxy distribution, the matter power spectrum $P(k)$ on the r.h.s. should be replace with the galaxy power spectrum, expressed, at large-scales, by $P_g(k)=b_1^2P(k)$, under the local bias assumption of \eqn{eq:bias}. \eqn{eq:Berror} constitutes the {\it Gaussian limit} to the bispectrum variance, as it neglects higher-order corrections dependent on the three-, four- and six-point, connected, correlation functions.

\subsubsection{Fisher matrix forecasts}

In this section we consider simple forecasts for the constraints on the non-Gaussian parameters from measurements of the galaxy bispectrum in future redshift surveys. Specifically, we will consider a Fisher matrix for reduced galaxy bispectrum $Q_g$ in terms of the non-Gaussian parameter $\fnl$ and the linear and quadratic bias parameters $b_1$ and $b_2$. These three parameters characterize the relative weight of the different non-Gaussian contributions to the galaxy bispectrum. Since the possibility to detect a primordial component relies on our ability to separate the three contributions, a robust result should, {\it at least}, involve a marginalization over bias. On the other hand, we will assume all cosmological parameters as known. This is in part justified by the weak dependence of the matter reduced bispectrum on cosmology discussed in the previous section. In this respect, it can be shown that the reduced bispectrum has the same signal to noise as the bispectrum. For a given triangular configurations, in fact,
\be
\left(\frac{S}{N}\right)_{(\kall)}\equiv\frac{Q_g(\kall)}{\D Q_g(\kall)}
\simeq \frac{B_g(\kall)}{\D B_g(\kall)}\,,
\ee
since the variance of $Q$ is dominated by the variance of $B$ \citep[see, for instance, ][]{ScoccimarroSefusattiZaldarriaga2004}.

The Fisher matrix can be written as
\be
F_{\a\b}\equiv\sum_{triangles}\frac{\pd Q_g}{\pd p_\a}\frac{\pd Q_g}{\pd p_\b}\frac{1}{\D Q_g^2}\,,
\ee
where the indeces $\a$ and $\b$ run over the parameters of interest $\fnl$, $b_1$ and $b_2$, while the reduced bispectrum variance, as mentioned above, can be expressed in first approximation as 
\be
\D Q_g^2(\kall)\simeq \frac{\D B_g^2(\kall)}{[P_g(k_1)P_g(k_2)+2{\rm ~perm.}]^2}\,,
\ee
with $\D B_g^2$ given by \eqn{eq:Berror}. Notice that $\D Q_g^2$ depends on the linear bias parameter $b_1$. The sum over the triangles configurations can be explicitly defined in terms of three sums over the wavenumber $k_1$, $k_2$ and $k_3$ in steps of $\D k$,
\be\label{eq:FisherSum}
\sum_{triangles} \equiv \sum_{k_1=k_{\min}}^{k_{\max}}\ \sum_{k_2=k_{\min}}^{k_1}\  \sum_{k_3=k_{\min}^*}^{k_2},
\ee
with $k_{\min}^* = \max(k_{\min},|k_1-k_2|)$ to ensure that a close triangle can be formed and with $\kmax$ representing the minimal physical scale included in the analysis. Clearly, larger values of $\kmax$ correspond to a much larger number of available configurations. For this reason, in fact, the cumulative signal to noise for the bispectrum, \ie the sum of the signal to noise over all measurable configurations, grows more rapidly with $\kmax$ than it does for the power spectrum. On the other hand, we expect the primordial component to decrease significantly at small scales (high-$k$). In practice, however, $\kmax$ can be defined as the smallest scale at which we can trust our model for the galaxy bispectrum, in our case, the tree-level expression in \eqn{eq:QgNG}. 

\begin{figure}
\begin{center}
\includegraphics[width=0.65\textwidth]{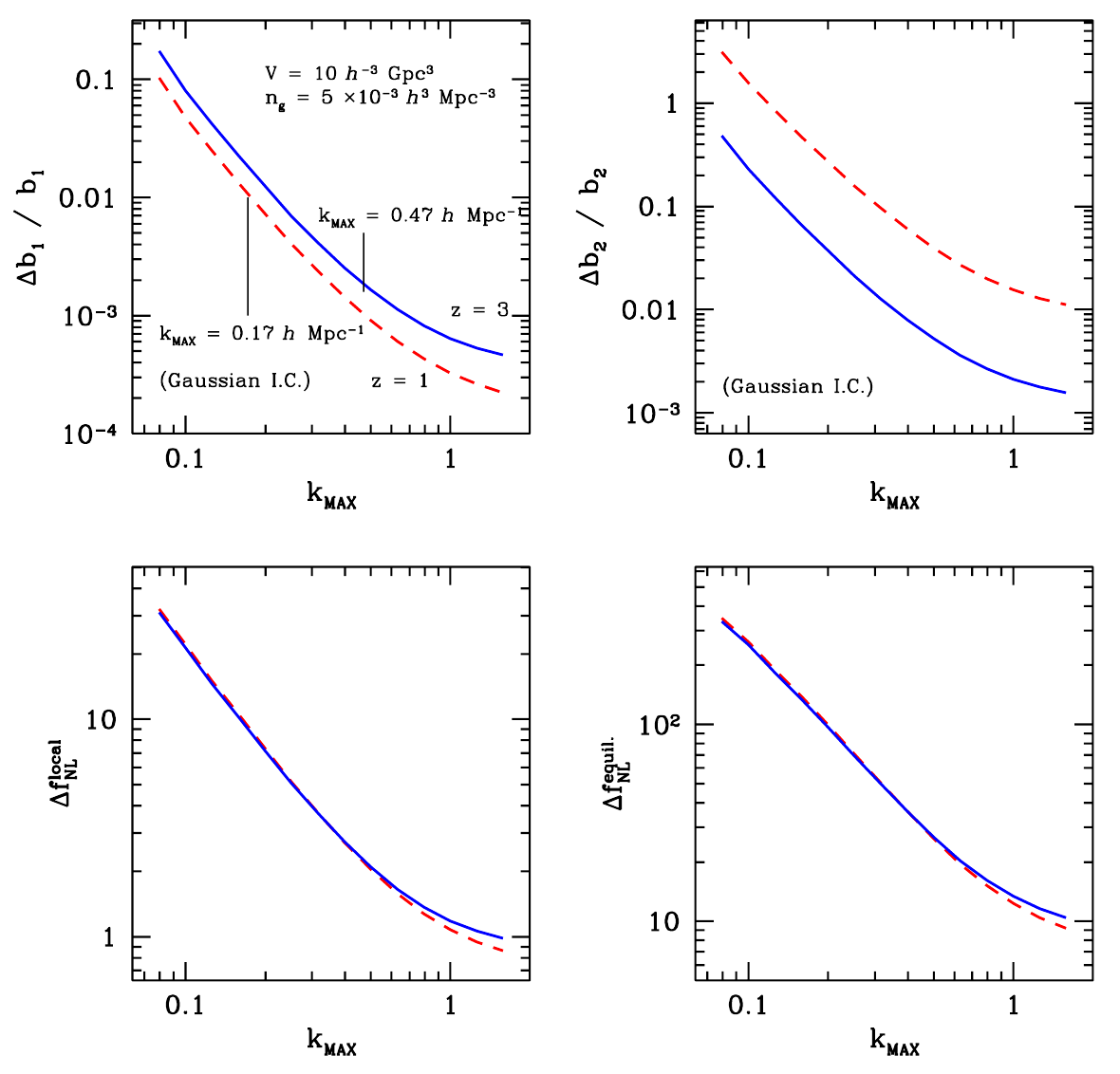}
\caption{{\it Upper panels}: predicted errors on galaxy bias parameters $b_1$ ({\it left}) and $b_2$ ({\it right}) as a function of the maximum wavenumber $\kmax$ considered for the sum defining the Fisher matrix, \eqn{eq:FisherSum}. The analysis corresponds to an ideal geometry survey of volume $V=10\cGpc$ and a galaxy number density of $n_g=5\times 10^{-3}\icMpc$. Dashed (red) lines assume a mean redshift of $z=1$, while continuous (blue) lines assume $z=3$. Both assume Gaussian initial conditions, \ie $\fnl=0$. The vertical lines correspond to the value of $\kmax$ determined as the inverse of the distance scale $R$  
defined by the condition $\sigma(R,z)=0.5$. {\it Lower panels}: predicted errors on the non-Gaussian parameters $\fnll$ ({\it left}) and $\fnle$ ({\it right}), marginalized over the bias parameters, as a function of $\kmax$. From \citet{SefusattiKomatsu2007}.}
\label{fig:SK07_kmax}
\end{center}
\end{figure}
In \fig{fig:SK07_kmax} the forecasted errors on bias parameters and non-Gaussian parameters as a function of $\kmax$ for an ideal geometry galaxy survey  of volume $V=10\cGpc$ and a galaxy number density of $n_g=5\times 10^{-3}\icMpc$ at redshift $z=1$ (dashed, red lines) and $z=3$ (continuous, blue lines). The negligible difference between the results for the non-Gaussian parameters at different redshift is a consequence of the fact that the signal to noise of the primordial component to the matter and galaxy bispectrum for a single triangular configuration, $B_0/\D B^2$, is, in our approximation, {\it constant}, both as a function of redshift and scale. This is not the case for the contributions due to gravitational instability and bias. It is clear that the choice of $\kmax$ significantly affects the final result. For instance, \citet{SefusattiKomatsu2007} define $\kmax$, for a given survey, from as the inverse of the scale $R$ given by the condition $\sigma(R,z)=0.5$ to ensure that the tree-level predictions is applied within the mildly nonlinear range. Notice that the choice of $\kmax$ depend on redshift, since at larger redshift we can expect a larger range of validity of perturbation theory predictions, both for matter and galaxy bispectrum. The dependence on the survey volume is simply given by $\sim 1/\sqrt{V}$. 

\citet{ScoccimarroSefusattiZaldarriaga2004}, from a Fisher matrix analysis as the one described above, have shown that the 2dF and SDSS surveys should be able to probe values of $\fnll\lesssim 100$, assuming $\kmax=0.3\kMpc$. They suggested as well that an all-sky survey with a galaxy number density of $n_g\sim 3\times 10^{-3}\icMpc$ up to redshift $z\sim 1$ can {\it probe values of $\fnll$ of order unity}.

\citet{SefusattiKomatsu2007} provided more specific predictions for a choice of planned and proposed high-redshift galaxy surveys, based on a similar Fisher approach, for the errors on non-Gaussian parameters both for the local and equilateral model.  It is found that, for equilateral non-Gaussianity, the degeneracy between the non-Gaussian parameter $\fnle$ and the bias parameters is severe, but it extends to unphysical regions of the $b_1$-$b_2$ plane and it can be severely reduced by introducing a correlation between linear and quadratic bias as the one predicted by the halo model. The marginalization over bias can be then replaced by a marginalization over the parameters of the Halo Occupation Distribution describing the galaxy population. 
\citet{SefusattiKomatsu2007} finds that future large-volume surveys ($V\sim 100\cGpc$ at $z\sim 1$, $2$), designed to accurately measure acoustic oscillations in the galaxy correlation function and thus map the late-time expansion of the Universe, should be able to probe $\fnll\sim 4$ and $\fnle\sim 20$, \ie values comparable to those expected from future CMB missions.  At that time they constituted they best forecasts for constraints on $\fnl$ from large-scale structure measurements. These results implied, in particular, that if Planck will indeed detect primordial non-Gaussianity, a confirmation by large-scale structure observations will be required.

\subsubsection{Effects of covariance and current results}
\label{SDSSF}

The simple Fisher matrix analysis described in the previous section makes several approximations, starting with the assumption of an ideal geometry for the survey under consideration, and the Gaussian variance for the galaxy bispectrum configurations. In fact we can expect a proper treatment of the survey selection function {\it and} of the bispectrum covariance to have a significant impact on the estimation of the non-Gaussian (and bias) parameters. For instance, triangular bispectrum configurations at the largest scales probed by a realistic redshift survey (where the initial component should provide the largest corrections) are indeed highly correlated, because of the limited number of measurable Fourier modes. 

The issue of bispectrum covariance has been studied in \citep{Scoccimarro2000B,ScoccimarroSefusattiZaldarriaga2004,SefusattiScoccimarro2005,SefusattiEtal2006}. For instance, \citet{ScoccimarroSefusattiZaldarriaga2004} compare the Fisher matrix results for an ideal survey with a volume and galaxy number density similar to those of the main sample of the SDSS, with the predictions resulting from a likelihood analysis of the same survey, {\it including} the effects of survey geometry and covariance. Such analysis involves all measurable triangular configurations defined by wavenumbers $k_1$, $k_2$, $k_3\le0.3\kMpc$, with $\D k=0.015\kMpc$, resulting in a total number of triangle bins, $N_T=1015$. The estimation of the corresponding, $1015\times 1015$, bispectrum covariance matrix clearly represents a challenging computational problem as it cannot be determined from a relatively small number of N-body simulations. This work uses instead a code \citep{Scoccimarro2000B} implementing particle displacements as predicted by second order Lagrangian perturbation theory \citep[2LPT, see, for instance, ][and references therein]{BernardeauEtal2002} to produce $6,000$ realizations of the density field. Such large number of realizations is in fact necessary for an accurate determination of the covariance matrix. In addition, the 2LPT results, including particle velocities, allow for exact redshift distortions. Each mock catalog, in redshift space, is then weighted according to the FKP procedure~\citep{FeldmanKaiserPeacock1994,MatarreseVerdeHeavens1997,Scoccimarro2000B} to take into account the SDSS selection function. The same covariance matrix is compared to analytic expressions in \citet{SefusattiEtal2006}.

\begin{figure}
\begin{center}
\includegraphics[width=0.6\textwidth]{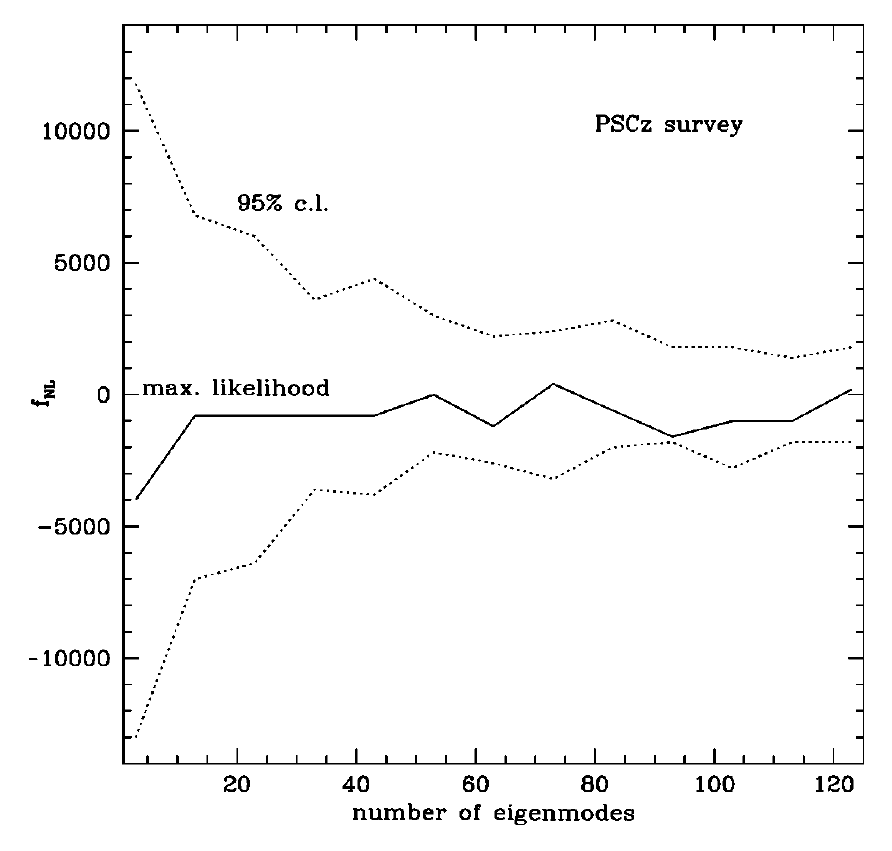}
\caption{$95\%$ confidence limits on $\fnl$ from the PSCz galaxy bispectrum after marginalization over bias parameters, as a function of the number of eigenmodes included in the likelihood analysis. From \citet{ScoccimarroSefusattiZaldarriaga2004}.}
\label{fig:PSCz}
\end{center}
\end{figure}

Given a proper estimate of the covariance matrix, a likelihood function for the reduced bispectrum $Q_n$ can be defined in terms of the normalized bispectrum eigenmodes $\hat{q}_n$ that diagonalize it \citep{Scoccimarro2000B}. These can be expressed as
\be
\hat{q}_n = \sum_{m=1}^{N_{\rm T}} \gamma_{mn} \frac{Q_m-\bar{Q}_m}{\Delta Q_m}\,,
\ee
where $\bar{Q}_m \equiv \la Q_m \ra$, $\Delta Q_m^2 \equiv \la (Q_m-\bar{Q}_m)^2 \ra$ and their signal to noise is given by
\be
\left(\frac{S}{N}\right)_n = \frac{1}{\l_n}  \left| \sum_{m=1}^{N_{\rm T}} \gamma_{mn} \frac{\bar{Q}_m}{\Delta Q_m}\right| \,,
\ee
where $\l_n$ represents the eigenvalue for $\hat{q}_n$, with $\la \hat{q}_n\, \hat{q}_m \ra = \l^2_n \, \d_{nm}$. The eigenmodes presenting the largest signal to noise can be easily interpreted by considering how they weight different bispectrum configurations. In fact, the largest signal to noise corresponds to an eigenmode defined by a nearly equal weighting of {\it all} triangles, and it therefore represents the overall bispectrum amplitude. The next eigenmode weights instead with opposite sign triangles close to the equilateral shape and nearly collinear triangle. Each eigenmode represents in fact a fraction of the information contained in the bispectrum configurations, and a crucial role in this respect is played by the shape and scale dependence. To illustrate this point, \fig{fig:PSCz} \citep[from][]{ScoccimarroSefusattiZaldarriaga2004} shows the 95\% C.L. limits on $\fnll$ from the likelihood analysis of the IRAS PSC$z$ catalog \citep{SaundersEtal2000} as a function of the number of eigenmodes included. 

Although the diagonalization of the covariance matrix does not ensure the exact independence of the eigenmodes, which can still present non-vanishing higher-order correlations, it has been shown that this is nevertheless a reasonable assumption in practice \citep{Scoccimarro2000B} . This allows us to write a likelihood function for the non-Gaussian and bias parameters, denoted generically as $p_\a$, in terms of the product of the probability distribution functions $P_n(x)$ for each individual eigenmode, that is 
\be
{\cal L}(\{p_\a\}) \propto \prod_{n=1}^{N_{\rm T}}\
P_{n}[\hat{q}_n(\{p_\a\})].
\label{eq:likelihood}
\ee
The probability distributions $P_n(x)$, which can be determined from the mock catalogs, are not expected in general to be Gaussian, although this can be in fact a good first-order approximation in the case of the SDSS main sample \citep{ScoccimarroSefusattiZaldarriaga2004}.

A direct implementation of this kind of analysis, taking into account all measurable bispectrum configurations and their covariance, has been performed by \citet{ScoccimarroEtal2001B} for different IRAS catalogs \citep{EfstathiouEtal1990, StraussEtal1992, FisherEtal1995} and by \citet{FeldmanEtal2001} for the IRAS PSC$z$ catalog \citep{SaundersEtal2000} considering the case of the $\chi^2$ model of primordial non-Gaussianity \citep{Scoccimarro2000A}. \citet{ScoccimarroSefusattiZaldarriaga2004} derives the limit $|\fnll|<1800$ at $95\%$ C.L. for the bispectrum measured in the PSCz catalog.

Along this lines, \citet{ScoccimarroSefusattiZaldarriaga2004} also studied the constraints on $\fnll$ for local non-Gaussianity, that could be obtained from measurements of the galaxy bispectrum in the SDSS main sample, including the effects of the survey geometry and bispectrum covariance, forecasting the 1-$\sigma$ error $\dfnll\simeq 150$, after marginalization over the bias parameters. This work compared this more realistic estimate of the predicted errors on $\fnll$ from the likelihood analysis of the SDSS bispectrum to the Fisher matrix forecast for an ideal geometry of nearly the same volume and galaxy density finding a worsening of a factor of 4-5. They point-out, however, that the realistic errors, which are an estimate from the north part of SDSS alone, should be taken as a an upper bound to the results actually achievable because of the FKP weighting scheme, not optimal at the largest scales where the primordial component is the largest and because of the fact that extra signal can be found as well in open configurations, not considered there, due to the broken translation invariance. We might add, based on the results of section \ref{sssec:lssmPT1l}, that nonlinear corrections present for non-Gaussian initial conditions might increase the overall signal due to a non-zero $\fnl$, particularly on small scales where a large number of triangular configurations can be measured. 

At this point we should remind the reader that all the results discussed so far on the galaxy bispectrum and its significance for constraining primordial non-Gaussianity, assume the expression \eqn{eq:QgNG} to be a reliable prediction. As we shall se in the remainder of this section, this is {\it not} the case, as additional effects of non-Gaussian initial conditions have to be taken into account. Nevertheless, the primordial component, whose direct detection has been the main target of the earlier works discussed above, is still expected to provide a contribution to the galaxy bispectrum, and there are good reasons to believe that these results can be still interpreted as a ``conservative estimate'' of the possibilities offered by bispectrum measurements in the large-scale structure to test the Gaussianity of the initial conditions.

\subsubsection{Primordial non-Gaussianity and {\it non-local} Galaxy Bias }
\label{ssec:bsgNGBias}

\begin{figure}[t]
\begin{center}
{\includegraphics[width=0.6\textwidth]{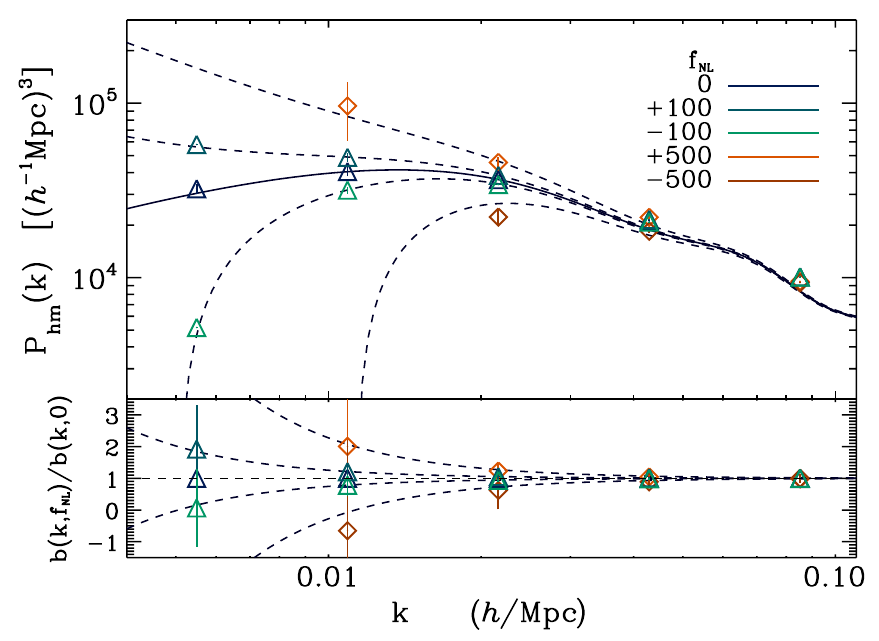}}
\caption{Matter-halo cross power spectrum measured in simulations with {\it local} non-Gaussian initial conditions for different values of $\fnl$. From \citet{DalalEtal2008}.}
\label{fig:bsgB1}
\end{center}
\end{figure}

  The constraints and forecasts discussed so far in this section are based on the tree-level expression for the galaxy bispectrum, \eqn{eq:QgNG}, derived under the assumption of {\it local} bias, \eqn{eq:bias}. As anticipated, our understanding of galaxy bias in presence of primordial  non-Gaussianity {\it radically changed} in the last two years, after \citet{DalalEtal2008} presented measurements of the halo power spectrum in simulations with non-Gaussian initial conditions of the local kind showing the presence of {\it large corrections at large scales}, {\it not captured by the local bias prescription!} \Fig{fig:bsgB1} shows the matter-halo cross-power spectrum for different values of $\fnl$ from these simulations, where the unexpected effect of non-Gaussianity at large scales is evident. 

Local bias, \eqn{eq:bias}), in fact, implies a leading contribution to the galaxy (or halo) power spectrum of the simple form 
\be
P_g(k)=b_1^2P_0(k)\,,
\ee
with no dependence on $\fnl$, while the simulations results of \citet{DalalEtal2008}, later confirmed by \citet{PillepichPorcianiHahn2008,DesjacquesSeljakIliev2009,GrossiEtal2009}, are consistent with a {\it scale-dependent} correction to the linear bias of the form
\be
P_g(k)=[b_1+\Delta b_1(k)]^2P_0(k)\,,
\ee
with
\be\label{eq:deltab1}
\Delta b_1 (k)=3\fnl (b_1-1)\d_c\frac{\Omega_m H_0^2}{k^2 T(k)D(z)}\,,
\ee
where $\d_c\simeq 1.68$ is the linear critical density  for spherical collapse, extrapolated at $z=0$. Such correction therefore increases with the scale, with redshift via the growth factor $D(z)$ and with the non-Gaussian parameter $\fnl$, and vanishes for unbiased populations ($b_1=1$). 
  
  A theoretical interpretation, based on the peak-background split \citep{ColeKaiser1989} has been assumed by \citet{DalalEtal2008, SlosarEtal2008, AfshordiTolley2008, GiannantonioPorciani2009}, and, with a somehow different derivation, by \citet{McDonald2008}. According to these works, the local relation between the galaxy density and the matter density in \eqn{eq:bias} is modified, in presence of {\it local} primordial non-Gaussianity, to include an explicit dependence on the primordial curvature perturbation, $\Phi$, \ie
\be\label{eq:biasNG}
\d_g(\xv) =  b_1\d(\xv)+c_1(\fnl)\Phi(\xv) +\frac{1}{2}b_2\d^2(\xv) +c_2(\fnl)\d(\xv)\Phi(\xv)+\dots
\ee
The galaxy two-point function, will be given by the following perturbative expansion,
\be
\la\d_g(\xv_1)\d_g(\xv_2)\ra =
b_1^2\la\d(\xv_1)\d(\xv_2)\ra+b_1c_1(\fnl)\la\d(\xv_1)\Phi(\xv_2)\ra+{\rm ~perm.}\dots,
\ee  
where the second term can be rewritten as the scale-dependent of the linear bias parameter of Eq.~(\ref{eq:deltab1}), with the $1/k^2$ behavior resulting from the relation between $\d_{\kv}$ and $\Phi_\kv$ given by $M(k)\sim k^2$. \citet{GiannantonioPorciani2009} describes, in fact, the galaxy distribution as {\it multivariate} distribution, although the matter density $\d$ and the curvature $\Phi$ are not two independent random field, but they are related by Poisson equation, \eqn{eq:Poisson}. It should be noted, that no derivation of a similar effect due to a different kind of primordial non-Gaussianity (if feasible) has been, so far, proposed. The derivations presented in the works cited above, in fact, all rely on the relatively simple expression defining the local model \eqn{eq:fnldefn} while their generalization to a model defined by a generic initial bispectrum is a quite challenging problem.

Following the results of \citet{DalalEtal2008}, moreover, an apparently different explanation, resulting in fact in a very similar {\it but distinct} effect on the galaxy power spectrum has been proposed by \citet{MatarreseVerde2008} and \citet{TaruyaKoyamaMatsubara2008}. \citet{TaruyaKoyamaMatsubara2008}, starting from the {\it local bias prescription} of \eqn{eq:bias}, point-out that the next-to-leading order correction to the galaxy two-point function in presence of local primordial non-Gaussianity, represents, in fact, a large correction, identical up to a constant factor, in the large-scale limit, to the bias correction of \eqn{eq:deltab1}. The perturbative expression for the galaxy power spectrum is given by
\be\label{eq:psTKM}
P_g(k)\simeq b_1^2P(k)+b_1b_2\!\!\int\!\! d^3q ~B(k,q,|\kv-\qv|),
\ee
where the second term, proportional to the quadratic bias parameter $b_2$ and dependent on the matter bispectrum $B$, corresponds to the lowest order, one-loop correction. Remarkably, for local non-Gaussianity, in the limit $k\rightarrow 0$, such correction presents the same scale and redshift dependence, and, for massive halos or highly biased populations ($b_1\gg 1$), even the same amplitude, as the one resulting from \eqn{eq:deltab1}. The expression, however, can be applied to {\it any} model of primordial non-Gaussianity, given the appropriate initial matter bispectrum \citep[see, for instance,][]{VerdeMatarrese2009}. In the case of equilateral non-Gaussianity, the correction is almost negligible, while local non-Gaussianity appears to be a limiting case leading to a particularly significant effect. The same correction has been considered already by \citet{Scoccimarro2000A} in the context of $\chi^2$ initial conditions, where it leads to a redefinition of the bias parameters, with no additional scale-dependence. 

\citet{MatarreseVerde2008} presents a different derivation of an expression similar to the one of \eqn{eq:psTKM}, based on earlier works on the density peak correlation function \citep{GrinsteinWise1986, MatarreseLucchinBonometto1986}. In this case, a specific prediction for the bias parameters, valid however only in the high density threshold limit,  is included. It is interesting to notice that the possibility of large-scale effects on the correlations of biased distributions has been explicitly pointed-out by \citet{GrinsteinWise1986}, although without further study. 

The two distinct corrections to the galaxy power spectrum, one corresponding to the modified bias relation of \eqn{eq:biasNG}, the other to the perturbative correction due to nonlinear bias of \eqn{eq:psTKM}, have been studied in a comprehensive framework recently by \citet{GiannantonioPorciani2009}, where the authors suggest that the effect measured in N-body simulations is mainly due to the multivariate nature of the galaxy distribution with local primordial non-Gaussianity, rather than the effect of nonlinear bias \eqn{eq:psTKM}. In addition, \citet{DesjacquesSeljakIliev2009} pointed-out that even the galaxy bias parameters $b_i$, related in the framework of the halo model to the halo bias parameters $b_{h,i}(M)$ for halo populations of mass $M$, present a dependence of $\fnl$ due to the effects of non-Gaussianity on the halo mass function. The picture that has been emerging in the last years is therefore quite complex and it should be stressed that a wide consensus in the community on a well defined model, even for the galaxy power spectrum, is still lacking. For instance, a discrepancy of the order of a $10\%$ between predictions and simulations results, did not find yet a unique interpretation \citep[see discussions in][]{PillepichPorcianiHahn2008,DesjacquesSeljakIliev2009,GrossiEtal2009,MaggioreRiotto2009C,GiannantonioPorciani2009}.

This rather surprising effect of local non-Gaussianity on the bias relation leads, remarkably, to the possibility of placing limits on $\fnll$ from {\it current} large-scale structure observations, {\it already comparable to limits from the CMB!} \citep{SlosarEtal2008, AfshordiTolley2008}. Specifically, \citet{SlosarEtal2008} derived from measurements of the cross-correlation of several large-scale structure data-sets and the CMB \citep{HoEtal2008} the 2-$\sigma$ constraints
\be
-29<\fnll<70\,,
\ee
leading to an marginal improvement of the WMAP results. Encouraging predictions for the constraints that can be derived in future spectroscopic as well as photometric redshift surveys can be found in \citet{CarboneVerdeMatarrese2008}. A fair comparison between these forecasts and those derived for the galaxy bispectrum in \citet{SefusattiKomatsu2007} is clearly not possible as the latter do not include the effect on the bias relation discussed above. Two observations, however, are in order. In the first place, these effects on the galaxy power spectrum are specific of the local model of non-Gaussianity, while the galaxy bispectrum is in principle sensitive to {\it any} initial component $B_0$. In the second place, robust results can be obtained from galaxy power spectrum measurements at large scales in {\it photometric} surveys. The degradation of the information that can be extracted from bispectrum measurements in photometric surveys with respect to spectroscopic ones is still to be properly studied. The impact of photometric errors on the accurate determination of the bispectrum dependence on the triangle shape can in fact be significant.

\subsubsection{The Galaxy Bispectrum {\rm after} Dalal {\rm et al.} (2008)}
\label{ssec:bsgNGBiasBisp}

First steps in the direction of an extension of the results discussed above to the galaxy bispectrum have been taken in \citet{JeongKomatsu2009B} and \citet{Sefusatti2009}. Specifically, \citet{JeongKomatsu2009B} considered an expression for the high-peak three-point function derived in \citet{MatarreseLucchinBonometto1986}, analogous to the one for the two-point function studied by \citet{MatarreseVerde2008}, and applied it to the case of local non-Gaussianity. \citet{Sefusatti2009} considered instead the perturbative approach of \citet{TaruyaKoyamaMatsubara2008} based on the local bias expansion of \eqn{eq:bias}, and applied it to local {\it and} equilateral non-Gaussianity. 

\begin{figure}[t]
\begin{center}
{\includegraphics[width=0.48\textwidth]{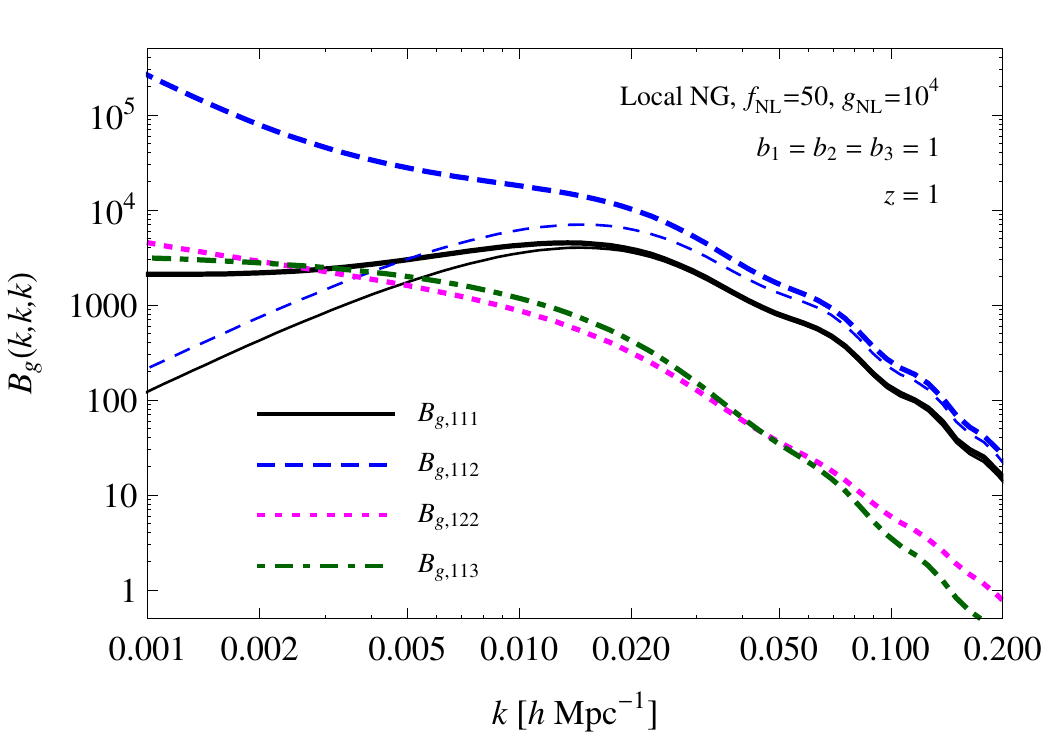}}
{\includegraphics[width=0.48\textwidth]{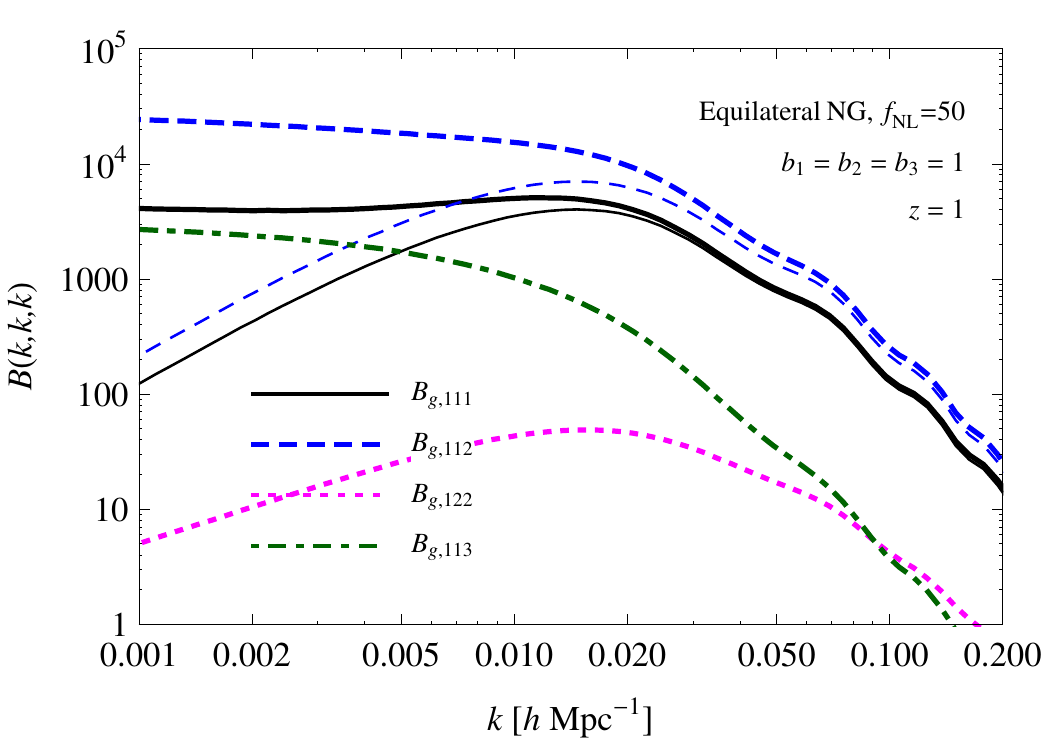}}
\caption{Large-scale contributions to the galaxy bispectrum due to primordial non-Gaussianity of the local ({\it left panel}) and equilateral ({\it right panel}) type described as one-loop corrections assuming a local bias prescription. Thin lines correspond to the contributions for Gaussian initial conditions. From \citet{Sefusatti2009}, see the reference for further details.}
\label{fig:bsgC1}
\end{center}
\end{figure}

These works show that the galaxy bispectrum is expected to be sensitive to both the initial matter bispectrum $B_0$, as well as to the initial matter {\it trispectrum} $T_0$, by means of a contribution analogous to \eqn{eq:psTKM} and given by
\be\label{eq:bsS}
B_g  \simeq  b_1^3B(k_1,k_2,k_3)
 + \frac{b_1^2b_2}{2}\!\!\int\!\! d^3q ~T(k_1,k_2,q,|\kv_3-\qv|),
\ee
which represents a large correction at large scales, with an asymptotic behavior characterized by an extra $1/k^2$ factor with respect to the primordial matter bispectrum component, $B_0$ and a dependence on $\fnl^2$. In addition, \citet{Sefusatti2009} points-out that, unlike the power spectrum, large-scale corrections due to nonlinear bias are present as well for {\it equilateral non-Gaussianity} (and virtually for any non-pathological form of the primordial bispectrum and trispectrum). \Fig{fig:bsgC1} shows the one-loop corrections to the galaxy bispectrum due to nonlinear bias and primordial non-Gaussianity under the assumption of local bias \citep{Sefusatti2009}. The left panel assumes local non-Gaussianity including a non-zero initial bispectrum and trispectrum, while the right panel assumes a non-zero initial bispectrum of the equilateral type. Thin lines correspond to Gaussian initial conditions. The black continuous line represents the matter bispectrum and therefore the first term on the r.h.s. of \eqn{eq:bsS}, while the blue dashed line correspond to the second term. Notice that at next-to-leading order in PT, the matter bispectrum $T$ depends on the initial trispectrum $T_0$ as well as the initial bispectrum $B_0$, so that an effect is present also for equilateral non-Gaussianity where the figure assumes $T_0=0$.

\begin{figure}[t]
\begin{center}
{\includegraphics[width=0.7\textwidth]{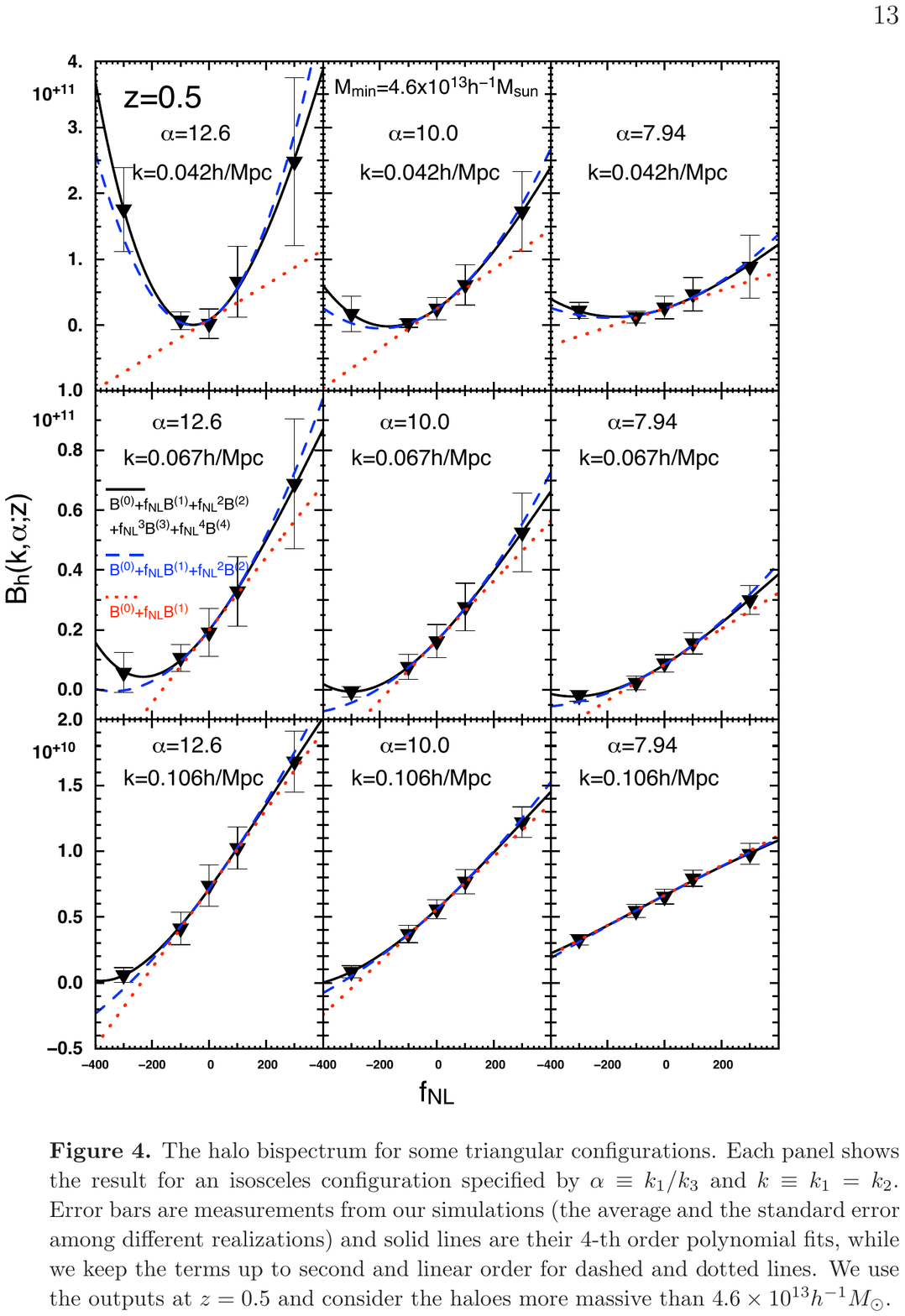}}
\caption{Measurements of a set of triangular configurations of the halo bispectrum in N-body simulations with local non-Gaussian initial conditions, as a function of the non-Gaussian parameter $\fnl$. Large values of the parameter $\a$ correspond to more squeezed configurations. In the upper panels, the dependence of the halo bispectrum on $\fnl^2$ is evident. From \citet{NishimichiEtal2009}.}
\label{fig:bsgC2}
\end{center}
\end{figure}
It should be noted, however, that these results ignore, at least for local non-Gaussianity, the modified bias relation of \eqn{eq:biasNG} \citep[see, in this respect, some comments in][]{GiannantonioPorciani2009},  and do not provide reliable predictions for the constant bias parameters. Furthermore, they have not been properly tested against measurements of the halo bispectrum in numerical simulations. The only work, at the time of writing, in this direction \citep{NishimichiEtal2009} shows, however that the dependence of the halo bispectrum on $\fnl$ is roughly consistent with the functional form resulting from the prediction of \eqn{eq:bsS}. The authors attempt as well, using a simple fit to their measurements, a preliminary forecast analysis for a future large-volume ($100\cGpc$), high-redshift survey, finding a detectable value of $\fnl\simeq 20$, using a {\it very limited} number of configurations. \Fig{fig:bsgC2} from \citet{NishimichiEtal2009} shows measurements of a set of triangular configurations of the halo bispectrum in simulations with local non-Gaussian initial conditions, as a function of the non-Gaussian parameter $\fnl$, where the dependence of the halo bispectrum on $\fnl^2$ is evident.

A simple but reasonable expectation would be that the inclusion of the effects of primordial non-Gaussianity on galaxy bias will improve the results of \citet{SefusattiKomatsu2007}, which are based on the detectability of the primordial component alone.  Our understanding of these phenomena is, however, evolving rapidly in these days, and these notes on recent developments are likely to become outdated relatively soon.

\subsection{Running non-Gaussianity}
\label{sec:runningNG}

\subsubsection{The case of a scale-dependent $\fnl$}

DBI models of inflation predict, as we have seen, a primordial curvature bispectrum very close to the equilateral model in its shape dependence. An additional but {\it quite generic} feature of these models is given by a significant departure from the hierarchical scaling $B_\Phi(k,k,k)\sim P_\Phi^2(k)$ \citep{Chen2005, ShanderaTye2006, ChenEtal2007, KhouryPiazza2008, ChenWang2009, RenauxPetel2009}. More recently, this possibility has been explored as well in models of local non-Gaussianity \citep{ByrnesChoiHall2008, ByrnesChoiHall2009, ByrnesTasinato2009, ByrnesEtal2009, KumarLeblondRajaraman2009}. 

 Under a phenomenological point of view, this extra scale-dependence can be described by a {\it running} $\fnl(k)$, or, more properly, in terms of an amplitude parameter $\fnl$ and a running parameter $\nng$, defined by
 \be
\fnl(K) \equiv \fnl\left( \frac{K}{\kp} \right)^{\nng},
\ee
where $\kp$ is a properly chosen pivot scale, while $K(k_1,k_2,k_3)=(k_1+k_2+k_3)/3$ defines an overall scale characteristic of the triangular configuration on which $B_\Phi(k_1,k_2,k_3)$ depends. In other terms, the $\fnl(K)$ defined above replaces the constant $\fnl$ in the definitions of the local and equilateral bispectra effectively introducing an extra dependence on scale. 

Observational consequences of a running $\fnl(K)$ have been explored in \citet{LoVerdeEtal2008} and \citet{SefusattiEtal2009}, while in \citet{TaruyaKoyamaMatsubara2008} this effect is included in the prediction for one-loop corrections to the matter and galaxy power spectrum.

\begin{figure}[t]
\begin{center}
{\includegraphics[width=0.48\textwidth]{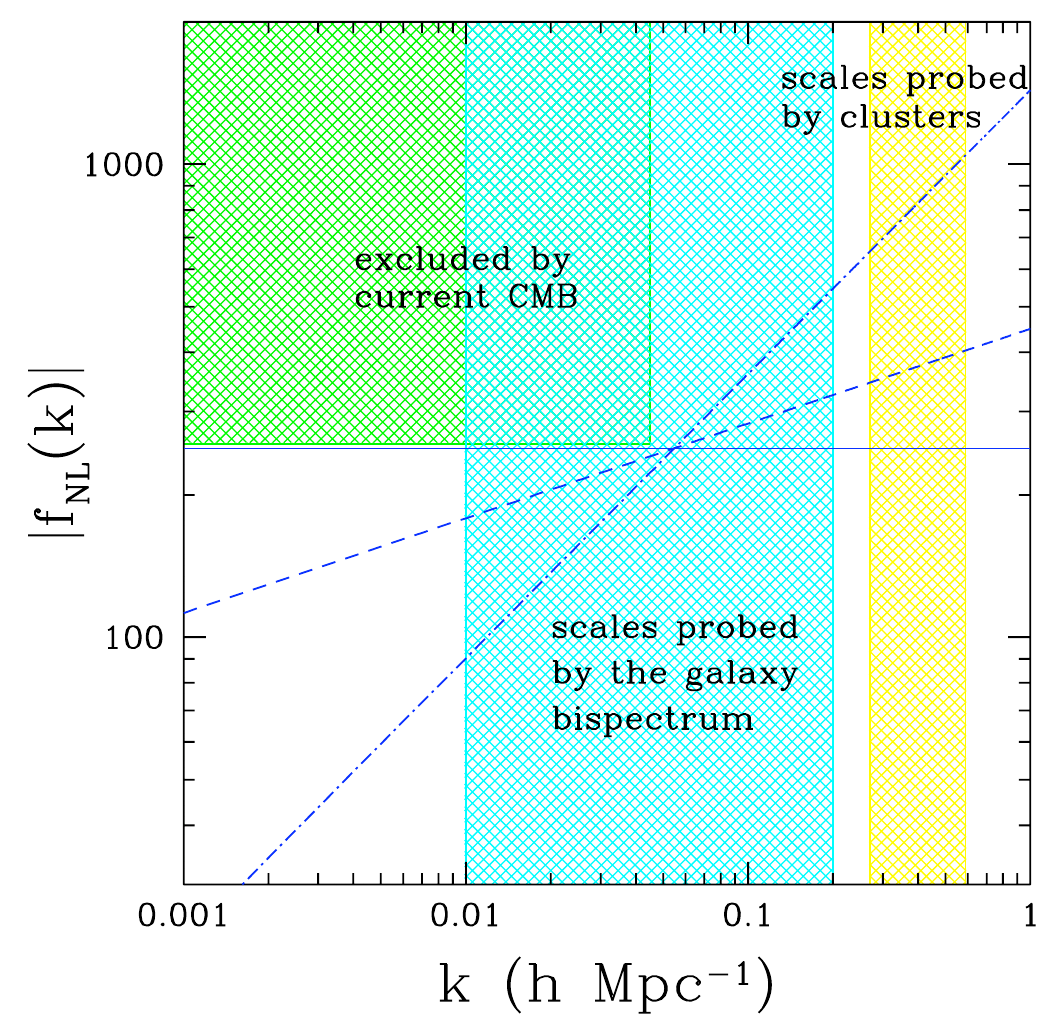}}
{\includegraphics[width=0.48\textwidth]{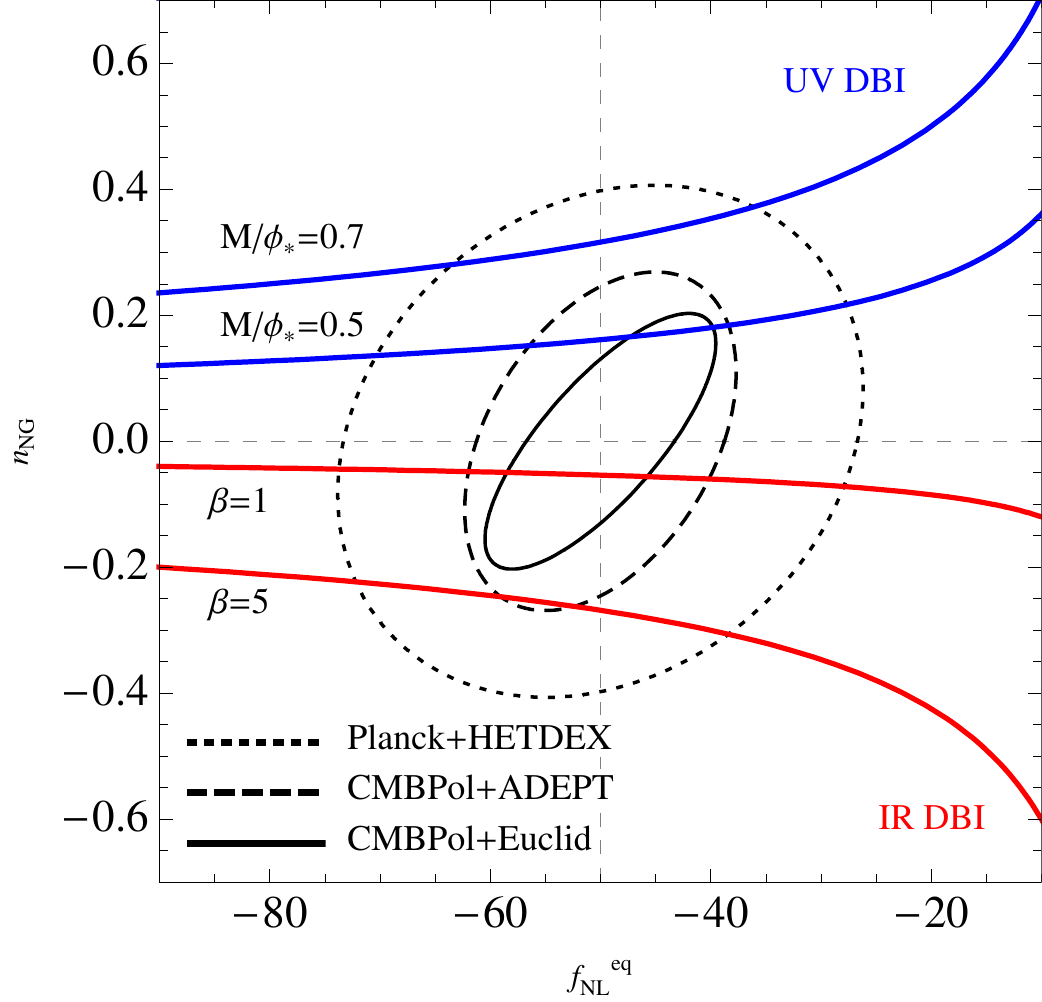}}
\caption{{\it Left panel}: Range of scales probed by different observables compared with $\fnl(k)$ for different values of the running parameter $\nng$. From \citet{LoVerdeEtal2008}.  {\it Right panel}: predictions of DBI models showing the peculiar relation between the amplitude $\fnle$ and the running $\nng$ for different values of the parameters of the inflaton Lagrangian. The figure includes the Fisher matrix forecasts for combined CMB and galaxy bispectrum measurements assuming a fiducial $\fnle=-50$. From \citet{SefusattiEtal2009}, see the reference for further details.}
\label{fig:running1}
\end{center}
\end{figure}

\citet{LoVerdeEtal2008} provided an analysis of the possibility of constraining the running parameter $\nng$ by combining current limits from the CMB on the amplitude parameter $\fnl$ at the pivot scale $\kp=0.04$ Mpc$^{-1}$ with future measurements of cluster abundance. The effect of a $\nng$ significantly different from $1$, can result in a much larger (or smaller) amount of non-Gaussianity on the smaller scales relevant for the cluster mass function. \Fig{fig:running1} ({\it left panel}, from \citealt{LoVerdeEtal2008}) illustrates the difference in the range of scales  probed by different observables. Focusing in particular on the equilateral model for the curvature bispectrum, this work assumes the amplitude of $\fnl(k)$ to be constrained by the CMB bispectrum {\it at} the pivot point scale $\kp$ and derives the expected constraints on its running by considering the effective amplitude of $\fnl(k)$ at the smaller scales ($k\sim 0.3-0.6\kMpc$) probed by cluster surveys. For an all-sky cluster survey up to redshift $z_{max}=1.3$ they find the 1-$\sigma$ constraints, marginalized over $\Omega_m$, $\sigma_8$ and $h$, assuming the fiducial values $\fnl=38$ and $\nng=0$, $\Delta\nng \simeq 2$ with a Planck prior $\Delta \fnl(k=\kp)=40$. Their analysis, however, does not include the {\it simultaneous} limits that measurement of the CMB bispectrum {\it alone} is expected to provide on both the amplitude $\fnl$ and running $\nng$. 

\subsubsection{Running non-Gaussianity and bispectrum measurements}

\citet{SefusattiEtal2009} performs a Fisher matrix analysis of the CMB bispectrum to obtain the sensitivity of this observable to the running of $\fnl(k)$. The results in the case of local non-Gaussianity, assuming the same pivot $\kp=0.04$ and marginalizing over the amplitude $\fnll$, are the 1-$\sigma$ uncertainties of $\dnng \simeq 0.68 ~(50/\fnll)~f_{sky}^{-1/2}$ for WMAP and $\dnng \simeq 0.1 ~(50/\fnll)~f_{sky}^{-1/2}$ for Planck, where $\fnll$ stands for the fiducial value of the amplitude parameter. In the case of equilateral non-Gaussianity, we have $\dnng \simeq 1.1 ~(100/\fnle)~f_{sky}^{-1/2}$ for WMAP and $\dnng \simeq 0.3 ~(100/\fnle)~f_{sky}^{-1/2}$ for Planck. Since it is always possible, given the observable of interest (\eg the CMB bispectrum for a specific experiment) {\it and} the non-Gaussian model, to choose the pivot point in such a way to remove any degeneracy between the amplitude and the running parameters, a measurement of the running parameter comes at {\it no cost}  with respect to the determination of the $\fnl(k=\kp)$. Notice, however, that, for reasons related to the numerical implementation of the CMB estimator, \citet{SefusattiEtal2009} assumes for the overall scale representative of a given triangular configuration, the {\it geometric} mean of the three wavenumber, \ie $K\equiv (k_1k_2k_3)^{1/3}$. While the difference with the more physically motivated definition in terms of the arithmetic mean $K=(k_1+k_2+k_3)/3$ is very small for equilateral non-Gaussianity, in the local model this is not the case. 

\citet{SefusattiEtal2009} considers as well the Fisher matrix from large-scale structure information, and specifically the galaxy power spectrum (including the effect on halo bias) for the local model and the galaxy bispectrum, but in terms of the simple description of \eqn{eq:QgNG}, therefore excluding halo bias effects. This different choice of observables with respect to the model of primordial non-Gaussianity assumes a  negligible effect of equilateral non-Gaussianity on the galaxy power spectrum (still to be confirmed by N-body simulations). It is shown, in particular, that future galaxy redshift surveys can significantly improve CMB results. \Fig{fig:running1} ({\it right panel}) shows the contours plots for the 1-$\sigma$ uncertainties resulting for a joint Fisher matrix analysis of CMB and large-scale structure information. The expected limits are plotted against the predictions for the relation between the amplitude $\fnll$ and the running $\nng$ from DBI inflationary models. It is interesting to notice how these models predict a stronger running for smaller values of the amplitude parameter. In this respect, constraining the value of $\nng$ can place additional limits on the parameters of the inflaton Lagrangian.

\section{Conclusions}

Weakly non-Gaussian initial conditions are {\it defined}, in most of the relevant inflationary models, by a non-vanishing bispectrum for the primordial curvature perturbations. The most direct observables of this primordial density correlator are, naturally, the bispectrum of the temperature fluctuations in the CMB and the bispectrum of the mass distribution at large scales as probed by galaxy surveys. In this review we presented an overview of the problems, results and expectations connected with the detection of (or constraints on) primordial non-Gaussianity specifically in bispectrum measurements of the CMB and LSS.

The CMB is an ideal observable for tests of primordial NG because temperature and polarization anisotropies can be described in the linear regime of cosmological perturbations. The statistical properties of the primordial curvature field are thus directly reflected in the pattern of CMB fluctuations. As we have seen, tests of primordial NG are formulated in terms of the estimation of the bispectrum amplitude $\fnl$ for each of the shapes predicted by different inflationary models. It was originally shown in the literature 
that a maximum likelihood estimator of the bispectrum optimally extracts {\em all} the $\fnl$ information from a CMB map. Extracting the primordial non-linear parameter from the bispectrum has subsequently become the standard way to test primordial NG in the CMB. The best $\fnl$ measurements to date come from analysis of the WMAP datasets, and roughly constrain the primordial bispectrum amplitude to be $\lesssim 100$ for the local, equilateral and orthogonal shapes. Despite already being very stringent (the NG part of the CMB temperature anisotropies is constrained at the level of $10^{-3}$ of the total fluctuation), these bounds are still far from the typical order of magnitude of primordial NG predicted by most inflationary models. As we have seen, Fisher matrix forecasts show that  future results from the Planck satellite (whose release date is predicted to be in 2012) will improve previous WMAP constraints by roughly one order of magnitude, thus impacting the range of some theoretical predictions. This significant improvement is due to the better sensitivity of Planck to many more bispectrum configurations in the analysis, and  the possibly of exploiting both temperature and polarization datasets.
 Another important limitation on current constraints is that inflationary predictions encompass more shapes than those that have been constrained so far. The reason why many shapes remain to be constrained is that they cannot be written as a {\em separable} product of one-dimensional functions of a single wavenumber. Separability, as we have seen, is a crucial property since it makes the actual analysis computationally affordable in terms of CPU time. We have reviewed recent work showing that this limitation can also be overcome in future analysis by means of a fully general, and mathematically well defined, eigenmode expansion of the bispectrum shape. Thanks to this, and in light of the significant improvement in sensitivity provided by Planck, better and more general CMB constraints on primordial NG models will be
 available in the near future. One caveat is that the high precision of the forthcoming CMB datasets makes them much more sensitive to other spurious (\ie non primordial) sources of NG, that could bias the $\fnl$ estimate. Achieving an accurate control on these contaminants is clearly a crucial goal for future analysis. 
As we have seen, much work is being done in order to predict, detect and isolate non-primordial NG effects, but some issues still have to be addressed. In particular a complete prediction of the total bispectrum generated by second order cosmological perturbations is not yet available, although a number of effects have been studied in detail. Accurate characterization of NG from diffuse foreground residuals is another important issue that will require further investigation.

For large-scale structure, many aspects of the general CMB scenario outlined above change, as should be evident from a comparison of the discussions in sections~\ref{sec:CMB} and \ref{sec:LSS}. In the first place, we cannot rely on a direct relation between the observed galaxy bispectrum and the primordial curvature bispectrum predicted by inflationary models. As we have seen, a small departure from Gaussian initial conditions should result in a correction to the galaxy bispectrum induced by gravitational instability and nonlinear bias, constituting the dominant contributions. The nature of this correction is a complex problem in its own right, since it is due to the linearly evolved initial matter bispectrum as well as to the effects of primordial non-Gaussianity on the galaxy bias relation. Such effects are still under investigations and we do not have, to date,  an accurate theoretical model.
On the other hand, early results from galaxy power spectrum measurements are very encouraging, albeit restricted at present to the local non-Gaussian model. Current data sets already appear to be able to confirm and improve CMB results. In this respect, it is evident that the ultimate goal is the implementation of a complete large-scale structure analysis in terms of all measurable correlators, including power spectrum, bispectrum and beyond, \ie an analysis that fully reflects the non-Gaussian nature of the mass and galaxy distributions even on large scales. 

There are several issues which remain to be resolved,  for which we can identify three main categories. First, we need to develop a robust {\it model for the galaxy correlators} accurately accounting for small-scale nonlinearities for both 
the matter and galaxy density fields, as well as in the presence of non-Gaussian initial conditions; this also must account describe {\it non-localities} in the bias relation. In this review we have briefly summarized the state of the art, noting that our understanding of these phenomena is evolving rapidly. Secondly, once a reliable model is available, it will be necessary to develop the machinery that will allow us, in the event of a future detection, to properly identify the effects of different models and their bispectrum shapes. In this respect, the CMB results presented in section~\ref{sec:CMB}, provide an important
benchmark. Finally, observational problems connected with redshift surveys such as the effects of redshift distortions and/or photometric errors, survey selection function, completeness, etc., will have to be addressed. We have not discussed these issues here as they are generic to all large-scale structure experiments, but they clearly represent a major challenge for the exploitation of future data-sets. Both the first and the last point are crucial for virtually all the science goals of future ground-based or satellite surveys, particularly dark energy studies. Although only partial results have been obtained so far, there is every indication that characterising non-Gaussianity in future galaxy surveys will result in a significant test of the initial conditions of the Universe.

To summarize, sufficient experimental sensitivity has been reached recently in CMB experiments (namely 
WMAP) to allow for meaningful constraints on the non-linear parameter $\fnl$ for several different
families of models. These results are already arguably the most stringent 
quantitative test of the predictions of standard inflation.  However, much tighter constraints on a 
broader range of models are expected from the future Planck data release.     Thus a dramatic confrontation is set to continue between the de facto standard model of inflation and observational datasets from both the CMB and large-scale structure.  Tests of primordial non-Gaussianity are rapidly becoming one of the most effective and promising approaches 
for gleaning important information about the physical processes that generated the primordial cosmological perturbations.

\acknowledgements

 J.~F., M.~L.\ and E.P.S.\ were supported by STFC grant ST/F002998/1 and the 
Centre for Theoretical Cosmology.  
E.~S.\ acknowledges support by the French Agence National de la Recherche under grant BLAN07-1-212615 and by the European Union under the Marie Curie Inter European Fellowship.

\appendix

\section{Basics of estimation theory}\label{sec:estimationbasics}

If a random variable $\mathbf{x}$ is characterized by a Probability Distribution Function (PDF) $p(\mathbf{x}|\l)$ dependent on a parameter $\l$ then an estimator for $\l$ is a function ${\curl{E}}(\mathbf{x})$ used to infer the value of the parameter. If a given data set $\{\mathbf{x}^{obs}\}$ is drawn from the distribution $p(\mathbf{x},\l)$, then $\hat{\l} = {\curl{E}}(\mathbf{x}^{obs})$ is the estimate of the parameter $\l$ from the given observations. Since ${\curl{E}}$ is a function of a random variable, it is itself a random variable. In the literature a random variable obtained as a function of another set of random variables is often referred to as a {\em statistic}.

 A general property usually required when building an estimator is its {\em unbiasedness}. 
An estimator for a parameter $\l$ is {\em unbiased} if its average value is equal to the true value of the parameter:
\be\label{eq:unbiasednessdef}
\langle \hat{\l} \rangle = \l.
\ee
The standard deviation is generally used to determine the error bars on $\l$ \ie 
\be\label{errbarsdef}
\sigma_\l = \sqrt{\left\langle \left( \hat{\l} - \langle \hat{\l} \rangle \right)^2 \right\rangle},
\ee
where $\langle . \rangle$ denotes statistical average and $\sigma^2$ is the variance of the 
inferred parameter. When we measure a parameter $\l$ from a set of observations drawn from the PDF $p(\mathbf{x}|\l)$, we clearly would like our estimate not only to be unbiased, but also to have as small error bars as possible. In other words, among all the possible unbiased estimators of $\l$ that can be built, we look for the one that minimizes $\sigma_\l$ defined in (\ref{errbarsdef}). If such an estimator exists, it is called an {\em optimal estimator}.

In this context a crucial role is played by the {\em Fisher information matrix}, defined as
\be\label{eq:fishermatrixdef}
F_{\l \l} = \left\langle \( \frac{ \partial \( \ln p(\mathbf{x}|\l) \) }{\partial \l} \) \right\rangle,
\ee
The Fisher matrix appears in an important theorem, known as the {\em Cramer-Rao} inequality, stating that {\em for any unbiased estimator of $\l$}
\be
\sigma_\l \geq \frac{1}{\sqrt{F_{\l \l}}}.
\ee
This theorem is then placing a {\em lower bound} on the error bars that can be attained when estimating a given parameter from a given set of observations. No matter which estimator is used, the smallest attainable error bars will be given by the square root of the inverse of the Fisher matrix. For a demonstration of this crucial result see \eg  \citet{KendallStuart1979} or, in relation to the CMB bispectrum, \citet{Babich2005}. It is then clear that the best estimator of a parameter is an unbiased estimator saturating the Rao-Cramer bound. If such an estimator is found, then it is impossible to obtain a better estimate using any other statistic. The question then becomes if, for a given the PDF 
$p(\mathbf{x}|\l)$, an estimator saturating the Rao-Cramer bound exists. 

It can be shown that a {\em necessary and sufficient condition} for an estimator ${\curl{E}}(\mathbf{x})$ of a parameter $\l$ to be optimal is the following:
\be\label{eq:optimalitycondition}
\frac{\partial{\ln p(\mathbf{x}|\l)}}{\partial \l} = F_{\l \l} ({\curl{E}}(\mathbf{x}) - \l) \; ,
\ee
where $F$ is the Fisher information matrix just introduced above.

Another crucial quantity in estimation theory is the so called {\em maximum-likelihood estimator}. In a maximum-likelihood (ML) approach we take the observed data set $\mathbf{x}^{obs}$ as fixed and we estimate 
$\l$ as the parameter that maximize the probability ({\em likelihood}) to observe the given data. In formulae, the 
ML-estimate of $\l$ is the value $\hat{\l}$ that satisfies:
\be\label{eq:MLcondition}
\frac{\partial \ln p(\mathbf{x}|\l)}{\partial \l} |_{\l = \hat{\l}} = 0 \; .
\ee
In this context the PDF $p(\mathbf{x}|\l)$ is often denoted as the {\em likelihood function} and indicated as $\cal{L}(\mathbf{x},\l)$. Two powerful theorems involving the likelihood have been proven:
\begin{enumerate}
\item If there is an optimal unbiased estimator (\ie an unbiased estimator saturating the Rao-Cramer bound) then it is the maximum-likelihood estimator or a function of it.
\item The maximum likelihood estimator is {\em asymptotically optimal}, \ie it saturates the Rao-Cramer bound when $N \rightarrow \infty$, N being the number of repeated observations in our data set $\mathbf{x}^{obs}_{(1)},\dots,
\mathbf{x}^{obs}_{(N)}$.
\end{enumerate}
These two theorems answer our initial question about the best estimator choice. The first theorem basically states that {\em if} a best method exist, then the ML-estimator is that method. Note that this result follows naturally from the optimality condition (\ref{eq:optimalitycondition}) introduced above. The second theorem says that for very large data sets the ML-estimator {\em is} the best method, \ie the one saturating the Rao-Cramer bound. In other words, when dealing with the practical problem of estimating a parameter from a given data set, we should in theory always choose a ML-likelihood approach. However in practice this is not always possible: for example, the PDF $p(\mathbf{x}|\l)$ might be too difficult to calculate or sample numerically, or the ML condition (\ref{eq:MLcondition}) (generally a complicated non-linear equation) too difficult to solve. In this case other approaches and different estimators have to be chosen.

An important role is played by the likelihood of Gaussian random variables. If a given observed variable  $O_{\a}$ is characterized by gaussianly distributed errors, then it is easy to see that its likelihood is
\be
{\cal{L}} = e^{\chi^2/2},
\ee
where the $chi^2$ statistic is defined as:
\be\label{eq:chi2stat}
\chi^2 = \sum_{\a} \frac{\[O_{\a}(\mathbf{\l}) - O^{obs}_{\a}(\l)\]^2}{\(\Delta O_\a\)^2},
\ee
where $O_{\a}^{obs}$ are the measured values of our observable. In the previous equation we made $O_{\alpha}$ dependent on a vector of parameters $\mathbf{\l}$, that we want to fit. Our observable cold be for example the CMB angular power spectrum $C_{\ell}$, the primordial power spectrum $P(k)$ or, like in our case the angular bispectrum $B_{\ell_1, \ell_2 \ell_£}$, and we might be interested in knowing the sensitivity of our observation to any cosmological parameter. Our statistical estimate of $\l$ will be obtained by minimizing $\chi^2$. That is clearly equivalent to maximize the likelihood. Let us now for simplicity work in the one-dimensional case (\ie our observable depends on a single parameter) and expand the $\chi^2$ about its minimum, that is about the best fit value of the parameter $\bar{\l}$
\be
\chi^2(\l) = \chi^2(\bar{\l}) + \frac{1}{2}\left.\frac{\partial^2 \chi^2}{\partial \l^2}\right|_{\l = \bar{\l}}(\l - \bar{\l})^2 \; .
\ee
The linear term vanishes here since we are in the minimum. The quadratic term represents the curvature and defines the error on $\l$. If the $\chi^2$ moves very quickly away from its minimum, then our determination of $\l$ will be more precise, while the error on $\l$ will be much larger otherwise. If we define
\be
F \equiv \frac{1}{2}\left.\frac{\partial^2 \chi^2}{\partial \l^2}\right|_{\l = \bar{\l}} \; ,
\ee
then we can estimate the minimum possible error on $\l$ as ${1/\sqrt{F}}$. It is easy to see that the curvature of the likelihood in the Gaussian case matches exactly the definition of Fisher matrix given above. The ${1/sqrt{F}}$ lower limit on the error bar then coincides, as it should, with the Rao-Cramer bound. This at the same time validates the choice of ${1/\sqrt{F}}$ as the error on the parameter, and also shows a simple way to interpret the Rao-Cramer bound. Since the Fisher matrix represents the curvature of the $\ln$ of the 
likelihood around its maximum, it also provide an intrinsic minimum error on the measurement of the parameter. A likelihood strongly peaked around its maximum for a given parameter will provide stronger constraints on that parameter and vice-versa. We have however to keep in mind that the curvature $F$ constructed above is the curvature of the likelihood {\em only} if the distribution of our observable $O_{\a}$ is Gaussian. This, strictly speaking, is in general not true, but it is a reasonably good approximation in most cases\footnote{A clarifying example is provided by the CMB angular power spectrum. We know that $C_{\ell}$ is distributed like a $\chi^2$ with $2 \ell +1$ degrees of freedom, that rapidly gets close to a Gaussian as $\ell$ grows.}. The Fisher matrix for any observable is then defined as the second derivative of the $\chi^2$ statistic \eqn{eq:chi2stat}. If we compute it explicitly we get
\be
F_{\l \l} = \sum_{\alpha} \frac{1}{\(\Delta O_{\alpha}\)^2}\[\(\frac{\partial O_{\a}}{\partial \l}\)^2 
+ \(O_{\a} - O_{\a}^{obs}\) \frac{\partial^2 O_{\a}}{\partial \l^2} \].
\ee 
The second term in the sum above is generally neglected. The idea, as explained in 
\citet{Dodelson2003} or in \citet{PressEtal1992} is that the observed $O_{\a}$ will oscillate around their real value, making the difference $\(O_{\a} - O_{\a}^{obs}\)$ oscillate around zero, resulting in cancellations. We are then left with the expression generally used in the literature,
\be\label{eq:fisher3}
F_{\l \l} = \sum_{\alpha} \frac{1}{\(\Delta O_{\alpha}\)^2}\[\(\frac{\partial O_{\a}}{\partial \l}\)^2\].
\ee

In the review we have applied the basic concepts described in this Appendix to the estimation of the non-Gaussian parameter $\fnlmodel$ from the bispectrum of CMB and LSS datasets. We would like to stress again that we have just very quickly sketched some essential concepts in estimation theory here. For excellent and much more comprehensive reviews of ideas and 
applications of estimation theory to cosmology we refer the reader to \citet{TegmarkTaylorHeavens1997,Dodelson2003,MartinezEtal2009}. The brief review provided here was actually largely inspired by these works. A detailed and complete book about statistical methods and estimation theory is \eg \citet{KendallStuart1979}.

\bibliography{Bibliography}

\end{document}